%% file: ms.tex
\documentclass[twoside,twocolumn,9pt]{article}
\usepackage{extsizes}
\usepackage[super,sort&compress,comma]{natbib}
\usepackage[version=3]{mhchem}
\usepackage[left=1.5cm, right=1.5cm, top=1.785cm, bottom=2.0cm]{geometry}
\usepackage{balance}
\usepackage{times,mathptmx}
\usepackage{sectsty}
\usepackage{graphicx}
\usepackage{lastpage}
\usepackage[format=plain,justification=justified,singlelinecheck=false,font={stretch=1.125,small,sf},labelfont=bf,labelsep=space]{caption}
\usepackage{float}
\usepackage{fnpos}
\usepackage[english]{babel}
\usepackage{array}
\usepackage{droidsans}
\usepackage{charter}
\usepackage[T1]{fontenc}
\usepackage[usenames,dvipsnames]{xcolor}
\usepackage{setspace}
\usepackage[compact]{titlesec}

\usepackage{comment,pgfplots}
\usepackage{subcaption}

\newcommand{\rd}[0]{\mathrm{d}}
\newcommand{\mr}[1]{\mathrm{#1}}
\usepackage[normalem]{ulem}	
\includecomment{mycomment}

\usepackage{hyperref}

\newcounter{Sequation}
\newenvironment{Sequation}
   {\stepcounter{Sequation}%
     \addtocounter{equation}{-1}%
     \equation}
   {\endequation}
\newcounter{Sfigure}
\newenvironment{Sfigure}
   {\stepcounter{Sfigure}%
     \addtocounter{figure}{-1}%
     \figure}
   {\endfigure}
\newcommand{\movie}[1]{\noindent #1 \\[\baselineskip]}

\usepackage{epstopdf}

\definecolor{cream}{RGB}{222,217,201}

\begin{document}

\thispagestyle{plain}



\renewcommand{\thefootnote}{\fnsymbol{footnote}}
\setcounter{secnumdepth}{5}

\renewcommand{\figurename}{\small{Fig.}~}
\sectionfont{\sffamily\Large}
\subsectionfont{\normalsize}
\subsubsectionfont{\bf}
\setstretch{1.125} 
\setlength{\skip\footins}{0.8cm}
\setlength{\footnotesep}{0.25cm}
\setlength{\jot}{10pt}
\titlespacing*{\section}{0pt}{4pt}{4pt}
\titlespacing*{\subsection}{0pt}{15pt}{1pt}


\makeatletter
\newlength{\figrulesep}
\setlength{\figrulesep}{0.5\textfloatsep}

\newcommand{\topfigrule}{\vspace*{-1pt}%
\noindent{\color{cream}\rule[-\figrulesep]{\columnwidth}{1.5pt}} }

\newcommand{\botfigrule}{\vspace*{-2pt}%
\noindent{\color{cream}\rule[\figrulesep]{\columnwidth}{1.5pt}} }

\newcommand{\dblfigrule}{\vspace*{-1pt}%
\noindent{\color{cream}\rule[-\figrulesep]{\textwidth}{1.5pt}} }

\makeatother

\twocolumn[
  \begin{@twocolumnfalse}
\sffamily

	\noindent\LARGE{\textbf{Theoretical study of vesicle shapes driven by coupling\\ curved proteins and active cytoskeletal forces}} \\

	\noindent\large{Miha Fo\v{s}nari\v{c},\textit{$^{a}$} Samo Peni\v{c},\textit{$^{b}$} Ale\v{s} Igli\v{c},\textit{$^{b}$}  Veronika Kralj-Igli\v{c},\textit{$^{a}$} Mitja Drab,\textit{$^{b}$} and Nir Gov$^{\ast}$\textit{$^{c}$}} \\

	\noindent\normalsize{Eukaryote cells have a flexible shape, which dynamically changes according to the function performed by the cell. One mechanism for deforming the cell membrane into the desired shape is through the expression of curved membrane proteins. Furthermore, these curved membrane proteins are often associated with the recruitment of the cytoskeleton, which then applies active forces that deform the membrane. This coupling between curvature and activity was previously explored theoretically in the linear limit of small deformations, and low dimensionality. Here we explore the unrestricted shapes of vesicles that contain active curved membrane proteins, in three-dimensions, using Monte-Carlo numerical simulations. The activity of the proteins is in the form of protrusive forces that push the membrane outwards, as may arise from the cytoskeleton of the cell due to actin or microtubule polymerization occurring near the membrane. For proteins that have an isotropic convex shape, the additional protrusive force enhances their tendency to aggregate and form membrane protrusions (buds). In addition, we find another transition from deformed spheres with necklace type aggregates, to flat pancake-shaped vesicles, where the curved proteins line the outer rim. This second transition is driven by the active forces, coupled to the spontaneous curvature, and the resulting configurations may shed light on the organization of the lamellipodia of adhered and motile cells.}


 \end{@twocolumnfalse} \vspace{0.6cm}

  ]

\renewcommand*\rmdefault{bch}\normalfont\upshape
\rmfamily
\section*{}
\vspace{-1cm}


\footnotetext{\textit{$^{a}$~Faculty of Health Sciences, University of Ljubljana, Ljubljana, Slovenia.}}
\footnotetext{\textit{$^{b}$~Faculty of Electrical Engineering, University of Ljubljana, Ljubljana, Slovenia}}%
\footnotetext{\textit{$^{c}$~Department of Chemical and Biological Physics, Weizmann Institute of Science, Rehovot 7610001, Israel. E-mail: nir.gov@weizmann.ac.il}}%


\footnotetext{\dag~MF and SP provided Monte-Carlo simulations; NG provided the model for active proteins and linear stability analysis; AI, VKI and MD provided the model of self-assembly in equilibrium.}



Curved membrane proteins \cite{zimmerberg2006proteins}, for example membrane embedded proteins with non-zero intrinsic curvature \cite{leibler86,FournierPRL1996,kralj-iglic96,kralj-iglic99,fovsnarivc2006influence,Baumgart2011,Gomez2016}, flexible nanodomains \cite{iglivc2007possible,Baumgart2011} or curved membrane-attached proteins \cite{zimmerberg2006proteins,iglivc2007elastic,Baumgart2011,noguchi2016membrane,MesarecCollSurf2016}, have been identified to play an important role in driving the formation of various membrane shapes \cite{markin81,Seifert1997,kralj-iglic96,mcmahon2005membrane,bovzivc2006coupling,Baumgart2011,Walani2014}. Coupling between non-homogeneous lateral distribution of membrane components and membrane shapes may be a general mechanism for the generation and stabilization of highly curved membrane structures \cite{chabanon2017systems}, such as spherical buds, membrane necks, thin tubular or undulated membrane protrusions \cite{markin81,HelfrichProst1988,seifert93,Seifert1997,KraljIglicPRE2000,kumar04,allain04,kraljiglicEBJ2005,IglHag2006JTB,kralj-iglicJStP2006,semrau2009membrane,Walani2014,MesarecEBJ2017,NoguchiSoftM2017}.
In addition, it was found in many cellular processes that the curved proteins, or complexes containing the curved proteins, are able to recruit the cytoskeleton of the cell to produce additional protrusive forces, for example due to actin polymerization \cite{gov2018guided,saha2018joining}. Such protrusive forces are active, meaning that they consume energy (ATP) and maintain the system out of thermal equilibrium \cite{alimohamadi2018role}. The resulting steady-state configurations of the system may therefore differ from those in thermal equilibrium. Note that in cellular membranes such active forces can originate also from other sources, such as ion pumps \cite{girard2005passive,gov2004membrane}.

Curved membrane proteins with a convex shape, such that they induce outwards bending of the membrane, that also recruit the cytoskeletal forces which push the membrane outwards, can serve as efficient initiators of membrane protrusions. This mechanism was first suggested theoretically \cite{gov2006dynamics}, and has since been found in experiments \cite{miki2000irsp53,krugmann2001cdc42,mattila2007missing,millard2005structural,disanza2006regulation,vaggi2011eps8,disanza2013cdc42,kuhn2015structure}. This coupling of convex curvature and recruitment of actin polymerization is therefore emerging as an efficient cellular mechanism for the production of actin-based protrusions \cite{prevost2015irsp53}. It also appears to be exploited by certain viruses during their budding from the infected cell \cite{gladnikoff2009retroviral,votteler2013virus}.
Previous studies of the coupling between curved membrane proteins and the cytoskeletal forces were mostly limited to the linear regime \cite{gov2006dynamics} or to simplified geometries \cite{kabaso2011theoretical}, and indicated that convex proteins can undergo phase separation and aggregation at lower concentrations (or higher temperatures) when the protrusive forces are present \cite{veksler2007phase}. We study here the membrane shapes and the aggregation properties of such systems using numerical simulations, which allow us to go beyond the linear deformations limit. We find that the presence of the active protrusive forces affects the phase-separation (budding) transition, as well as induces transitions into new classes of shapes that are not accessible in the equilibrium (passive) systems. The ability of active processes, associated with curved proteins, to lead to global shape transitions was previously found in numerical simulations \cite{ramakrishnan2015organelle}, where activity was in the form of proteins with fluctuating spontaneous curvature.

\section{Theoretical model}
\label{sec:theoretical_model}

The energy of the membrane is expressed as the sum of contributions of membrane bending, direct interactions between membrane proteins and outward protrusive cytoskeletal forces,
\begin{equation}
	W = W_\mr{b} + W_\mr{d} + W_\mr{F},
	\label{eq:W}
\end{equation}
respectively.


For membrane bending energy the standard Helfrich expression\cite{Helfrich1973} is used,
\begin{equation}
	W_\mr{b} = \frac{\kappa}{2} \int_A (C_1+C_2 - C_0)^2 \rd A,
	\label{eq:Wb}
\end{equation}
 where the integral runs over the whole area $A$ of the membrane with bending stiffness $\kappa$, $C_1$ and $C_2$ are principal curvatures and $C_0$ is the spontaneous curvature of the membrane. The proteins on the membrane are modeled as patches of the membrane with given spontaneous curvature $c_0$ \cite{iglivc2007elastic,iglivc2007possible,fovsnarivc2006influence}. On the patches occupied by the curved proteins we therefore set $C_0 = c_0$ and elsewhere we assume a symmetric membrane $C_0=0$.

For direct interactions between neighboring proteins we assume the step potential,
\begin{equation}
	W_\mr{d} = - w \sum_{i < j} \mathcal{H}(r_0 - r_{ij}),
	\label{eq:Wd}
\end{equation}
where $w$ is a direct interaction constant, the sum runs over all protein-protein pairs, $r_{ij}$ are their mutual in-plane distances,  $\mathcal{H}(r)$ is the Heaviside step function and $r_0$ is the range of the direct interaction.


Finally, the energy contribution of the local protrusive forces due to the cytoskeleton is
\begin{equation}
	W_\mr{F} = - F \sum_{i} \hat n_i \cdot \vec{x}_i,
	\label{eq:Wf}
\end{equation}
where $F$ is the size of the force, the sum runs over all proteins, $\hat n_i$ is the outwards facing normal to the membrane at the location of the protein $i$ and $\vec{x}_i$ is the position vector of the protein $i$.

Note that the system with the active forces does not generally has global force balance, since the proteins are not symmetrically distributed on the surface of the vesicle. However, we are interested in shape changes, so disregard any center-of-mass motion.

\section{Monte-Carlo simulations}
\label{sec:mcsims}

The membrane is represented by a set of $N$ vertices that are linked by tethers of variable length $l$ to form a closed, dynamically triangulated, self-avoiding two-dimensional network \cite{Gompper_NelsonStatMech,gompper96} of approximately $2N$ triangles and with the topology of a sphere. The lengths of the tethers can vary between a minimal value, $l_\mr{min}$, and a maximal value, $l_\mr{max}$. Self-avoidance of the network is ensured by choosing the appropriate values for $l_\mr{max}$ and the maximal displacement of the vertex $s$ in a single updating step. In this work we used $s/l_\mr{min}=0.15$ and $l_\mr{max}/l_\mr{min}=1.7$. The dynamically triangulated network acquires its lateral fluidity from a bond flip mechanism. A single bond flip involves the four vertices of two neighboring triangles. The tether connecting the two vertices in diagonal direction is cut and reestablished between the other two, previously unconnected, vertices. The treatment is within the Rouse description, as it ignores the effects of hydrodynamics.

The microstates of the membrane are sampled according to the Metropolis algorithm. The probability of accepting the change of the microstate due to vertex move or bond flip is $\min{\left[1,\exp{(-\Delta E/kT)}\right]}$, where $\Delta E$ is the energy change, $k$ is the Boltzmann constant and $T$ is absolute temperature. The energy for a given microstate is specified in Eq.~\ref{eq:W}. The bending energy is discretized as described by Gompper and Kroll \cite{gompper96,Gompper_NelsonStatMech,Ramakrishnan2011}. For each set of parameters, the system is initially thermalized. Ensemble averaging is done over 200 statistically independent microstates.

In this work we set $N_c$ of the total $N=3127$ vertices to represent proteins and have spontaneous curvature $c_0=1/l_\mr{min}$ for curved proteins, unless stated otherwise, or $c_0=0$ for flat proteins. All other vertices represent symmetric membrane and have zero spontaneous curvature. The positive sign of $c_0$ for curved proteins means that the proteins have the tendency to curve the membrane outwards. If the two vertices representing proteins are nearest neighbors, there is an addition energy term $-w$ assigned to their bond (direct protein-protein interaction). Direct interaction constant $w$ is assumed to be of the order of the thermal energy $kT_0$, where $T_0 \approx 300$~$K$ is ``room temperature'', and membrane bending stiffness $\kappa$ is of the order of $20~kT_0$. In the following we fix the ratio $\kappa/w=20$, unless stated otherwise.  For the size of the protrusive force $F$ we expect the order of thermal energy $kT_0$ per minimal bond length $l_\mr{min}$.

Note that since the volume of the vesicle is not conserved, it can adjust to accommodate any shape of the membrane, and therefore the membrane is in a tensionless regime for the passive system. The active forces, pointing outwards, induce a finite membrane stretch and tension. In addition, we can introduce a non-zero pressure difference across the membrane, which can act to inflate the vesicle and induce a finite tension (Fig.~S8$^{\dag}$).

\section{Results and discussion}
\label{sec:results_and_discussion}


Using our simulations we aim to improve the understanding of clustering of curved and active proteins on the membrane and how this process of demixing is coupled with membrane shape changes. Especially we are interested in the budding of curved protein clusters. We expect the demixing and budding to be enhanced by attractive direct interaction between the proteins and the additional membrane deformation induced by the protrusive (cytoskeletal) forces recruited by the proteins.

In our Monte-Carlo simulations the whole membrane is the triangulated surface.
 To quantitatively analyse the demixing of the proteins in our system, we define the ensemble averaged mean cluster size as
\begin{equation}
	\left<\bar{N}_\mr{vc}\right> =\left< \frac{\sum_i{{N^{(i)}_\mr{vc} N^{(i)}_\mr{cl}}}}{\sum_i {N^{(i)}_\mr{cl}}}\right>,
	\label{eq:eamcs}
\end{equation}
where the angle brackets denote the canonical ensemble average. Inside the brackets, {\em i.e.} for a given microstate, $\bar{N}_\mr{vc}$ is the  mean cluster size and the sums run over all clusters of vertices representing proteins. In the sums, $N^{(i)}_\mr{vc}$ is the number of vertices in cluster $i$ and $N^{(i)}_\mr{cl}$ is the number of clusters of size $N^{(i)}_\mr{vc}$.

\subsection{Phase transition in thermal equilibrium}
\label{subsec:f=0}

In Fig.~\ref{fig:snaps} we plot the cluster size distribution and snapshots of typical microstates of vesicles with curved proteins, in the absence of active protrusive forces. The system is in thermal equilibrium\footnote[4]{Note that in some cases, for example when large necklace-like clusters form, thermal equilibrium is not always easily obtainable. However, after monitoring different measures for membrane shape change and demixing, we expect that the presented results correspond to the correct phase behaviour.} %
 at different temperatures and densities (area coverage fraction, $\rho=N_c/N$) of curved proteins.

At low average protein densities the equilibrium vesicle shapes remain quasi-spherical, with clusters that increase in size with decreasing temperature (in the far left column of Fig.~\ref{fig:snaps}, the largest clusters are composed of 5 proteins at $T/T_0=1.33$ and of 8 proteins at $T/T_0=0.63$). At higher average protein densities, cluster sizes increase and curved protein buds burst on the membrane.

At even larger average protein densities, the vesicle shapes deviate drastically from quasi-spherical and large necklace-like protein clusters often form. The size of these necklace-like clusters and the number of ``beads'' they contain increase with decreasing temperature. These necklace-like structures form since the isotropically curved proteins can not form flat aggregates, due to their spontaneous curvature, and resemble aggregates calculated for membrane-adsorbed spherical particles \cite{deserno2007nature}. The theory of self-assembly of curved proteins can approximately explain observed necklace-like structures (see SI1 text$^{\dag}$), while anisotropic curved proteins may form aggregates with other geometries on vesicles \cite{simunovic2013linear,helle2017mechanical}. Necklace-like membrane protrusions have been observed in cellular membranes, under different conditions \cite{heinrich2014reversible}, and in many in-vitro experiments \cite{tsafrir2001pearling,yu2009pearling}.

\begin{mycomment}
\begin{figure*}
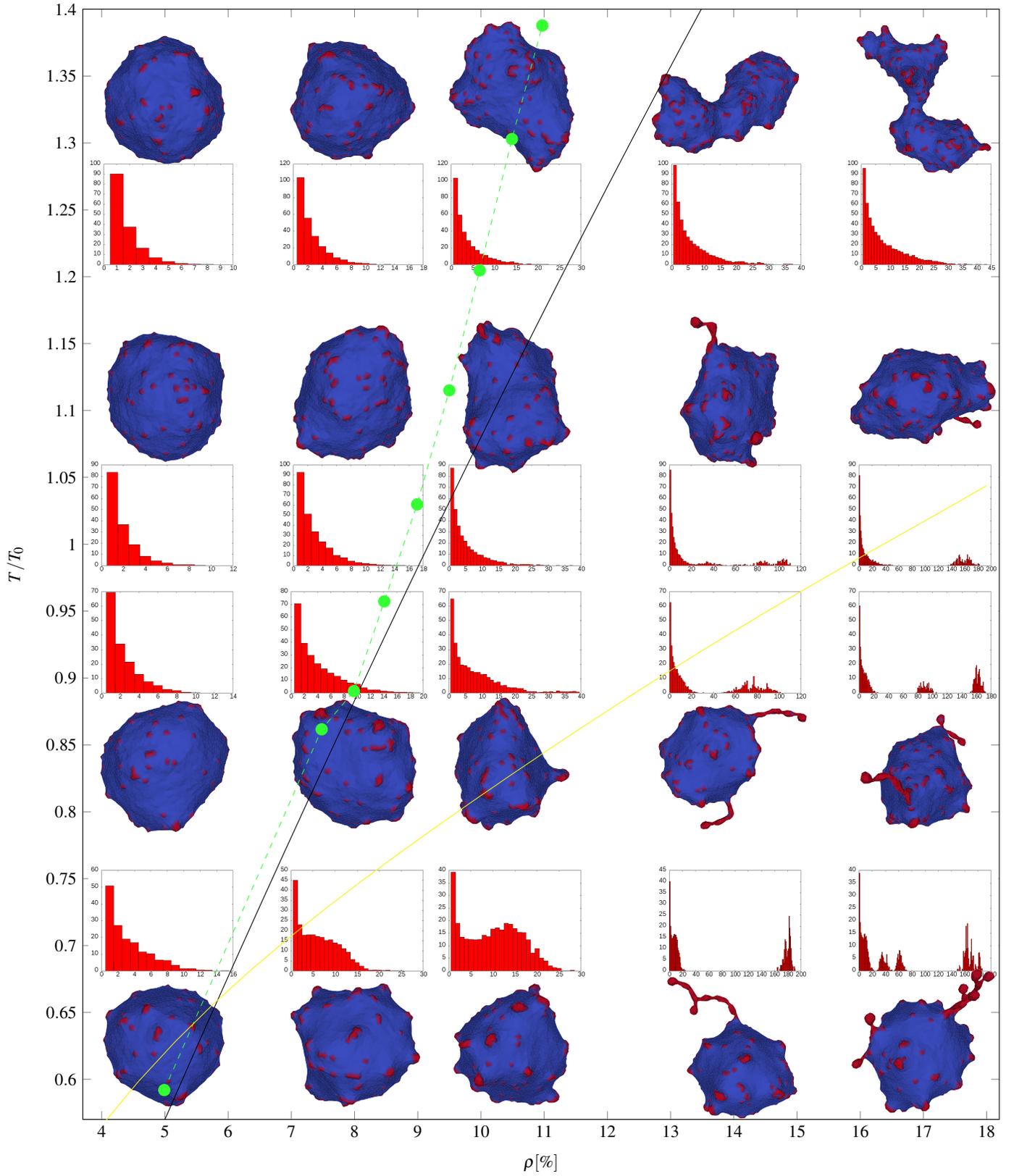

	\centering
	\include{snaps_fig}
	\caption{Microstates of the vesicles in thermal equilibrium (absence of protrusive forces), for different average densities of curved proteins $\rho$ and relative temperatures $T/T_0$. The blue vertices represent the protein-free bilayer and have zero spontaneous curvature; red vertices denote the curved proteins and have spontaneous curvature $c_0$. In the corresponding cluster-size distributions, the $y$-axis is the ensemble averaged number of protein clusters of each size and the $x$-axis is the protein cluster size. Solid black line denotes the prediction of the critical temperature by the linear stability analysis (Eq.~\ref{eq:tc0}), while green points denote the line where $\left<\bar{N}_\mr{vc}\right>=2$. The yellow line marks the critical boundary of the phase space below which the self-assembly theory predicts aggregate growth (for $r=1.35 l_{min}$, $R_{0}=10 r$; Eq.~\ref{eq.mi8}).}
	\label{fig:snaps}
\end{figure*}

\end{mycomment}

\begin{figure}
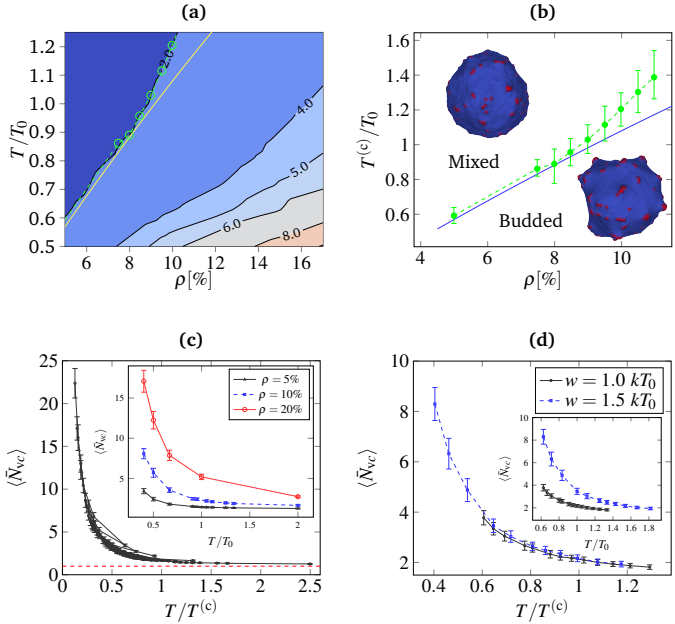

	\begin{subfigure}{0.5\columnwidth}
		\centering \include{contourplot_c1_f0}
	\end{subfigure}
	\begin{subfigure}{0.5\columnwidth}
		\centering \include{figs/testlatexfigf0_4panel}
		\label{subfig:passive b}
	\end{subfigure}
	\begin{subfigure}{0.5\columnwidth}
		\centering \include{figs/eamcsvstdivtc_F0only}
	\end{subfigure}
	\begin{subfigure}{0.5\columnwidth}
		\centering \include{figs/eamcsvstdivtc_w_comparison_F0only}
	\end{subfigure}
	\caption{Results for the equilibrium system with curved proteins ($c_0=1/l_\mr{min}$) without active forces $F=0$. {\bf (a)} Contour plot of the ensemble averaged mean cluster size $\left<\bar{N}_\mr{vc}\right>$ (Eq.~\ref{eq:eamcs}) as a function of protein density $\rho$ and relative temperature $T/T_0$. The prediction of the critical temperature $T^\mr{(c)}$ by the linear stability analysis (Eq.~\ref{eq:tc0}) is shown (solid yellow line), as well as the points where $\left<\bar{N}_\mr{vc}\right>=2$ (green). {\bf (b)} Critical temperatures as predicted by the linear stability analysis (solid line) and from simulations ($\left<\bar{N}_\mr{vc}\right>=2$, green points). Representative snapshots are added for the mixed phase (for $\rho=0.05$, $T/T_0=4/3$) and the budded phase (for $\rho=0.08$, $T/T_0=0.625$). {\bf (c)} Mean cluster size $\left<\bar{N}_\mr{vc}\right>$ as function of the temperature normalized by the critical temperature $T/T^\mr{(c)}$, for $\rho=5$, $7.5$, $8$, $8.5$, $9$, $9.5$, $10.5$, $11$, $11.5$, $12$, $12.5$, $14$, $15$, $15.5$, $17$, $20$ and $25\%$. The horizontal dashed line indicates where $\left<\bar{N}_\mr{vc}\right>=1$. INSET:  $\left<\bar{N}_\mr{vc}\right>$ as a function of $T/T_0$ for three values of $\rho$. {\bf (d)} $\left<\bar{N}_\mr{vc}\right>$ as a function of $T/T^\mr{(c)}$ for two different values of the direct interaction constant $w$. The average protein density is $\rho=9.5\%$ (INSET: the same but as a function of $T/T_0$).}
	\label{fig:passive}
\end{figure}

We compare these simulation results with the prediction of the linear stability analysis (corresponding to the spinodal) of Gladnikoff et al.\cite{gladnikoff2009retroviral}. The linear stability analysis yields the critical thermal energy, $k T^\mr{(c)}$, below which the instability occurs and buds start to form:%
\footnote[5]{We express the binding interaction between viral proteins $J$ introduced in Gladnikoff et al.\cite{gladnikoff2009retroviral} (see the forth term in the expression for the membrane free energy, Eq.~(S1), in its Supporting Material) with our direct interaction constant $w$ (see Eq.~\ref{eq:Wb}). We get that the aggregation interaction energy contribution per viral protein is $J \phi (1-\phi)/2$, where $\phi$ is the local concentration of viral proteins. Change of $\phi$ from $0$ to $50~\%$ gives an energy change of $J/8$. On the other hand in our simulations the energy contribution due to direct interaction between curved proteins is defined on bonds between vertices, where each bond carries energy $-w$, if the vertices connected by the bond both represent curved proteins. In a hexagonal mesh the concentration $\phi=1/2$ for a given vertex corresponds the case where 3 out of 6 bonds carry the energy $-w$ and the energy per vertex is therefore $-3w/2$. Comparing the energy changes per protein when $\phi$ changes from $0$ to $50~\%$ in both models, we get $J=-12w$. We also identify $\phi_0$ in\cite{gladnikoff2009retroviral} with our normalized average densities of curved proteins $\rho=N_c/N$, $H$ with one half of our spontaneous curvature $c_0$ and $f_c$ with our $F$ per $a^2$, where $a$ is protein lateral size. We use $a=l_\mr{min}$ and take into account that in our simulations we assume tensionless membrane ($\sigma = 0$). In our model the membrane is not flat. For a membrane with curvature $1/R_0$ at the site of the curved protein, the spontaneous curvature $c_0$ (or $2H$ in Eq.~(S8) in the Supporting Material of \cite{gladnikoff2009retroviral}) effectively changes to $c_0 (1-1/\rho R_0)$. We can then express from Eq.~(S8) the critical temperature $T^\mr{(c)}$.}
\begin{equation}
	kT^\mr{(c)} = 12 w (1-\rho) \rho \left(1 + \sqrt{\frac{l_\mr{min}^2 F c_0}{12 w}\left(1-\frac{1}{\rho R_0}\right)}\right),
	\label{eq:tc0}
\end{equation}
where $1/R_0$ is mean membrane curvature at the site of a curved membrane protein. In the following we approximate $R_0$ with the radius of a spherical vesicle with the same membrane area $A$%
\footnote[6]{To estimate $R_0$ we assume a network of equilateral triangles with sides of lengths $l=(l_\mr{min}+l_\mr{max})/2$. This gives $R_0 \approx \sqrt{\sqrt{3}\: l^2 \: N/8\pi} \approx 0.35\: l_\mr{min} \sqrt{N}$.}
.

In thermal equilibrium, in the absence of the protrusive forces $F=0$, this simplifies to: $T^\mr{(c)} = 12 w (1-\rho) \rho/k$, which is plotted in Fig.~\ref{fig:snaps}. It can be seen that the prediction of the linear stability analysis qualitatively agrees with our simulation results: above the critical temperature line the protein budding is weak and the cluster size distribution is highly peaked at the size of isolated proteins, while below it buds are larger (together with the corresponding membrane deformation) and the size distribution exhibits a secondary peak at aggregates containing $8$ or more proteins (which is the number required to form the smallest closed spherical cluster of proteins). For a more quantitative description, $T^\mr{(c)}$ is defined from the simulations as the temperature where $\left<\bar{N}_\mr{vc}\right>=2$ (Eq.~\ref{eq:eamcs})\footnote[7]{The points $\left<\bar{N}_\mr{vc}\right>$ as a function of temperature (see Fig.~\ref{fig:passive}b) where fitted using a function $y(x)=a_1/x^{a_2} + a_3$, where $a_1$, $a_2$ and $a_3$ are free parameters and then $y(2)$ was taken as the estimate for $T^\mr{(c)}$. For non-zero $F$ only points at temperatures above the transition into a pancake-like shapes are taken into account when fitting. The error bars in Fig.~\ref{fig:passive} were obtained by fitting points $\left<\bar{N}_\mr{vc}\right>\pm \varepsilon$, where $\varepsilon$'s are standard deviations of $\left<\bar{N}_\mr{vc}\right>$.}, which agrees quite well with the predicted budding transition line, as plotted in Figs.~\ref{fig:snaps},~\ref{fig:passive}a,b. We find that the mean aggregate size collapses to a universal curve when the temperature is scaled by $T^\mr{(c)}$ (Fig.~\ref{fig:passive}c). The agreement with Eq.~\ref{eq:tc0} is also found as function of protein interaction strength (Fig.~\ref{fig:passive}d), and as function of vesicle radius (Fig.~S7$^{\dag}$). We conclude that the critical temperature for budding in the passive system agrees very well with the prediction of the linear stability model \cite{gladnikoff2009retroviral} (Eq.~\ref{eq:tc0}).

The phase separation of the passive system was additionally approximated with a two-dimensional model of self-assembly of curved proteins (see SI1$^{\dag}$ for details).
The total free energy of the model reads
\begin{eqnarray}
	{F} &=& M \left[ \tilde x_1 ~\tilde{\mu}_1 + kT {\tilde x_1}
(\ln {\tilde x_1} -1) \right]+ \nonumber \\
	&+& M \sum \limits_{i=1}^\infty \left[x_i ~{\mu}_i
+ kT \frac{x_i}{i} \left(\ln \frac{x_i}{i}-1\right)\right] -\nonumber\\
	&-& \mu M \,(\tilde x_1 + \sum \limits_{i=1}^\infty x_i).
\label{eq.miha1}
\end{eqnarray}
Here, $\tilde{x}_{1}$ and $x_{i}$ are the number densities of nanodomains in the weakly curved region and highly curved aggregates, respectively. The energy contributions come not only from the free energies per nanodomain ($\mu_{i}$ and $\tilde{\mu}_{1}$), but also from configurational entropy, while the Lagrange multiplier $\mu$ assures a constant number of nanodomains in the system through a conservation relation. We minimize $F$ with respect to the number densities, arriving at the equilibrium distributions for aggregate size (Eq.~S10$^{\dag}$). This model predicts that the critical density beyond which the growth of aggregates is energetically favourable is given by (Eq.~S11$^{\dag}$)
\begin{equation}
\tilde x_{c} \approx \exp \left(~ \frac{ \Delta f - w
}{kT}\right)\,,
\label{eq.mi8}
\end{equation}
where $\Delta f = f_{c} - f_{sp}$ is the difference between the energy of a single protein nanodomain on the highly curved necklace-like aggregates ($f_{c}$) and on the weakly curved membrane region ($f_{sp}$). Comparison between the model phase transition and MC simulations of aggregate distributions shows good overall agreement, with large necklace-like aggregates appearing below the calculated transition line (Fig.~\ref{fig:snaps}).


\subsection{Phase transitions in the presence of active protrusive forces}
\label{subsec:F}

Next, we consider the effects of active protrusive forces on the system. In Fig.~\ref{fig:snapsf1} we plot the typical shapes and aggregate size distribution as in Fig.~\ref{fig:snaps}, but with $F=1\: k T_0/l_\mr{min}$. We can see that the active protrusive forces promote demixing and budding of the convex curved proteins, such that the transition temperature $T^\mr{(c)}$ is shifted to higher temperatures and lower densities, as expected and predicted by the linear stability analysis \cite{gladnikoff2009retroviral} (Eq.~\ref{eq:tc0}).

A more dramatic effect of the active forces is seen in Fig.~\ref{fig:snapsf1} at low temperatures, where below the budding transition there is now a second transition to a new class of shapes that was not seen in the equilibrium system (Fig.~\ref{fig:snaps}). Namely, below the red dashed curve we find that the vesicles change from deformed-spherical to flattened pancake-like shapes, where all or nearly all the proteins aggregate at the rim, forming one large cluster in the form of a closed ring. We locate the transition into the pancake-like shapes regime from the sharp change of the slope of the mean cluster size $\left<\bar{N}_\mr{vc}\right>$ as function of $T$ (see Fig.~\ref{fig:compareactive}b,d), as shown in Figs.~\ref{fig:snapsf1},\ref{fig:compareactive}a,c.%
\footnote[8]{The error bars on the pancake-transition curve in Fig.~\ref{fig:compareactive}c mark the region in which the transition occurs, {\i.e.} the region between observed no-pancake states with lowest $T$ and pancake states with largest $T$ (Fig.~\ref{fig:snapsf1}).}
The transition is sharp (see Fig.~\ref{fig:snapsf1_llc}a), which suggests similarity to a first-order phase transition (although the system is out-of-equilibrium). However, we did not find any significant hysteresis (Fig.~S5$^{\dag}$).

Note that in Fig.~\ref{fig:compareactive}b we plot $\left<\bar{N}_\mr{vc}\right>$ as a function of $T/T^\mr{(c)}$, where $T^\mr{(c)}$ is the critical temperature from Eq.~\ref{eq:tc0}. In the presence of the actin protrusive force the curves collapse well at temperatures above the pancake-like shape transition, as in the passive case (Figs.\ref{fig:passive}b,c). However, at lower temperatures, where this shape transition dominates the system behavior, there is no such collapse as function of $T/T^\mr{(c)}$. This is expected, since the linear analysis of Gladnikof et al. \cite{gladnikoff2009retroviral} does not take into account large-scale global shape changes, which are highly non-linear. A similar behavior is found for different protein-protein interaction strengths (Fig.~S4$^{\dag}$). As expected, a stronger force promotes the transition to pancake-like shapes at higher temperatures (Fig.~\ref{fig:compareactive}d).

The organization of the proteins into a circular cluster around the rim of a flat vesicle is highly effective in stretching out the flat membrane parts. We indeed find that these regions are almost devoid of proteins (Figs.~\ref{fig:snapsf1},\ref{fig:snapsf1_llc}a), since these regions are energetically unfavorable for the curved proteins. The stretching of the membrane in these regions also acts to suppress aggregation of the curved proteins \cite{gladnikoff2009retroviral}, and the rim aggregate is highly stable (see Supporting Movie S1$^{\dag}$).

Below a critical density, we find that there are simply not enough proteins to form a continuous cluster around the rim of the flattened vesicle, and the system changes to a different class of shapes (Fig.~\ref{fig:snapsf1_llc}a). In this regime the curved proteins form arc-like clusters that line the flattened ends of an elongated vesicle, and remain highly dynamic (see Supporting Movie S2$^{\dag}$). Since the curved proteins are isotropic, the curvature of the rim of the flattened vesicles along circumferential direction is much smaller compared to $c_0$. The curved protein therefore tend to bend the rim and cause it to undulate along this direction (Fig.~\ref{fig:snapsf1_llc}b). At large protein densities we find that the excess proteins crowd the rim and cause it to undergo buckling and curling, so as to be able to accommodate more curved proteins (Fig.~\ref{fig:snapsf1}). At the highest densities, the excess proteins extend from the rim cluster as spherical and necklace-like clusters (Fig.~\ref{fig:snapsf1}).

\begin{mycomment}
\begin{figure*}
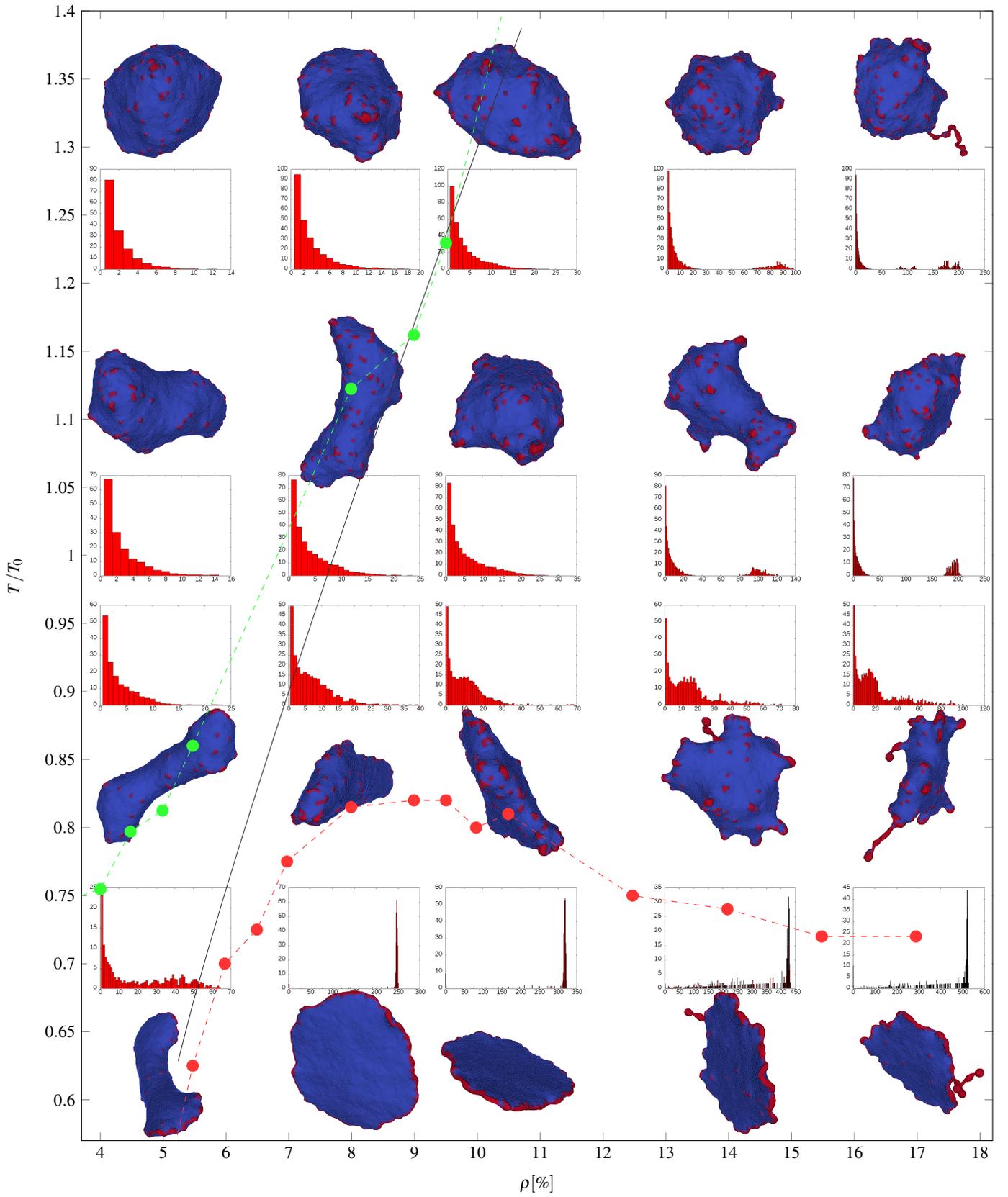

	\centering
	\include{snaps_fig_f1}
	\caption{Same as on Fig.~\ref{fig:snaps} but for the system with active protrusive forces $F=1\: k T_0/l_\mr{min}$. Approximate temperatures below which a transition into a pancake-like shapes is observed are indicated (red).}
	\label{fig:snapsf1}
\end{figure*}

\end{mycomment}

\begin{figure}
	\begin{subfigure}{0.5\columnwidth}
		\centering \include{contourplot_c1_f1}
	\end{subfigure}
	\begin{subfigure}{0.5\columnwidth}
		\centering \include{figs/eamcsvstdivtc_F1only}
	\end{subfigure}

	\begin{subfigure}{0.5\columnwidth}
		\centering \include{figs/testlatexfig_4panel}
	\end{subfigure}
	\begin{subfigure}{0.5\columnwidth}
		\centering \include{figs/eamcsvst_f_comparison}
	\end{subfigure}
	\caption{Results for the system with active curved proteins ($c_0=1/l_\mr{min}$, $F=1\: k T_0/l_\mr{min}$). {\bf (a)} Contour plot of the ensemble averaged mean cluster size $\left<\bar{N}_\mr{vc}\right>$ (Eq.~\ref{eq:eamcs}) as a function of protein density $\rho$ and relative temperature $T/T_0$. The prediction of the critical temperature $T^\mr{(c)}$ by the linear stability analysis (Eq.~\ref{eq:tc0}) is show (solid yellow curve), as well as the points where $\left<\bar{N}_\mr{vc}\right>=2$ (green) and approximate temperatures below which a transition into a pancake-like shapes is observed (red). {\bf (b)} $\left<\bar{N}_\mr{vc}\right>$ as a function of temperature normalized by the critical temperature $T/T^\mr{(c)}$, for $\rho=5$, $7.5$, $8$, $8.5$, $9$, $9.5$, $10.5$, $11$, $11.5$, $12$, $12.5$, $14$, $15$, $15.5$, $17$, $20$ and $25\%$. The horizontal dashed line indicates where $\left<\bar{N}_\mr{vc}\right>=1$.  INSET: $\left<\bar{N}_\mr{vc}\right>$ as a function of $T/T_0$ for three values of $\rho$. {\bf (c)} Critical temperatures as predicted by the linear stability analysis (solid line) and from simulations (green points). Approximate temperatures below which a transition into a pancake shapes is observed in the simulations are shown in red. Representative snapshots are added for the mixed phase (for $\rho=0.05$, $T/T_0=4/3$), the budded phase (for $\rho=0.14$, $T/T_0=4/3$) and the pancake phase (for $\rho=0.105$, $T/T_0=0.625$). {\bf (d)} $\left<\bar{N}_\mr{vc}\right>$ as a function of temperature for three different values of the actin protrusive force (legend). The average protein density is $\rho=9.5\%$. For $F=1.5\: k T_0/l_\mr{min}$, the snapshots of steady-state microstates are show for $T/T_0 = 0.625$ (pancake), $1$ (prolate) and $1.33$ (quasi-spherical).}
	\label{fig:compareactive}
\end{figure}

\begin{figure}

	\centering \include{snaps_fig_f1_llc}

	\centering
	\begin{tikzpicture}
		\node[anchor=center] (zoomededges) at (0,0) {\frame{\includegraphics[width=0.8\columnwidth]{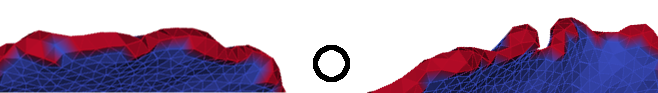}}};
		\node[above] at (0,0.74) {\bf (b)};
	\end{tikzpicture}

	\caption{{\bf (a)} Zoom-in of the lower left corner of Fig.~\ref{fig:snapsf1}, showing system snapshots at low protein densities. Approximate temperatures below which a transition into pancake-like shapes is observed are also indicated (red). {\bf (b)} Zoom-in on the edge of the pancake shapes for $\rho=10.5\%, T/T_0=0.625$ (LEFT) and $\rho=15.5\%, T/T_0=0.625$ (RIGHT), to highlight their convoluted shape. The circle in the middle has radius $l_{min} = 1/c_0$ and indicates the length-scale of a single spherical bud of proteins.}
	\label{fig:snapsf1_llc}
\end{figure}

\begin{figure}
	\centering
		\centering \includegraphics[width=\columnwidth]{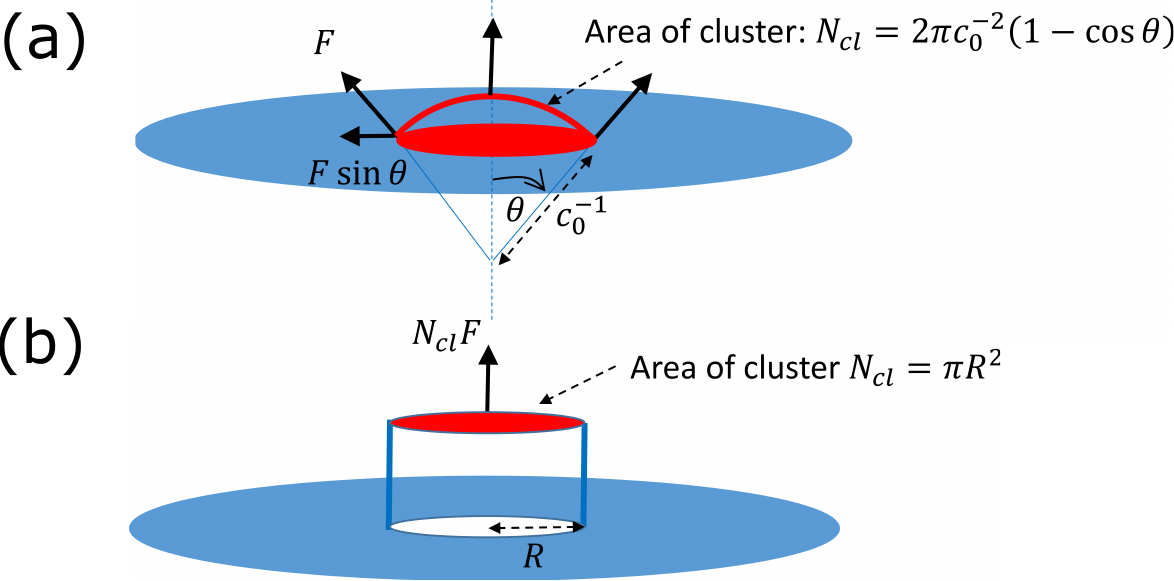}
%
	\caption{{\bf (a)} Schematic illustration of the side-ways active protrusive forces (black arrows) that act to destabilize spherical aggregates (red) of curved proteins, on a flat membrane (blue). {\bf (b)} Schematic illustration of the forces when a flat protein aggregate (red) drives the growth of a tubular protrusion from a flat membrane.}
	\label{fig:schematic}
\end{figure}

We can propose the following mechanism that drives the transition from deformed spherical vesicles with small buds to pancake-shapes in the presence of the active protrusive forces. When the proteins are isolated, or in very small clusters, the membrane is rather flat and the protrusive force promotes the aggregation of small clusters \cite{gladnikoff2009retroviral}, since the active force is directed at the outwards normal at each protein, thereby enhancing the outwards deformation that the curved protein induces. However, for larger clusters that are highly curved, the active forces point in different directions which acts to inflate and deform the clusters (Fig.~\ref{fig:schematic}a). We can estimate the critical cluster size at which the in-plane projection of the active forces are large enough to compete with the direct attraction between the proteins ($w$), and can destabilize spherical aggregates. The cluster size for an angle $\theta$ is: $N_{cl}=2\pi c_0^{-2}\left(1-\cos{(\theta)}\right)$. The critical angle can be estimated by the following force balance: $F\sin{(\theta)}\simeq w/l_\mr{min}$. From this estimate we expect that the active forces will destabilize spherical aggregates above a critical cluster size. The observed transition line is indeed found to follow a critical cluster size contour (Fig.~\ref{fig:compareactive}a). As the temperature decreases and the mean cluster size increases beyond the critical size, the system transitions to another global configuration where the proteins form a rim cluster that is highly stable and contains almost all the proteins. In this configurations there are no side-ways active forces that act to destabilize the protein aggregate, but rather the active forces now act to stretch the whole vesicle in the same direction as the deformation due to the proteins' curvature, and thereby stabilize the pancake configuration.

\subsection{Dependence on the spontaneous curvature of the active proteins }

We next explore the role of the spontaneous curvature of the active proteins in driving the shape transition discussed above. We start by calculating the phase diagram for flat active proteins, i.e. with $c_0=0$. The result is shown in Fig.~\ref{fig:snapsc0}a. We find that the budding transition line is still well described by the linear stability expression (Eq.~\ref{eq:tc0}). At low temperatures we find a global shape transition from quasi-spherical vesicles to shapes with highly elongated protrusions, which are driven by the protrusive force provided by a cluster of proteins at the protrusion's tip. An approximate transition curve, which marks the transition into such hydra-like shapes regime (Fig.~\ref{fig:snapsc0}a,b) can be obtained from locating the sharp change of the slope of the asphericity's\footnote[9]{Deviation of the vesicle from quasi-spherical shapes can be characterized conveniently in terms of the {\em asphericity}. To this end, we make use of the Gyration tensor, whose components are defined for a discrete object through
\begin{equation}
S_{ij} = \frac{1}{N} \sum \limits_{k=1}^{N} r_{k,i}\: r_{k,j}
\qquad (i,j = 1, 2 , 3).
\end{equation}
Here, $r_{k,i}$ is the $i$-th Cartesian coordinate of the position vector
$\vec{r}_k$ of the $k$-th particle. The origin of the coordinate system
is located at the center of mass and the sum runs over all particles
of the object. From the principal moments (i.e., the eigenvalues)
$\lambda_1 \ge \lambda_2 \ge \lambda_3$ of $S_{ij}$ (calculated
using the algorithm outlined by Smith \cite{Smith:1961:ESM}),
we obtain the asphericity of the object \cite{rudnick87}
\begin{equation} \label{asphericity}
Asph = \frac{\left< \left( \lambda_1 - \lambda_3 \right)^2 + \left( \lambda_2 - \lambda_3 \right)^2 + \left( \lambda_1 - \lambda_2 \right)^2 \right>}{2 \left< \left(  \lambda_1 + \lambda_2 + \lambda_3 \right)^2 \right>}
\end{equation}
with $\langle \dots \rangle$ denoting ensemble averages. We point out that
a one-dimensional object (where $\lambda_2=\lambda_3=0$) leads to $A=1$,
a two-dimensional axisymmetric disk (where
$\lambda_1=\lambda_2$ and $\lambda_3=0$) entails $A=1/4$,
and a sphere (where $\lambda_1=\lambda_2=\lambda_3$)
gives rise to $A=0$. The asphericity $A$ has frequently
been used in the past to characterize polymers and membranes \cite{fovsnarivc2013monte,luo98,ostermeir10}.} dependence on $T$ (Fig~\ref{fig:snapsc0}c).

This shape transition can be understood by calculating the conditions that allow a tube-like protrusion to start growing from the vesicle, driven by the protrusive force induced by a circular protein aggregate (Fig.~\ref{fig:schematic}b). This occurs when there is a force balance between the protrusive force provided by the protein cluster at the tip and the elastic restoring force due to membrane bending (as in tether pulling): $N_{cl}F\simeq 2\pi\kappa/\sqrt{N_{cl}}$. From this force balance we derive the radius of the protrusions (details in the SI$^{\dag}$, Eq.~S17$^{\dag}$): $R_c=\left(\frac{2\kappa a}{F}\right)^{1/3}$ (where $a$ is the average area per protein), which is in very good agreement with the simulated widths of the protrusions (Fig.S3). As the temperature decreases, the mean cluster size increases until it is larger than the threshold size (Fig.~\ref{fig:schematic}b) for the elongation of tube-like protrusions (Fig.~\ref{fig:snapsc0}b). In this phase the shapes are highly dynamic and unstable, with protrusions merging and growing (see Supporting Movie S3$^{\dag}$).

By fixing a low temperature and large enough density, we explore the dependence of the global vesicle shape transitions on the spontaneous curvature of the active proteins. This is shown in Fig.~\ref{fig:snapsc0}d, where we change the spontaneous curvature of the active proteins from $c_0=0$ (flat proteins) to $c_0=1/l_\mr{min}$ (the spontaneous curvature used in the previous section). We find that as $c_0$ is increased, the multi-tube-like shapes transform continuously into shapes that have a single tube-like part and flattened arc-like clusters at the two tips. Above a critical spontaneous curvature there is a sharp transition into the flattened shapes with continuous rim cluster. From our simple estimates (Figs.~\ref{fig:schematic}a,b) we predict that the critical cluster size that enables the pancake shape transition decreases with increasing $c_0$, while the tube-like shape transition does not depend on this parameter. We therefore expect that above a critical value of $c_0$ the pancake shape will dominate, as we observe (Fig.~\ref{fig:snapsc0}d).

\begin{mycomment}

\begin{figure}
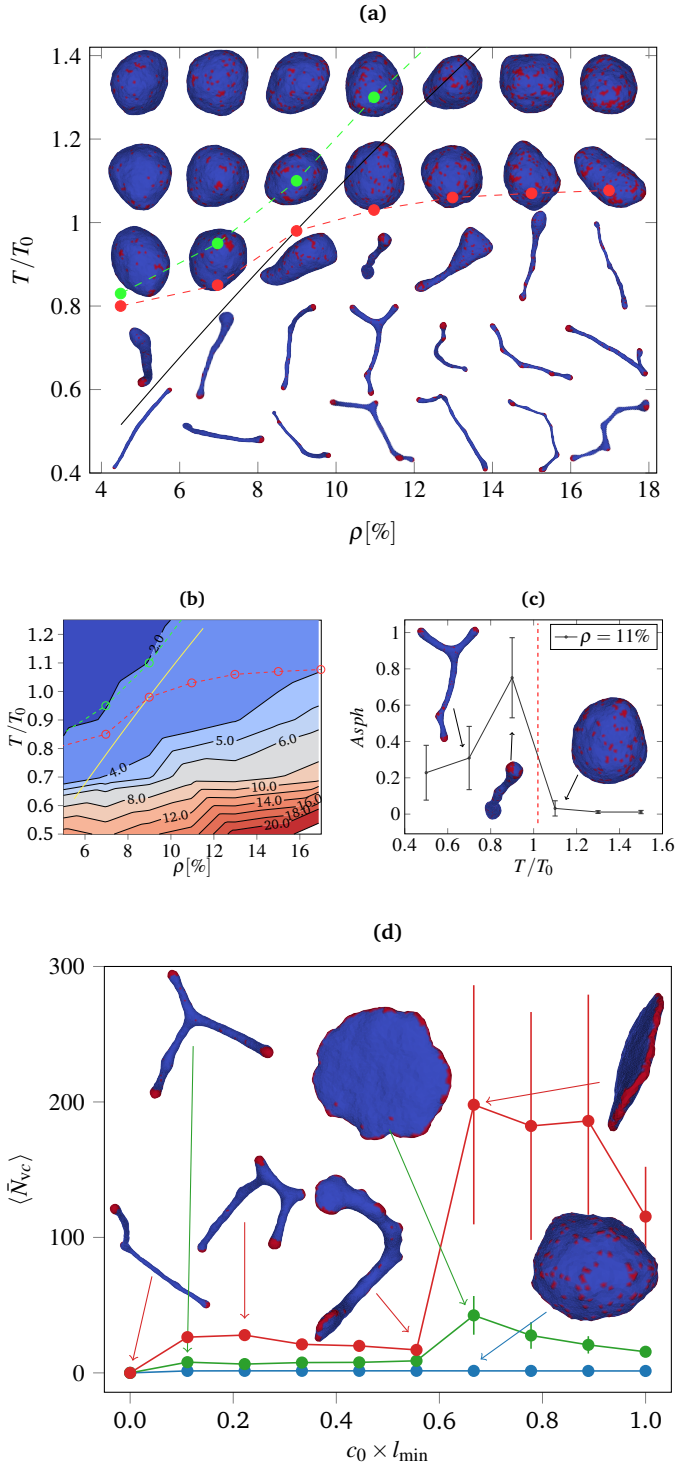

	\begin{subfigure}{\columnwidth}
		\centering \include{snapsc0_fig}
	\end{subfigure}
	\begin{subfigure}{0.5\columnwidth}
		\centering \include{contourplot_c0_f1}
	\end{subfigure}
	\begin{subfigure}{0.5\columnwidth}
		\centering \include{figs/asph_c0}
	\end{subfigure}
	\begin{subfigure}{\columnwidth}
		\centering \include{figs/pancaketohydra}
	\end{subfigure}
	\caption{{\bf (a)} Same as Fig.~\ref{fig:snapsf1}, but for flat active proteins ($c_0=0,F=1\: k T_0/l_\mr{min}$). Transition curve for hydra-like shapes is shown in red. {\bf (b)} Contour plot of the ensemble averaged mean cluster size $\left<\bar{N}_\mr{vc}\right>$ as a function of protein density $\rho$ and temperature $T$. Transition curve for hydra-like shapes is shown in red. {\bf (c)} Asphericity (Eq.~\ref{asphericity}) as a function of temperature. Red vertical dashed line indicates the transition from quasi-sphere to tube-like shapes. {\bf (d)} $\left<\bar{N}_\mr{vc}\right>$ as a function of spontaneous curvature for interaction constant $w = 0\: kT_0$ (blue), $1\: kT_0$ (green) and $1.5\: kT_0$ (red); for $\rho=11\%$ and $T/T_0=0.7$. Error bars denote standard deviations. Some corresponding snapshots are shown, where colors of the arrows match the color of the corresponding data points.}
	\label{fig:snapsc0}
\end{figure}

\end{mycomment}

\section{Conclusions}

We have explored here the coupling of convex proteins (such as complexes that contain IRSp53) that recruit the protrusive force of the cytoskeleton (most commonly due to actin polymerization), using Monte-Carlo computer simulations. We found that the presence of the protrusive forces gives rise to the formation of protein aggregates and budding at a higher temperature and lower average protein density, compared to the passive system that is in thermal equilibrium (no active forces). This is in an agreement with analytic linear-stability analysis \cite{gladnikoff2009retroviral}. The more robust aggregation and budding due to the recruited active forces of the cytoskeleton has important consequences for a variety of biological processes, such as budding of viruses \cite{gladnikoff2009retroviral,goff2007host,votteler2013virus} and initiation of cellular protrusions (such as filopodia) during development and cell motility.

Beyond the budding transition, we found a new and unexpected global shape transition that occurs only in the presence of the active forces. In this transition the spherical vesicle is transformed into a flat, pancake-like shape, with all the curved proteins and associated cytoskeleton forces along the circular rim. This structure resembles the lamellipodia and ruffles of spreading and motile cells \cite{fritz2017actin} (see Supporting Movies S1$^{\dag}$ and S2$^{\dag}$), where the actin polymerization is localized to the highly curved leading edge. Our results show that lamellipodia-like structures spontaneously form when convex proteins recruit the protrusive force of the cytoskeleton, on a closed membrane. Future studies could test these predictions, by exploring the spontaneous curvature properties of the actin nucleators at the leading edges of these cellular structures.

Let us note that flat oblate membrane shapes can be obtained without the presence of active protrusive forces, for vesicles with low volume to area ratios \cite{iglic2000torocyte,bovzivc2006coupling,iglivc2007elastic,mesarec2014numerical}. However, in our simulations the volume of the vesicle is free to relax, and under such conditions we find that the active forces are essential for the curved proteins to drive the global flattened vesicle shape transition.

In addition, a variety of tubular and flattened shapes may be stabilized in the absence of active forces, due to the presence of anisotropic curved membrane proteins (or protein complexes) \cite{ramakrishnan2018biophysics}. Tubular protrusions can occur due to accumulation of anisotropic membrane components, as was shown theoretically \cite{FournierGalatola1998,KraljIglicPRE2000,KraljIglicJPA2002,Bobrovska2013,MesarecEBJ2017,NoguchiSoftM2017,natesan2015phenomenology,NoguchiSoftM2017} and supported by experiments \cite{KraljIglicPRE2000,kraljiglicEBJ2005,iglivc2007possible}. Vesicles with flattened edges can be stabilized by anisotropic (arc-like) proteins that form a cluster at the edge of the flattened regions \cite{natesan2015phenomenology}.
We also note that the edge of the disc-shaped vesicles that we obtained here are highly convoluted (Fig.~\ref{fig:snapsf1_llc}b), due to the isotropic curvature of the proteins. Lamellipodia in cells may avoid this and maintain a smooth edge by using anisotropic proteins that recruit the actin polymerization. All of these considerations motivate the exploration of the vesicle shapes induced by coupling active forces with anisotropic membrane constituents in our future studies.

To conclude, our study highlights the rich variety of membrane shapes that may be induced by curved membrane proteins that recruit the active forces of the cytoskeleton \cite{gov2018guided}. These include steady-state shapes that are not possible for the passive, equilibrium system. Future computer simulations could explore further the space of these non-equilibrium, and dynamic membrane shapes.

\section*{Acknowledgements}

NSG is the incumbent of the Lee and William Abramowitz Professorial Chair of Biophysics. The study was supported by the  grants No. P2-0232, P3-0388, J5-7098, J2-8166 and J2-8169 of Slovenian Research Agency (ARRS).

\clearpage


\bibliography{refs} 
\bibliographystyle{rsc} 


\clearpage
\onecolumn
\section*{SUPPLEMENTAL INFORMATION}

\section*{S1: A theoretical model of self-assembly of curved nanodomains in a two-component membrane}

We use the theory of self-assembly to describe the accumulation of  curved membrane nanodomains composed of lipids and  proteins  into spherical or necklace membrane protrusions. The curved nanodomains (of total number $N$) are initially distributed in the weakly curved spherical membrane surface of
constant mean curvature $H \mathrm{=} 1/R_{0}$. We assume that
the nanodomains  are laterally mobile over the
membrane surface. For isotropic curved membrane protrusion of  constant high mean curvature
$H \mathrm{=} 1/r$. Here, $r$ is the radius of curvature everywhere on the membrane protrusion
which may be a sphere or necklace formation (see Fig.~\ref{fig.figAAB6}) and assume $R_{0} > r$.
\\
\begin{Sfigure}[h]
\begin{center}
\includegraphics[width=0.5\textwidth]{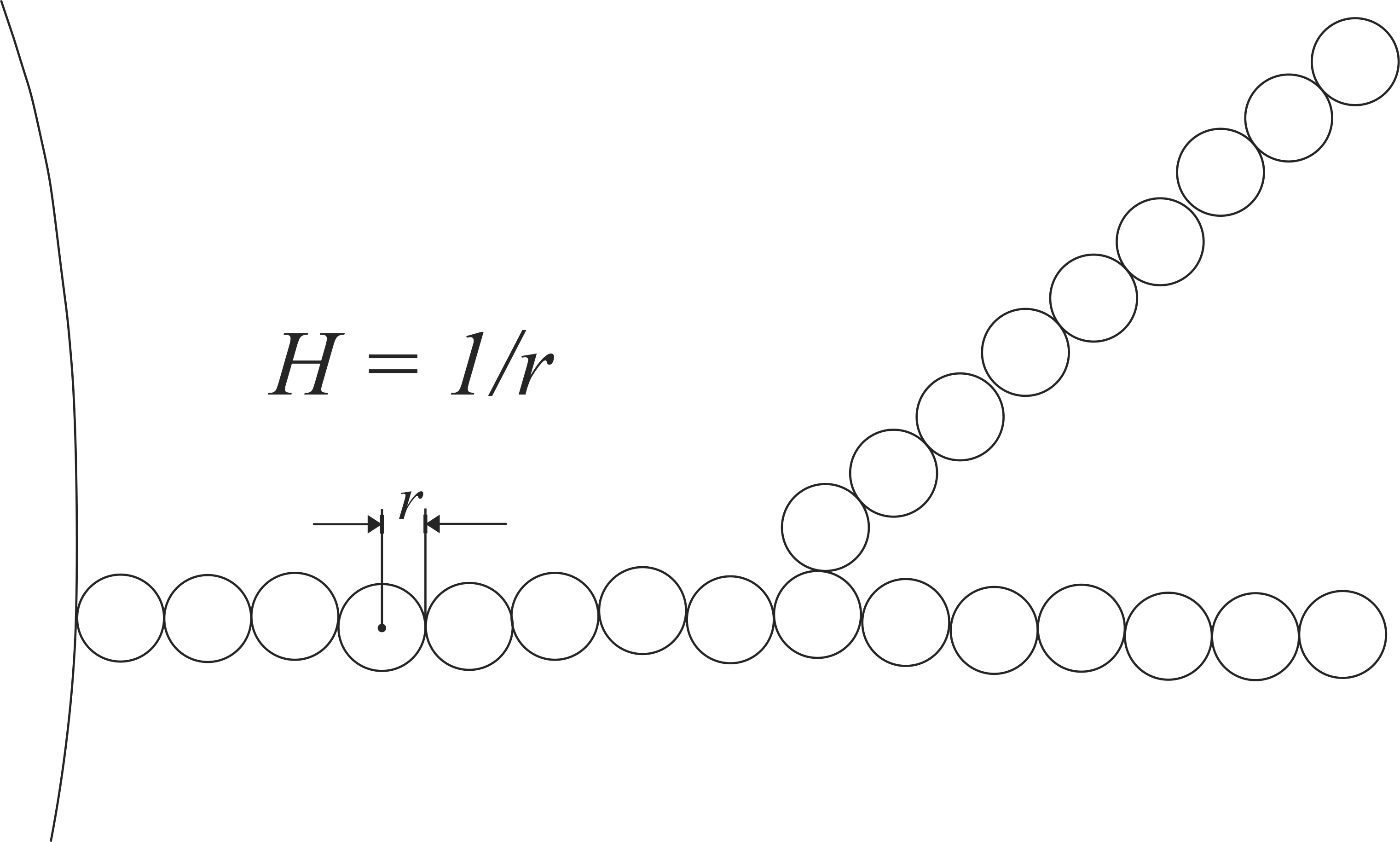}
\end{center}
\caption{Growth of necklace-like protrusions is energetically favorable when critical concentration $\tilde x_{c}$ is surpassed.}
\label{fig.figAAB6}
\end{Sfigure}

For the sake of simplicity we assume that  the free energy of a single flexible membrane  nanodomain  can be written in the form [V. Kralj-Igli\v{c} et al. Deviatoric elasticity as a possible physical mechanism explaining
collapse of inorganic micro and nanotubes, Physics letters A, 2002]:
\begin{Sequation}
f = \frac{\xi}{2}(H - H_{0})^2 a_{o} \,.
\label{eq.freene1}
\end{Sequation}
where $H_{0}$ is the intrinsic mean curvature of an isotropic  membrane nanodomain, $\xi$ is the elastic constant and $a_{0}$
is the area per single nanodomain.
In aggregates of  curved flexible membrane nanodomains the local membrane bending
constant is $k_{\mathrm{c}}=\xi/4$ and  the membrane spontaneous curvature  $c_{0}=2 H_{0}$.

Curved flexible membrane  nanodomains in aggregates interact with neighbouring  membrane nanodomains. We denote the corresponding interaction energy per curved flexible  membrane nanodomain (monomer) in an  aggregate composed of $i$ nanodomains
  as $w(i)$ where we assume that the energy $w(i)$ depends on the size of
the aggregate composed of $i$ nanodomains. The mean free energy per nanodomain in
a  curved  aggregate (where $H=D=1/r$) composed of $i$
nanodomains can be written as:
\begin{Sequation}
{\mu}_i = f_{c} - w(i)\,,
\label{eq.mi1}
\end{Sequation}
where $f_{c} = {f(H \mathrm{=} 1/r)}$ and $w(i) > 0$.
We assume that in the weakly curved spherical regions of the membrane (having
$H \mathrm{=} 1/R_{0}$)  the concentration of nanodomains
is always below the critical aggregation concentration and
therefore nanodomains  cannot form two-dimensional flat
aggregates. The mean energy per nanodomain  in the weakly curved  membrane
regions is $\tilde{\mu}_1 = f_{sp}~$, where $f_{sp} = f(H \mathrm{=}1/R_{o})$. The number density of curved proteins in the weakly curved  membrane regions is
\begin{Sequation}
\tilde{x}_1= \frac{\tilde{N}_1}{M}\,,
\label{eq.mol1}
\end{Sequation}
where $\tilde{N}_1$ is the number of monomeric curved nanodomains  in the weakly curved membrane
regions and $M$ is the number of lattice sites in the whole
system. The distribution of highly curved aggregates in the membrane
protrusions on the
scale of number density is expressed as
\begin{Sequation}
x_i= \frac{i N_i}{M}\,,
\label{eq.mol2}
\end{Sequation}
where $N_i$ denotes the number of  aggregates with
aggregation number $i$. The number densities $\tilde{x}_1$ and $x_i$ must fulfil the conservation
condition for the total number of flexible nanodomains in or on the membrane:
\begin{Sequation}
{\tilde{x}_1} ~+ \sum_{i=1}^\infty x_i = N/M\,.
\label{eq.mol2bb}
\end{Sequation}
The free energy $F$ of all nanodomains in or on the membrane can be written as:
\begin{Sequation}
{F} = M \left[ \tilde x_1 ~\tilde{\mu}_1 + kT {\tilde x_1}
(\ln {\tilde x_1} -1) \right] + M \sum \limits_{i=1}^\infty \left[x_i ~{\mu}_i
+ kT \frac{x_i}{i} \left(\ln \frac{x_i}{i}-1\right)\right] - \mu M \,(\tilde x_1 + \sum \limits_{i=1}^\infty x_i)\,~,
\label{eq.mi3}
\end{Sequation}
where $\mu$ is the Lagrange parameter assuring conservation of protein concentrations. The above expression for the free energy
 also involves  the contributions of configurational entropy. We minimize $F$ with respect to $\tilde x_1$ and $x_{i}$:
\begin{Sequation}
\frac{\partial F}{\partial \tilde x_{i}} = 0\,, \frac{\partial F}{\partial x_{i}} = 0,\, i = 1,\,2,\,3,\, ...\,,
\label{eq.min18}
\end{Sequation}
which leads to equilibrium distributions:
\begin{Sequation}
\tilde{x}_1 = \exp \left(- ~\frac {f_{sp} - \mu}{kT} \right)\,,
\label{eq.mi4}
\end{Sequation}
\begin{Sequation}
{x}_i = i~\exp \left(-\frac{i}{k\,T} \, \left[f_{c} - w - \mu \right] \right)\,,
\label{eq.mi5}
\end{Sequation}
where we assumed for simplicity that $w(i)=w$ is independent of aggregate size. The
quantity $ \mu$ can be expressed from Eq.~\ref{eq.mi4} and substituted
in Eq.~\ref{eq.mi5} to get:
\begin{Sequation}
{x}_i = i~ \left[ \tilde x_{1} \cdot \exp \left( \frac{ f_{sp} +
w - f_{c}}{kT} \right) \right]^{i}\,.
\label{eq.mi7}
\end{Sequation}
We see that if the concentration $\tilde{x}_{1}$ is small, aggregate growth will not be favourable, since $x_{1} > x_{2} > x_{3}\, ...$. Furthermore, $x_{i}$ can never exceed unity, leading to the maximal possible value of the number density of monomeric
curved flexible nanodomains in the weakly curved parts of the membrane when $\tilde x_{1}$ approaches $\exp \left[(f_{c}-f_{sp}-w)/{k\,T}\right]$. The critical concentration is therefore
\begin{Sequation}
\tilde x_{c} \approx \exp \left(~ \frac{ \Delta f - w
}{kT}\right)\,,
\label{eq.mi8}
\end{Sequation}
where $\Delta f = f_{c} - f_{sp}$ is the difference between the energy of
a  single nanodomain  on the  highly curved membrane  protrusion and the energy
of the single nanodomain  in the weakly curved membrane region with:
\begin{Sequation}
\Delta f = \frac{\xi a_{o}}{2 r}\left(\frac{1}{r}-2H_{0}\right) - \frac{\xi a_{o}}{2 R_{0}}\left(\frac{1}{R_{0}}-2H_{0}\right)    \,.
\label{eq.mi9}
\end{Sequation}
If $\tilde x_{1}$ is above $\tilde x_{c}~$, the formation of a very long necklace membrane protrusions composed of curved membrane proteins  is energetically favourable. It can be seen from Eq.~\ref{eq.mi8} that longitudinal growth of the
necklace membrane protrusions is dependent on the  energy
difference $\Delta f$  (Eq.~\ref{eq.mi9}) and  the strength of the direct
interaction between nanodomains   $w$. The critical
concentration $\tilde x_{c}$ strongly depends on
$H_{0}$.

In the approximation limit  $R_{0} \gg r$ we can rewrite Eq.~\ref{eq.mi9} as:
\begin{Sequation}
\Delta f \simeq \frac{\xi}{2r}\left(\frac{1}{r}-2H_{0}\right) = \frac{2k_{c}}{r}\left(\frac{1}{r}-c_{0}\right)\,,
\label{eq.mi10}
\end{Sequation}
where $k_{\mathrm{c}}$ and $c_{0}$ are the local bending constant and spontaneous curvature
of  aggregates of nanodomains, respectively. We may rewrite Eq. \ref{eq.mi8}:
\begin{Sequation}
\tilde x_{c} \approx \exp \left(~ {2 ~\frac{k_{c} }{kT }~ \frac{a_{o}}{r^{2}}   \left( 1-c_{0}r  \right) - \frac{w}{kT}
}\right)\,.
\label{eq.mi11}
\end{Sequation}
For $1 < c_{0}r$ the value of $\Delta f$ is always negative. The theoretically predicted existence  of  necklace  membrane protrusions (without application of  the local forces) within the self-assembly theory is in line with our MC predictions.


Since the density of nanodomains in or on the membrane is defined with the conservation condition (Eq. \ref{eq.mol2bb}), this also gives us the relation between normalized temperature $T/T_{0}$ and total curved nanodomains concentrations $\rho=N/M$. Using the parameters from the MC simulations, we may graph dependencies $x_{i}(i)$, as seen in Fig.~\ref{fig.figAAB7}. Above small concentrations and especially above $\tilde x_{c}$, aggregates start to form, where the peaks of the distributions are strongly dependent on the total protein concentration in the lattice. We see that the critical line beyond which aggregate growth is favourable agrees well with the results of MC simulations.

\begin{Sfigure}[h]
\begin{center}
\includegraphics[width=0.5\textwidth]{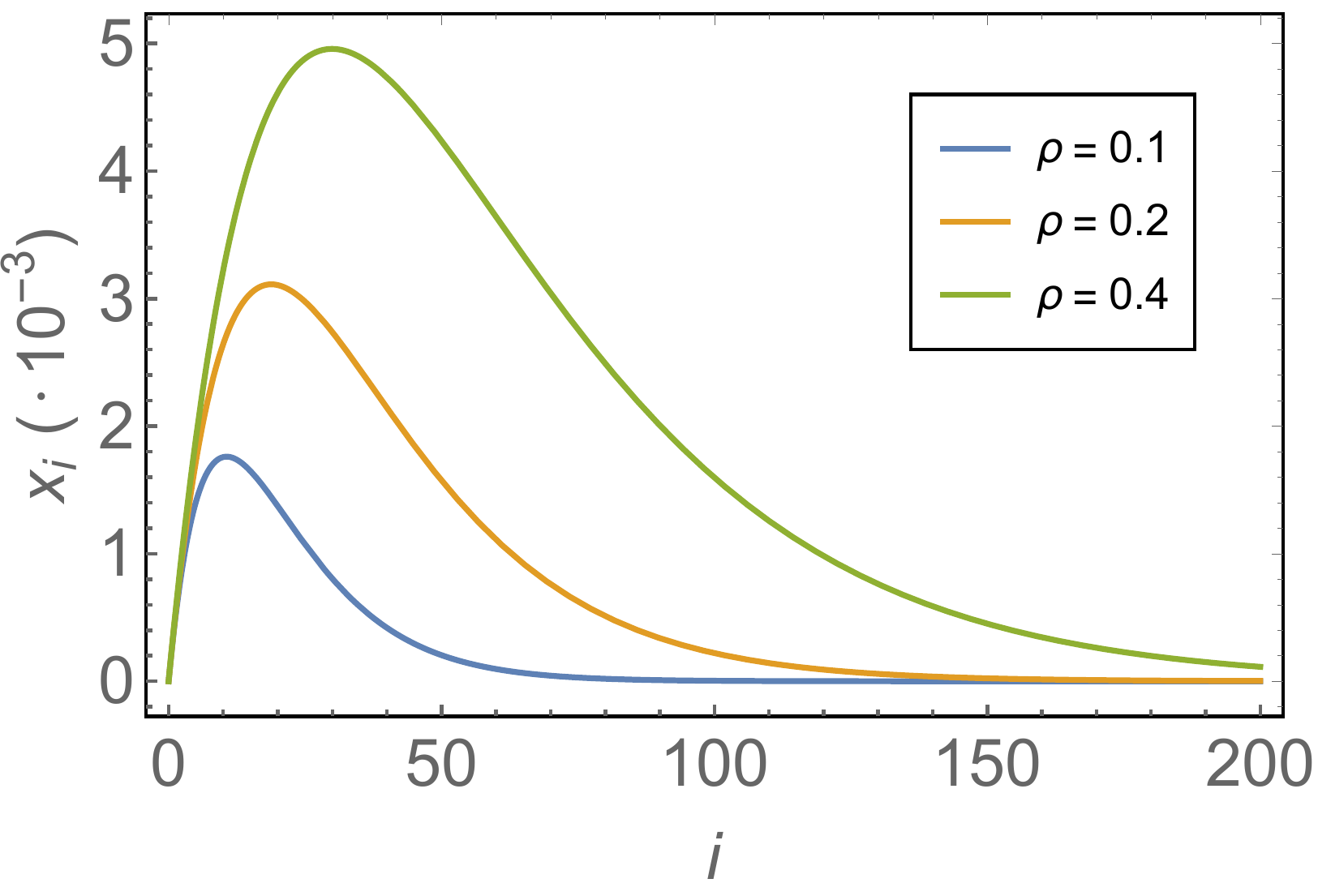}
\end{center}
\caption{Aggregate concentrations in dependence on number of nanodomains in the aggregate for different number of flexible nanodomains on the membrane.}
\label{fig.figAAB7}
\end{Sfigure}



\clearpage
\section*{S2: A theoretical analysis of the critical cluster size that enables tubular shapes for flat active proteins}
The conditions that trigger the transition into the tubular-shapes (Fig.7) are given by the following force balance:

The force applied at the tip of the cylindrical protrusion by the cluster of active proteins is
\begin{Sequation}
	F_a=F\frac{\pi R^2}{a}
\label{ftip}
\end{Sequation}
where $F$ is the force per active protein, $R$ is the radius of the cylinder, and $a$ is the area of a protein on the membrane.

This is balanced by the restoring force of the membrane bending energy
\begin{Sequation}
	F_b=\kappa\frac{2\pi}{R}
\label{fbend}
\end{Sequation}
with $\kappa$ the bending modulus.
The force balance gives the radius of the cylindrical protrusions in this phase of the vesicle shapes
\begin{Sequation}
	R_c=\left(\frac{2\kappa a}{F}\right)^{1/3}
\label{rc}
\end{Sequation}

The prediction of Eq.~\ref{rc} is in good agreement with simulations (see Fig.~\ref{fig:S2}), where we took for $a$ the area that correspons to one vertex in a hexagonal mesh, $a=\sqrt{3}\: l_0^2/2$, where $l_0=(l_{min}+l_{max})/2$.


In the phase of tubular shapes, there are several protrusions (typically 2-3) that pull in opposite directions to provide an overall force balance, and maintain the relative stability of this shape. Some fusions of protrusions do occur, especially for cases with a larger number of thinner protrusions, so that their number fluctuates.


\begin{Sfigure}[h!]
	\centering \include{figs/hydra_radius_dependence}
	\caption{Radius of cylindrical protrusions as a function of the $\kappa$ to $F$ ratio for the system withwith almost flat active proteins with parameters $c_0=1/(0.9 l_\mr{min})$, $\rho=11\%$, $w = 1 kT_0$ and $T/T_0=0.7$ (see top-left hydra-like snapshot on Fig.~7d). Black solid curve is the prediction of Eq.~\ref{rc}, while red dots are the results of the simulations with error bars indicating standard errors.}
	\label{fig:S2}
\end{Sfigure}


\clearpage
\section*{SI3: Cluster size dependence on the strength of the direct interaction for active system}

See Fig.~\ref{fig:Swdepf1}.

\begin{Sfigure}[h!]
	\centering \include{figs/eamcsvstdivtc_w_comparison_F1only}
	\caption{Mean cluster size $\left<\bar{N}_\mr{vc}\right>$ as a function of $T/T^\mr{(c)}$ for two different values of the direct interaction constant (legend), for an active system with $F=1\: k T_0/l_\mr{min}$. The average protein density is $\rho=9.5\%$. The graphs do not collapse, unlike in the passive system (Fig.~2d).}
	\label{fig:Swdepf1}
\end{Sfigure}

\clearpage
\section*{SI4: Testing for hysteresis of the pancake transition}

See Fig.~\ref{fig:Syhsteresis}.

\begin{Sfigure}[h!]
	\centering \include{./figs/hysteresis}
	\caption{Hysteresis test for the transition into the pancake shape (corresponding to the system shown in Figs.~3,4), showing ensemble averaged mean cluster size for active curved proteins as a function of temperature for two different initial states -- above (blue) and below (red) pancake transition. Average protein density is $\rho=11\%$. Error bars denote standard deviations.}
	\label{fig:Syhsteresis}
\end{Sfigure}


\clearpage
\section*{SI5: Cluster size dependence on the density}

The activity-driven transition is clearly seen in Fig.~4b of the main text -- in the mean cluster size $\left<\bar{N}_\mr{vc}\right>$ as a function of temperature $T/T_0$ for different average densities of proteins $\rho$. Without the active protrusive force, $\left<\bar{N}_\mr{vc}\right>$ monotonically increases with $\rho$ and decreases with $T$, while the protrusive force gives rise to the sharp transition into pancake-like shapes. The lower stability of the rim aggregate at high protein densities, that we already noticed in Fig.3, is manifested in the non-monotonic dependencies of $\left<\bar{N}_\mr{vc}\right>$ on $\rho$ and $T$ (Fig.~\ref{S5}).

\begin{Sfigure}[h!]
	\centering \include{figs/eamcsvsrhoSI}
	\caption{Ensemble averaged mean cluster size as a function of the average density of curved proteins with $c_0=1/l_\mr{min}$. Results with active protrusive force $F=1\: k T_0/l_\mr{min}$ are shown for $T/T_0 = 0.625$ (solid) and without it for $T/T_0 = 0.4$ (dashed).}
	\label{S5}
\end{Sfigure}


\clearpage
\section*{SI6: Vesicle size dependence of the budding and pancake transition}

The dependence of the pancake transition on the vesicle radius mirrors the effect on the overall cluster size distribution: a smaller vesicle has smaller protein clusters and a lower transition temperature (Fig.~\ref{S6}).

\begin{Sfigure}[h!]
		\centering \include{figs/pancaketranstempvsn}
		\caption{Dependence of the budding (green) and pancake (red) transition curves as functions of the number of vertices composing the vesicle with $F=1\: k T_0/l_\mr{min}$,$c_0=1/l_\mr{min}$, $\rho=9.5\%$. Spherical vesicle with the same membrane area $A$ has the radius $R_0 \approx 0.35 \sqrt{N}$ (in units of $l_\mr{min}$). Black solid curve is the prediction for the budding transition line from the linear stability analysis.}
	\label{S6}
\end{Sfigure}


\clearpage
\section*{SI7: Pressure difference dependence of the pancake transition}

See Fig.~\ref{S7}.

\begin{Sfigure}[h!]
	\begin{subfigure}{\textwidth}
		\begin{center}
			\include{snapspressure001_fig}
			\caption{Difference of pressures between inside and outside of the vesicle $\Delta p = 0.01\:kT_0/l_\mr{min}$}
		\end{center}
	\end{subfigure}
	\begin{subfigure}{\textwidth}
		\centering \include{snapspressure_fig}
		\centering \caption{Difference of pressures between inside and outside of the vesicle $\Delta p = 0.1\:kT_0/l_\mr{min}$}
	\end{subfigure}
	\caption{Pressurization snapshots: Same system as in Fig.3, but with a finite pressure difference $\Delta p = p_\mr{inside} - p_\mr{outside}$. At high pressure, the membrane tension inhibits the transition to the pancake shape.}
	\label{S7}
\end{Sfigure}

\clearpage
\section*{MOVIES}

\movie{{\bf \href{https://www.dropbox.com/s/cpvtph4skeclk37/movieS1_Fosnaric_f1c1w1t6rho7.ogv?dl=0}{Movie S1}}: Animation of snapshots in steady-state of the system with $\rho=7\,\%$ of curved active proteins with $F=1\,k T_0/l_{min}$, $c_0=1/l_{min}$, $w=1\,kT_0$ at $T/T_0=0.6$ (see the last snapshot in the second line from below on Fig.~5a).
}

\movie{{\bf \href{https://www.dropbox.com/s/czq0p2kwsjfzryp/movieS2_Fosnaric_f1c1w1t6rho5.ogv?dl=0}{Movie S2}}: Animation of snapshots in steady-state of the system with $\rho=5\,\%$ of curved active proteins with $F= 1\,k T_0/l_{min}$, $c_0=1/l_{min}$, $w=1\,kT_0$ at $T/T_0=0.6$ (see the second snapshot in the second line from below on Fig.~5a).
}

\movie{{\bf \href{https://www.dropbox.com/s/lnvt7cxs9gsq4fa/movieS3_Fosnaric_f1c1p9w1t7rho11.ogv?dl=0}{Movie S3}}: Animation of snapshots in steady-state of the system with $\rho=11\,\%$ of almost flat active proteins with $F= 1\,k T_0/l_{min}$, $c_0=1/(9 l_{min})$, $w=1\,kT_0$ at $T/T_0=0.7$ (see the top-left shape on Fig.~7d).
}


\end{document}

%% file: snaps_fig.tex
\begin{tikzpicture}
\begin{axis}[
	width=0.99\textwidth,height=0.91\textheight,
	xlabel={$\rho[\%]$},
	ylabel=$T/T_0$,ymin=0.57,ymax=1.4,xmin=0.037,xmax=0.182,
	scaled x ticks=manual:{}{\pgfmathparse{(#1)*100}},
	tick label style={/pgf/number format/fixed}]

	\node[inner sep=0pt] at (axis cs:0.04988807163415414,1.245) {\includegraphics[width=0.15\textwidth]{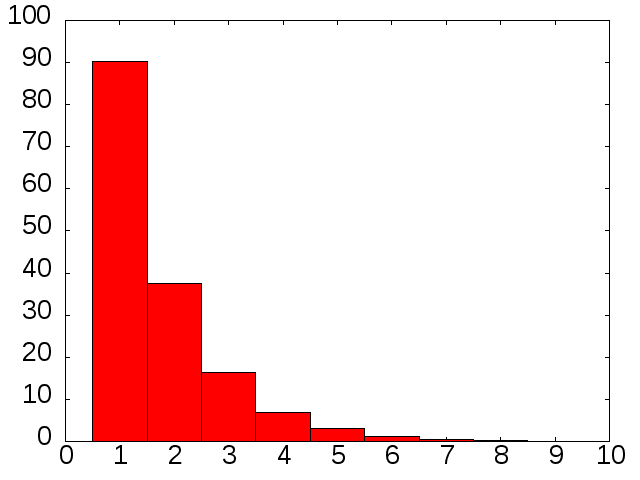}};
	\node[inner sep=0pt] at (axis cs:0.04988807163415414,1.3333333333333333) {\includegraphics[width=0.22\textwidth]{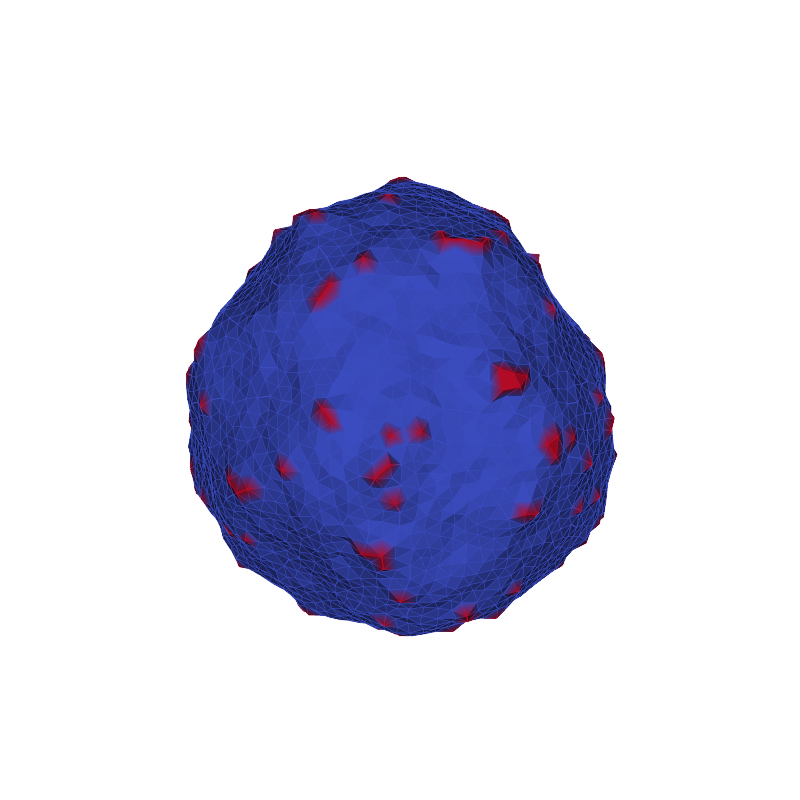}};
	\node[inner sep=0pt] at (axis cs:0.0799488327470419,1.245) {\includegraphics[width=0.15\textwidth]{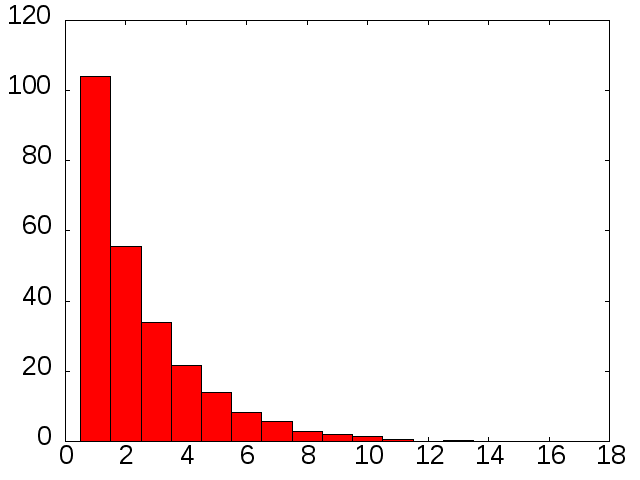}};
	\node[inner sep=0pt] at (axis cs:0.0799488327470419,1.3333333333333333) {\includegraphics[width=0.22\textwidth]{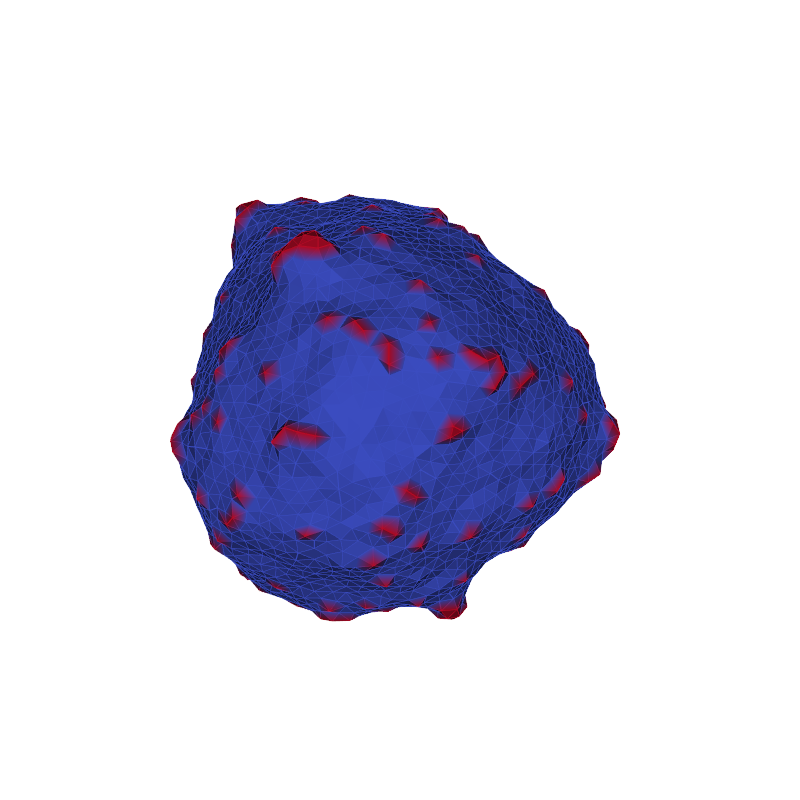}};
	\node[inner sep=0pt] at (axis cs:0.10489286856411896,1.245) {\includegraphics[width=0.15\textwidth]{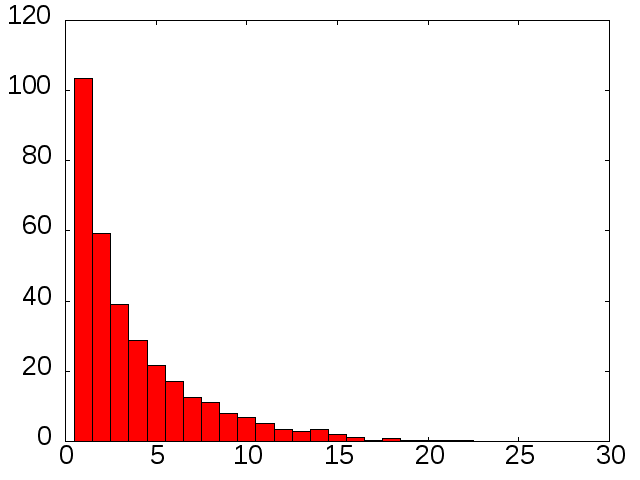}};
	\node[inner sep=0pt] at (axis cs:0.10489286856411896,1.3333333333333333) {\includegraphics[width=0.22\textwidth]{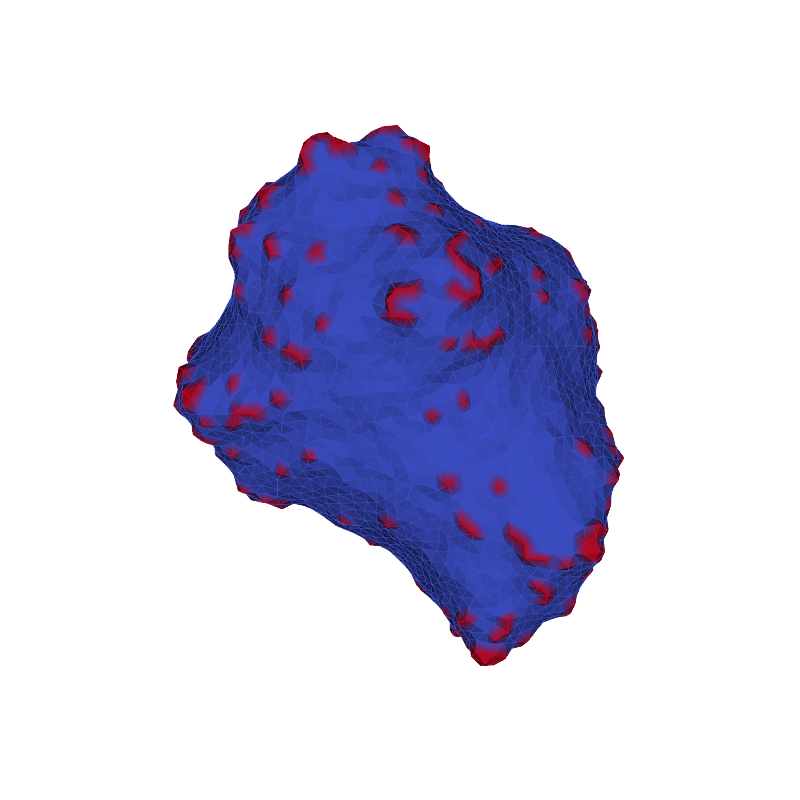}};
	\node[inner sep=0pt] at (axis cs:0.13975055964182923,1.245) {\includegraphics[width=0.15\textwidth]{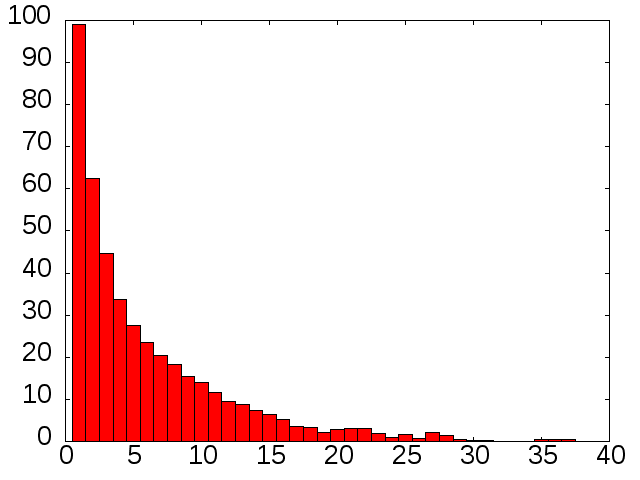}};
	\node[inner sep=0pt] at (axis cs:0.13975055964182923,1.3333333333333333) {\includegraphics[width=0.22\textwidth]{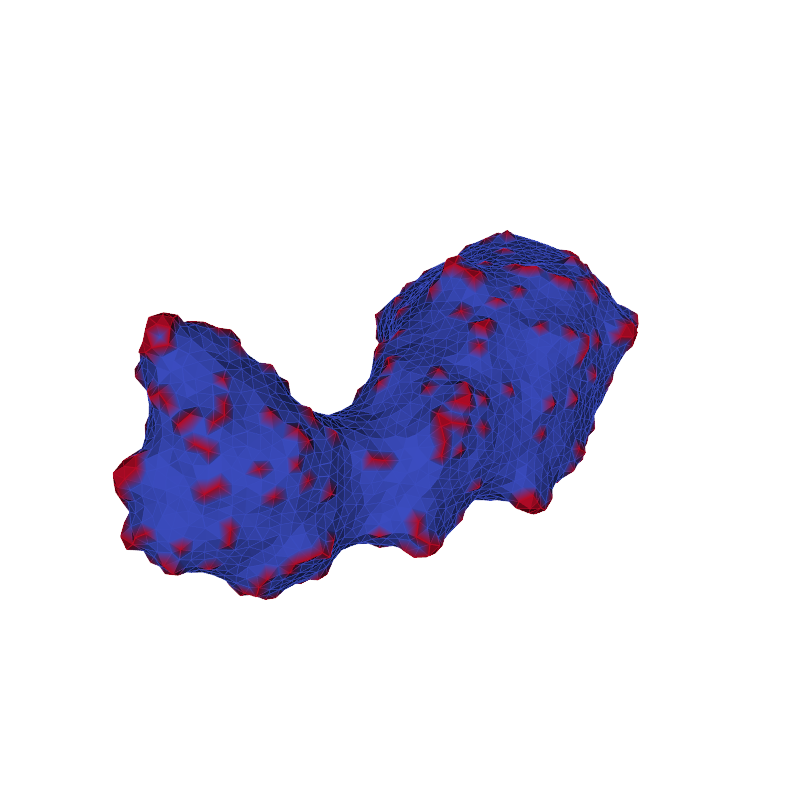}};
	\node[inner sep=0pt] at (axis cs:0.16981132075471697,1.245) {\includegraphics[width=0.15\textwidth]{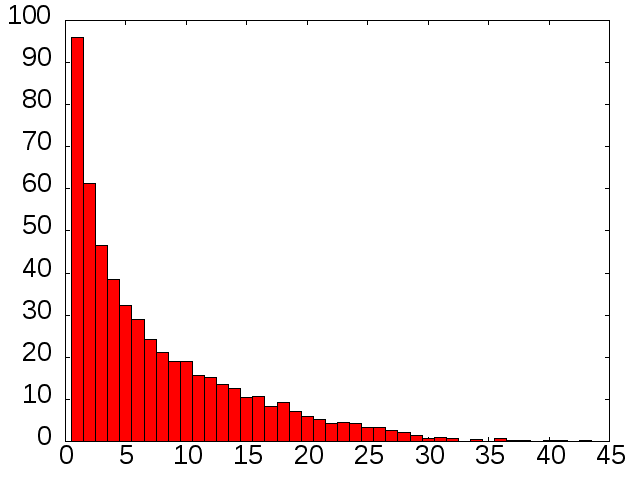}};
	\node[inner sep=0pt] at (axis cs:0.16981132075471697,1.3333333333333333) {\includegraphics[width=0.22\textwidth]{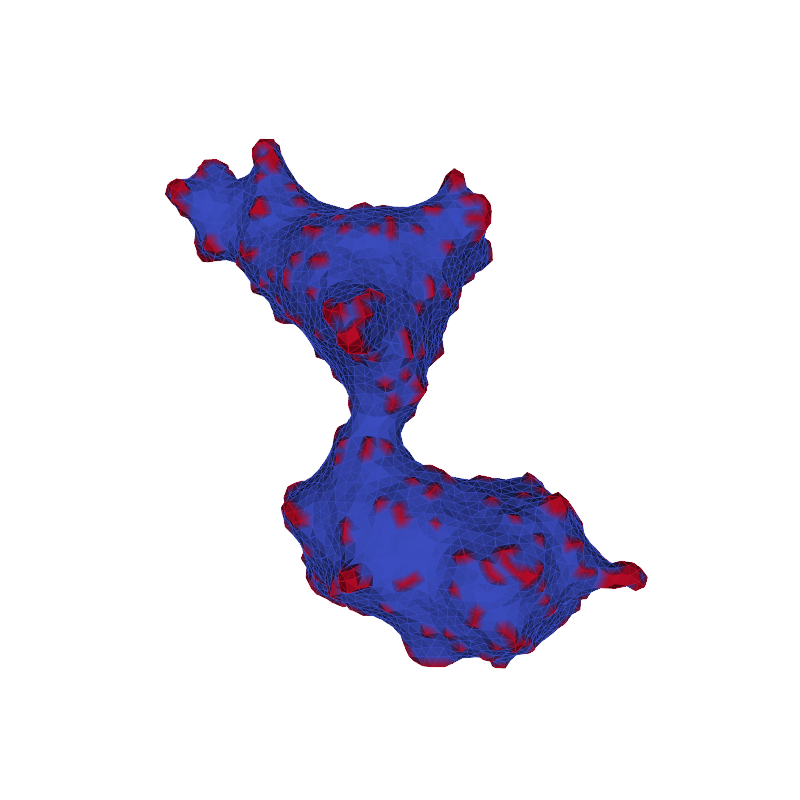}};
	\node[inner sep=0pt] at (axis cs:0.04988807163415414,1.02) {\includegraphics[width=0.15\textwidth]{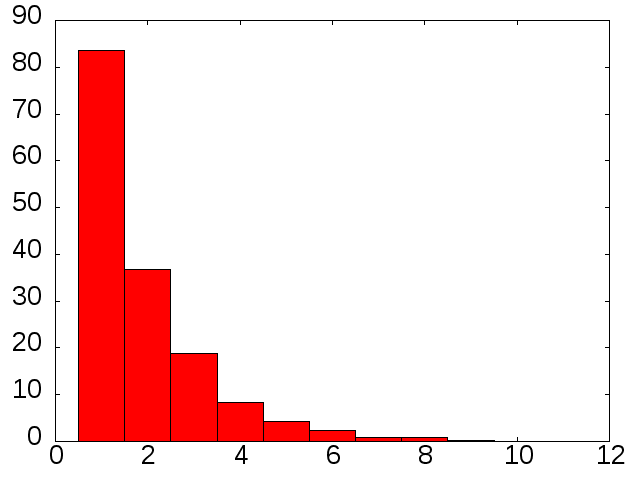}};
	\node[inner sep=0pt] at (axis cs:0.04988807163415414,1.1111111111111112) {\includegraphics[width=0.22\textwidth]{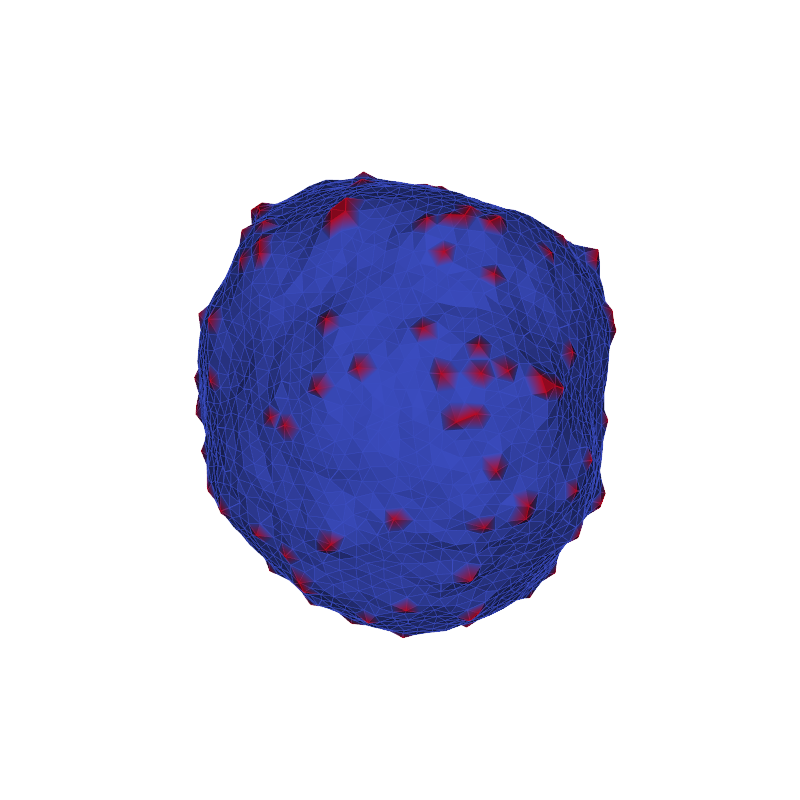}};
	\node[inner sep=0pt] at (axis cs:0.0799488327470419,1.02) {\includegraphics[width=0.15\textwidth]{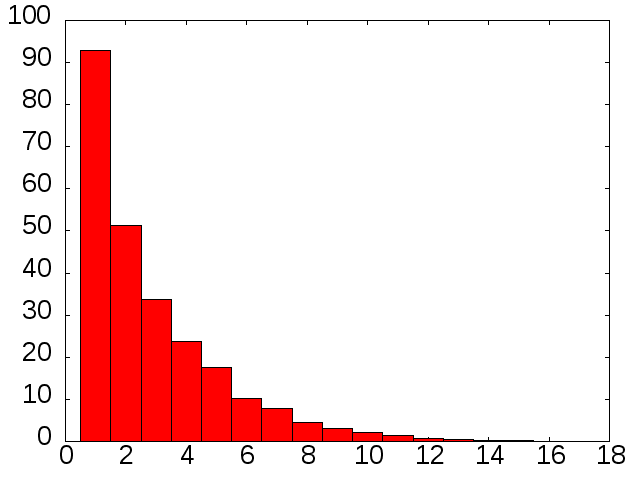}};
	\node[inner sep=0pt] at (axis cs:0.0799488327470419,1.1111111111111112) {\includegraphics[width=0.22\textwidth]{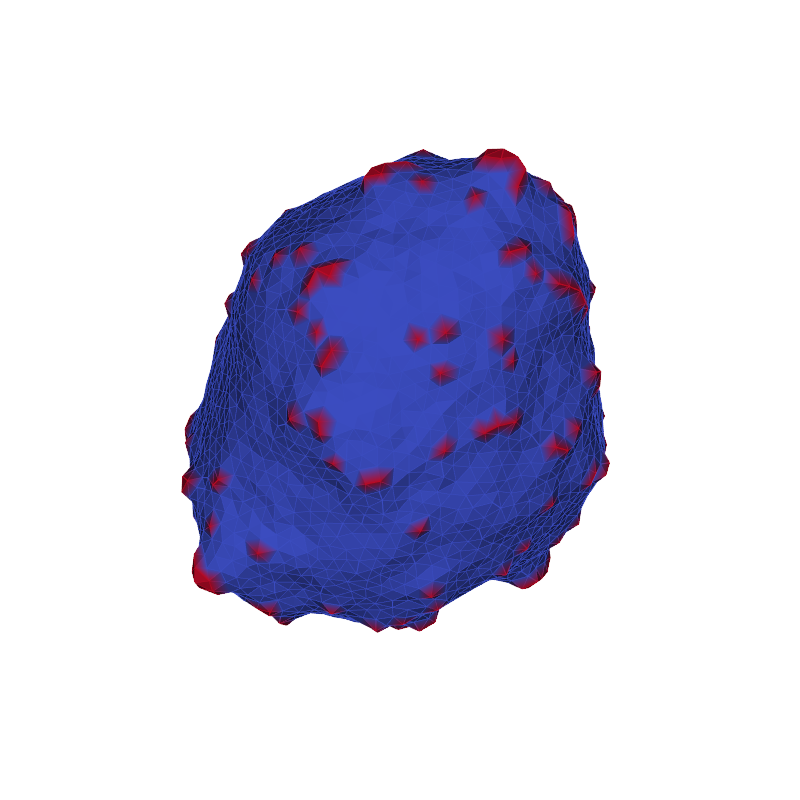}};
	\node[inner sep=0pt] at (axis cs:0.10489286856411896,1.02) {\includegraphics[width=0.15\textwidth]{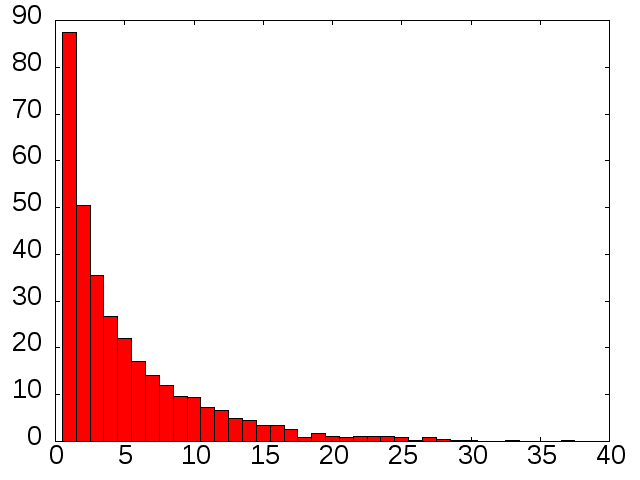}};
	\node[inner sep=0pt] at (axis cs:0.10489286856411896,1.1111111111111112) {\includegraphics[width=0.22\textwidth]{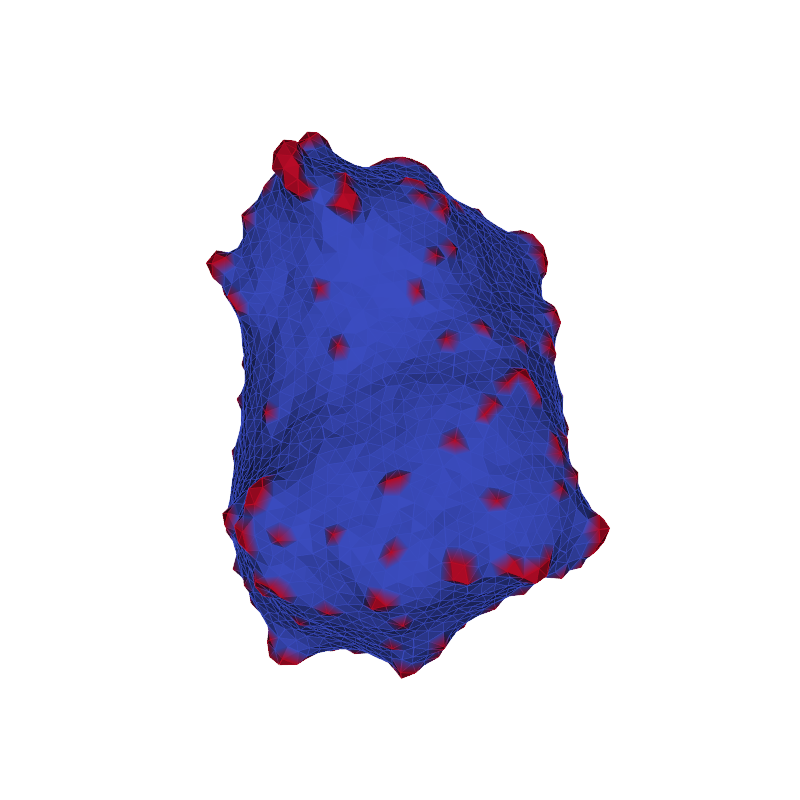}};
	\node[inner sep=0pt] at (axis cs:0.13975055964182923,1.02) {\includegraphics[width=0.15\textwidth]{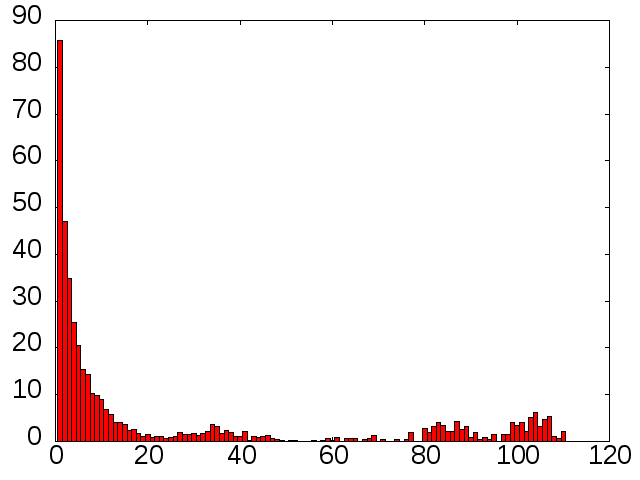}};
	\node[inner sep=0pt] at (axis cs:0.13975055964182923,1.1111111111111112) {\includegraphics[width=0.22\textwidth]{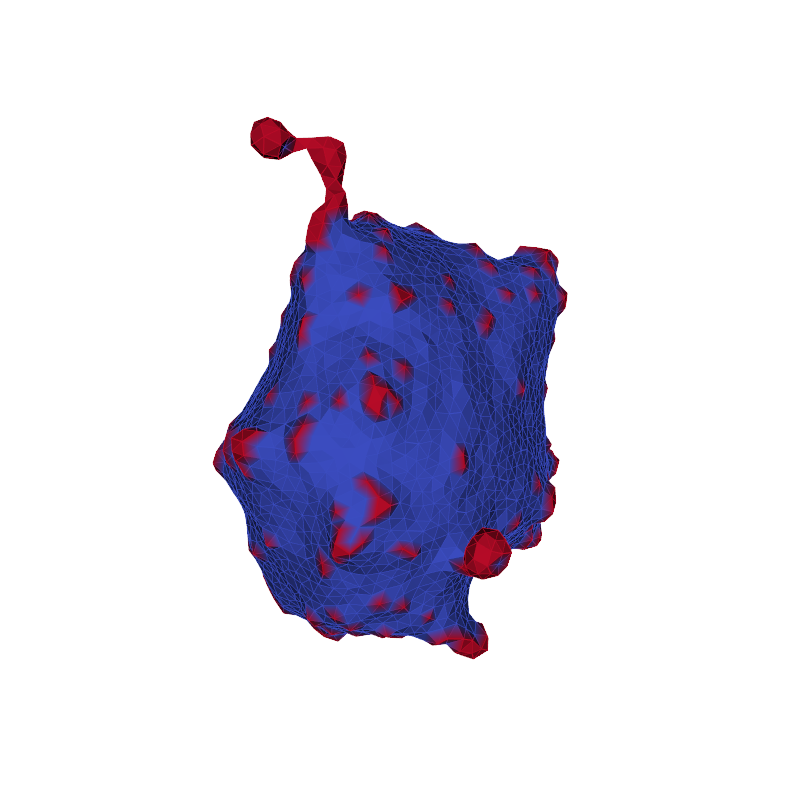}};
	\node[inner sep=0pt] at (axis cs:0.16981132075471697,1.02) {\includegraphics[width=0.15\textwidth]{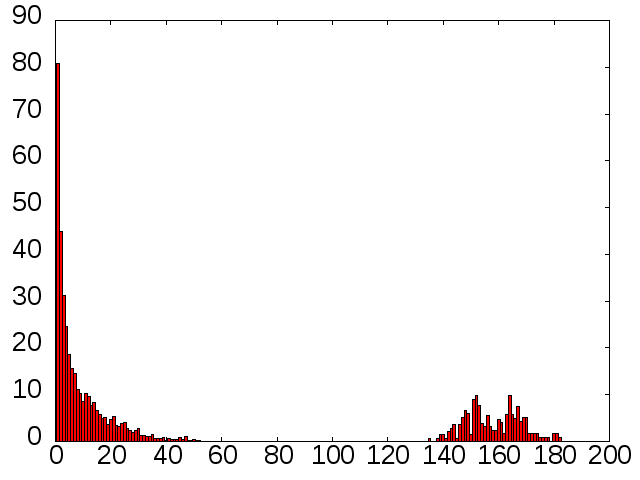}};
	\node[inner sep=0pt] at (axis cs:0.16981132075471697,1.1111111111111112) {\includegraphics[width=0.22\textwidth]{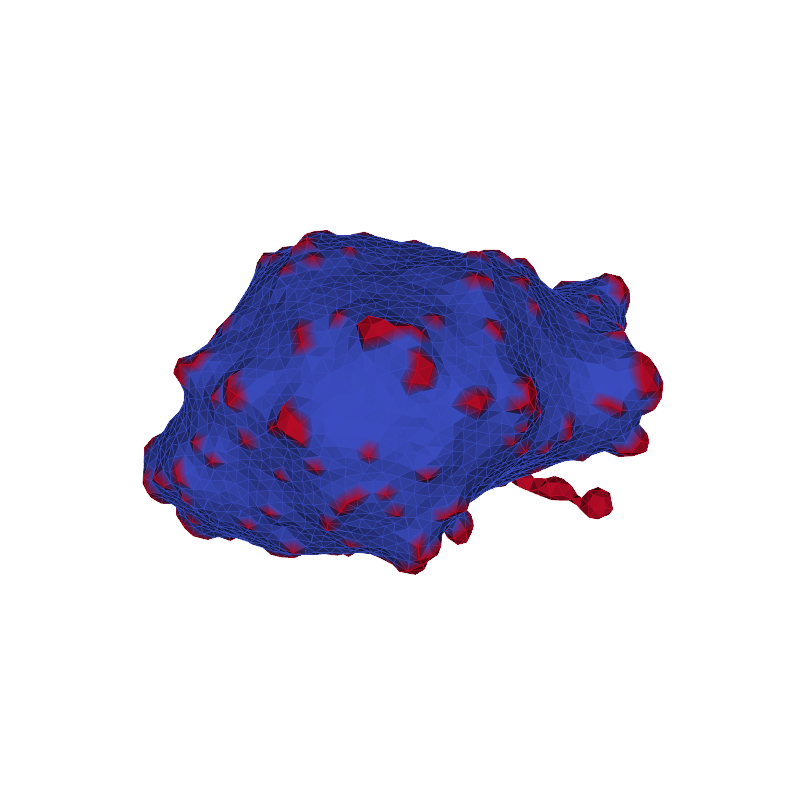}};
	\node[inner sep=0pt] at (axis cs:0.04988807163415414,0.925) {\includegraphics[width=0.15\textwidth]{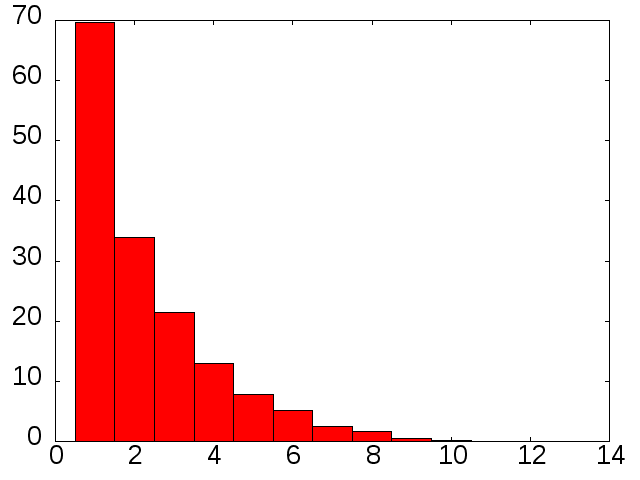}};
	\node[inner sep=0pt] at (axis cs:0.04988807163415414,0.8333333333333334) {\includegraphics[width=0.22\textwidth]{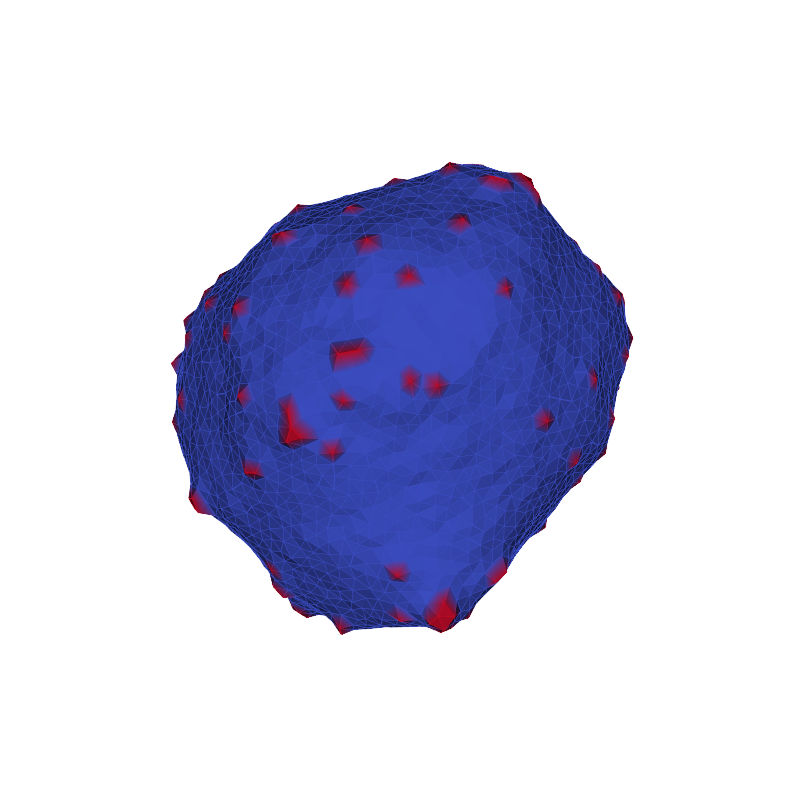}};
	\node[inner sep=0pt] at (axis cs:0.0799488327470419,0.925) {\includegraphics[width=0.15\textwidth]{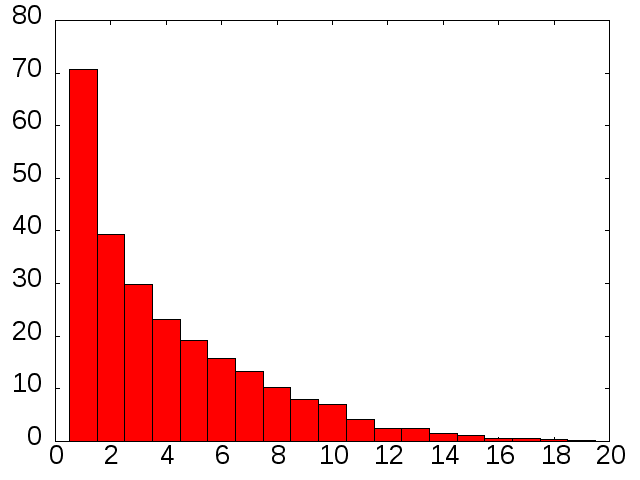}};
	\node[inner sep=0pt] at (axis cs:0.0799488327470419,0.8333333333333334) {\includegraphics[width=0.22\textwidth]{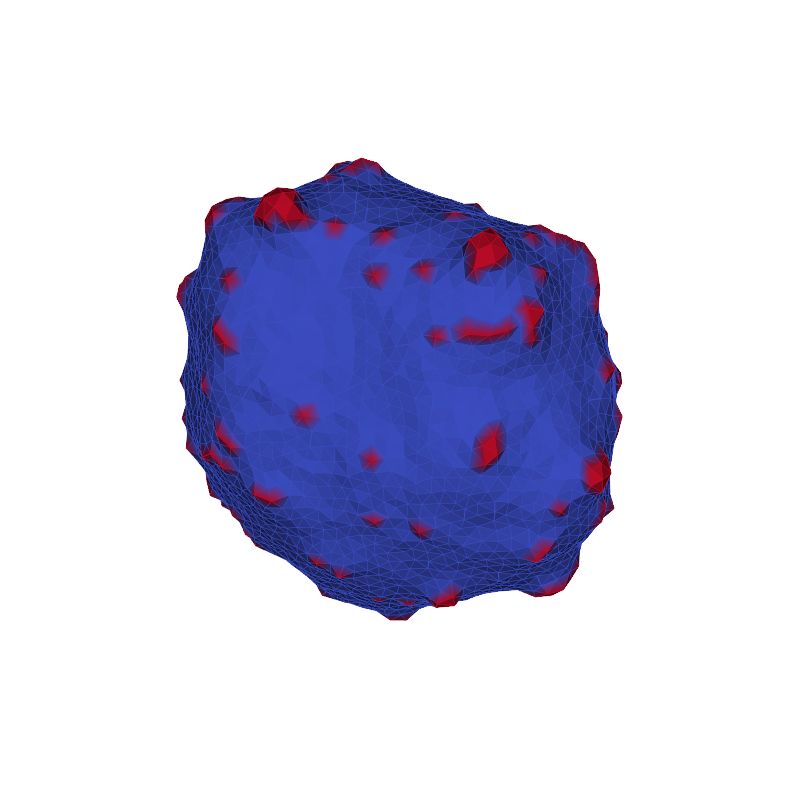}};
	\node[inner sep=0pt] at (axis cs:0.10489286856411896,0.925) {\includegraphics[width=0.15\textwidth]{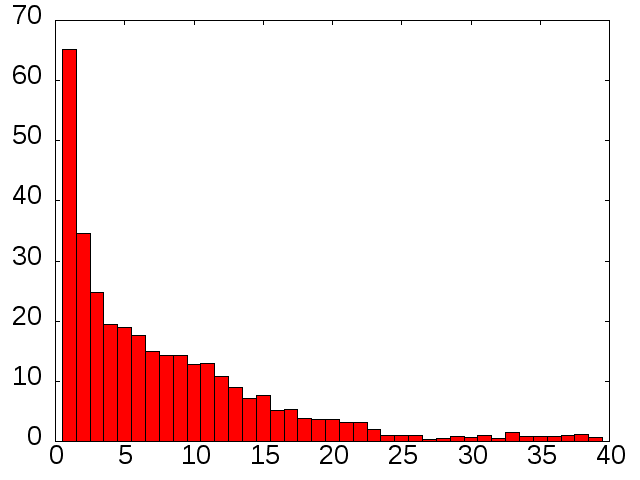}};
	\node[inner sep=0pt] at (axis cs:0.10489286856411896,0.8333333333333334) {\includegraphics[width=0.22\textwidth]{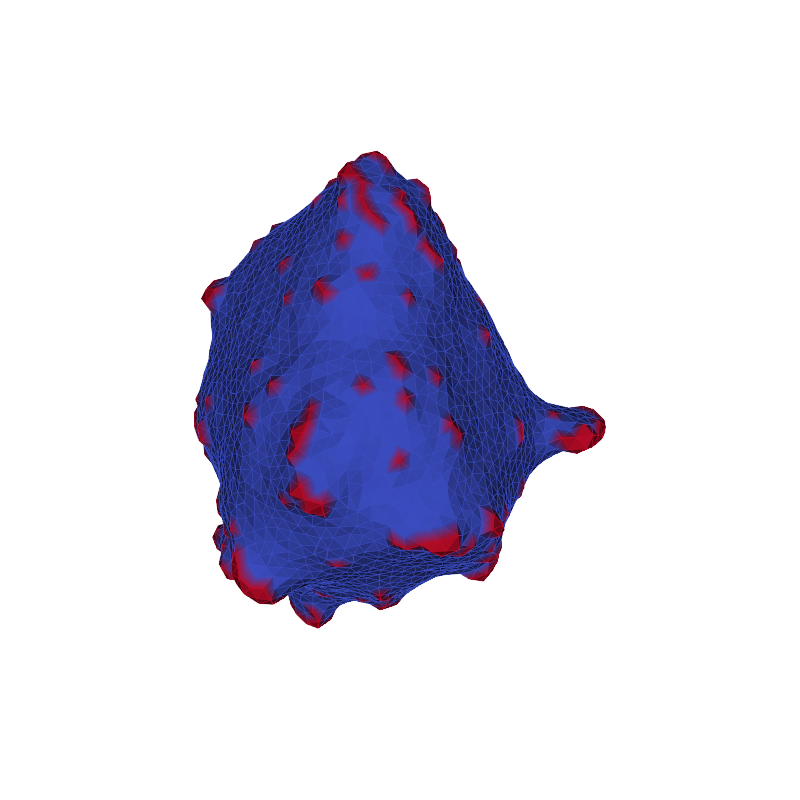}};
	\node[inner sep=0pt] at (axis cs:0.13975055964182923,0.925) {\includegraphics[width=0.15\textwidth]{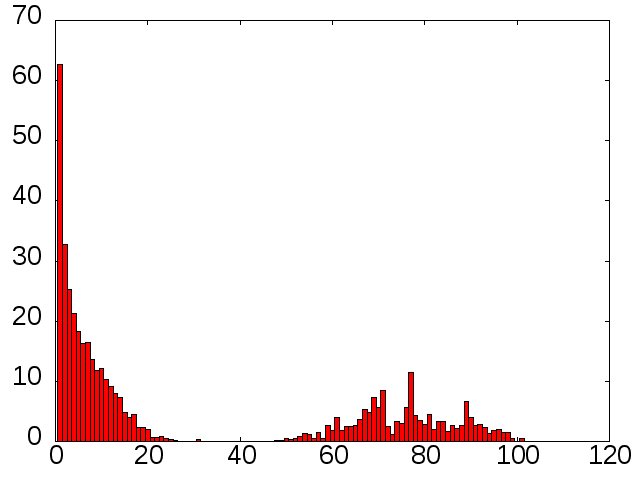}};
	\node[inner sep=0pt] at (axis cs:0.13975055964182923,0.8333333333333334) {\includegraphics[width=0.22\textwidth]{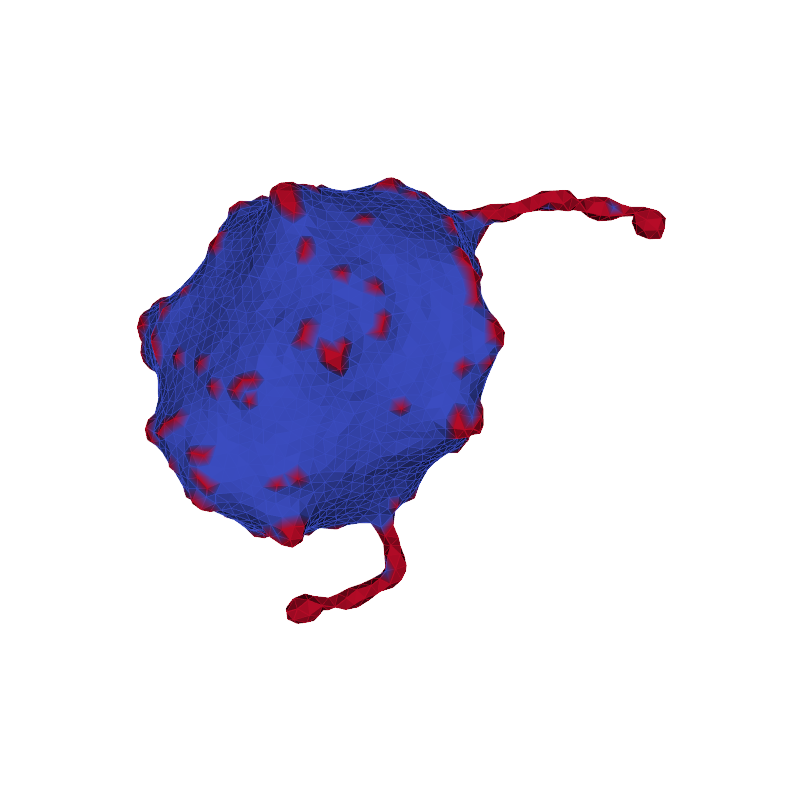}};
	\node[inner sep=0pt] at (axis cs:0.16981132075471697,0.925) {\includegraphics[width=0.15\textwidth]{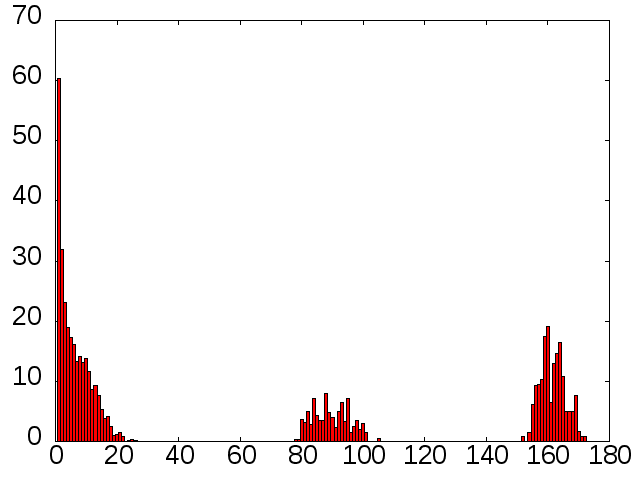}};
	\node[inner sep=0pt] at (axis cs:0.16981132075471697,0.8333333333333334) {\includegraphics[width=0.22\textwidth]{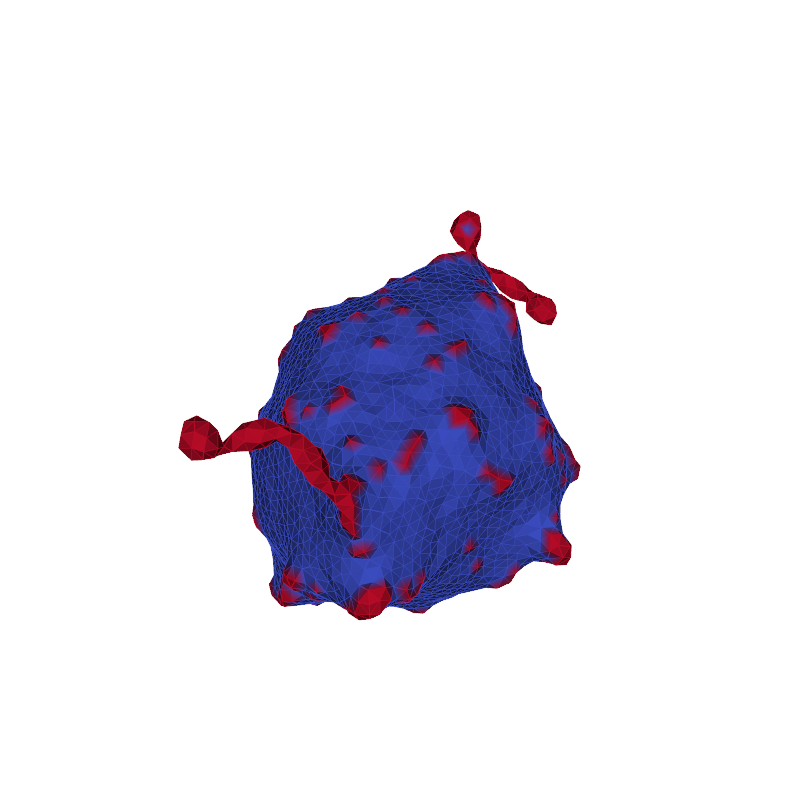}};
	\node[inner sep=0pt] at (axis cs:0.04988807163415414,0.717) {\includegraphics[width=0.15\textwidth]{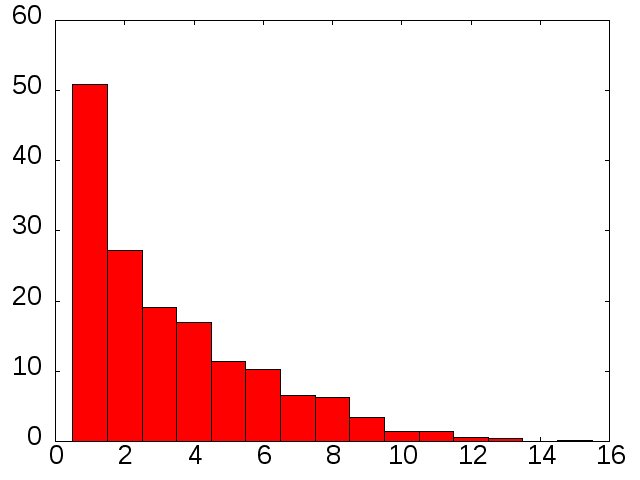}};
	\node[inner sep=0pt] at (axis cs:0.04988807163415414,0.625) {\includegraphics[width=0.22\textwidth]{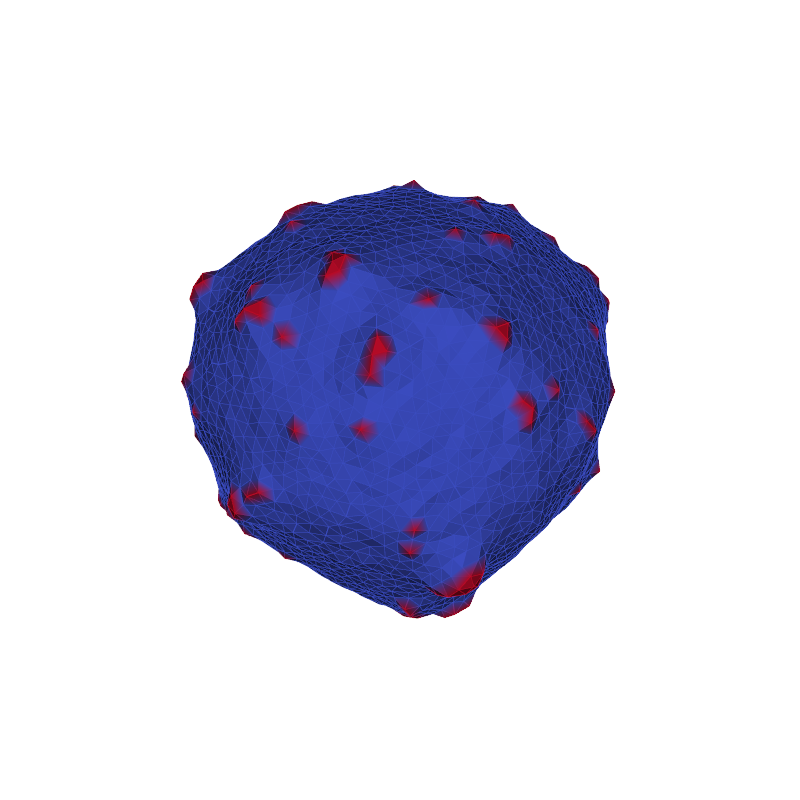}};
	\node[inner sep=0pt] at (axis cs:0.0799488327470419,0.717) {\includegraphics[width=0.15\textwidth]{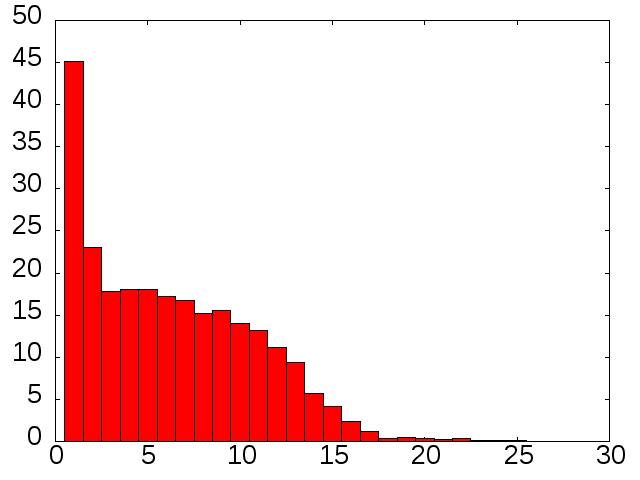}};
	\node[inner sep=0pt] at (axis cs:0.0799488327470419,0.625) {\includegraphics[width=0.22\textwidth]{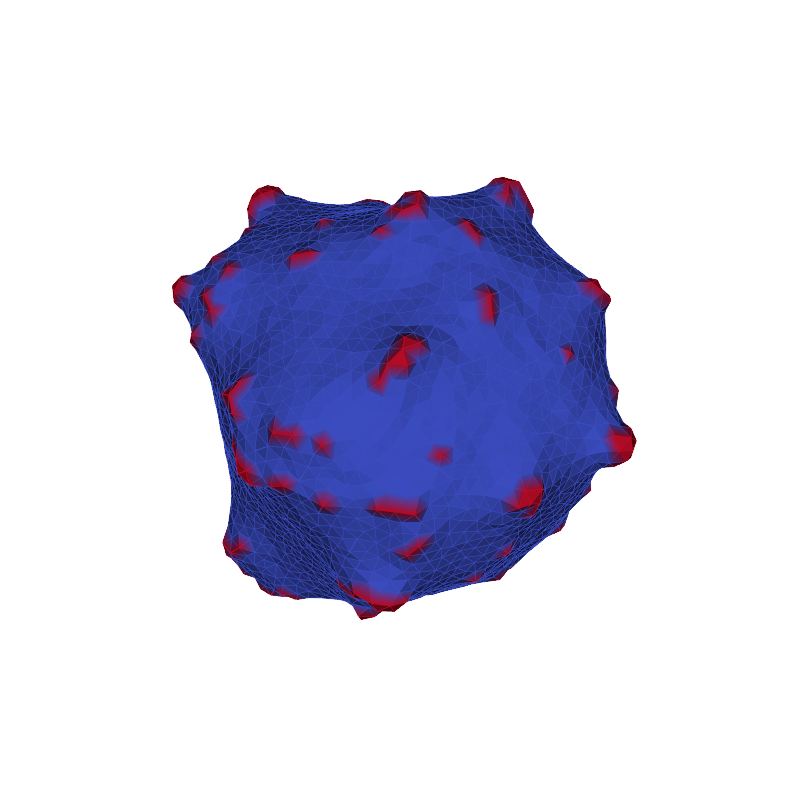}};
	\node[inner sep=0pt] at (axis cs:0.10489286856411896,0.717) {\includegraphics[width=0.15\textwidth]{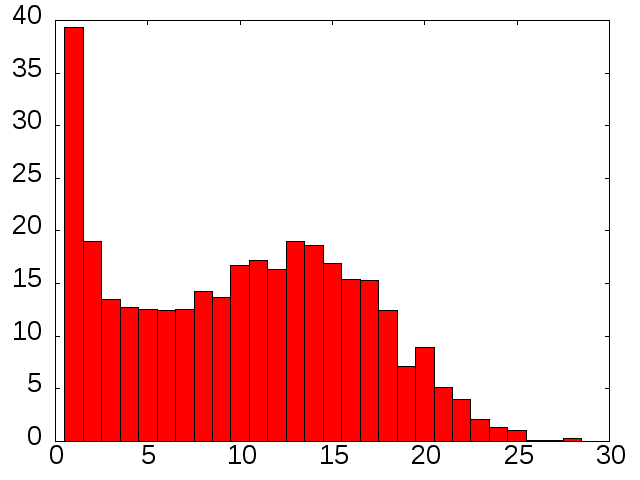}};
	\node[inner sep=0pt] at (axis cs:0.10489286856411896,0.625) {\includegraphics[width=0.22\textwidth]{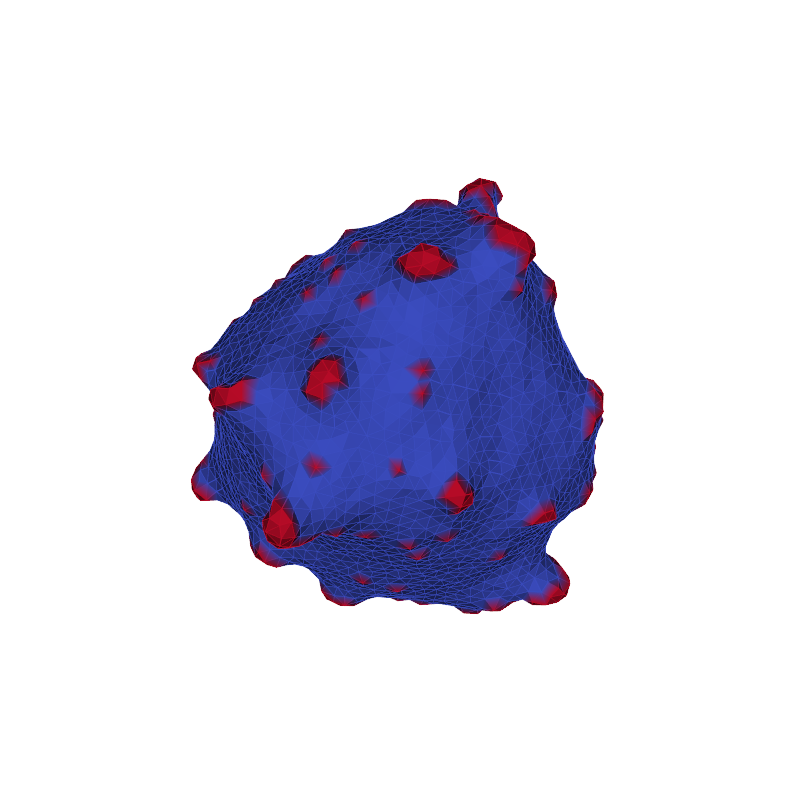}};
	\node[inner sep=0pt] at (axis cs:0.13975055964182923,0.717) {\includegraphics[width=0.15\textwidth]{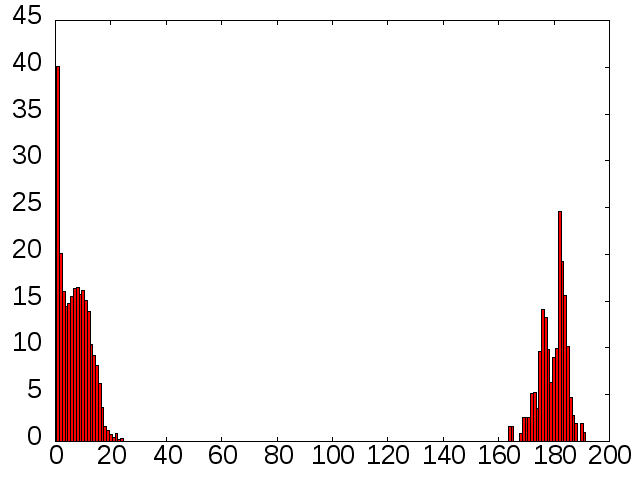}};
	\node[inner sep=0pt] at (axis cs:0.13975055964182923,0.625) {\includegraphics[width=0.22\textwidth]{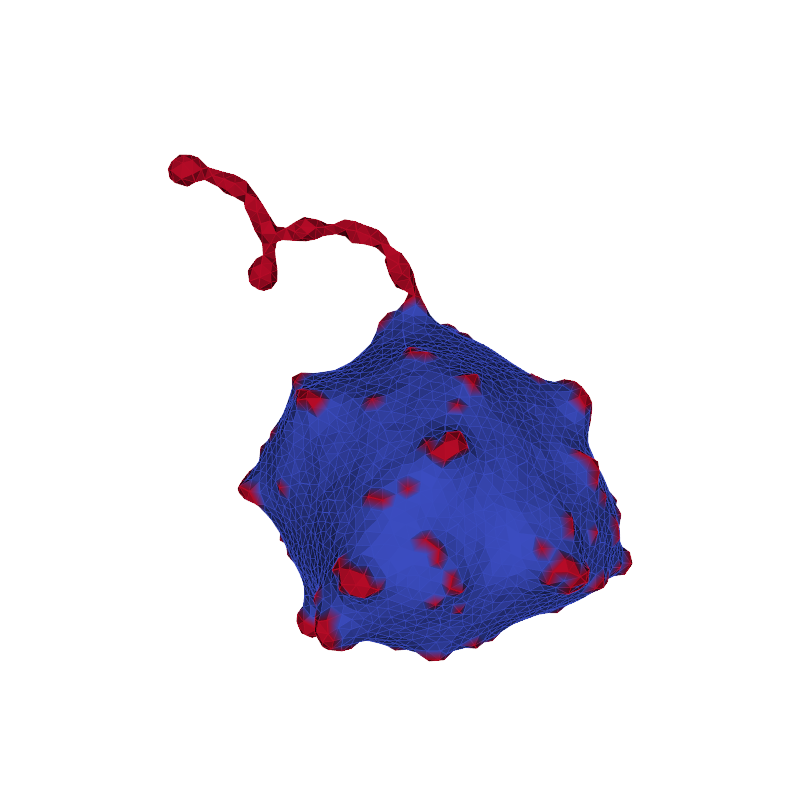}};
	\node[inner sep=0pt] at (axis cs:0.16981132075471697,0.717) {\includegraphics[width=0.15\textwidth]{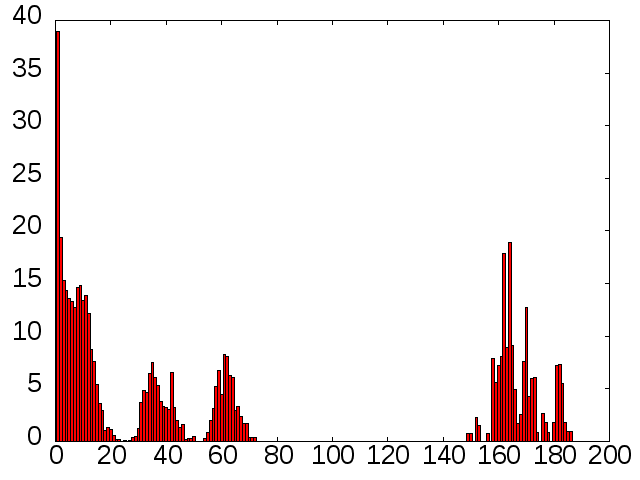}};
	\node[inner sep=0pt] at (axis cs:0.16981132075471697,0.625) {\includegraphics[width=0.22\textwidth]{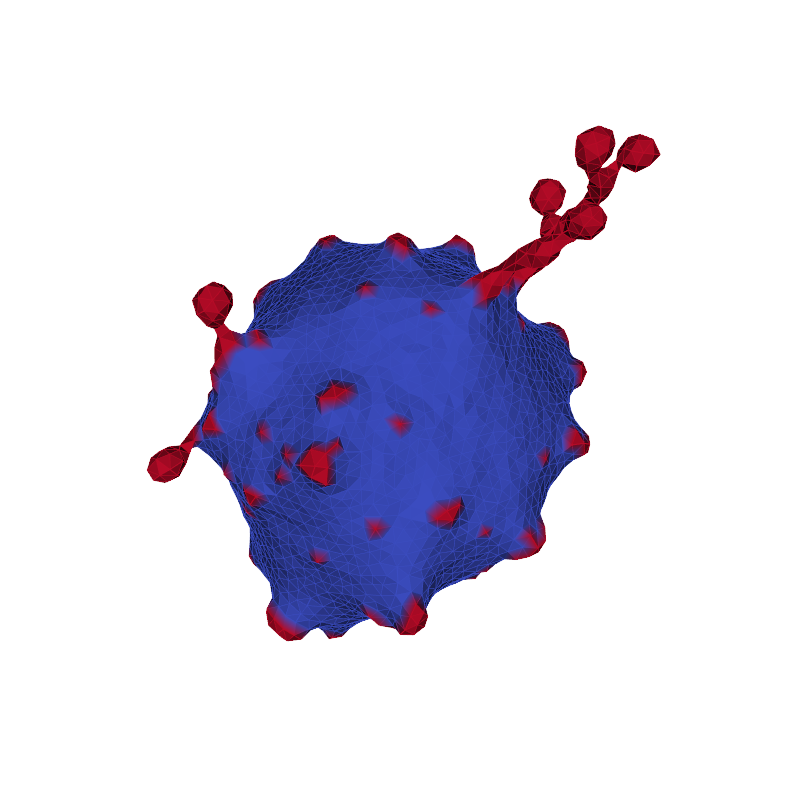}};

	\addplot[domain=0.045:0.14]{12*x*(1-x)};
color=black!80] table {./figs/data/testf0latexfig.dat};
\addplot [mark=*,mark options={solid, scale=1.5}, style={dashed}, color=green!80
] table[y error minus index=2, y error plus index=3] {./figs/data/transitionplot_errorbars.dat};

	\addplot[domain=0.04:0.18, color=yellow]{0.175734 + 15.1481*x - 185.248*x^2 + 1475.51*x^3 - 5968.79*x^4 + 9547.55*x^5};

\end{axis}
\end{tikzpicture}

%% file: figs/testlatexfigf0_4panel.tex
\begin{tikzpicture}[scale=0.5]
\begin{axis}[
 	title={\bf (b)},
	font=\huge,
	xlabel={$\rho[\%]$},
	ylabel={$T^\mr{(c)}/T_0$},
	grid=none,
	scaled x ticks=manual:{}{\pgfmathparse{(#1)*100}},
	tick label style={/pgf/number format/fixed},
    xmax=0.115,
]
\addplot[style={solid}, color=blue!80, domain=0.045:0.140]{12*x*(1-x)};
\addplot [mark=*,mark options={solid, scale=1}, style={dashed}, color=green, error bars/.cd, y dir=both, y explicit, error bar style={solid}] table[y error minus index=2, y error plus index=3] {./figs/data/transitionplot_errorbars.dat};

\node[anchor=center] (mixed snap) at (axis cs:0.06,1.3) {\includegraphics[width=0.9\columnwidth]{figs/snaps/snap_156_15.png}};
\node[anchor=center] (mixed snap) at (axis cs:0.1,0.69) {\includegraphics[width=0.9\columnwidth]{figs/snaps/snap_250_32.png}};

\node[above] at (axis cs:0.056,0.84) {Mixed};
\node[above] at (axis cs:0.073,0.5) {Budded};

\end{axis}
\end{tikzpicture}

%% file: figs/eamcsvstdivtc_F0only.tex
\begin{tikzpicture}[scale=0.5]
\begin{axis}[
    title={\bf (c)},
	font=\huge,
    xlabel={$T/T^\mr{(c)}$},
    ylabel={$\left<\bar{N}_{\mr vc}\right>$},
    xmin=0,xmax=2.6,
    ymin=0,ymax=25,
    grid=none,
]
	\input ./figs/eamcsvstdivtc_addplots.tex
	\coordinate (insetPosition) at (rel axis cs:0.37,0.37);
	\addplot[domain=0.0:2.6, style={dashed,red}]{1};
\end{axis}

\begin{axis}[at={(insetPosition)},anchor={north east},small,
	ymin=0.18,ymax=19,
	xlabel near ticks,
	ylabel near ticks,
	xlabel={$T/T_0$},
	ylabel={$\left<\bar{N}_{\mr vc}\right>$},
	legend entries={
$\rho=5\%$,
$\rho=10\%$,
$\rho=20\%$,
},
]
	\input ./figs/eamcsvst_addplots.tex
\end{axis}

\end{tikzpicture}

%% file: figs/eamcsvstdivtc_w_comparison_F0only.tex
\begin{tikzpicture}[scale=0.5]
\begin{axis}[
    title={\bf (d)},
	font=\huge,
    xlabel={$T/T^\mr{(c)}$},
    ylabel={$\left<\bar{N}_{\mr vc}\right>$},
    legend entries={$w=1.0~kT_0$,$w=1.5~kT_0$},
    legend pos=north east,
     ymin=1.5,	
     ymax=10,
]
\input ./figs/eamcsvstdivtc_w_comparison_addplots.tex
\coordinate (insetPosition) at (rel axis cs:0.364,0.25);
\end{axis}

\begin{axis}[at={(insetPosition)},anchor={north east},footnotesize,
	xlabel near ticks,
	ylabel near ticks,
	xlabel={$T/T_0$},
	ylabel={$\left<\bar{N}_{\mr vc}\right>$},
	ymax=10,
]
	\input ./figs/eamcsvst_w_comparison_addplots.tex
\end{axis}

\end{tikzpicture}

%% file: snaps_fig_f1.tex
\begin{tikzpicture}
\begin{axis}[
	width=0.99\textwidth,height=0.93\textheight,
	xlabel={$\rho[\%]$},
	ylabel=$T/T_0$,ymin=0.57,ymax=1.4,xmin=0.037,xmax=0.182,
	scaled x ticks=manual:{}{\pgfmathparse{(#1)*100}},
	tick label style={/pgf/number format/fixed}]

	\node[inner sep=0pt] at (axis cs:0.04988807163415414,1.245) {\includegraphics[width=0.15\textwidth]{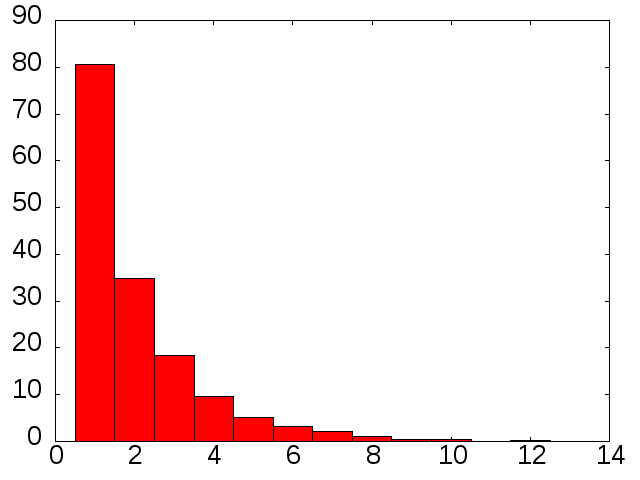}};
	\node[inner sep=0pt] at (axis cs:0.04988807163415414,1.3333333333333333) {\includegraphics[width=0.22\textwidth]{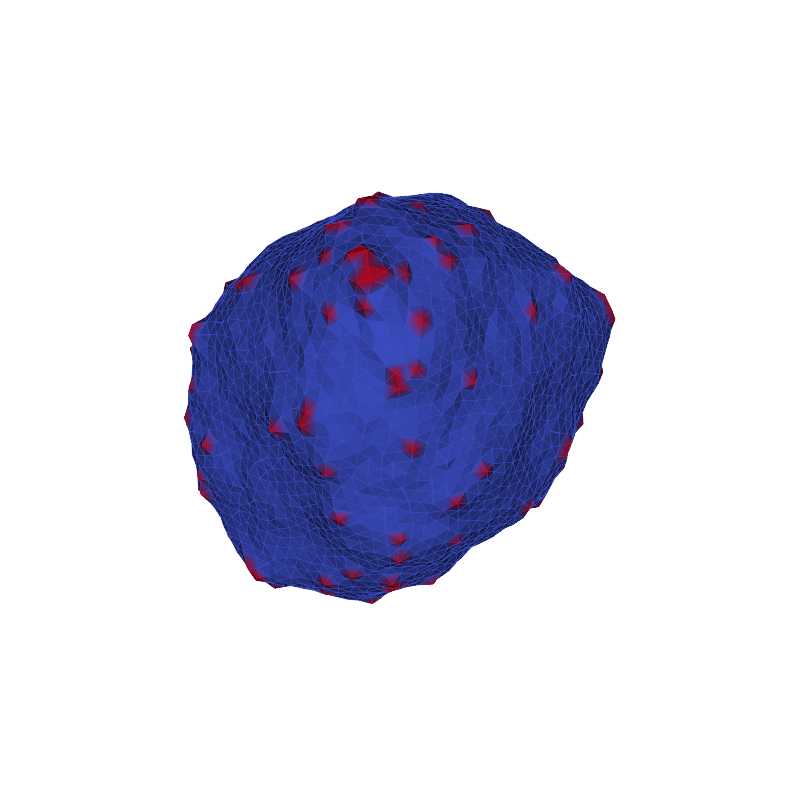}};
	\node[inner sep=0pt] at (axis cs:0.0799488327470419,1.245) {\includegraphics[width=0.15\textwidth]{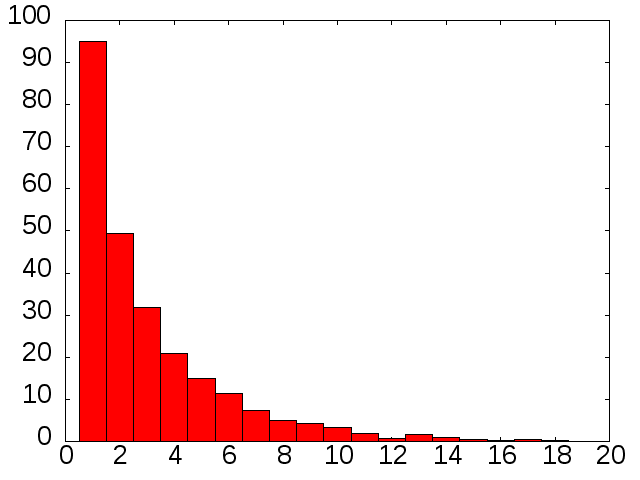}};
	\node[inner sep=0pt] at (axis cs:0.0799488327470419,1.3333333333333333) {\includegraphics[width=0.22\textwidth]{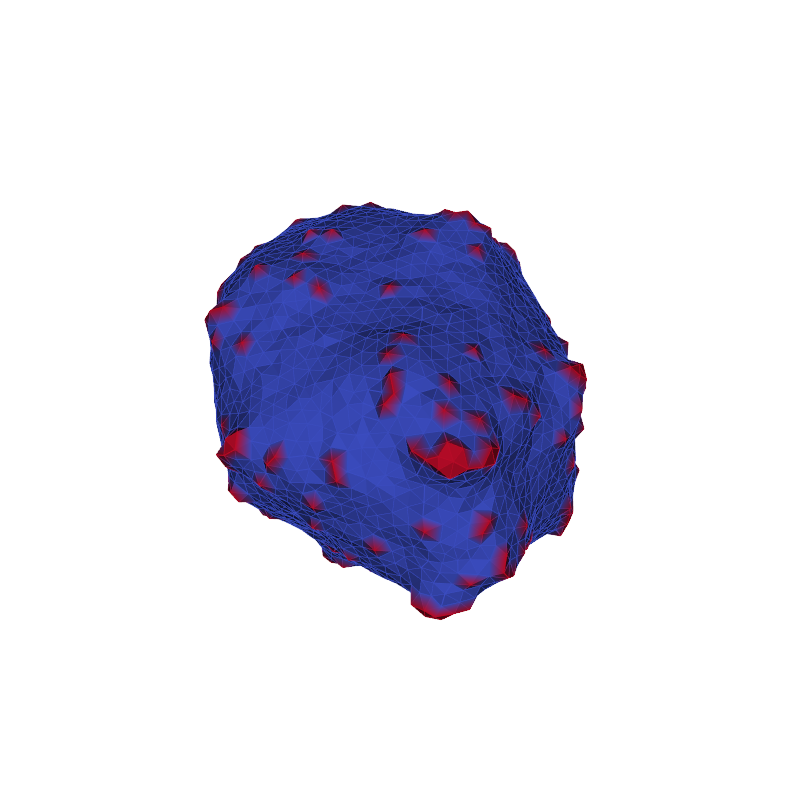}};
	\node[inner sep=0pt] at (axis cs:0.10489286856411896,1.245) {\includegraphics[width=0.15\textwidth]{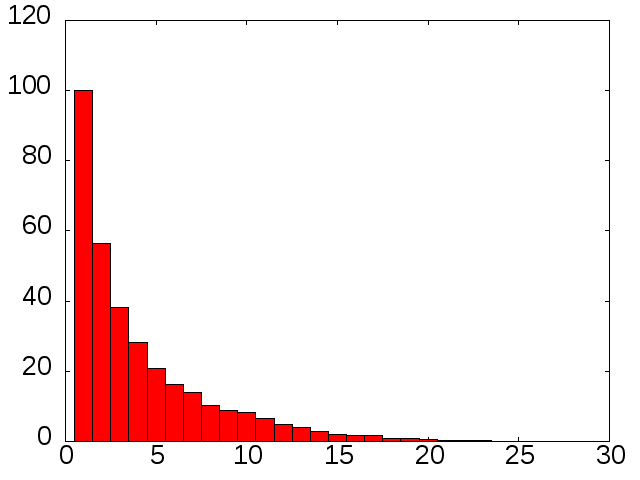}};
	\node[inner sep=0pt] at (axis cs:0.10489286856411896,1.3333333333333333) {\includegraphics[width=0.22\textwidth]{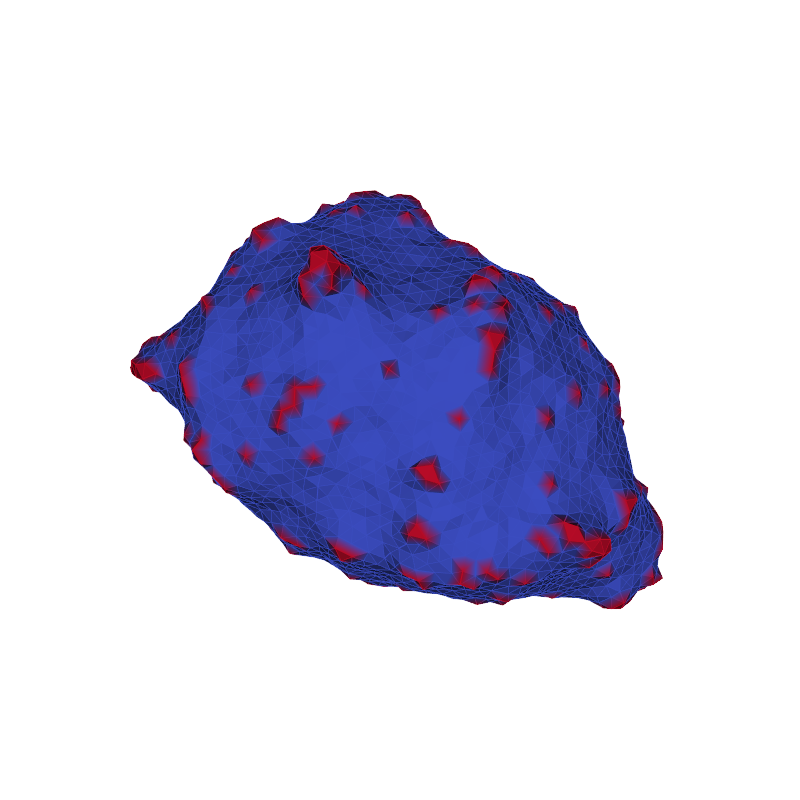}};
	\node[inner sep=0pt] at (axis cs:0.13975055964182923,1.245) {\includegraphics[width=0.15\textwidth]{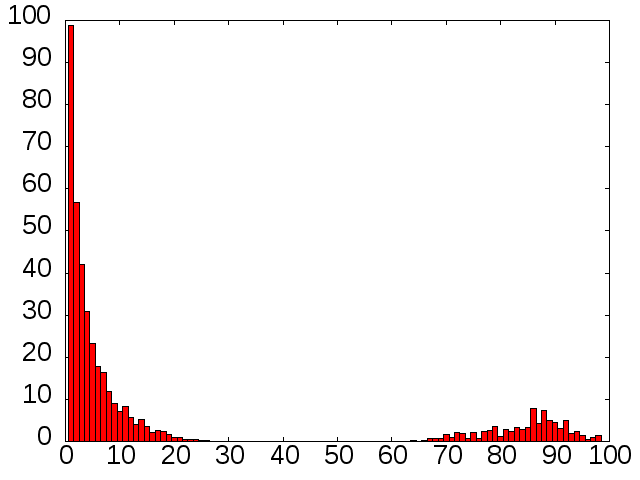}};
	\node[inner sep=0pt] at (axis cs:0.13975055964182923,1.3333333333333333) {\includegraphics[width=0.22\textwidth]{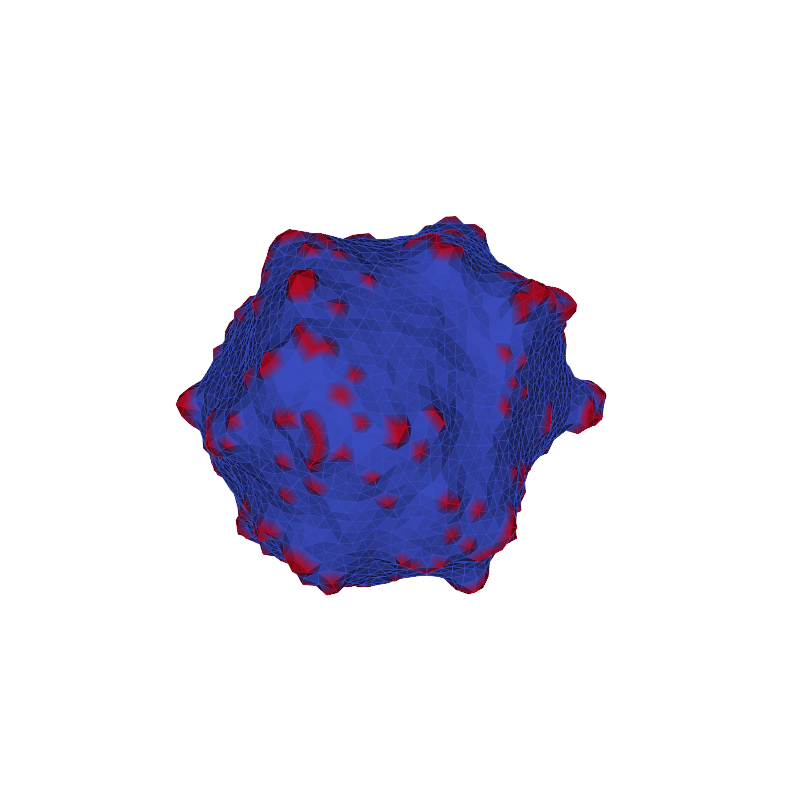}};
	\node[inner sep=0pt] at (axis cs:0.16981132075471697,1.245) {\includegraphics[width=0.15\textwidth]{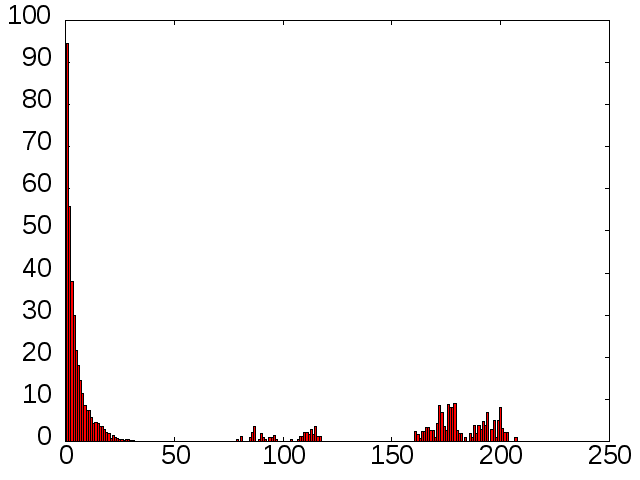}};
	\node[inner sep=0pt] at (axis cs:0.16981132075471697,1.3333333333333333) {\includegraphics[width=0.22\textwidth]{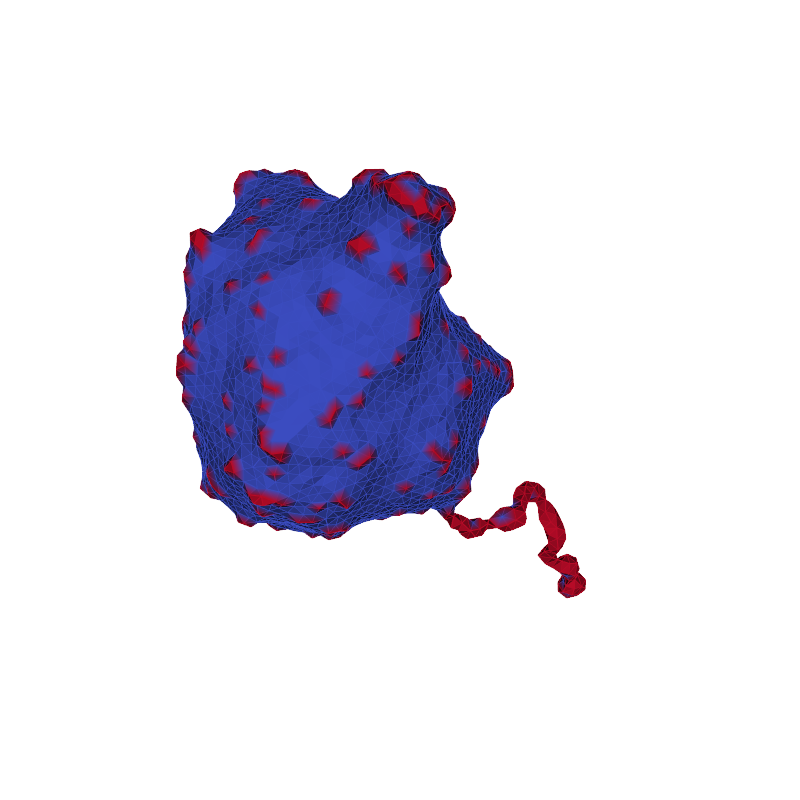}};
	\node[inner sep=0pt] at (axis cs:0.04988807163415414,1.02) {\includegraphics[width=0.15\textwidth]{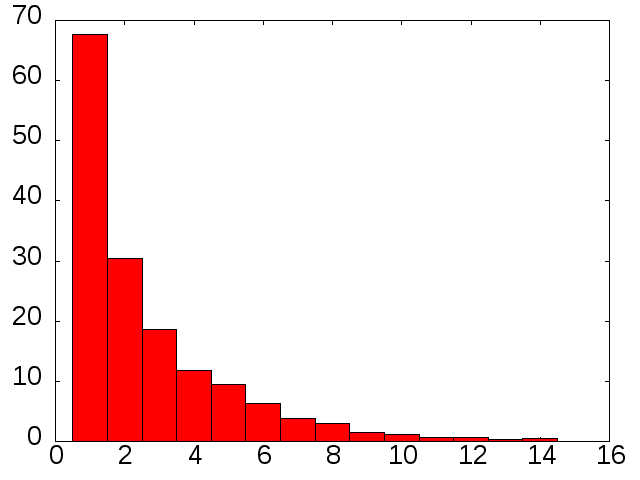}};
	\node[inner sep=0pt] at (axis cs:0.04988807163415414,1.1111111111111112) {\includegraphics[width=0.22\textwidth]{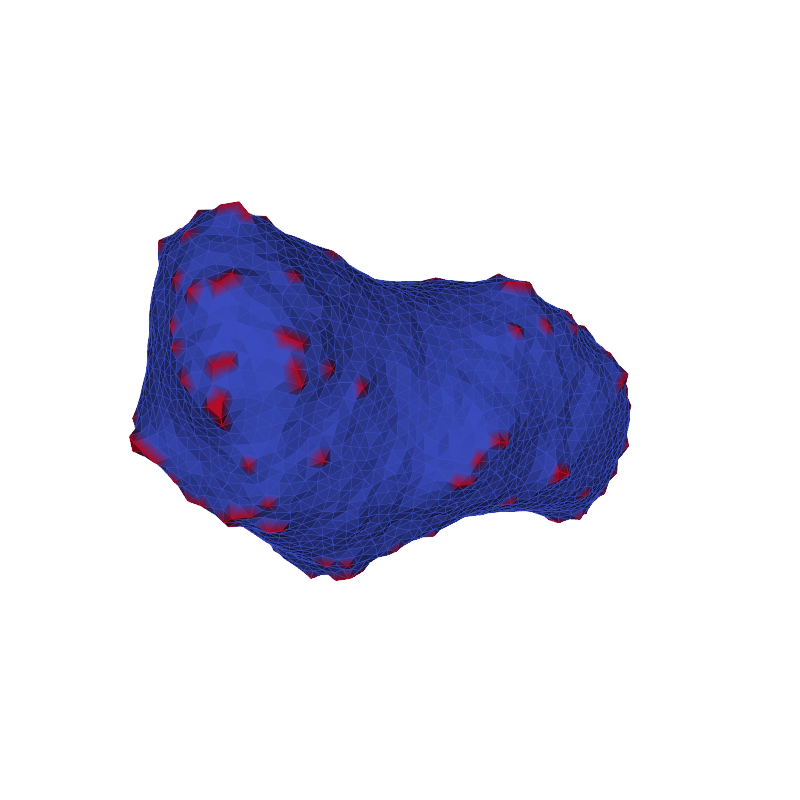}};
	\node[inner sep=0pt] at (axis cs:0.0799488327470419,1.02) {\includegraphics[width=0.15\textwidth]{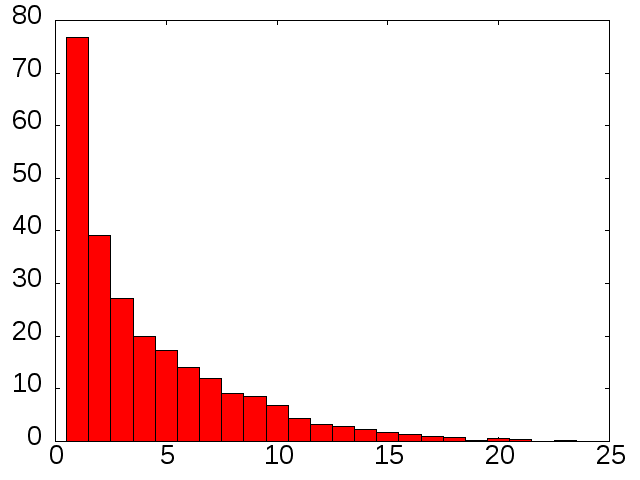}};
	\node[inner sep=0pt] at (axis cs:0.0799488327470419,1.1111111111111112) {\includegraphics[width=0.22\textwidth]{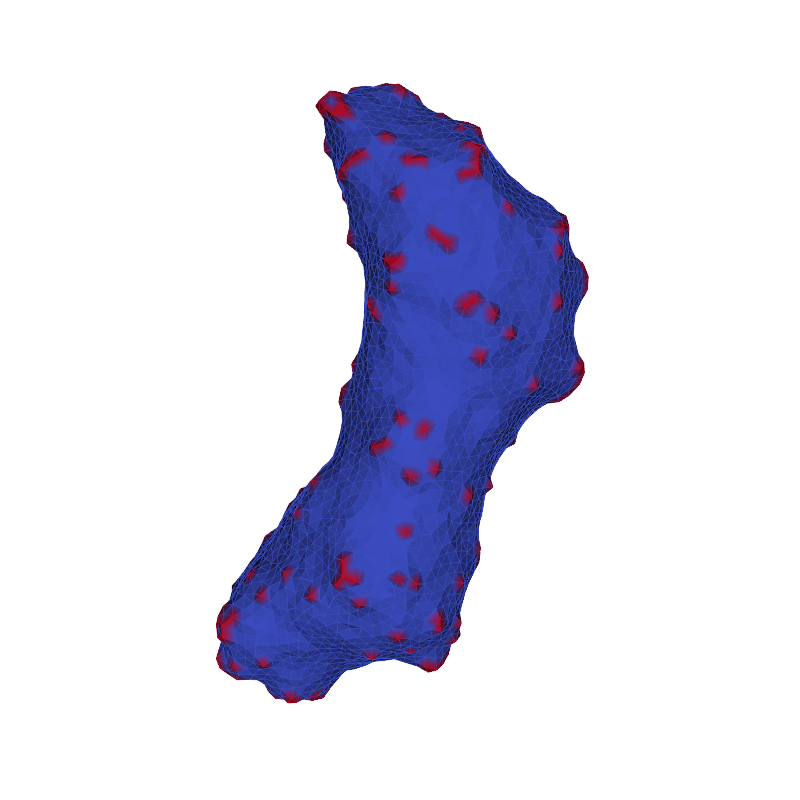}};
	\node[inner sep=0pt] at (axis cs:0.10489286856411896,1.02) {\includegraphics[width=0.15\textwidth]{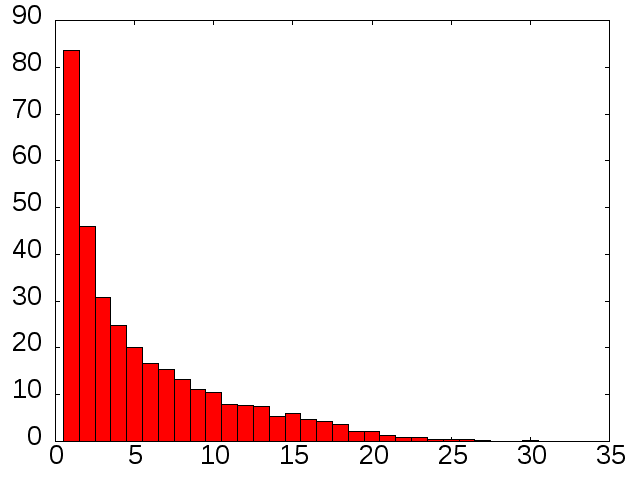}};
	\node[inner sep=0pt] at (axis cs:0.10489286856411896,1.1111111111111112) {\includegraphics[width=0.22\textwidth]{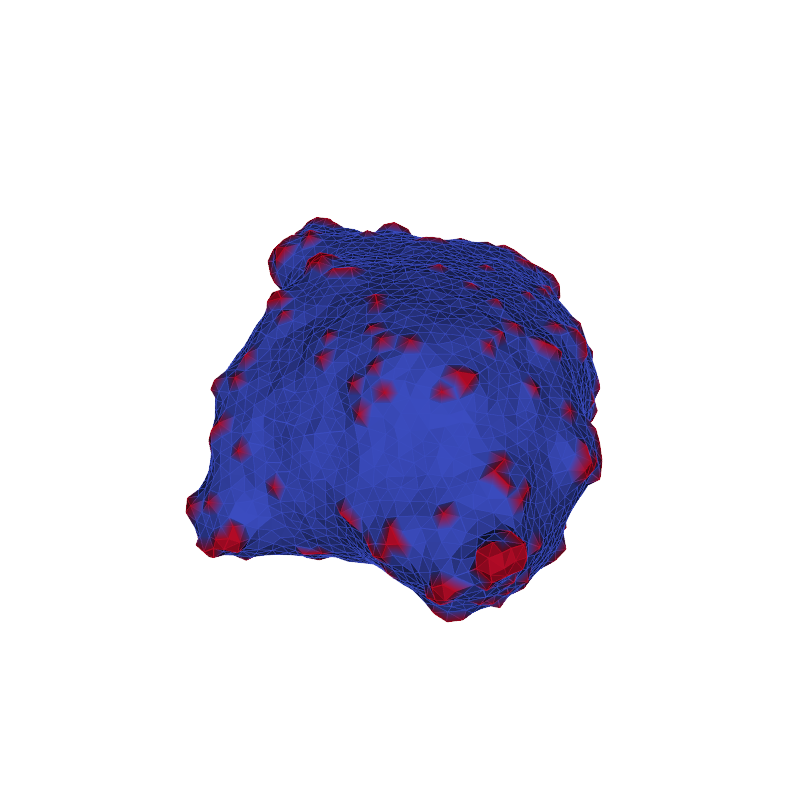}};
	\node[inner sep=0pt] at (axis cs:0.13975055964182923,1.02) {\includegraphics[width=0.15\textwidth]{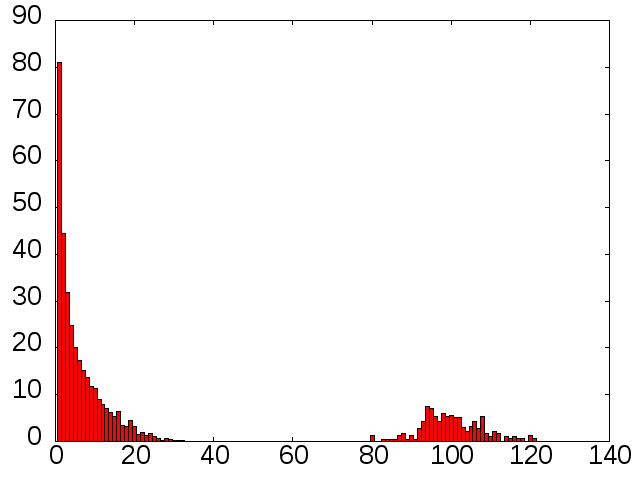}};
	\node[inner sep=0pt] at (axis cs:0.13975055964182923,1.1111111111111112) {\includegraphics[width=0.22\textwidth]{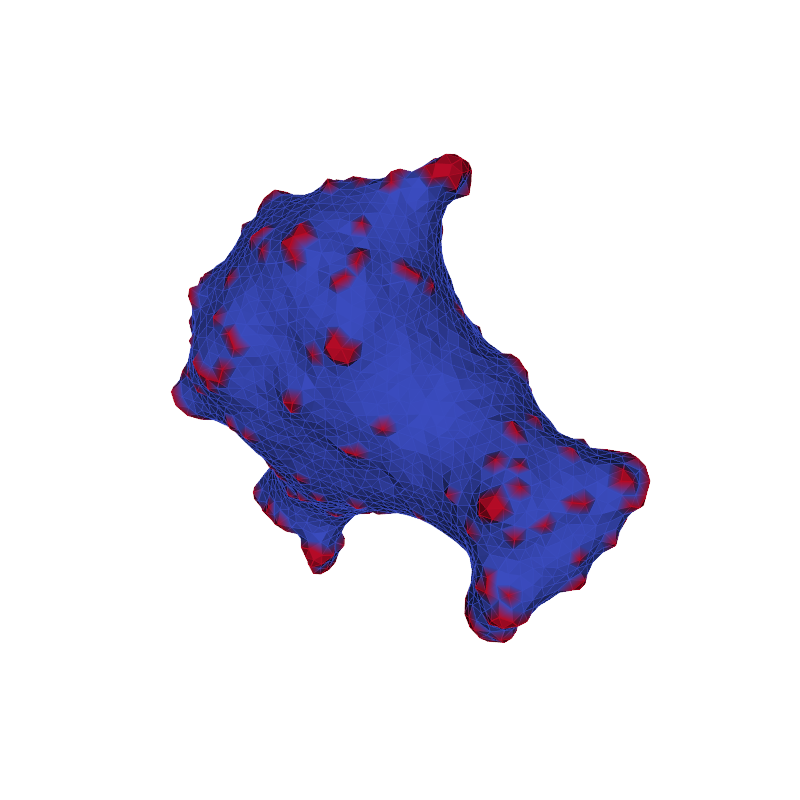}};
	\node[inner sep=0pt] at (axis cs:0.16981132075471697,1.02) {\includegraphics[width=0.15\textwidth]{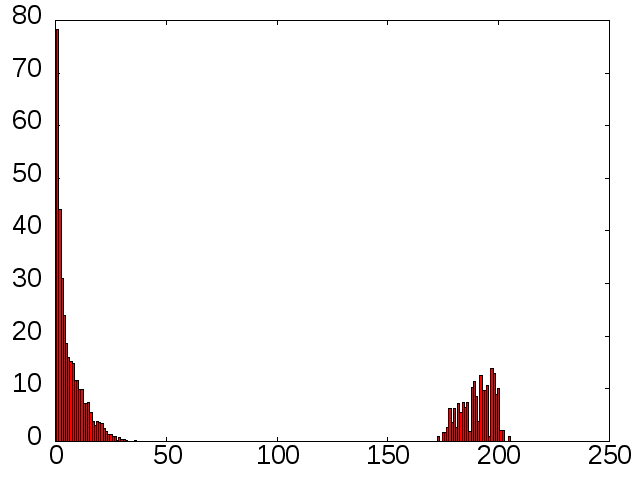}};
	\node[inner sep=0pt] at (axis cs:0.16981132075471697,1.1111111111111112) {\includegraphics[width=0.22\textwidth]{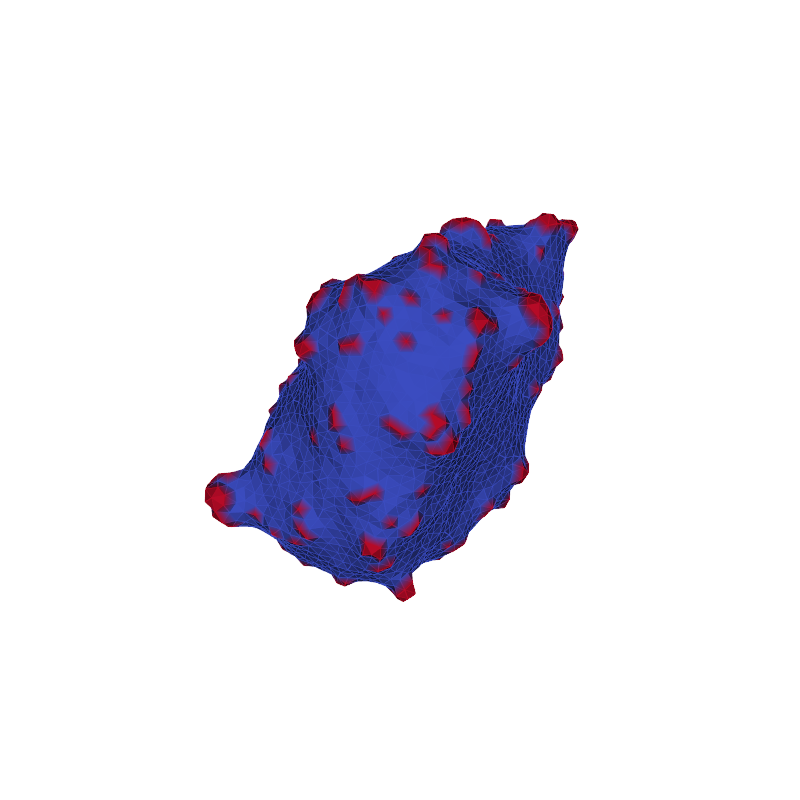}};
	\node[inner sep=0pt] at (axis cs:0.04988807163415414,0.925) {\includegraphics[width=0.15\textwidth]{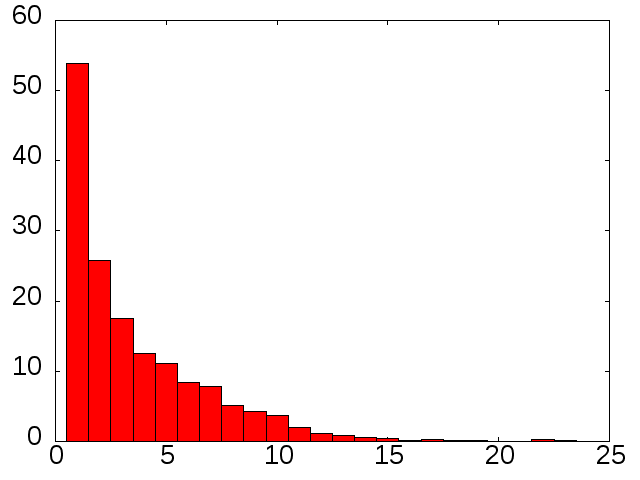}};
	\node[inner sep=0pt] at (axis cs:0.04988807163415414,0.8333333333333334) {\includegraphics[width=0.22\textwidth]{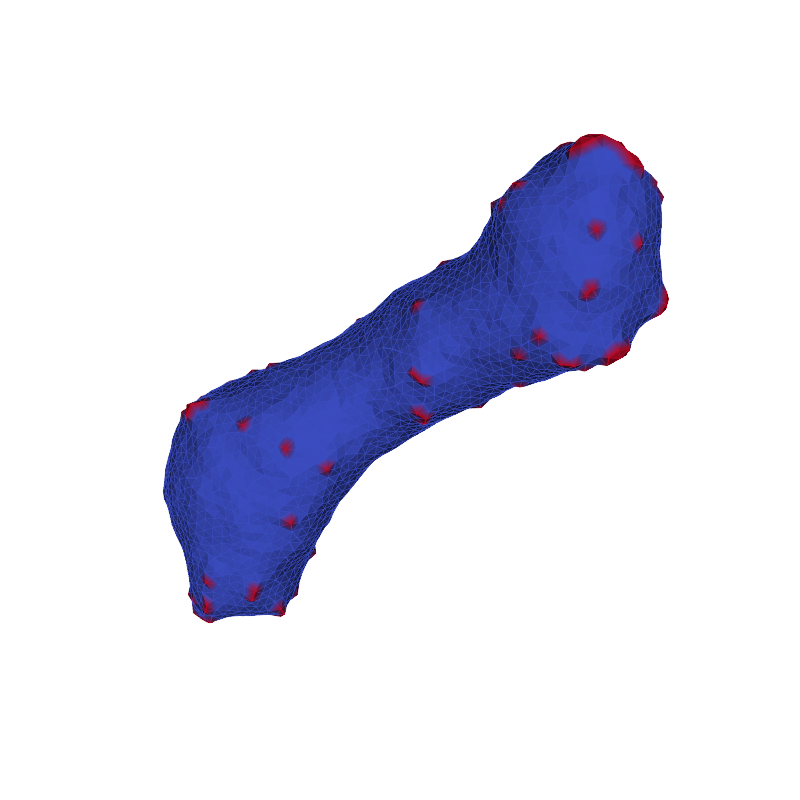}};
	\node[inner sep=0pt] at (axis cs:0.0799488327470419,0.925) {\includegraphics[width=0.15\textwidth]{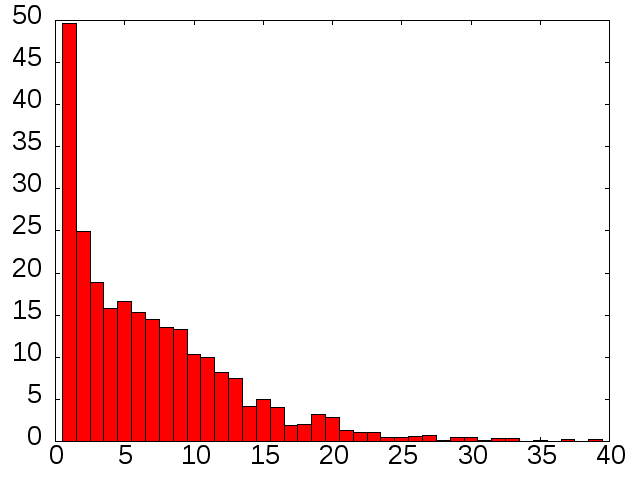}};
	\node[inner sep=0pt] at (axis cs:0.0799488327470419,0.8333333333333334) {\includegraphics[width=0.22\textwidth]{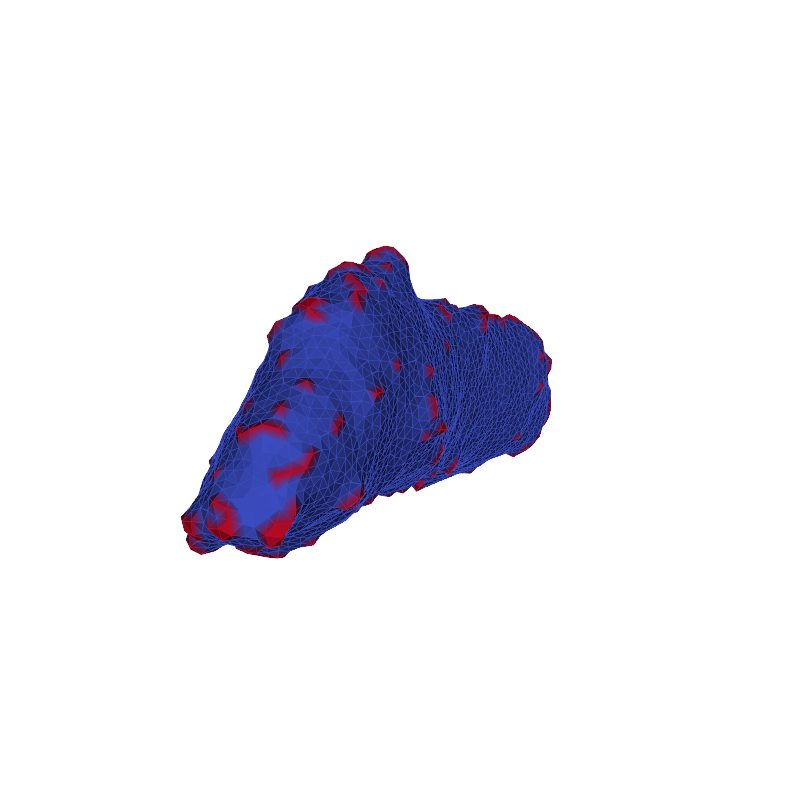}};
	\node[inner sep=0pt] at (axis cs:0.10489286856411896,0.925) {\includegraphics[width=0.15\textwidth]{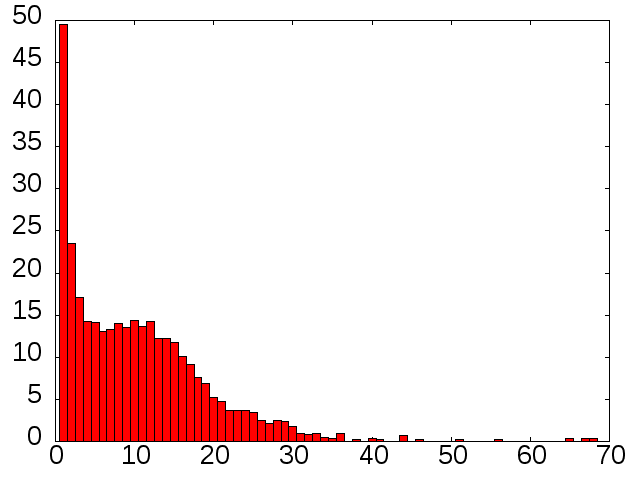}};
		\node[inner sep=0pt] at (axis cs:0.10489286856411896,0.8333333333333334) {\includegraphics[width=0.22\textwidth]{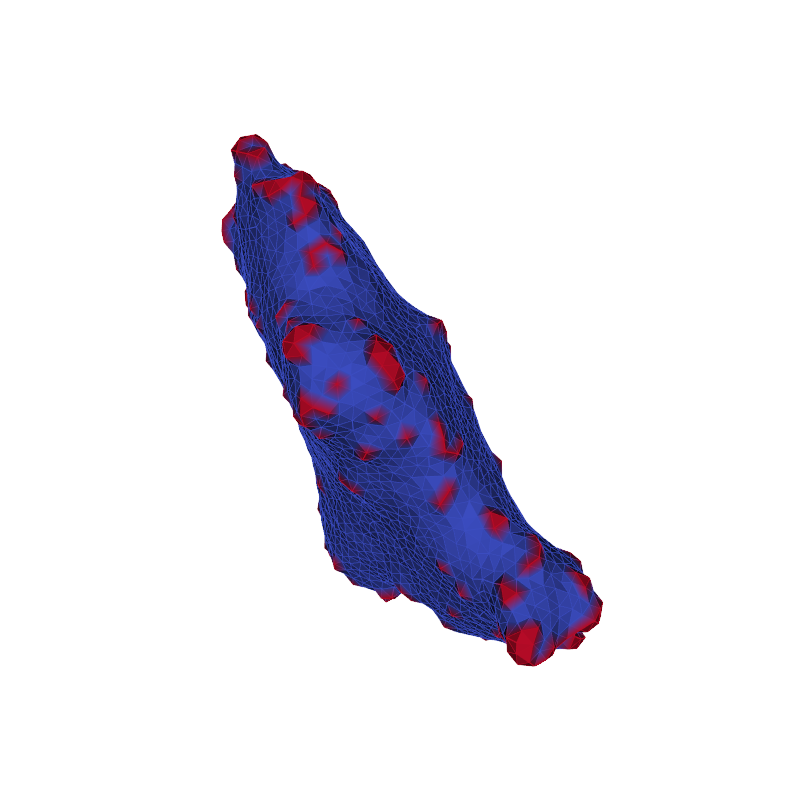}};
	\node[inner sep=0pt] at (axis cs:0.13975055964182923,0.925) {\includegraphics[width=0.15\textwidth]{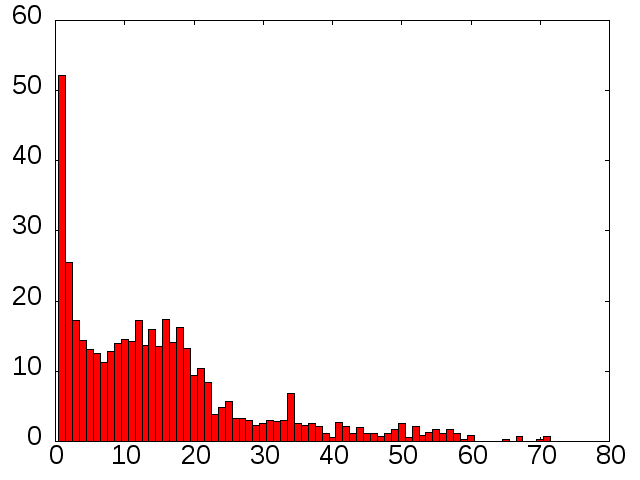}};
	\node[inner sep=0pt] at (axis cs:0.13975055964182923,0.8333333333333334) {\includegraphics[width=0.22\textwidth]{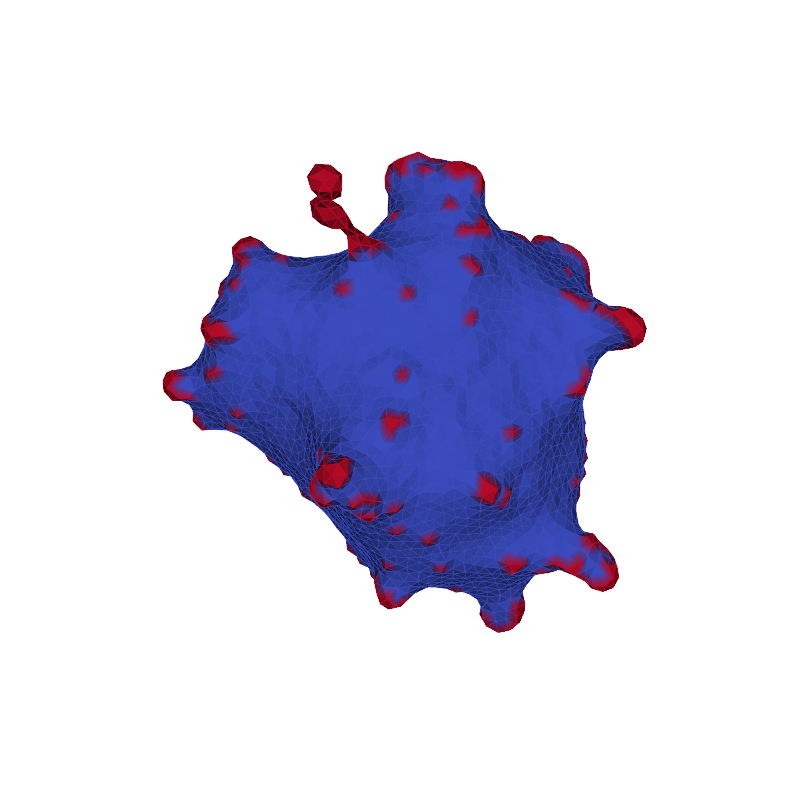}};
	\node[inner sep=0pt] at (axis cs:0.16981132075471697,0.925) {\includegraphics[width=0.15\textwidth]{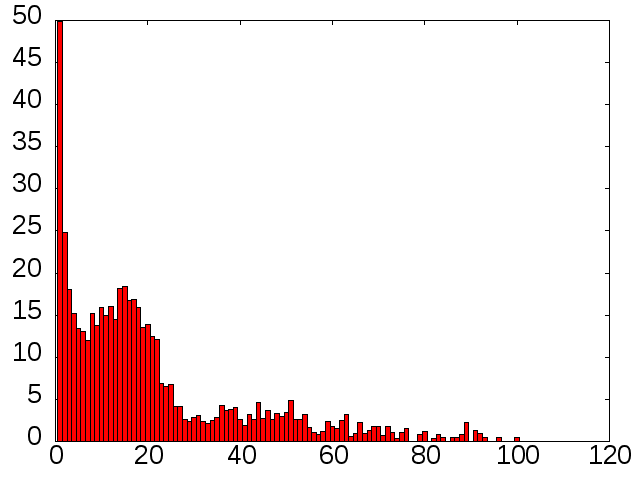}};
	\node[inner sep=0pt] at (axis cs:0.16981132075471697,0.8333333333333334) {\includegraphics[width=0.22\textwidth]{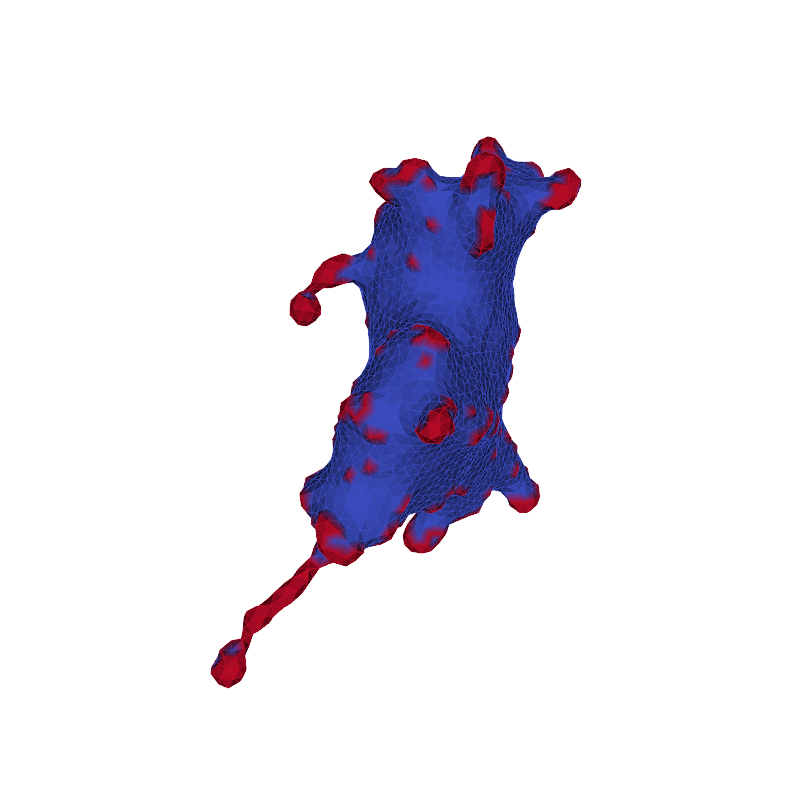}};
	\node[inner sep=0pt] at (axis cs:0.04988807163415414,0.717) {\includegraphics[width=0.15\textwidth]{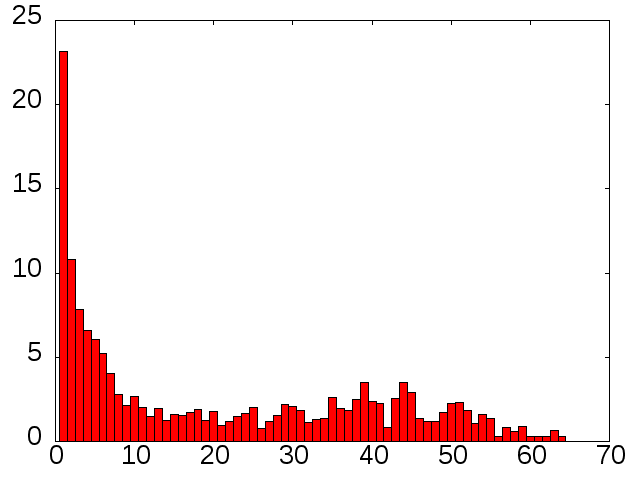}};
	\node[inner sep=0pt] at (axis cs:0.04988807163415414,0.625) {\includegraphics[width=0.22\textwidth]{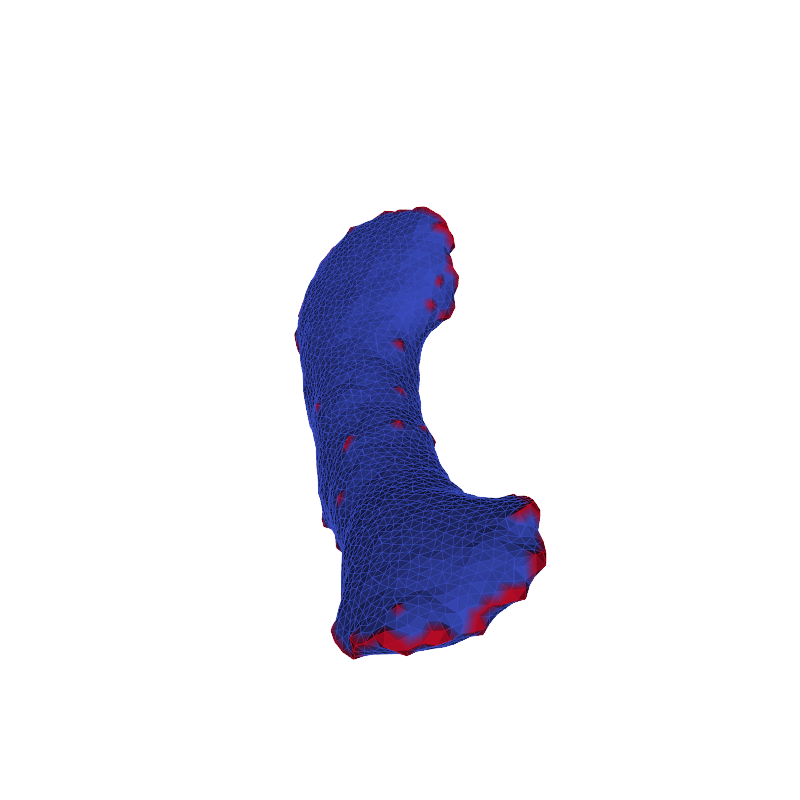}};
	\node[inner sep=0pt] at (axis cs:0.0799488327470419,0.717) {\includegraphics[width=0.15\textwidth]{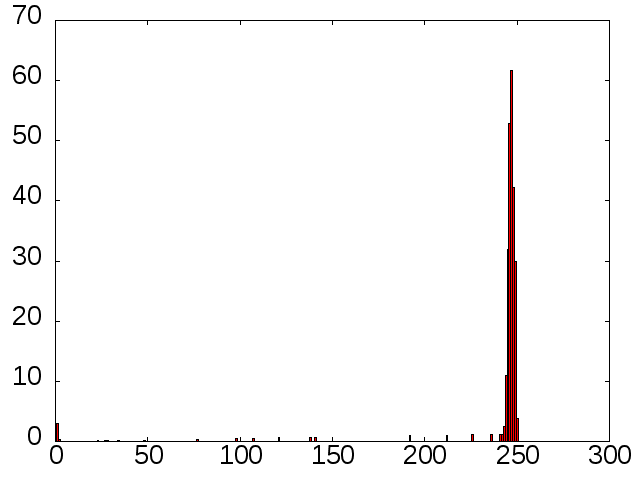}};
	\node[inner sep=0pt] at (axis cs:0.0799488327470419,0.625) {\includegraphics[width=0.22\textwidth]{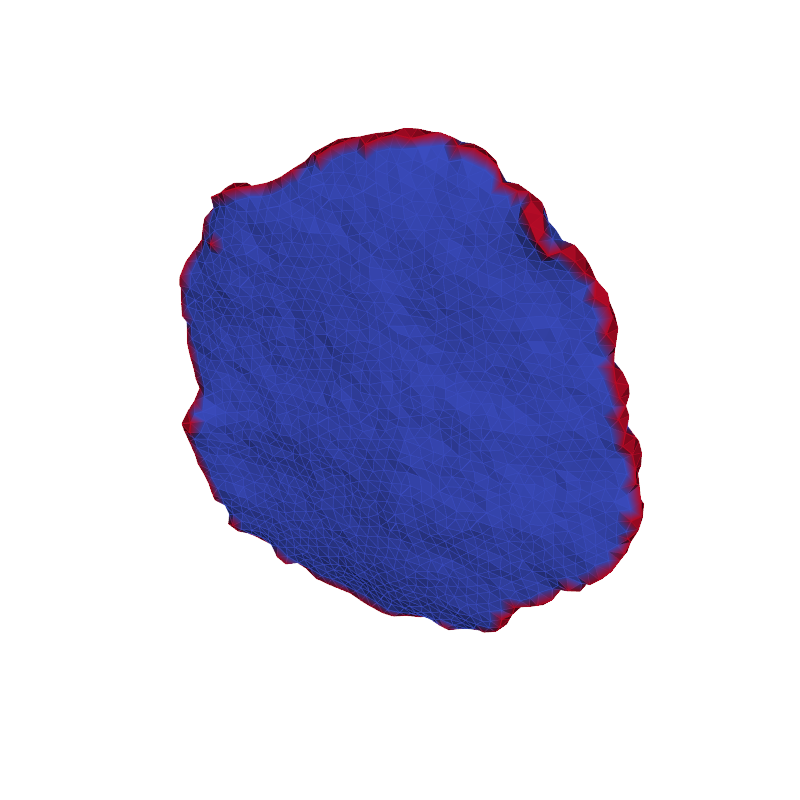}};
	\node[inner sep=0pt] at (axis cs:0.10489286856411896,0.717) {\includegraphics[width=0.15\textwidth]{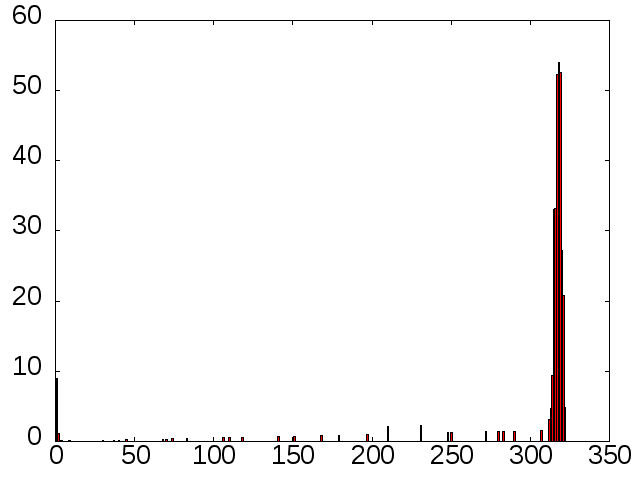}};
	\node[inner sep=0pt] at (axis cs:0.10489286856411896,0.625) {\includegraphics[width=0.22\textwidth]{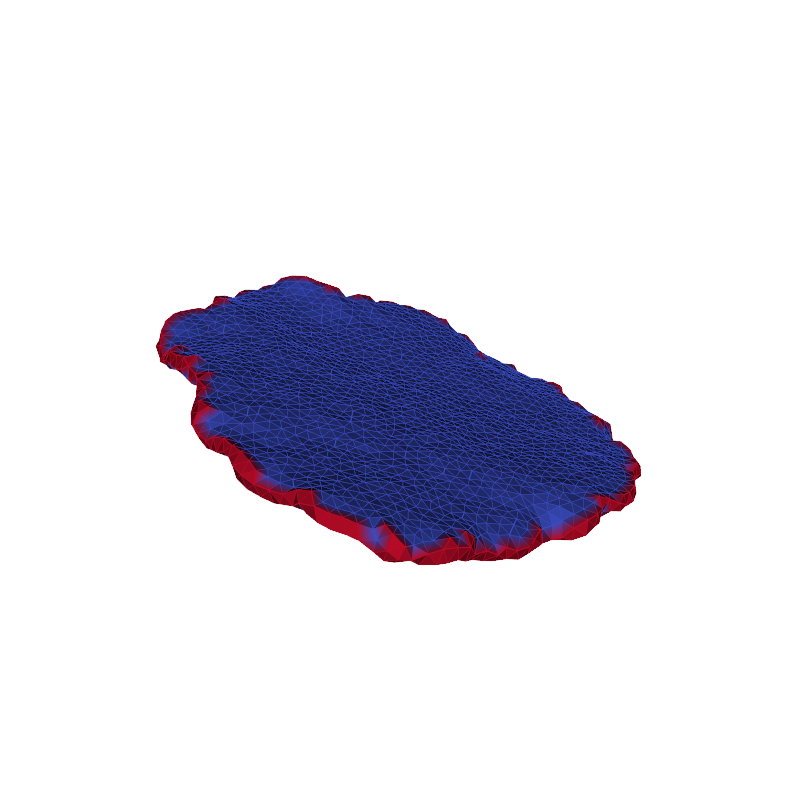}}; 
	\node[inner sep=0pt] at (axis cs:0.13975055964182923,0.717) {\includegraphics[width=0.15\textwidth]{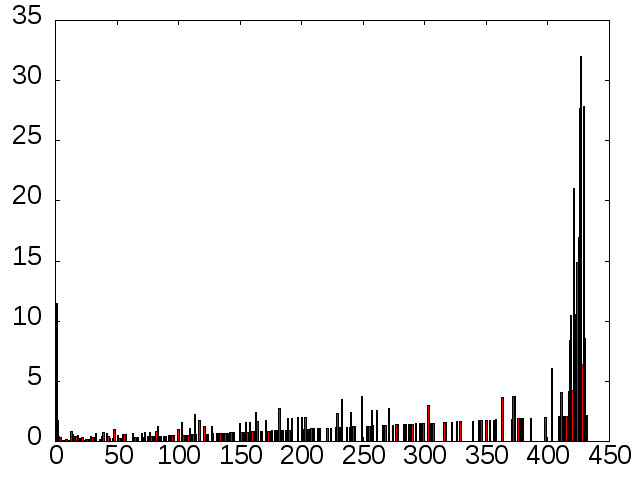}};
	\node[inner sep=0pt] at (axis cs:0.13975055964182923,0.625) {\includegraphics[width=0.22\textwidth]{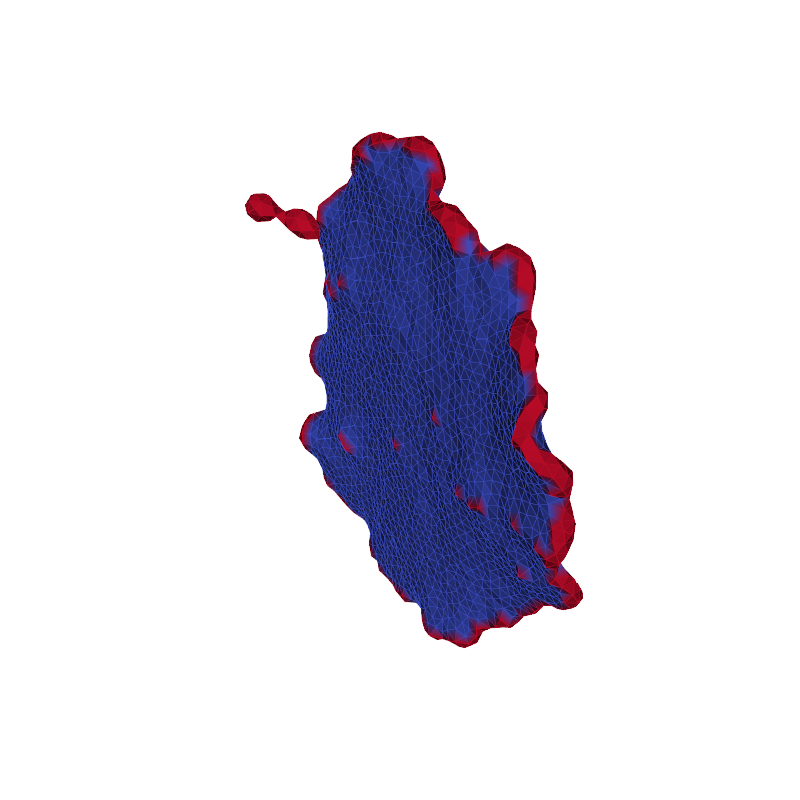}};
	\node[inner sep=0pt] at (axis cs:0.16981132075471697,0.717) {\includegraphics[width=0.15\textwidth]{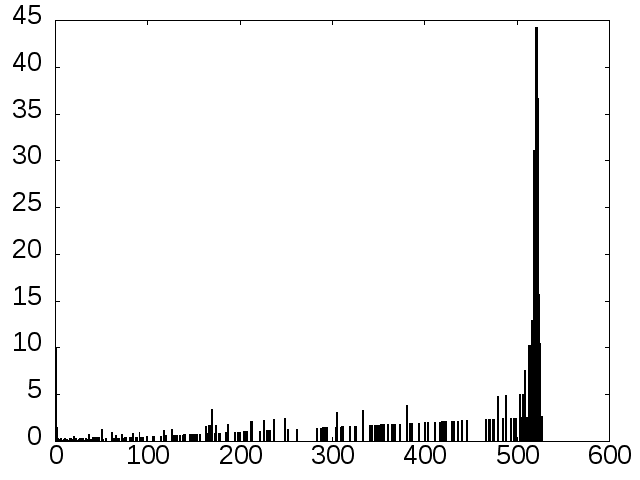}};
	\node[inner sep=0pt] at (axis cs:0.16981132075471697,0.625) {\includegraphics[width=0.22\textwidth]{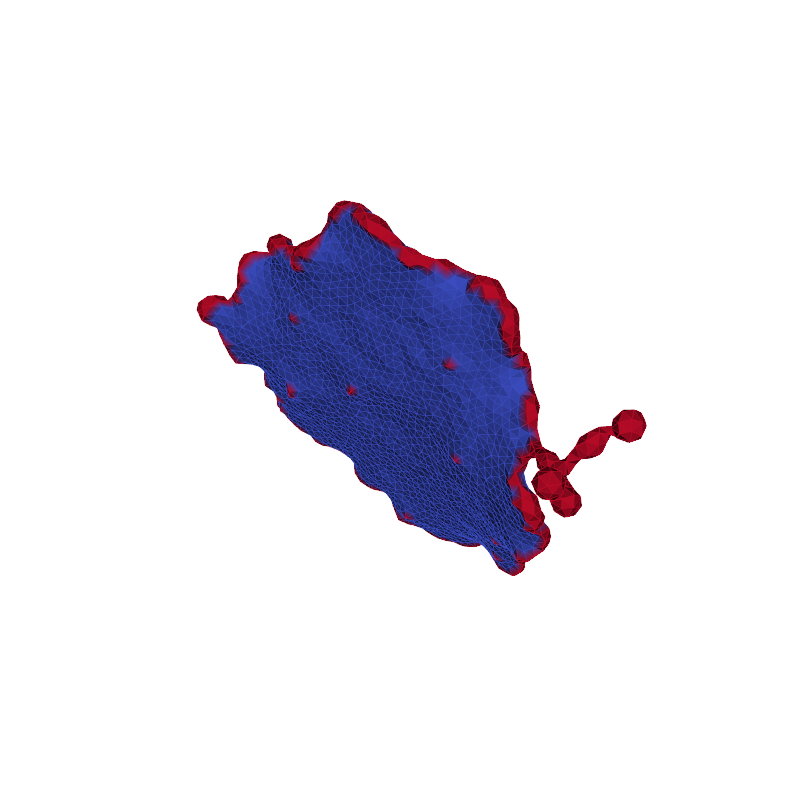}};

	\addplot[style={solid}, color=black!80,domain=0.05:0.107]{12*x*(1-x)*(1+sqrt((1-1/(x*0.3544*sqrt(3127)))/12))};
	\addplot [mark=*,mark options={solid, scale=1.5}, style={dashed}, color=green!80
] table[y error minus index=2, y error plus index=3] {./figs/data/transitionplot_f1_errorbars.dat};
	\addplot [mark=*,mark options={solid, scale=1.5}, style={dashed}, color=red!80
] table[y error minus index=2, y error plus index=3] {./figs/data/pancake_phase_transition_f1_error_bars.dat};

\end{axis}
\end{tikzpicture}

%% file: figs/eamcsvstdivtc_F1only.tex
\begin{tikzpicture}[scale=0.5]
\begin{axis}[
    title={\bf (b)},
	font=\huge,
    xlabel={$T/T^\mr{(c)}$},
    ylabel={$\left<\bar{N}_{\mr vc}\right>$},
    xmin=0.2,xmax=1.67,
    grid=none,
]
	\input ./figs/eamcsvstdivtc_f1_addplots.tex
	\coordinate (insetPosition) at (rel axis cs:0.35,0.35);
	\addplot[domain=0.0:2.6, style={dashed,red}]{1};
\end{axis}

\begin{axis}[at={(insetPosition)},anchor={north east},footnotesize,
	xlabel near ticks,
	ylabel near ticks,
	xmin=0.6,xmax=1.35,
	ymin=0,ymax=60,
	xlabel={$T/T_0$},
	ylabel={$\left<\bar{N}_{\mr vc}\right>$},
	legend entries={
$\rho=5\%$,
$\rho=10\%$,
$\rho=17\%$
},
	grid=none,
]
	\input ./figs/eamcsvst_mixred_addplots.tex
\end{axis}

\end{tikzpicture}

%% file: figs/testlatexfig_4panel.tex
\begin{tikzpicture}[scale=0.5]
\begin{axis}[
 	title={\bf (c)},
	font=\huge,
	xlabel={$\rho[\%]$},
	ylabel={$T^\mr{(c)}/T_0$},
	scaled x ticks=manual:{}{\pgfmathparse{(#1)*100}},
	grid=none,
	tick label style={/pgf/number format/.cd,fixed,precision=5},
	xmin=0.032,
    xmax=0.165,
	ymin=0.25,
]
	\addplot[style={solid}, color=black!80,domain=0.045:0.155]{12*x*(1-x)*(1+sqrt((1-1/(x*0.3544*sqrt(3127)))/12))};
	\addplot [mark=*,mark options={solid, scale=1}, style={dashed}, color=green, error bars/.cd, y dir=both, y explicit, error bar style={solid}] table[y error minus index=2, y error plus index=3] {./figs/data/transitionplot_f1_errorbars.dat};
	\addplot [mark=*,mark options={solid, scale=0.5}, style={dashed}, color=red!80, error bars/.cd, y dir=both, y explicit, error bar style={solid}] table[y error minus index=2, y error plus index=3] {./figs/data/pancake_phase_transition_f1_error_bars.dat};

	\addplot [mark=*,mark options={solid, scale=0.4}, style={dashed}, color=white] table {
0.0499 0.5
};

\node[anchor=center] (mixed snap) at (axis cs:0.06,1.7) {\includegraphics[width=0.9\columnwidth]{figs/snaps/snap_156_15f1.png}};
\node[anchor=center] (budded snap) at (axis cs:0.144,1.3) {\includegraphics[width=0.9\columnwidth]{figs/snaps/snap_437_15f1.png}};
\node[anchor=center] (budded snap) at (axis cs:0.092,0.58) {\includegraphics[width=0.9\columnwidth]{figs/snaps/snap_328_32f1.png}};

\node[above] at (axis cs:0.056,1.1) {Mixed};
\node[above] at (axis cs:0.11,0.9) {Budded};
\node[above] at (axis cs:0.14,0.4) {Pancake};

\end{axis}
\end{tikzpicture}

%% file: figs/eamcsvst_f_comparison.tex
\begin{tikzpicture}[scale=0.5]
\begin{axis}[
	title={\bf (d)},
	font=\huge,
	xlabel={$T/T_0$},
	ylabel={$\left<\bar{N}_{\mr vc}\right>$},
	legend entries={$F=0.5~kT_0$,$F=1.0~kT_0$,$F=1.5~kT_0$},
]
\input ./figs/eamcsvst_f_comparison_addplots.tex
\node[anchor=center] (2zdesne) at (axis cs:1.25,30) {\includegraphics[width=0.7\textwidth]{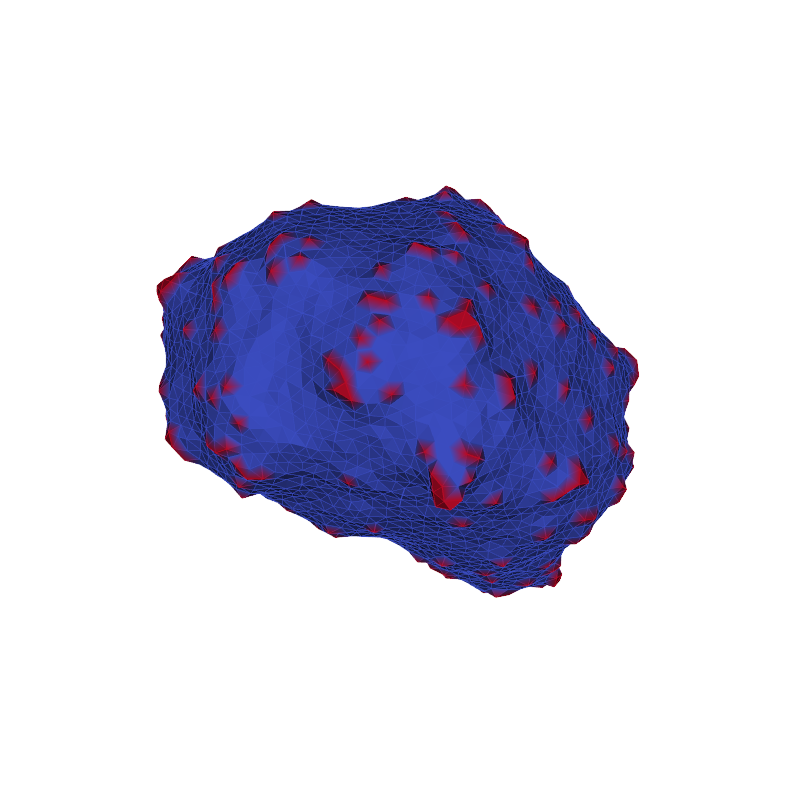}};
\node[anchor=center] (4zdesne) at (axis cs:1,35) {\includegraphics[width=0.7\textwidth]{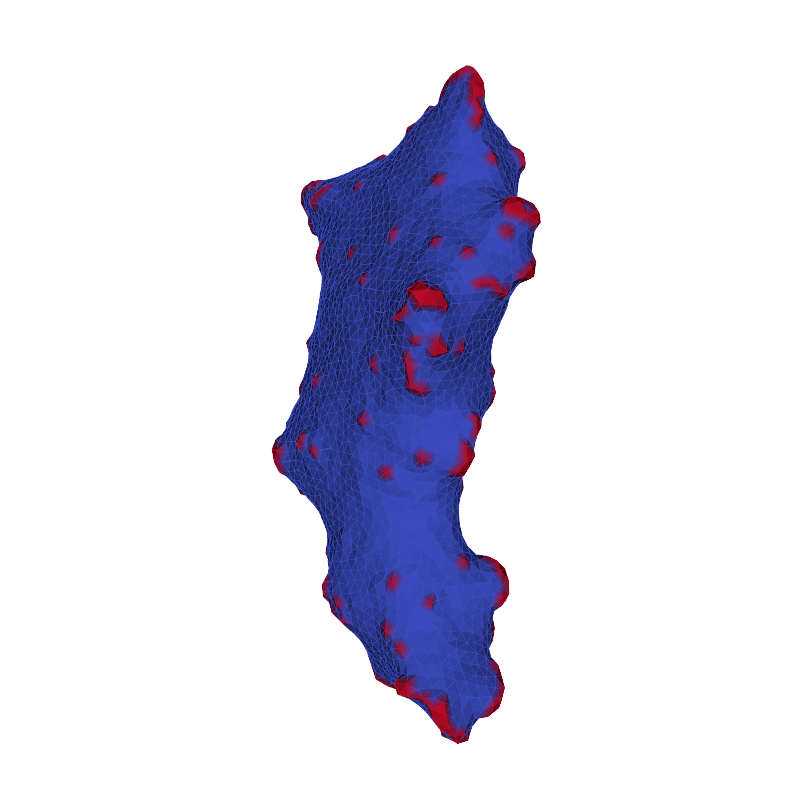}};
\node[anchor=center] (zadnja) at (axis cs:0.8,70) {\includegraphics[width=0.7\textwidth]{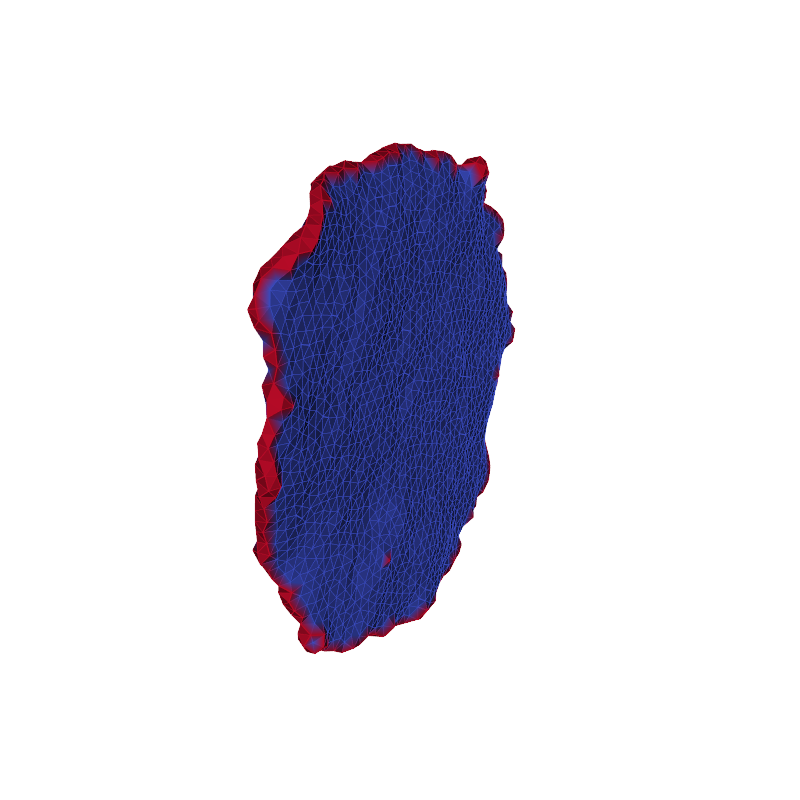}};
\draw[->] ([xshift=-0.0em,yshift=-2em]2zdesne.center) -- (axis cs:1.25,5);
\draw[->] ([yshift=-3.7em]4zdesne.center) -- (axis cs:1,5);
\draw[->] ([xshift=-2em,yshift=-1em]zadnja.center) -- (axis cs:0.64,58);
\end{axis}
\end{tikzpicture}

%% file: snaps_fig_f1_llc.tex
\begin{tikzpicture}
\begin{axis}[
	width=\columnwidth,
	title={\bf (a)},
	ylabel near ticks,
	xlabel={$\rho[\%]$},
	scaled x ticks=manual:{}{\pgfmathparse{(#1)*100}},
	tick label style={/pgf/number format/fixed},
	ylabel=$T/T_0$,
	ylabel near ticks,
	xmin=0.0375,
	xmax=0.0725,
	ymin=0.525,
	ymax=0.77,
]

	\node[inner sep=0pt] at (axis cs:0.03997441637352095,0.5555555555555556) {\includegraphics[width=0.08\textwidth,angle=30]{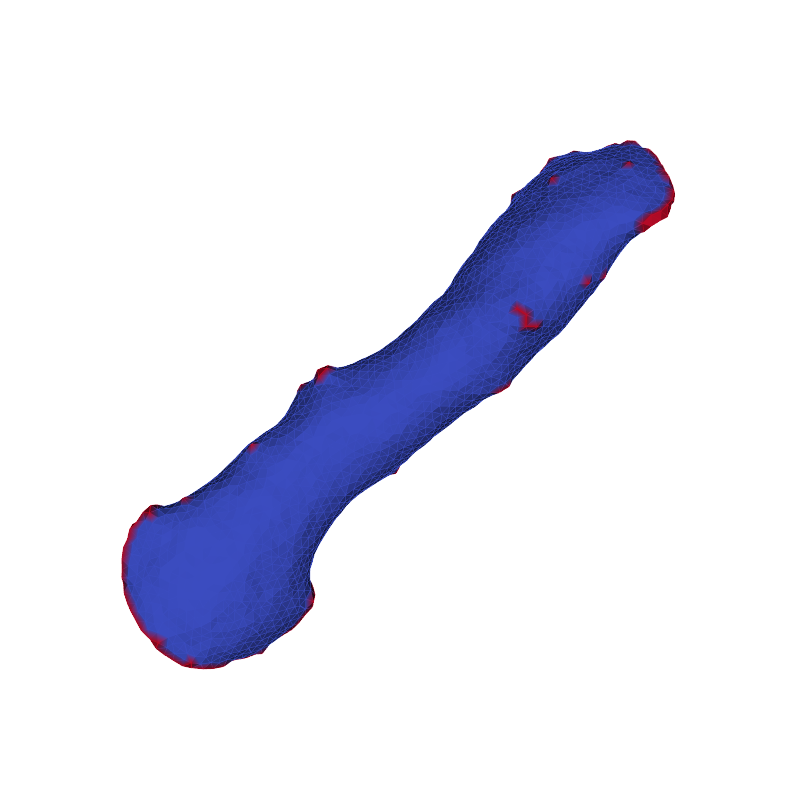}};
	\node[inner sep=0pt] at (axis cs:0.03997441637352095,0.6060606060606061) {\includegraphics[width=0.08\textwidth,angle=-50]{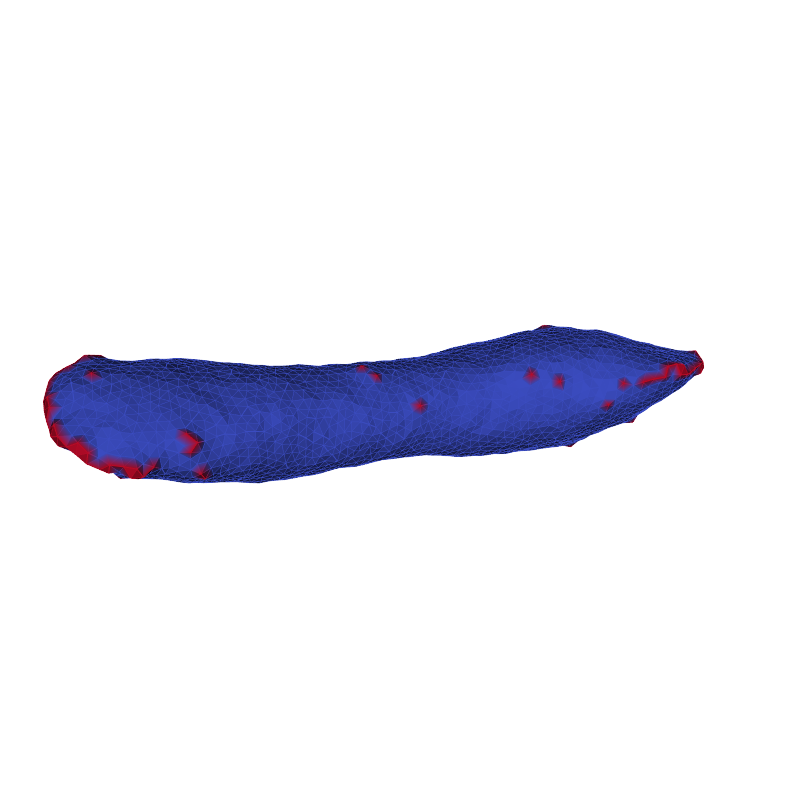}};
	\node[inner sep=0pt] at (axis cs:0.03997441637352095,0.6451612903225806) {\includegraphics[width=0.08\textwidth]{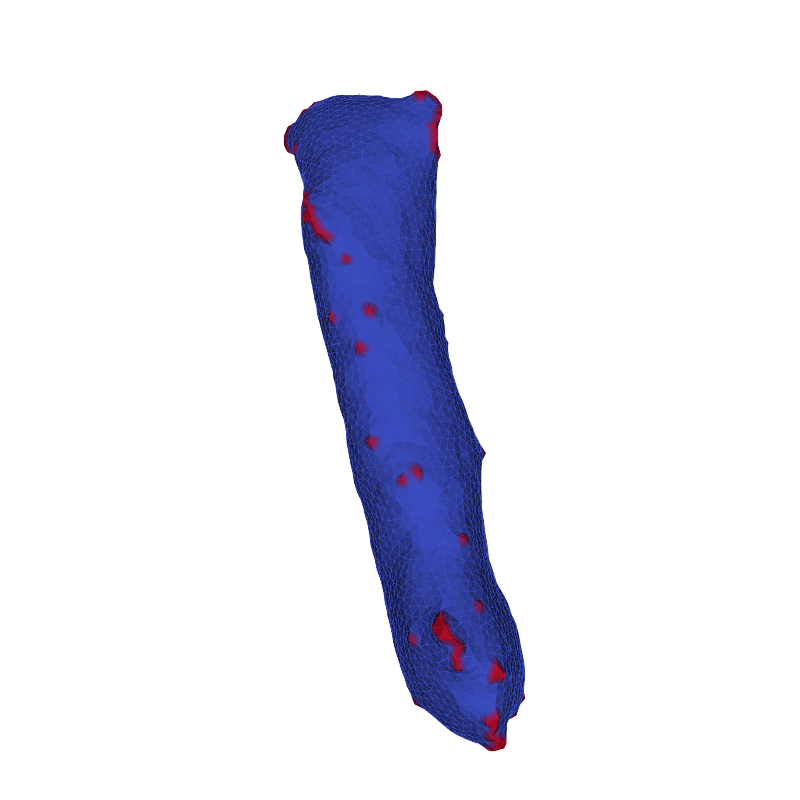}};
	\node[inner sep=0pt] at (axis cs:0.03997441637352095,0.6896551724137931) {\includegraphics[width=0.08\textwidth]{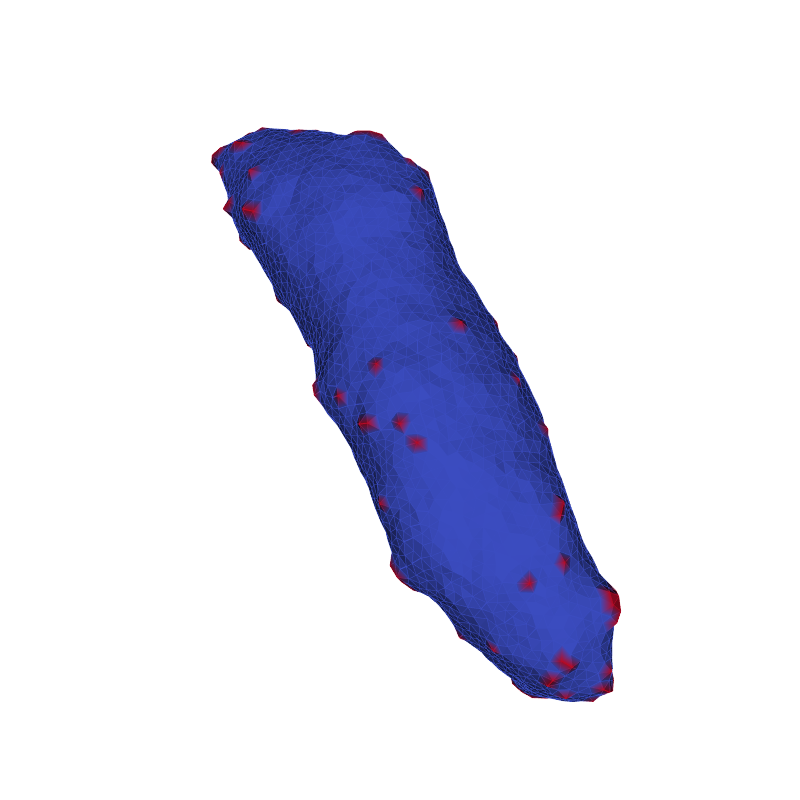}};
	\node[inner sep=0pt] at (axis cs:0.03997441637352095,0.7407407407407407) {\includegraphics[width=0.08\textwidth]{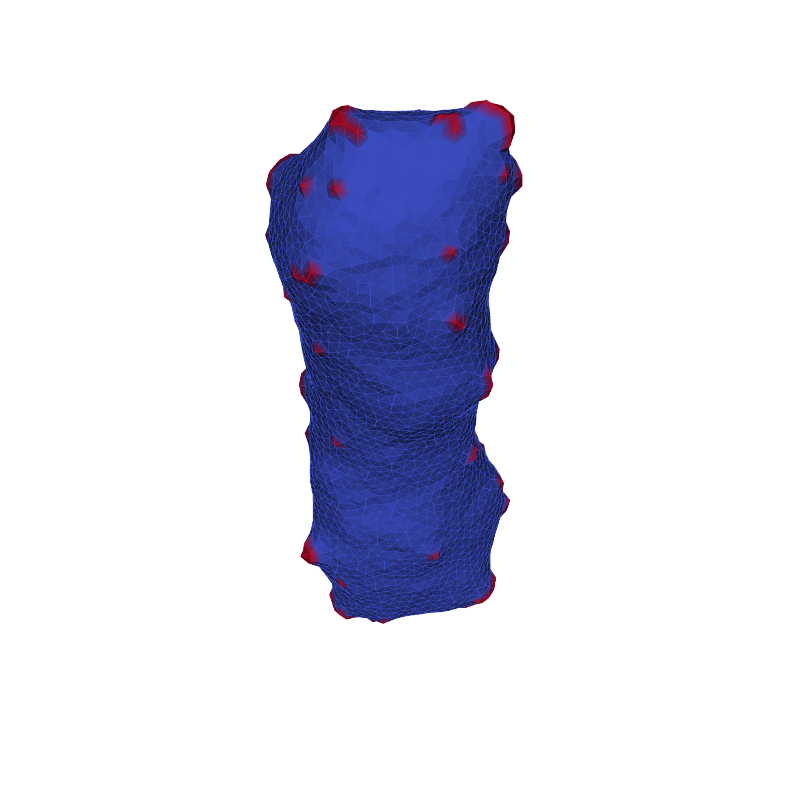}};
	\node[inner sep=0pt] at (axis cs:0.04477134633834346,0.5555555555555556) {\includegraphics[width=0.08\textwidth]{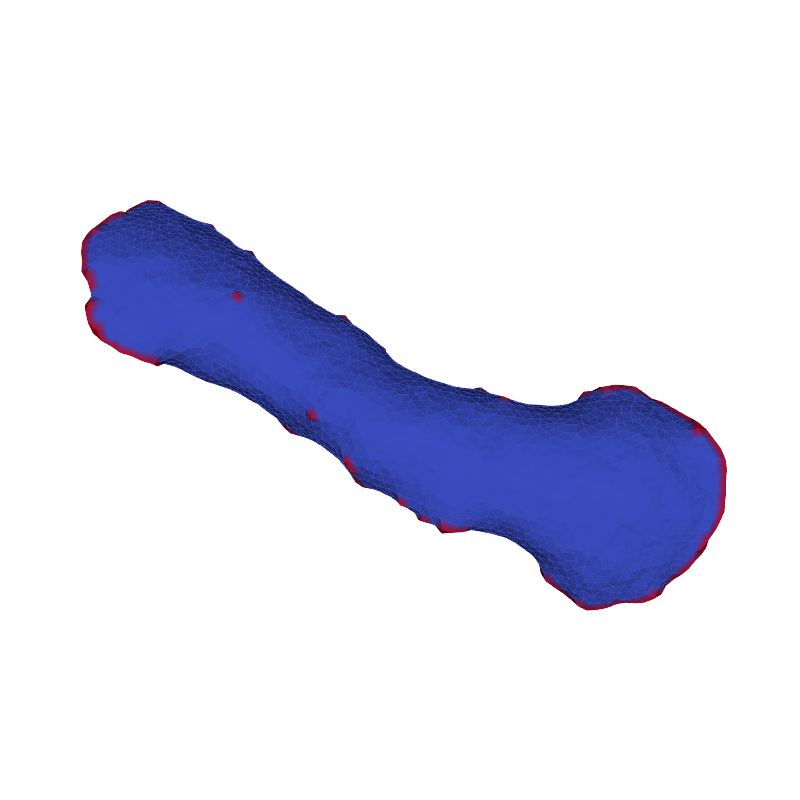}};
	\node[inner sep=0pt] at (axis cs:0.04477134633834346,0.6060606060606061) {\includegraphics[width=0.08\textwidth]{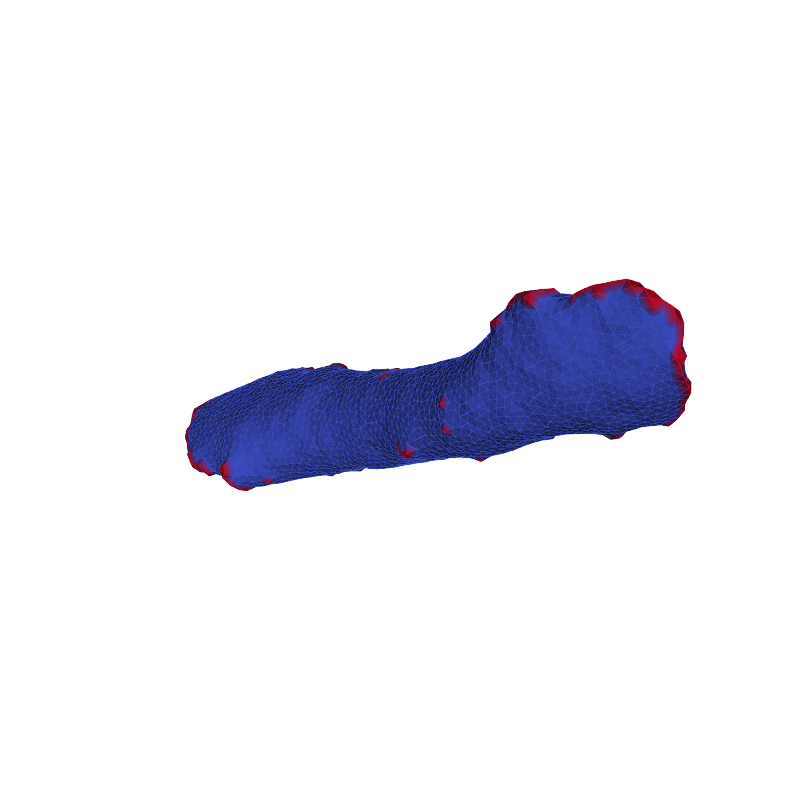}};
	\node[inner sep=0pt] at (axis cs:0.04477134633834346,0.6451612903225806) {\includegraphics[width=0.08\textwidth]{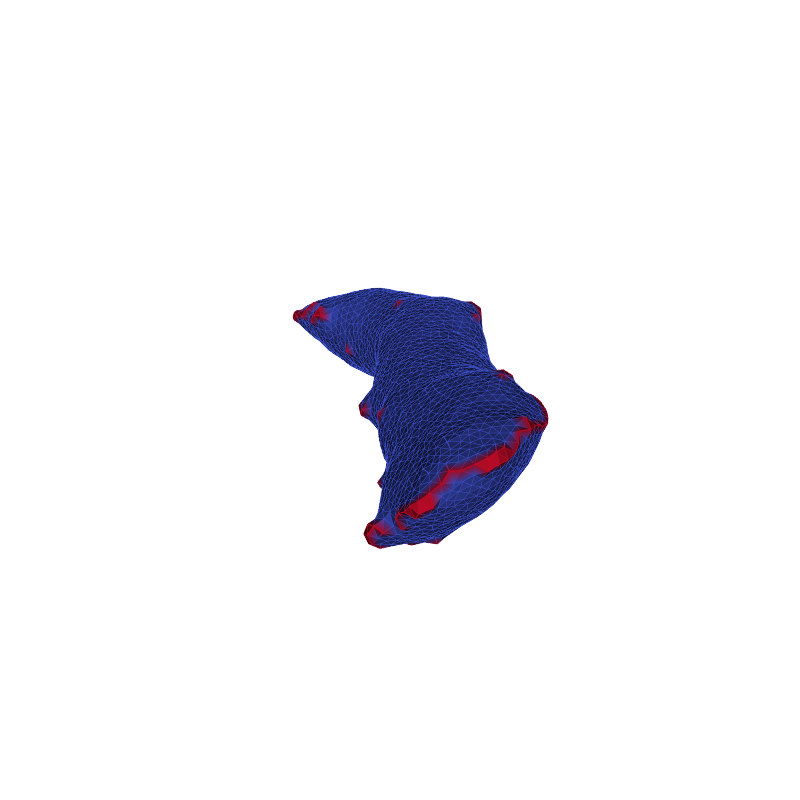}};
	\node[inner sep=0pt] at (axis cs:0.04477134633834346,0.6896551724137931) {\includegraphics[width=0.08\textwidth]{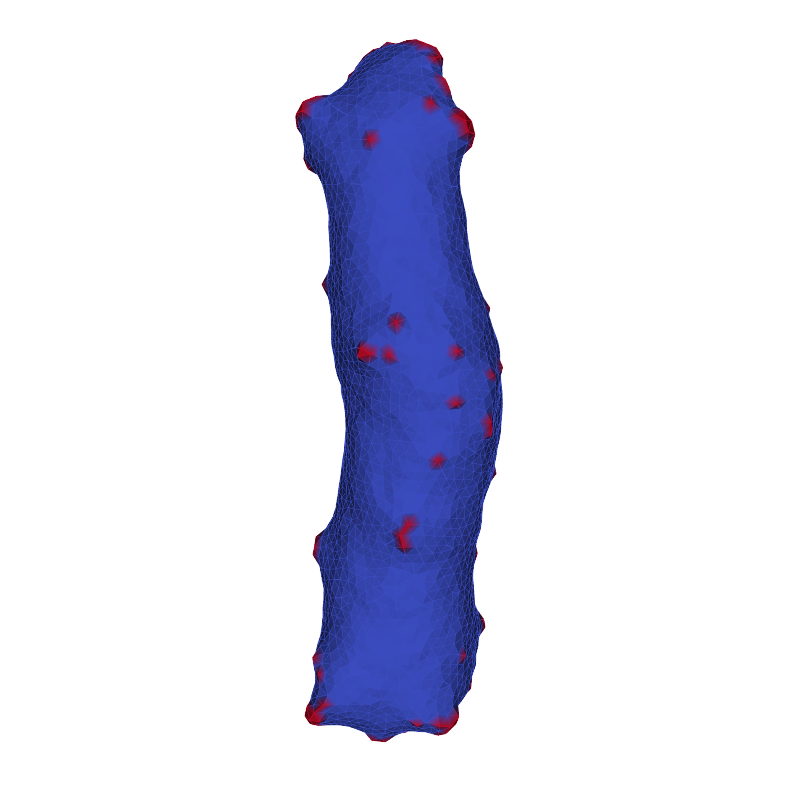}};
	\node[inner sep=0pt] at (axis cs:0.04477134633834346,0.7407407407407407) {\includegraphics[width=0.08\textwidth,angle=-20]{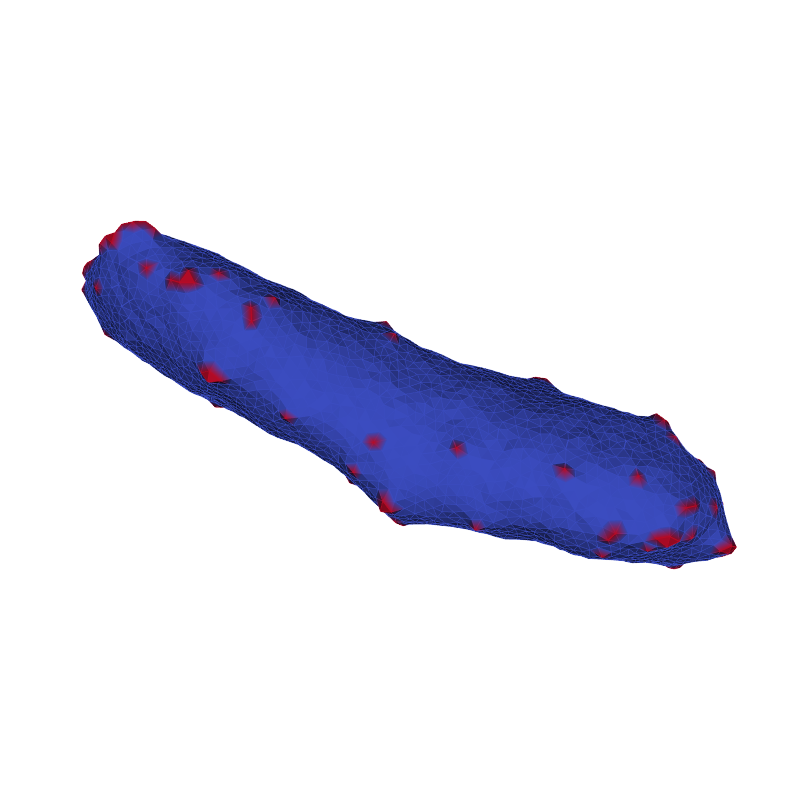}};
	\node[inner sep=0pt] at (axis cs:0.04988807163415414,0.5555555555555556) {\includegraphics[width=0.08\textwidth,angle=90]{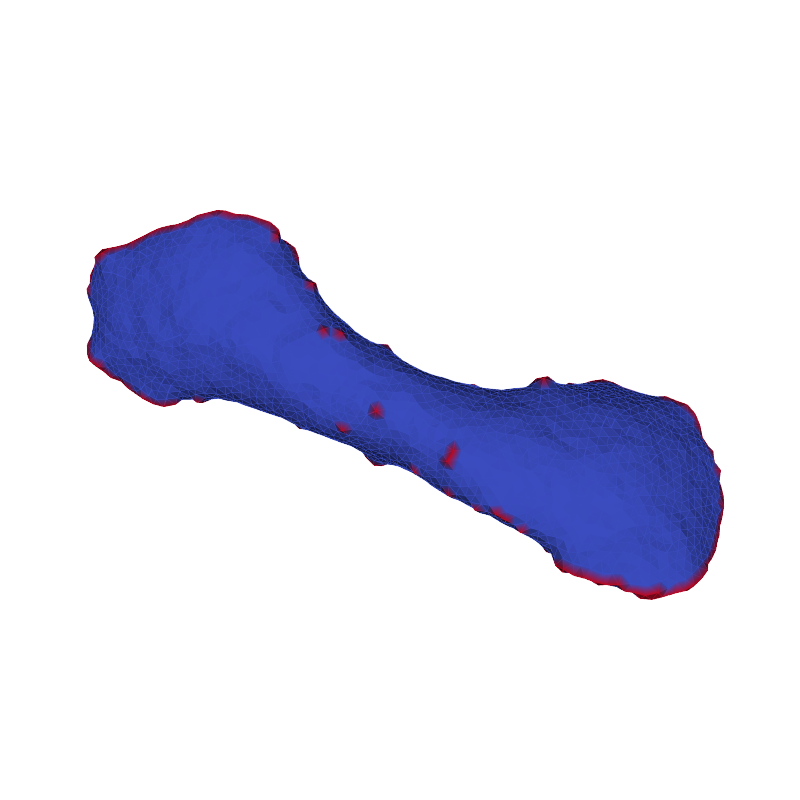}};
	\node[inner sep=0pt] at (axis cs:0.04988807163415414,0.6060606060606061) {\includegraphics[width=0.08\textwidth]{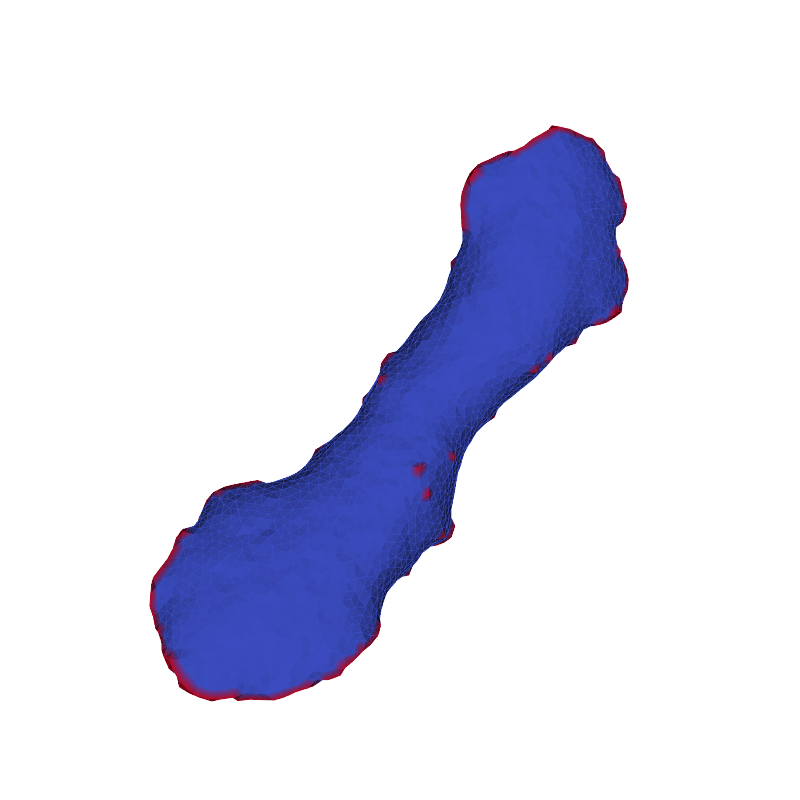}};
	\node[inner sep=0pt] at (axis cs:0.04988807163415414,0.6451612903225806) {\includegraphics[width=0.08\textwidth]{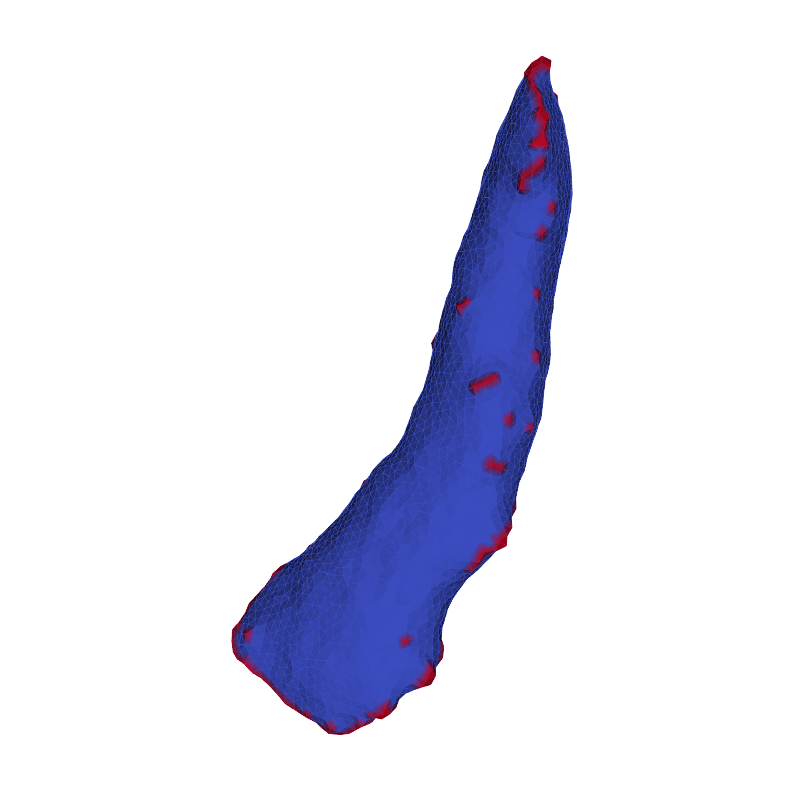}};
	\node[inner sep=0pt] at (axis cs:0.04988807163415414,0.6896551724137931) {\includegraphics[width=0.08\textwidth,angle=90]{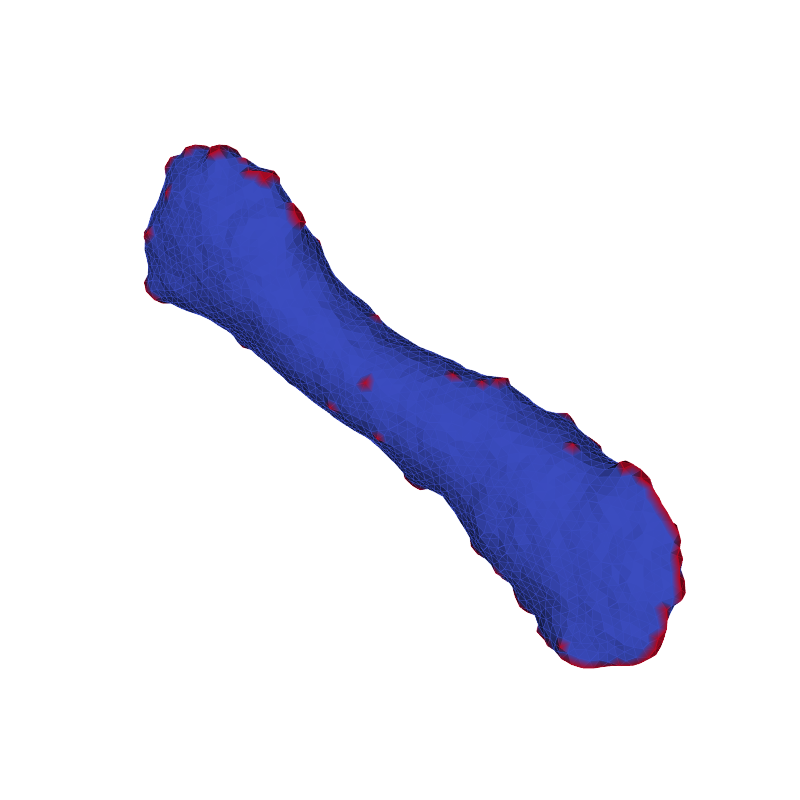}};
	\node[inner sep=0pt] at (axis cs:0.04988807163415414,0.7407407407407407) {\includegraphics[width=0.08\textwidth]{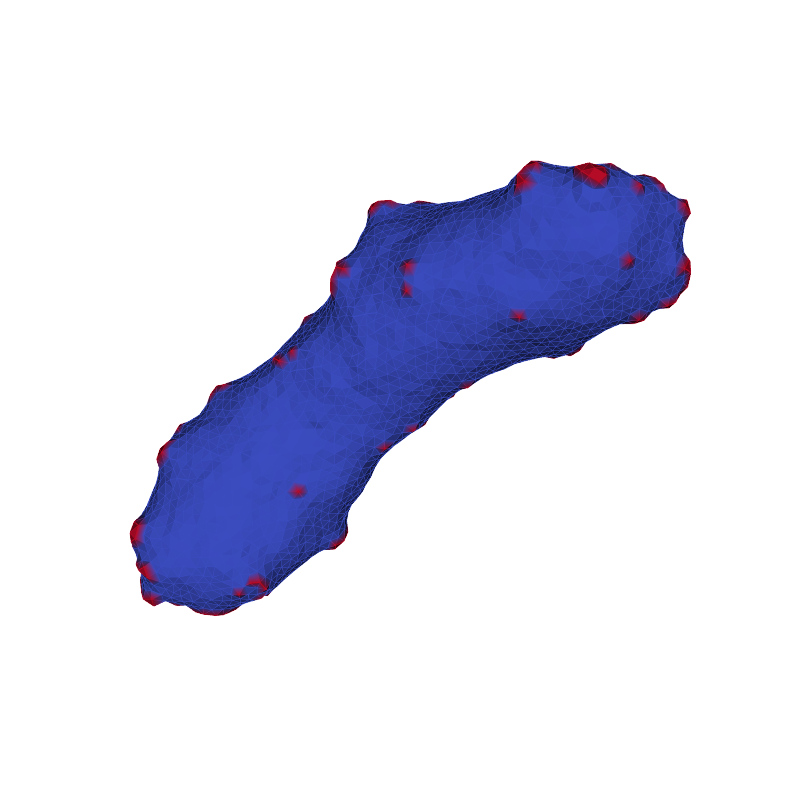}};
	\node[inner sep=0pt] at (axis cs:0.05468500159897666,0.5555555555555556) {\includegraphics[width=0.08\textwidth]{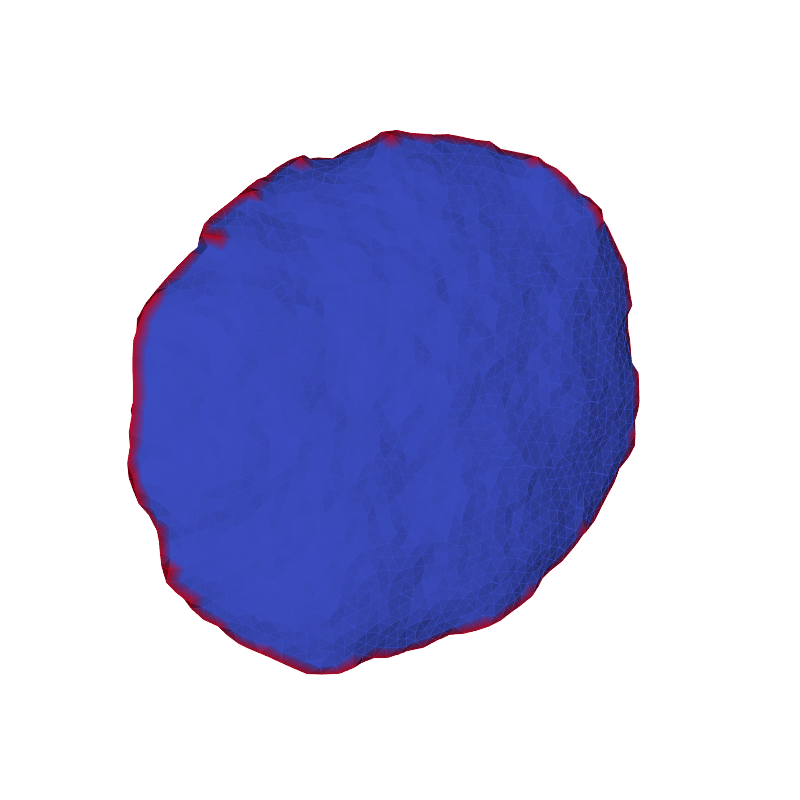}};
	\node[inner sep=0pt] at (axis cs:0.05468500159897666,0.6060606060606061) {\includegraphics[width=0.08\textwidth]{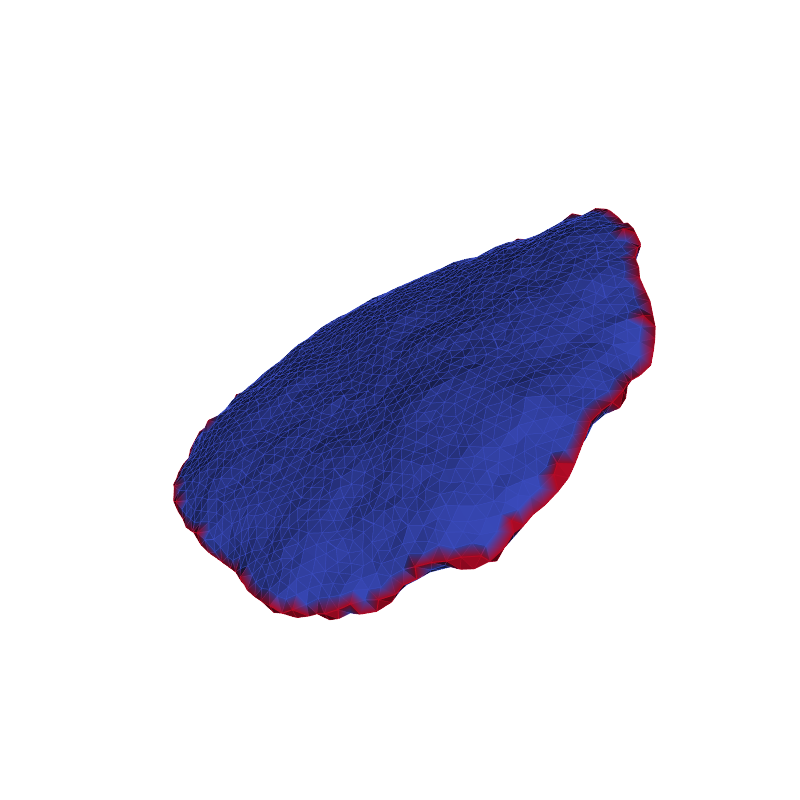}};
	\node[inner sep=0pt] at (axis cs:0.05468500159897666,0.6451612903225806) {\includegraphics[width=0.08\textwidth]{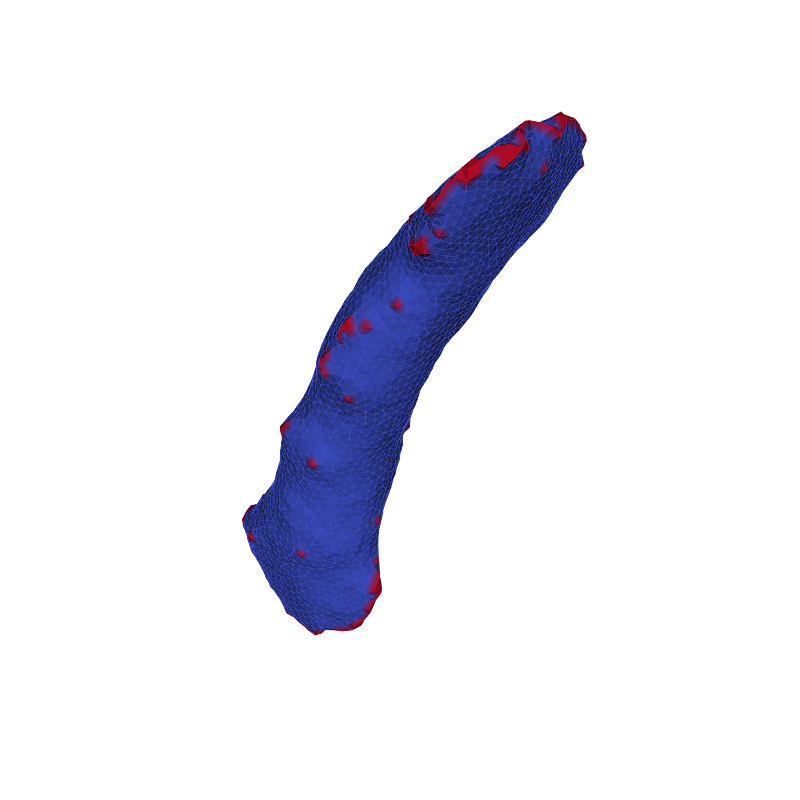}};
	\node[inner sep=0pt] at (axis cs:0.05468500159897666,0.6896551724137931) {\includegraphics[width=0.08\textwidth,angle=15]{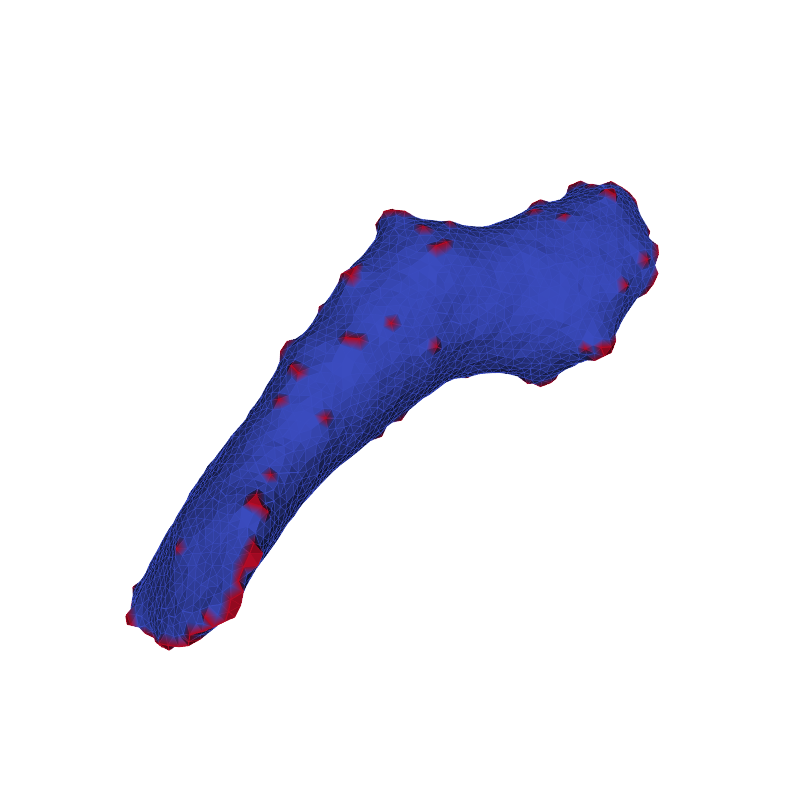}};
	\node[inner sep=0pt] at (axis cs:0.05468500159897666,0.7407407407407407) {\includegraphics[width=0.08\textwidth,angle=90]{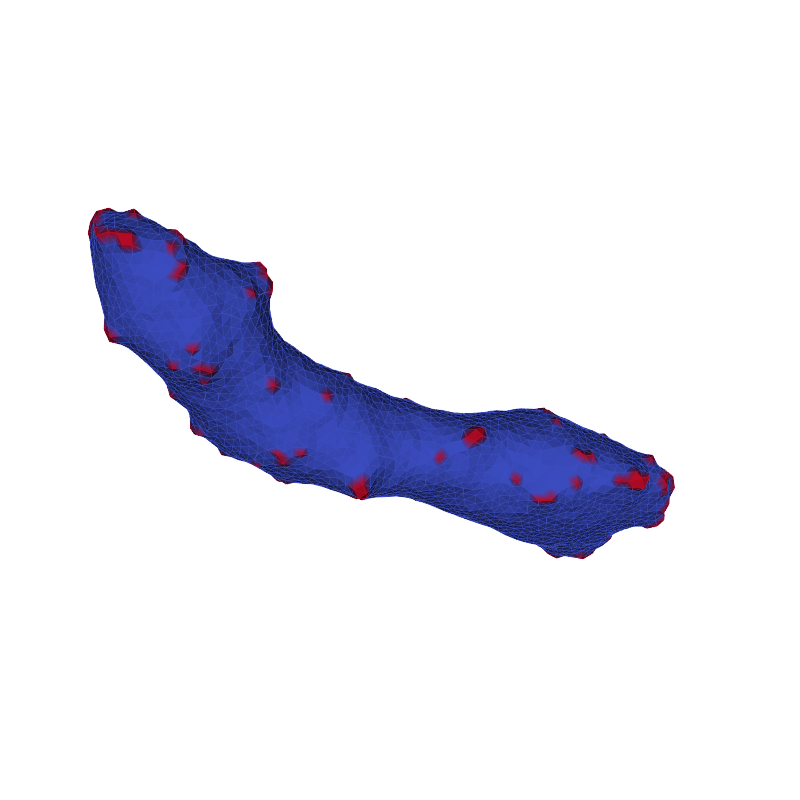}};
	\node[inner sep=0pt] at (axis cs:0.059801726894787334,0.5555555555555556) {\includegraphics[width=0.08\textwidth]{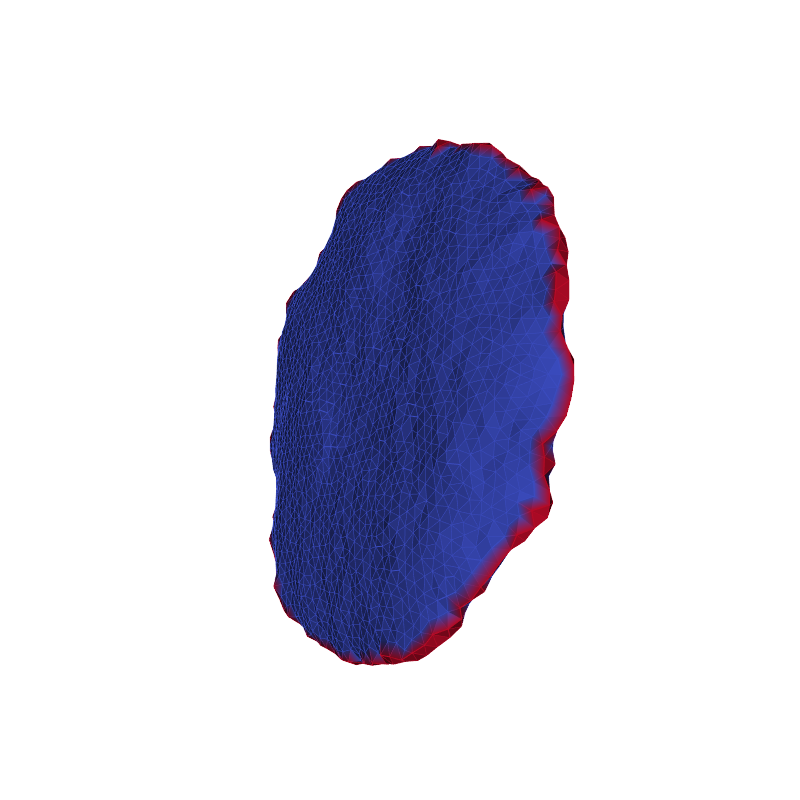}};
	\node[inner sep=0pt] at (axis cs:0.059801726894787334,0.6060606060606061) {\includegraphics[width=0.08\textwidth]{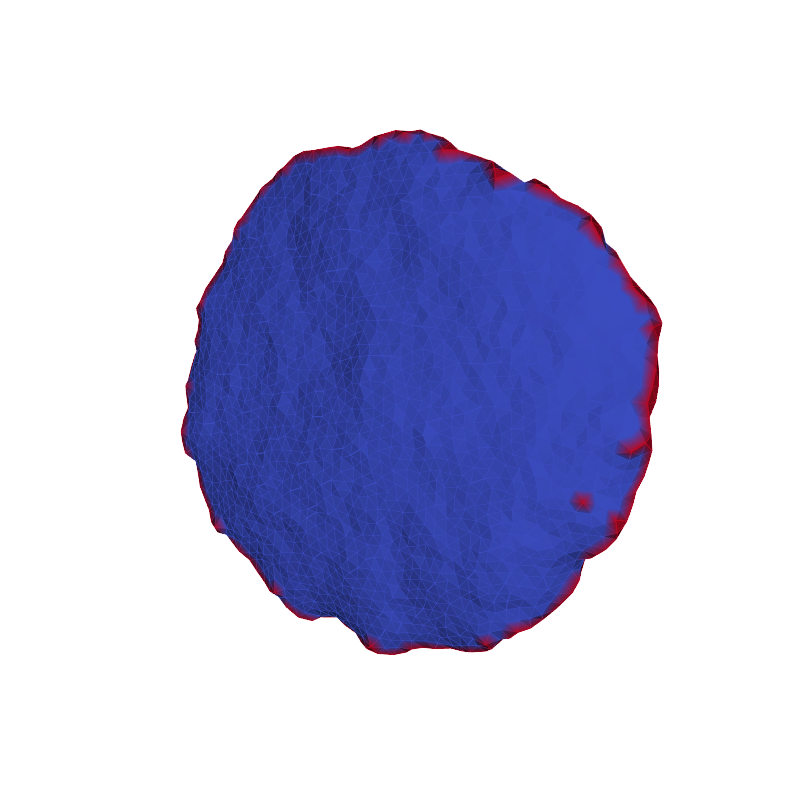}};
	\node[inner sep=0pt] at (axis cs:0.059801726894787334,0.6451612903225806) {\includegraphics[width=0.08\textwidth]{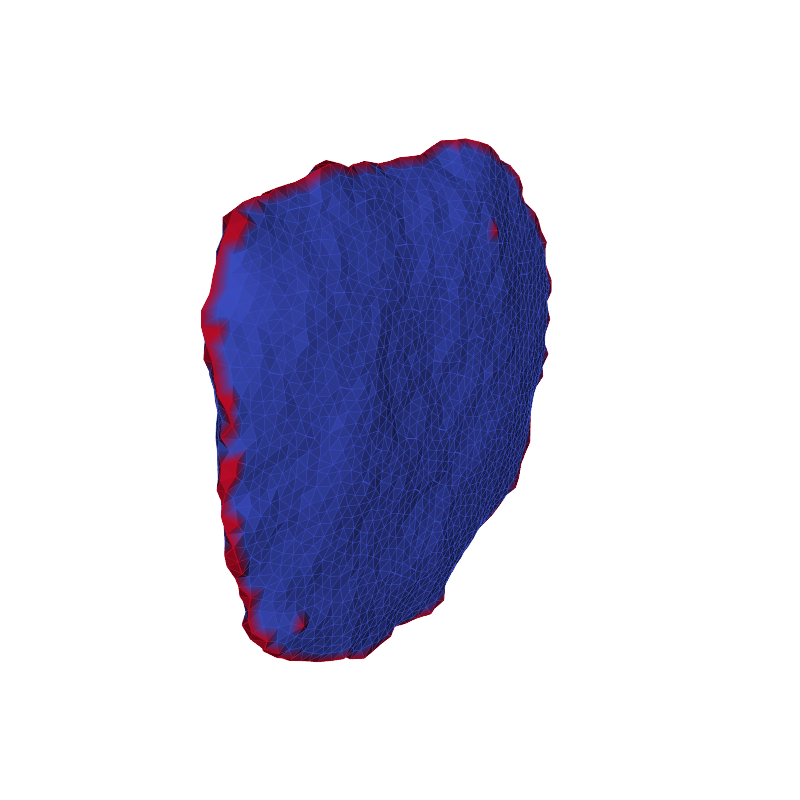}};
	\node[inner sep=0pt] at (axis cs:0.059801726894787334,0.6896551724137931) {\includegraphics[width=0.08\textwidth]{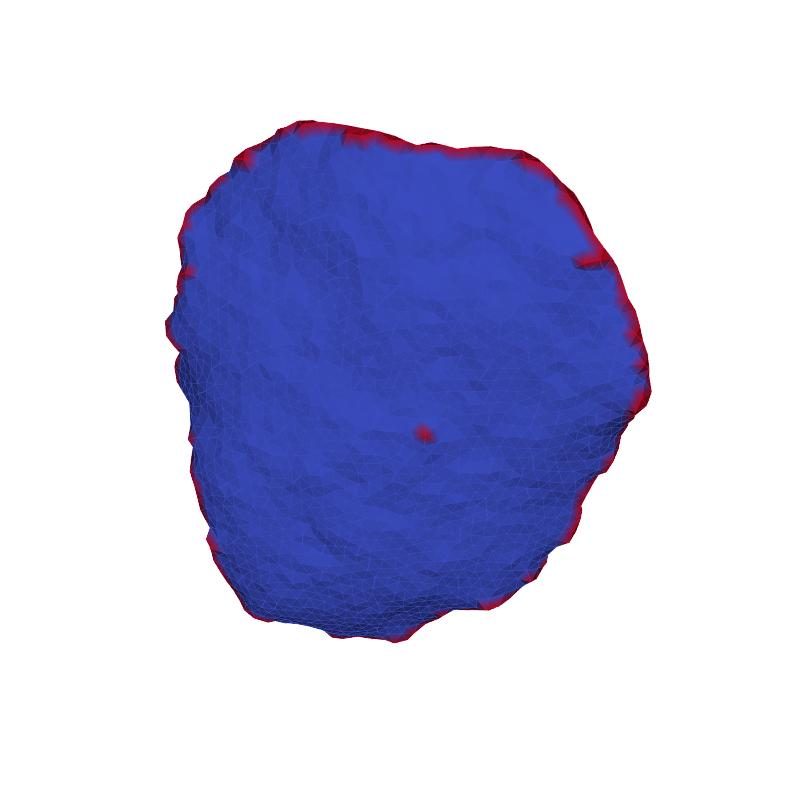}};
	\node[inner sep=0pt] at (axis cs:0.059801726894787334,0.7407407407407407) {\includegraphics[width=0.08\textwidth,angle=0]{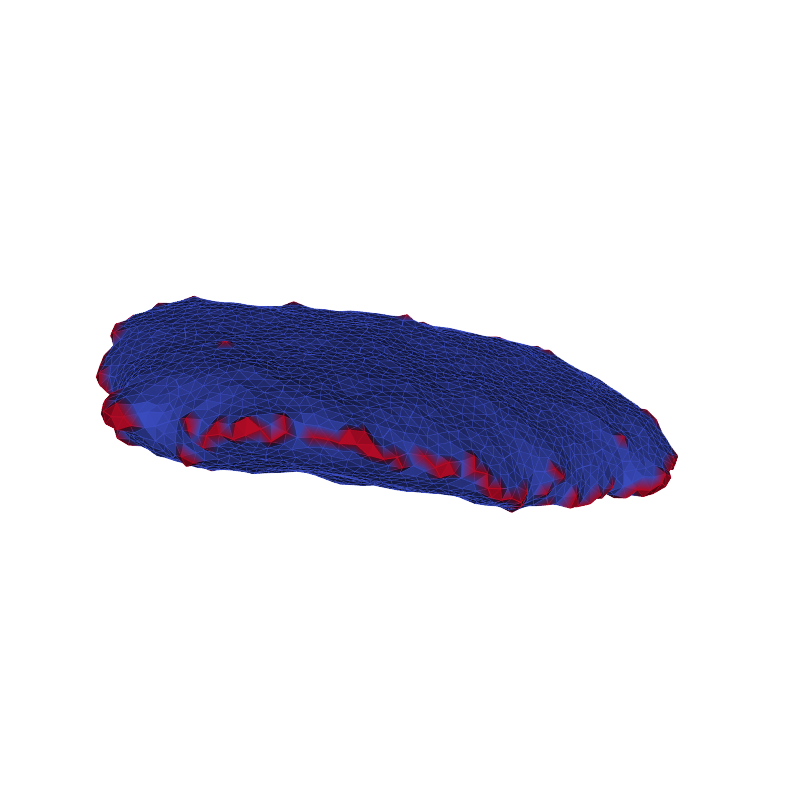}};
	\node[inner sep=0pt] at (axis cs:0.06491845219059801,0.5555555555555556) {\includegraphics[width=0.08\textwidth]{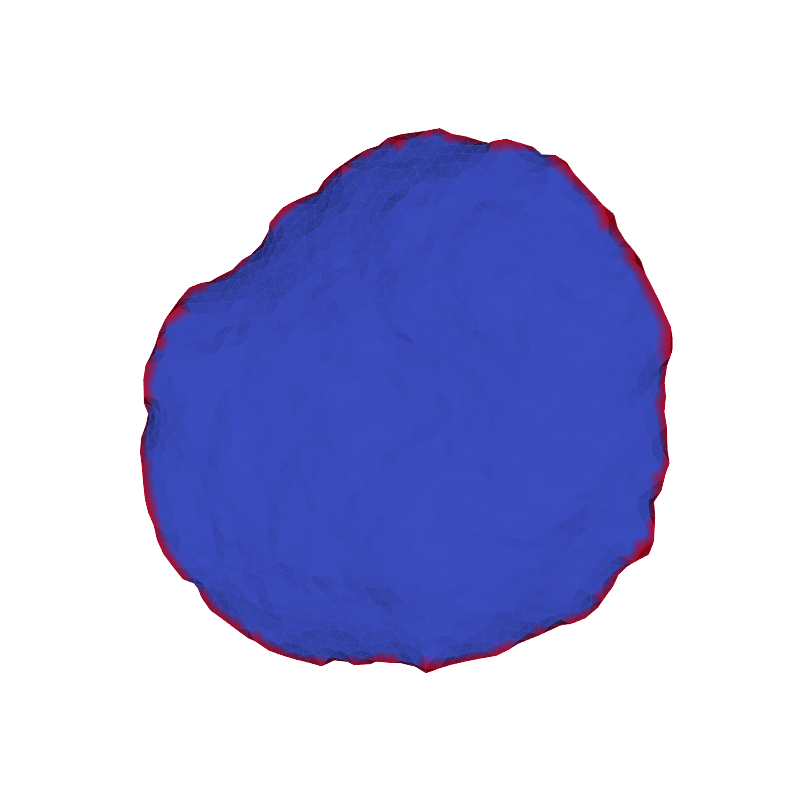}};
	\node[inner sep=0pt] at (axis cs:0.06491845219059801,0.6060606060606061) {\includegraphics[width=0.08\textwidth]{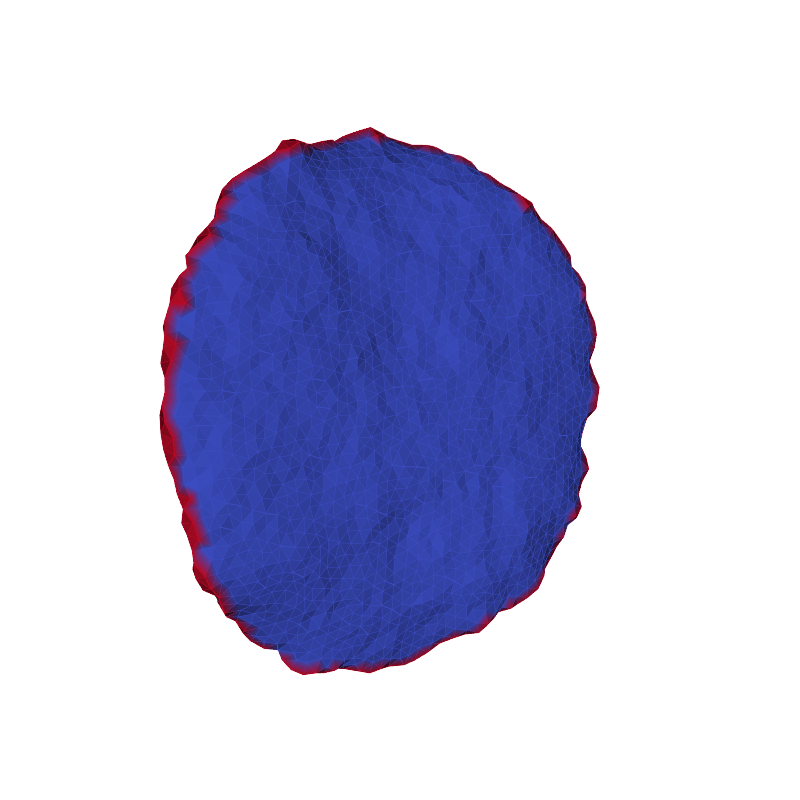}};
	\node[inner sep=0pt] at (axis cs:0.06491845219059801,0.6451612903225806) {\includegraphics[width=0.08\textwidth]{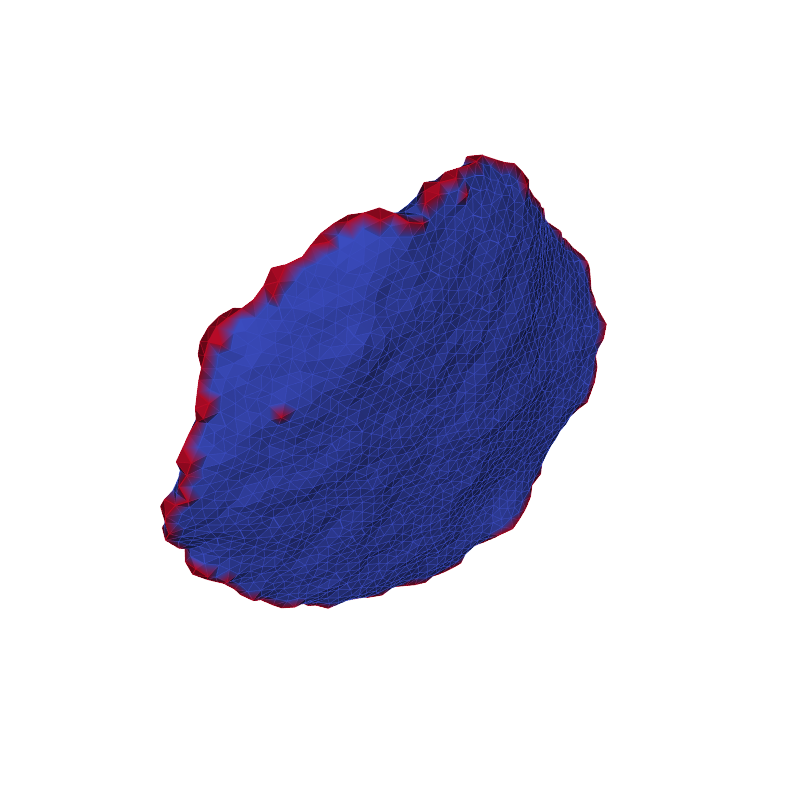}};
	\node[inner sep=0pt] at (axis cs:0.06491845219059801,0.6896551724137931) {\includegraphics[width=0.08\textwidth]{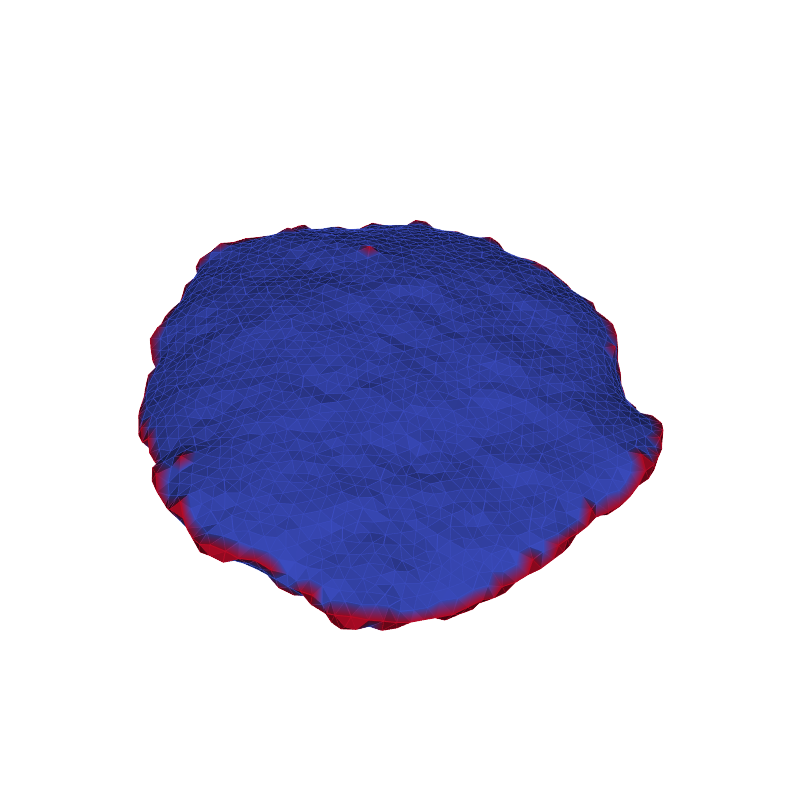}};
	\node[inner sep=0pt] at (axis cs:0.06491845219059801,0.7407407407407407) {\includegraphics[width=0.08\textwidth,angle=-105]{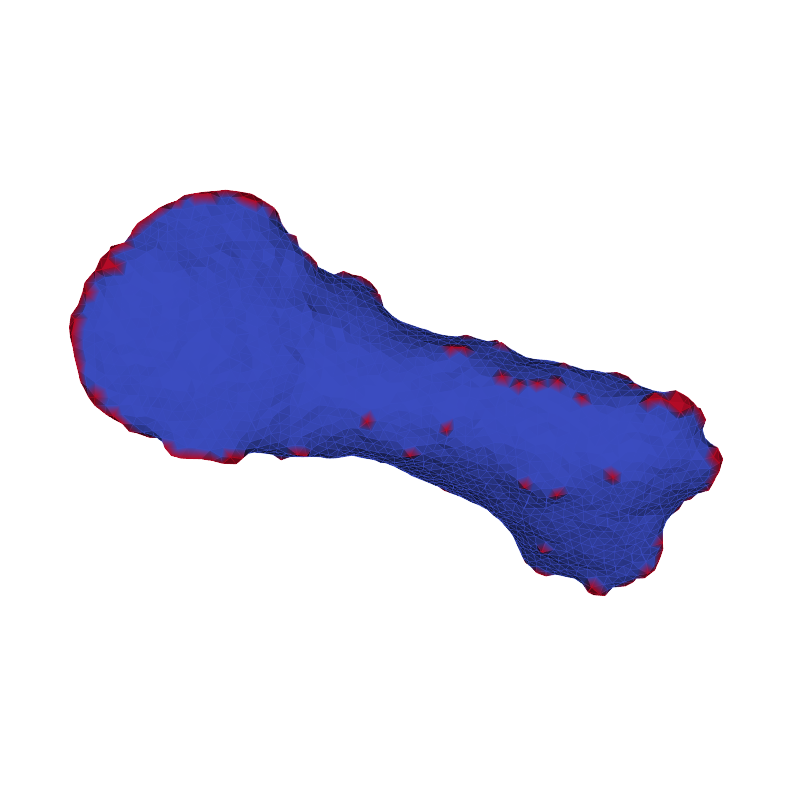}};
	\node[inner sep=0pt] at (axis cs:0.06971538215542053,0.5555555555555556) {\includegraphics[width=0.08\textwidth]{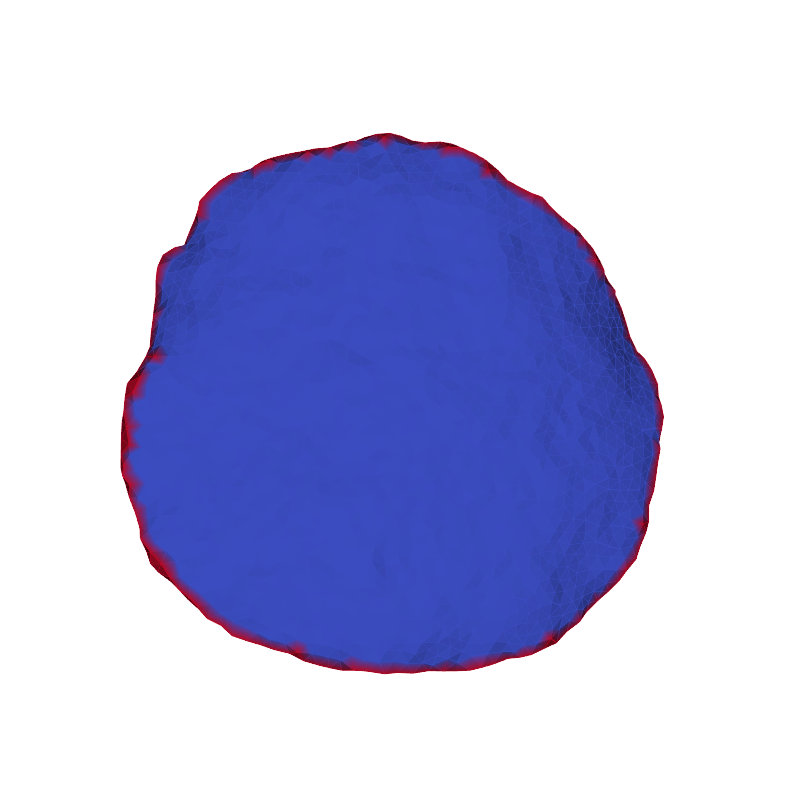}};
	\node[inner sep=0pt] at (axis cs:0.06971538215542053,0.6060606060606061) {\includegraphics[width=0.08\textwidth]{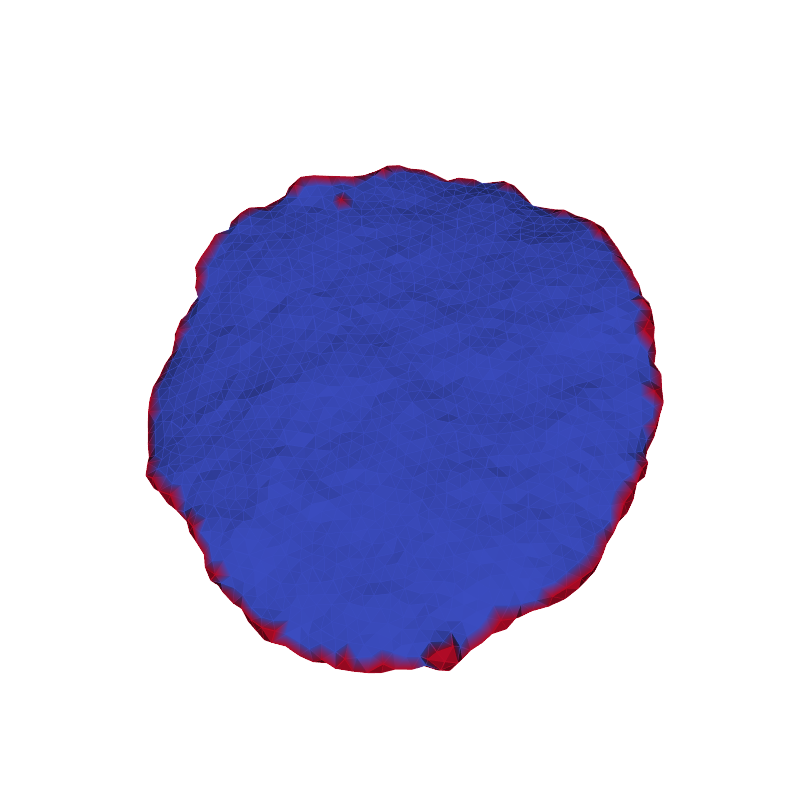}};
	\node[inner sep=0pt] at (axis cs:0.06971538215542053,0.6451612903225806) {\includegraphics[width=0.08\textwidth]{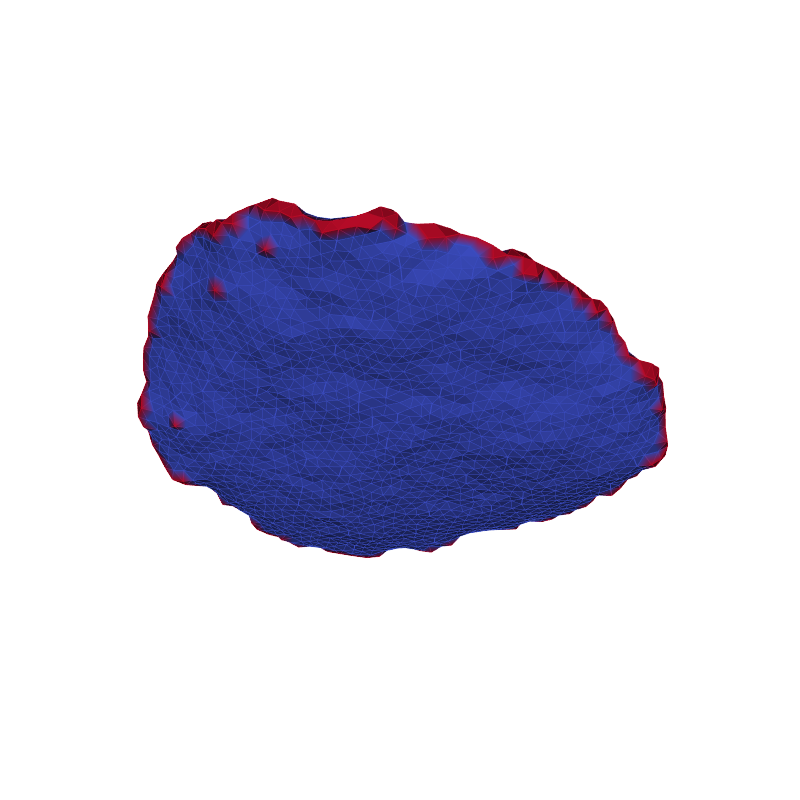}};
	\node[inner sep=0pt] at (axis cs:0.06971538215542053,0.6896551724137931) {\includegraphics[width=0.08\textwidth]{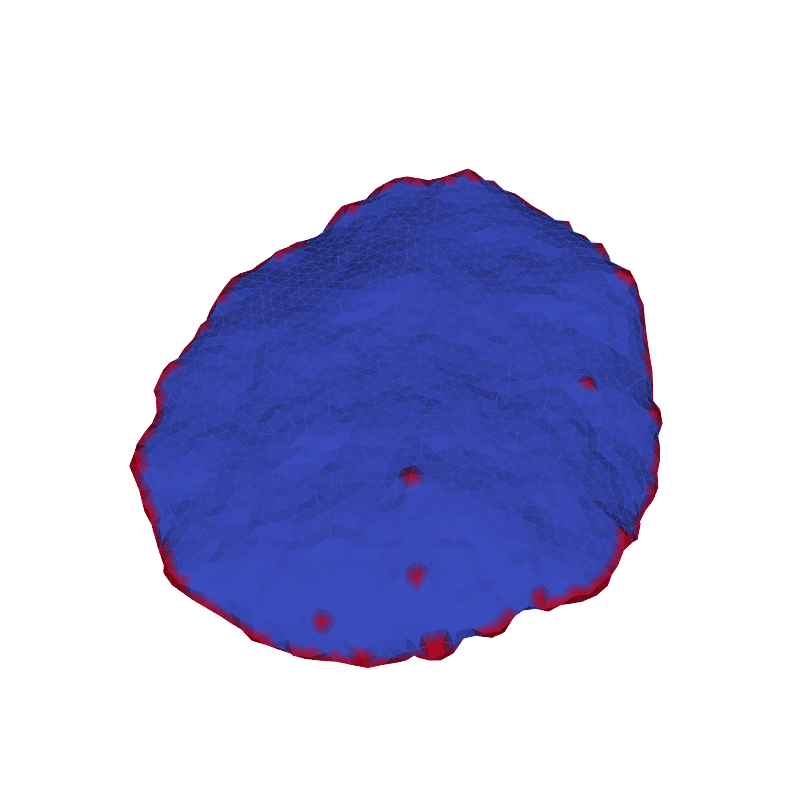}};
	\node[inner sep=0pt] at (axis cs:0.06971538215542053,0.7407407407407407) {\includegraphics[width=0.08\textwidth]{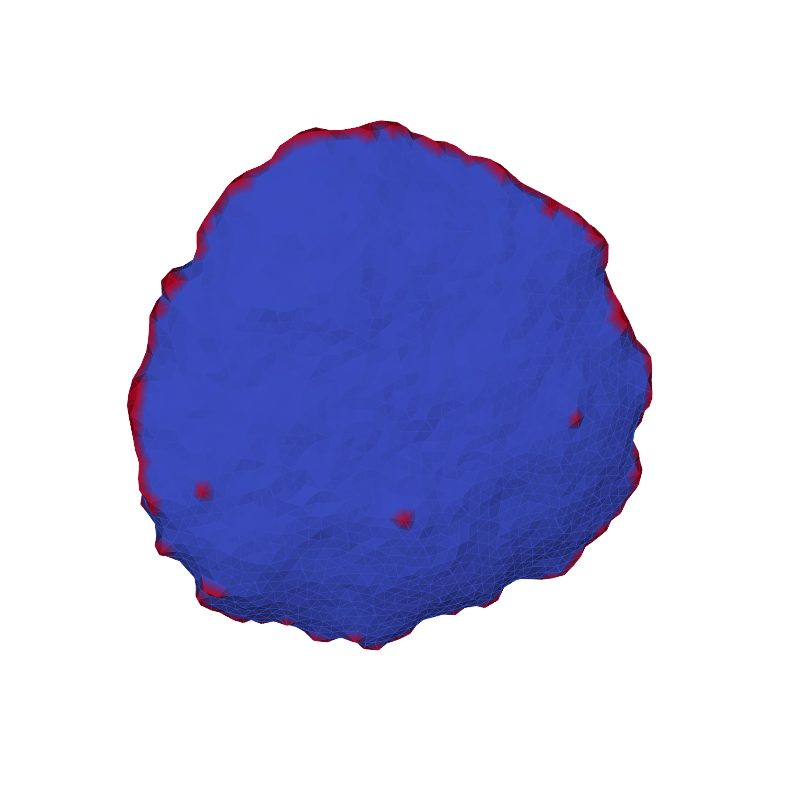}};

	\addplot [mark=*,mark options={solid, scale=1.0}, style={dashed}, color=red!80
] table[y error minus index=2, y error plus index=3] {./figs/data/pancake_phase_transition_f1_error_bars.dat};

\end{axis}
\end{tikzpicture}

%% file: snapsc0_fig.tex
\begin{tikzpicture}
\begin{axis}[
	width=\columnwidth,
	height=0.30\textheight,
	ylabel near ticks,
	title={{\small \bf (a)}},
	xlabel={$\rho[\%]$},
	scaled x ticks=manual:{}{\pgfmathparse{(#1)*100}},
	tick label style={/pgf/number format/fixed},
	ylabel=$T/T_0$,
	ymin=0.4,
	ymax=1.42,
	xmin=0.037,
	xmax=0.182,
	tick label style={/pgf/number format/.cd, fixed, precision=2}]

	\node[inner sep=0pt] at (axis cs:0.04988807163415414,0.5) {\includegraphics[width=0.16\textwidth]{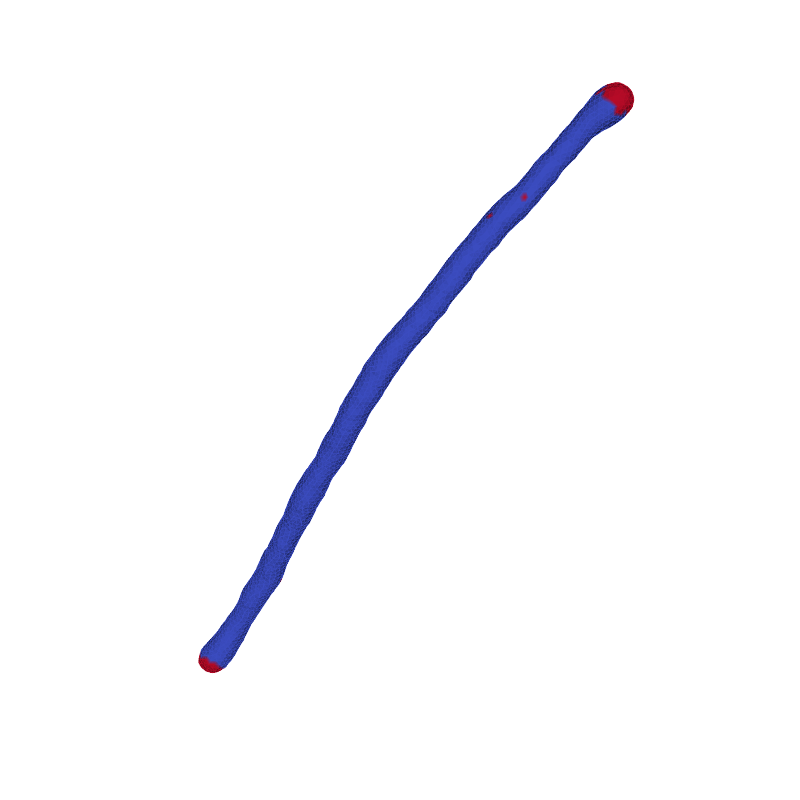}};
	\node[inner sep=0pt] at (axis cs:0.06971538215542053,0.5) {\includegraphics[width=0.16\textwidth]{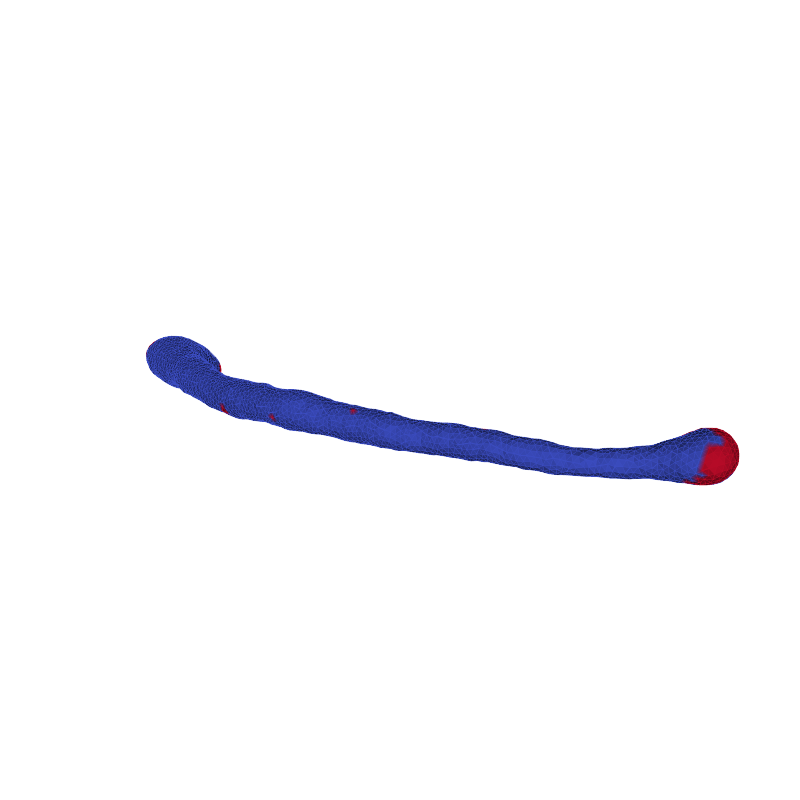}};
	\node[inner sep=0pt] at (axis cs:0.08986248800767509,0.5) {\includegraphics[width=0.16\textwidth]{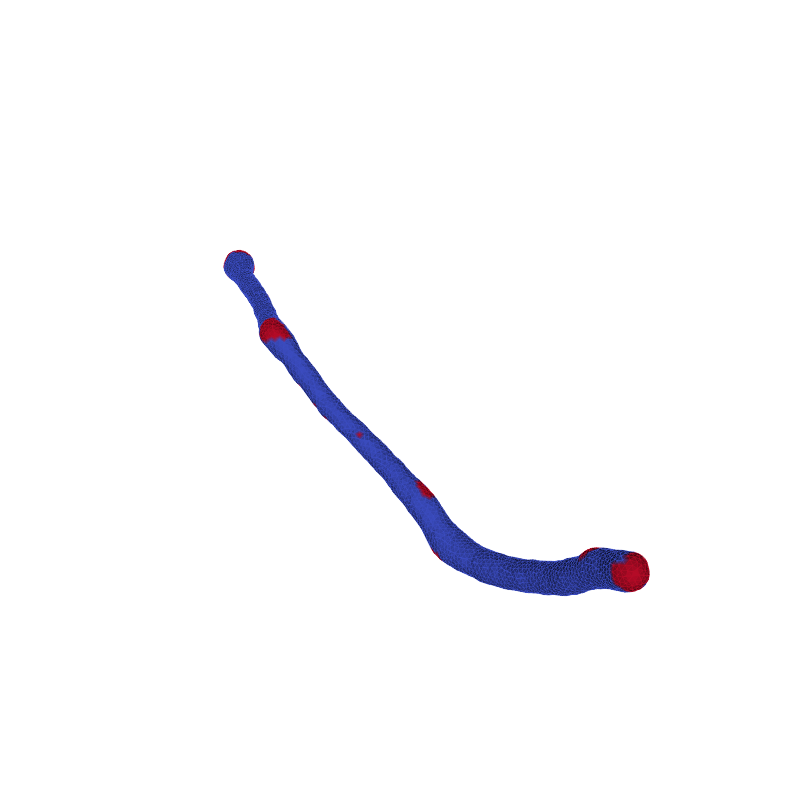}};
	\node[inner sep=0pt] at (axis cs:0.10968979852894148,0.5) {\includegraphics[width=0.16\textwidth,angle=30]{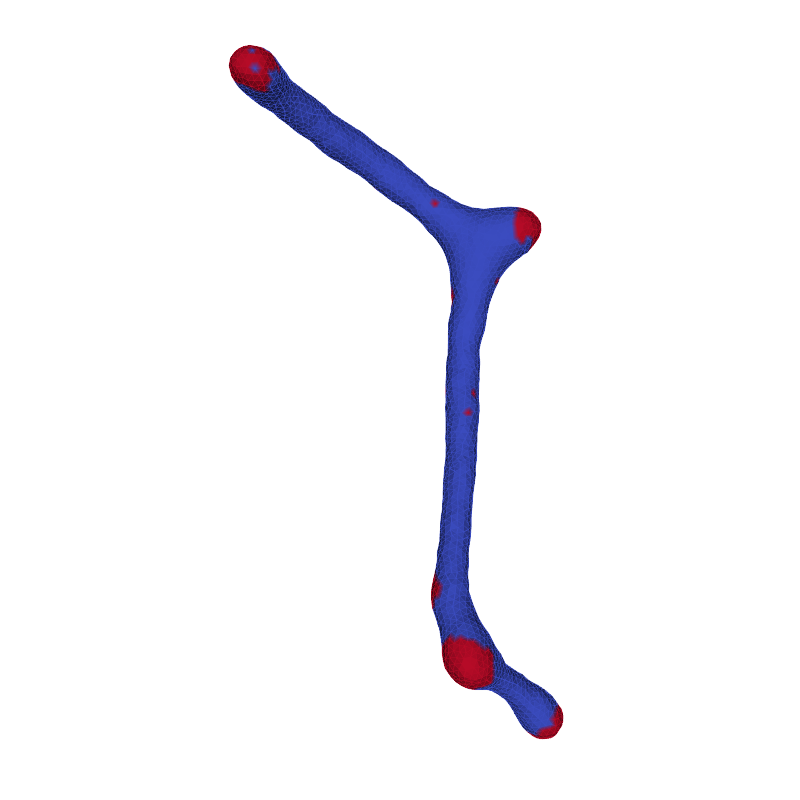}};
	\node[inner sep=0pt] at (axis cs:0.12983690438119602,0.5) {\includegraphics[width=0.16\textwidth]{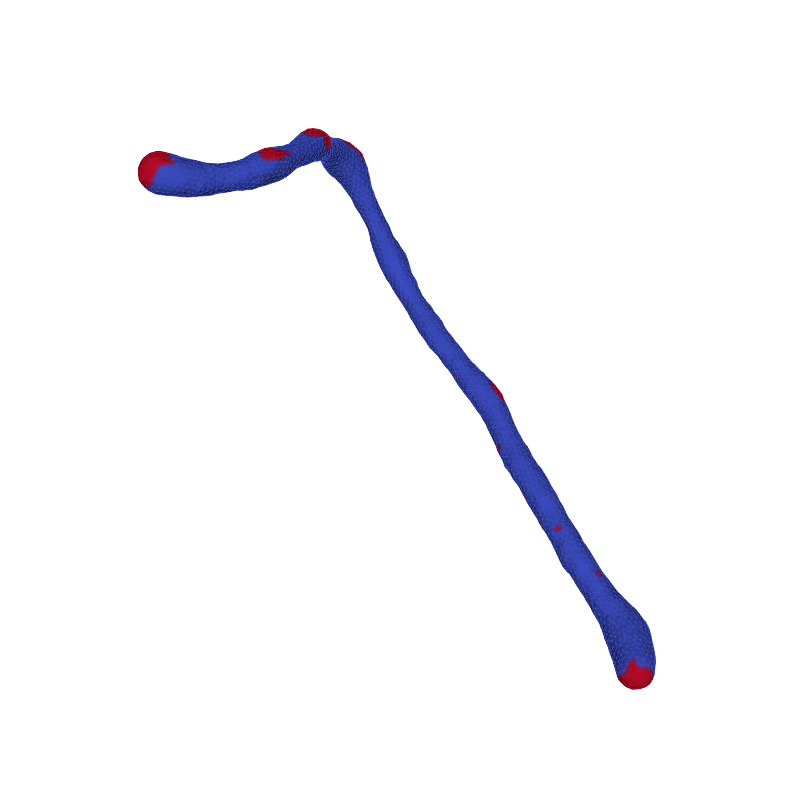}};
	\node[inner sep=0pt] at (axis cs:0.1499840102334506,0.5) {\includegraphics[width=0.16\textwidth]{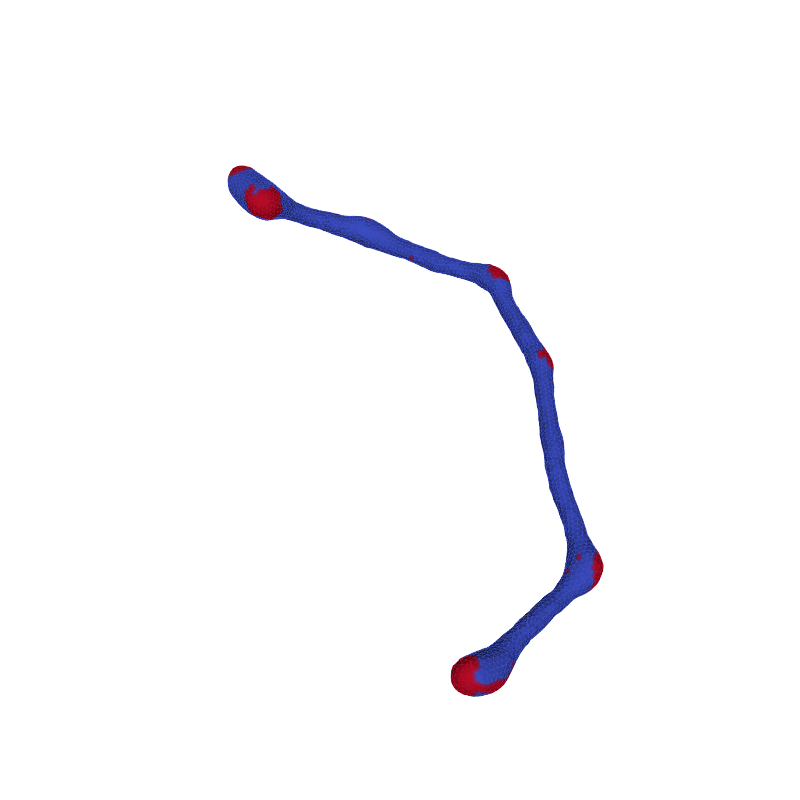}};
	\node[inner sep=0pt] at (axis cs:0.16981132075471697,0.5) {\includegraphics[width=0.16\textwidth,angle=30]{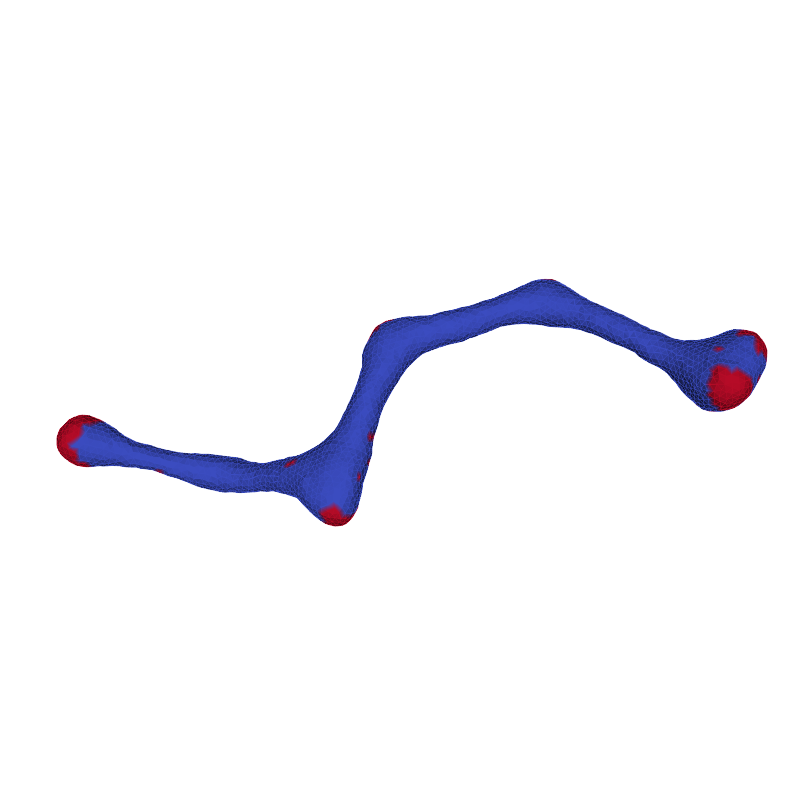}};
	\node[inner sep=0pt] at (axis cs:0.04988807163415414,0.6896551724137931) {\includegraphics[width=0.16\textwidth]{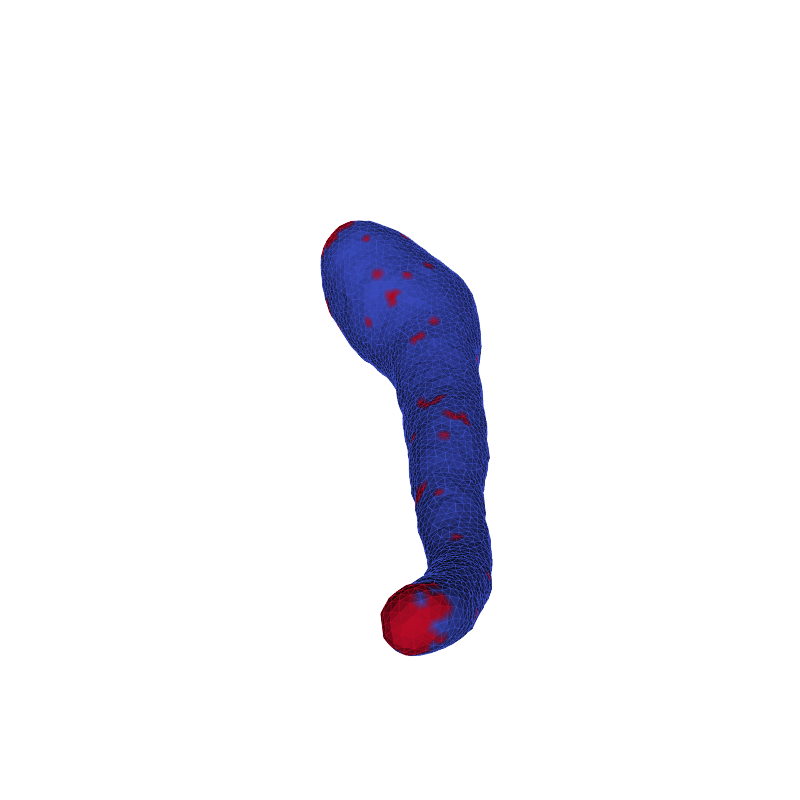}};
	\node[inner sep=0pt] at (axis cs:0.06971538215542053,0.6896551724137931) {\includegraphics[width=0.16\textwidth,angle=-10]{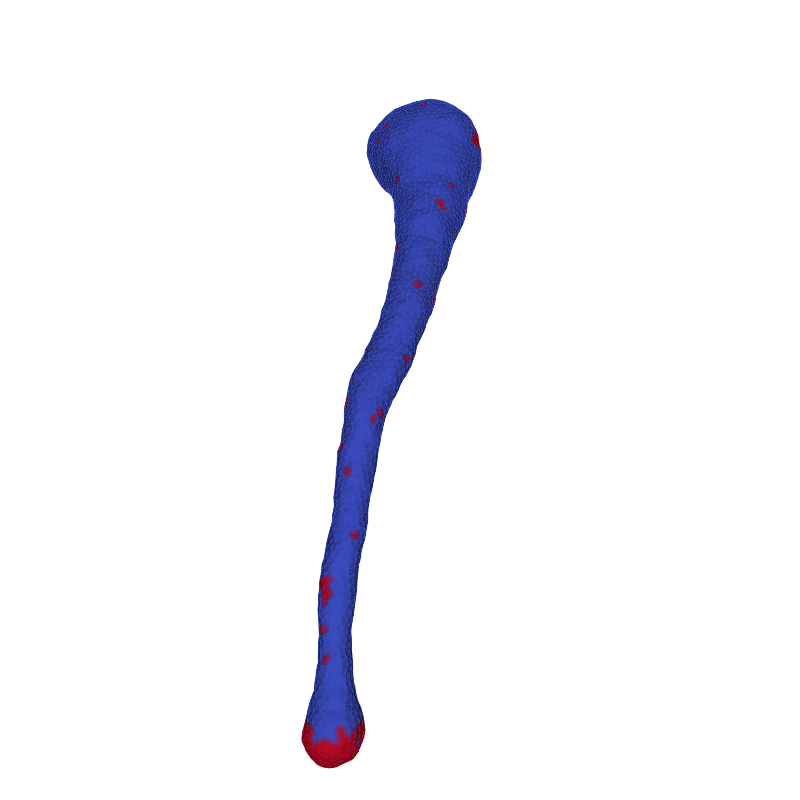}};
	\node[inner sep=0pt] at (axis cs:0.08986248800767509,0.6896551724137931) {\includegraphics[width=0.16\textwidth]{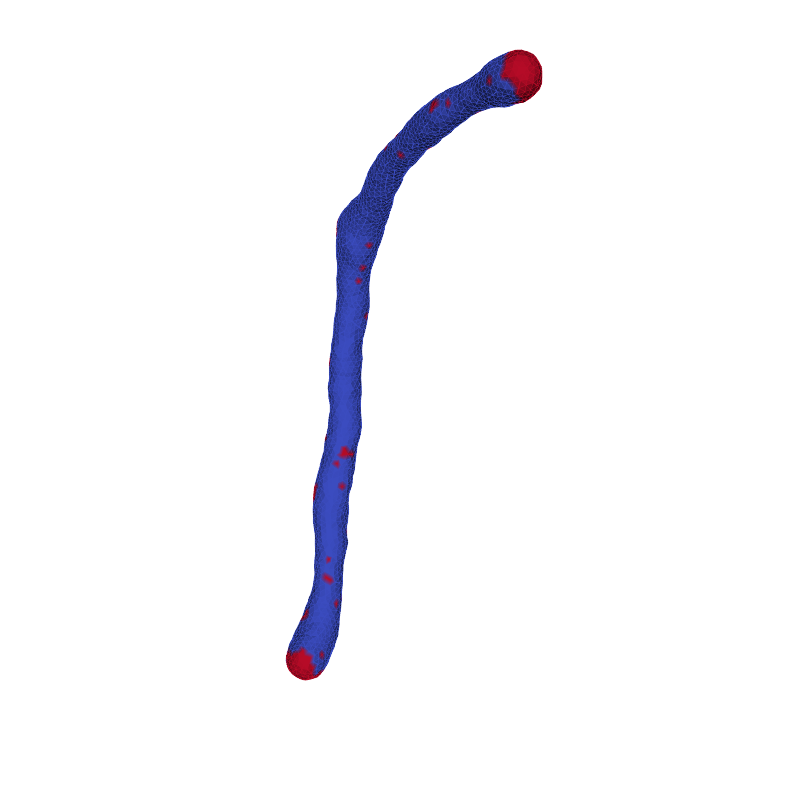}};
	\node[inner sep=0pt] at (axis cs:0.10968979852894148,0.6896551724137931) {\includegraphics[width=0.16\textwidth]{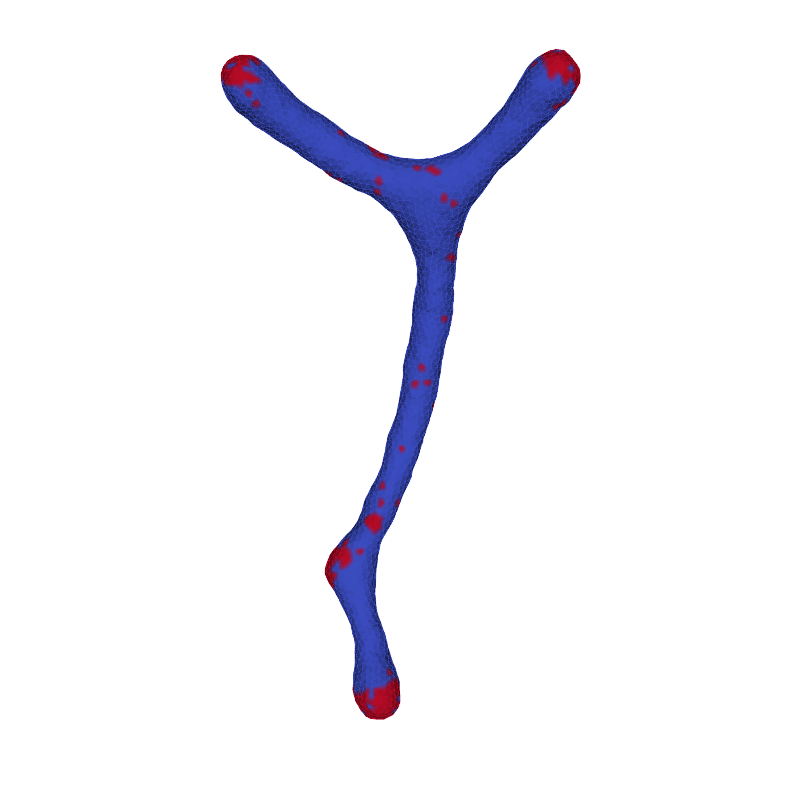}};
	\node[inner sep=0pt] at (axis cs:0.12983690438119602,0.6896551724137931) {\includegraphics[width=0.16\textwidth]{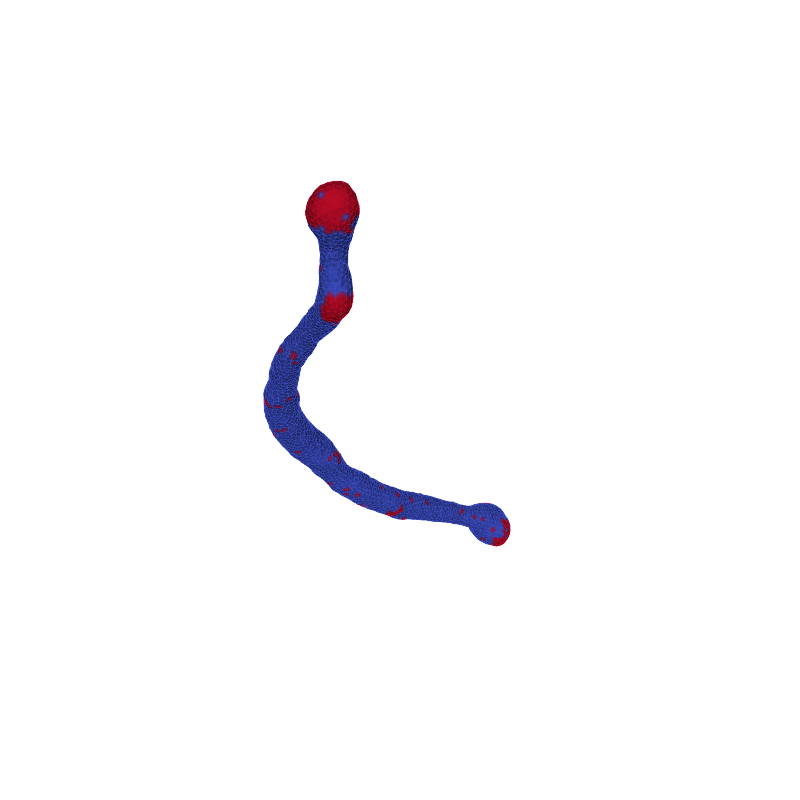}};
	\node[inner sep=0pt] at (axis cs:0.1499840102334506,0.6896551724137931) {\includegraphics[width=0.16\textwidth]{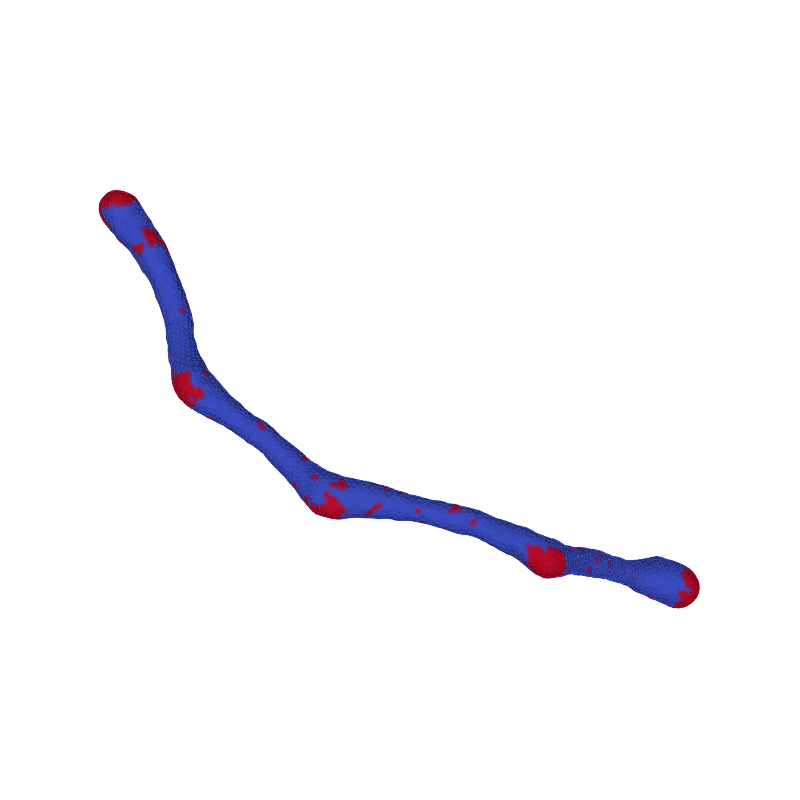}};
	\node[inner sep=0pt] at (axis cs:0.16981132075471697,0.6896551724137931) {\includegraphics[width=0.16\textwidth]{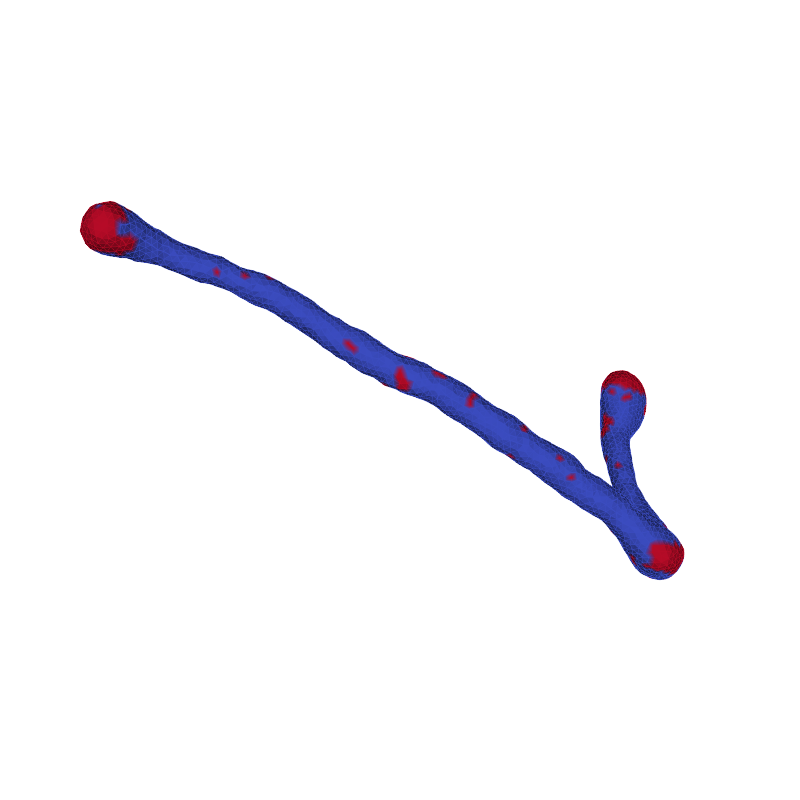}};
	\node[inner sep=0pt] at (axis cs:0.04988807163415414,0.9090909090909091) {\includegraphics[width=0.16\textwidth]{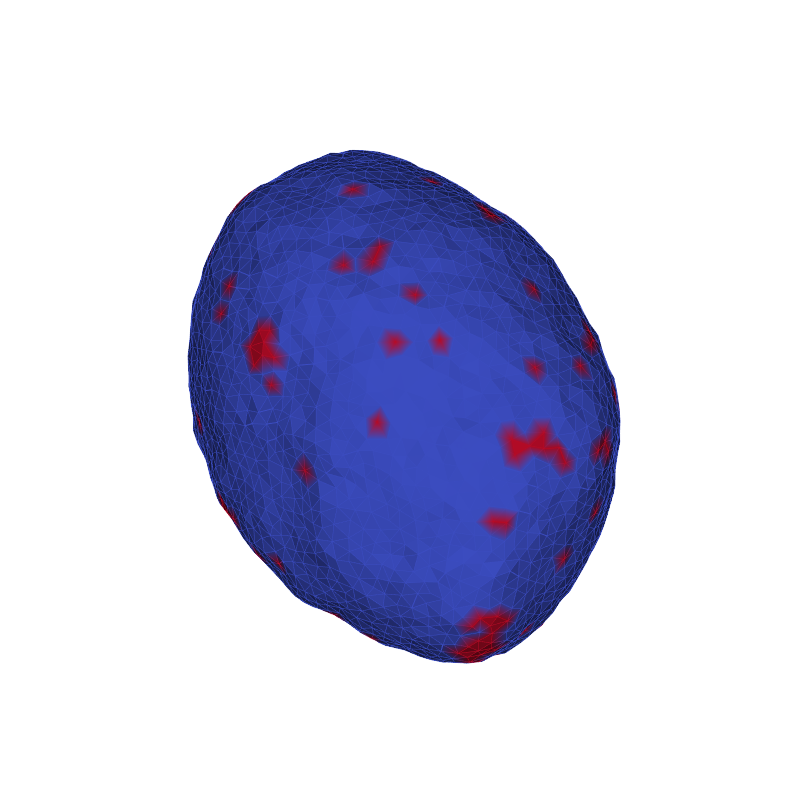}};
	\node[inner sep=0pt] at (axis cs:0.06971538215542053,0.9090909090909091) {\includegraphics[width=0.16\textwidth]{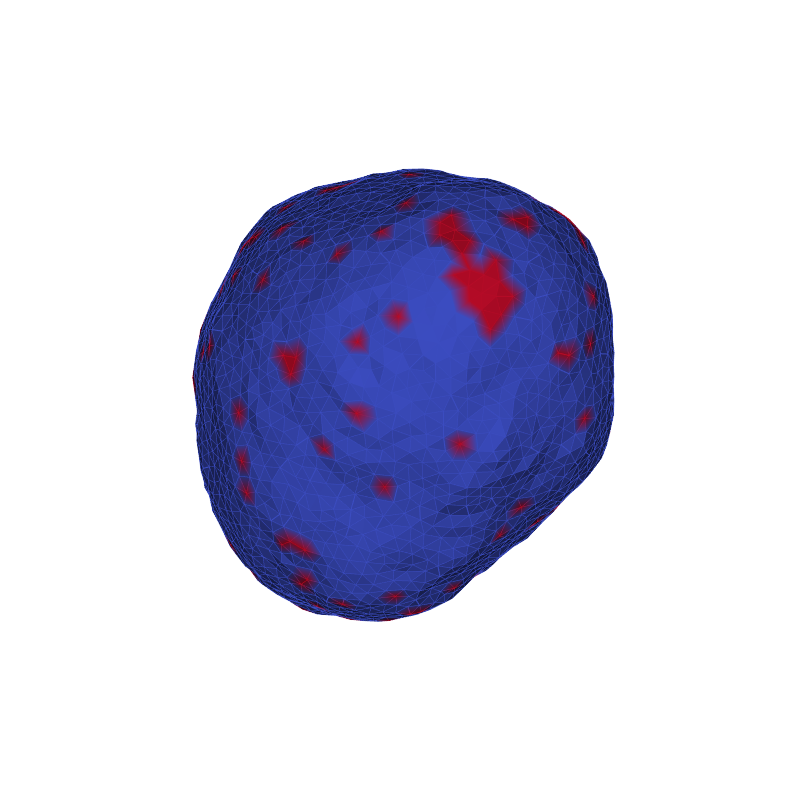}};
	\node[inner sep=0pt] at (axis cs:0.08986248800767509,0.9090909090909091) {\includegraphics[width=0.16\textwidth]{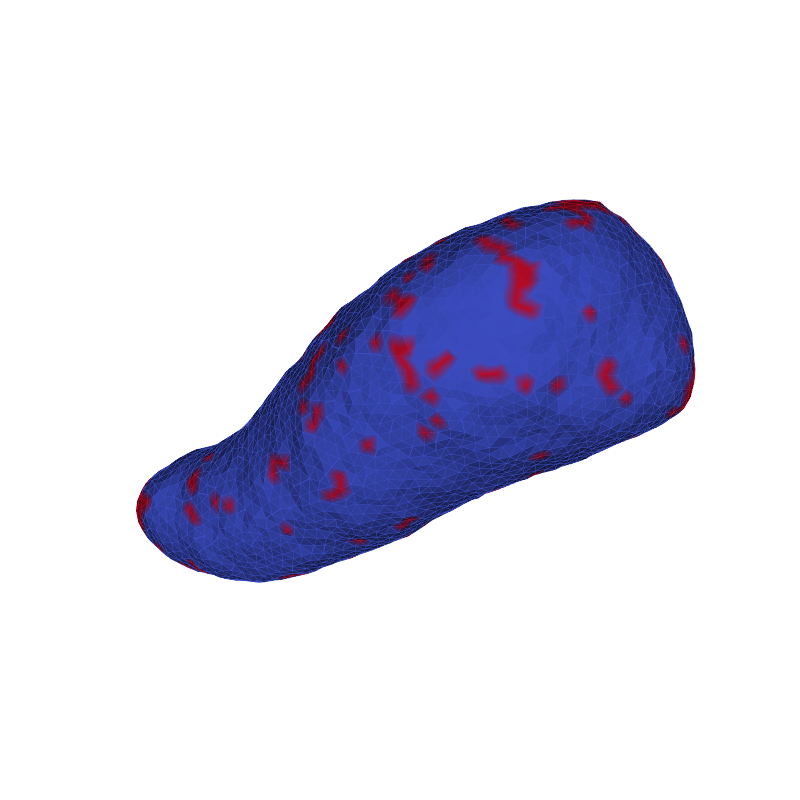}};
	\node[inner sep=0pt] at (axis cs:0.10968979852894148,0.9090909090909091) {\includegraphics[width=0.16\textwidth]{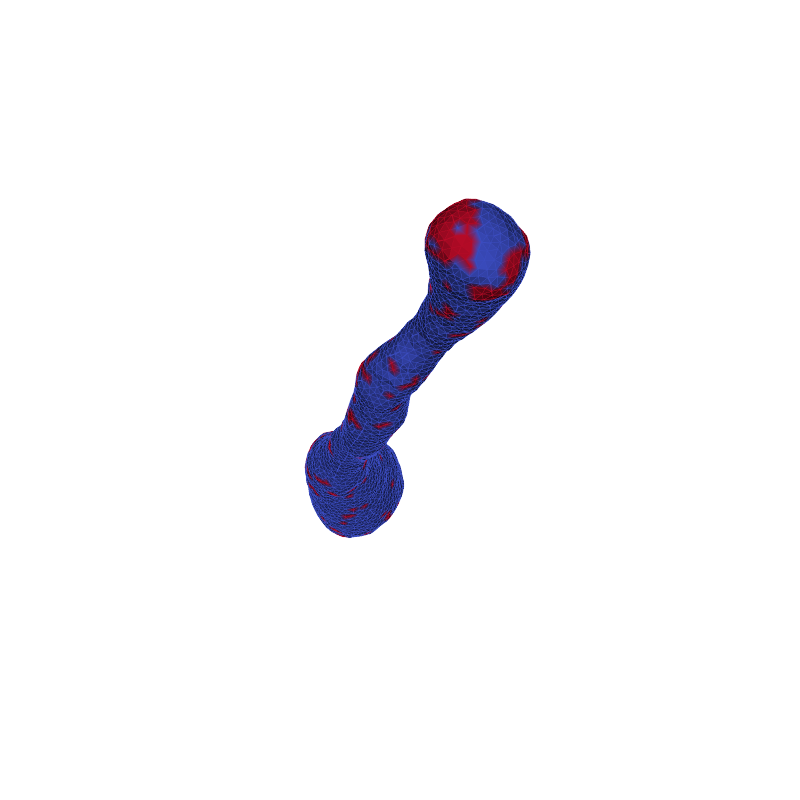}};
	\node[inner sep=0pt] at (axis cs:0.12983690438119602,0.9090909090909091) {\includegraphics[width=0.16\textwidth]{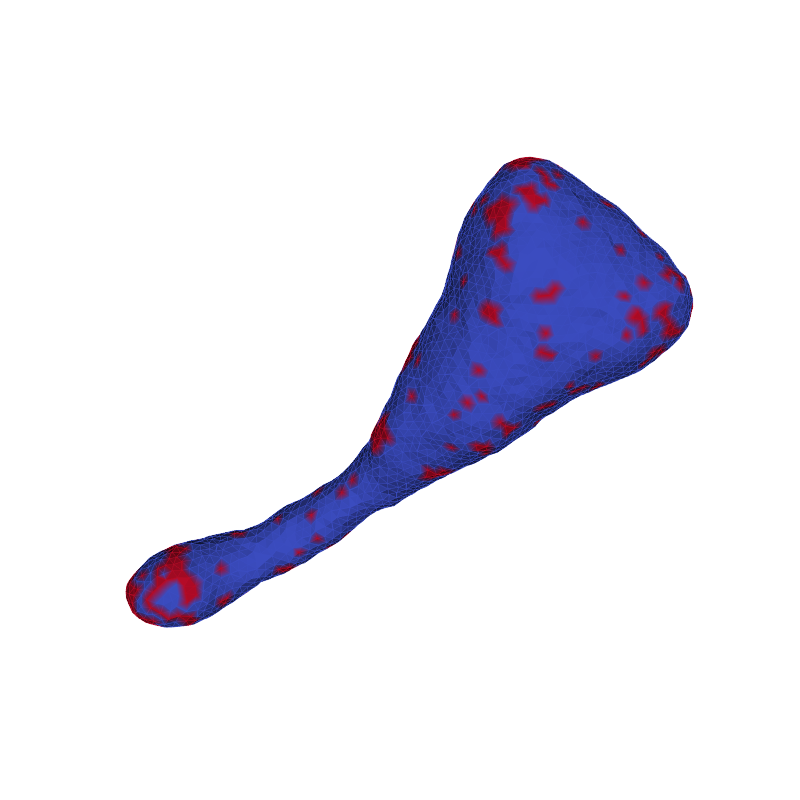}};
	\node[inner sep=0pt] at (axis cs:0.1499840102334506,0.9090909090909091) {\includegraphics[width=0.16\textwidth]{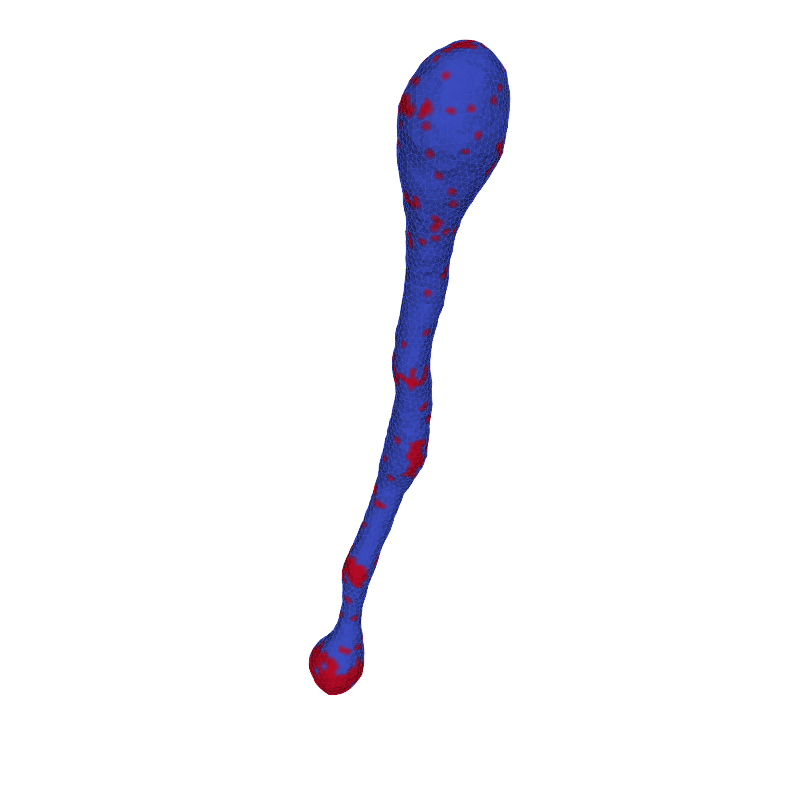}};
	\node[inner sep=0pt] at (axis cs:0.16981132075471697,0.9090909090909091) {\includegraphics[width=0.16\textwidth]{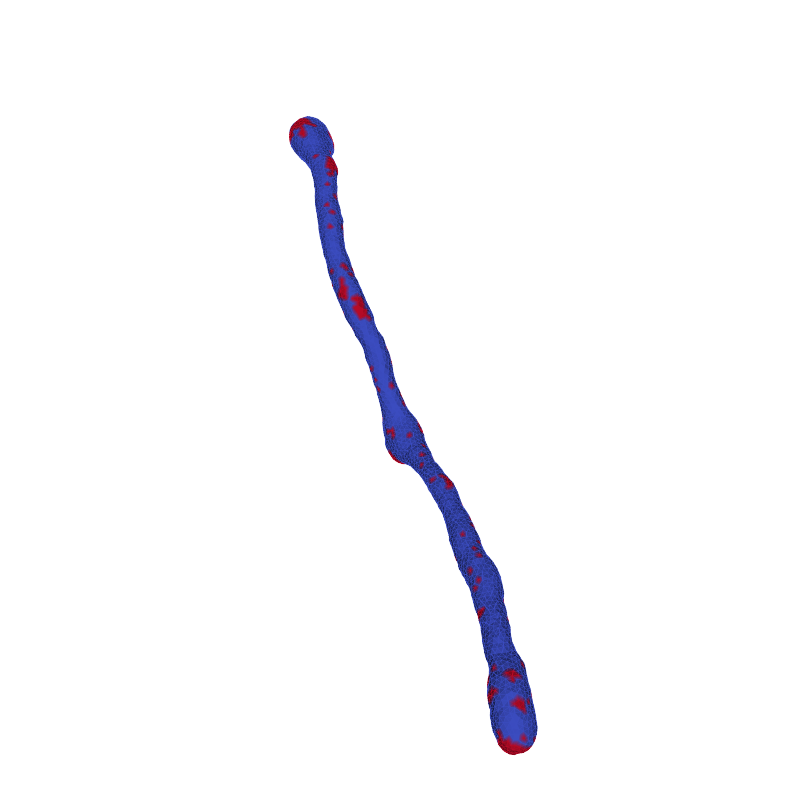}};
	\node[inner sep=0pt] at (axis cs:0.04988807163415414,1.1111111111111112) {\includegraphics[width=0.16\textwidth]{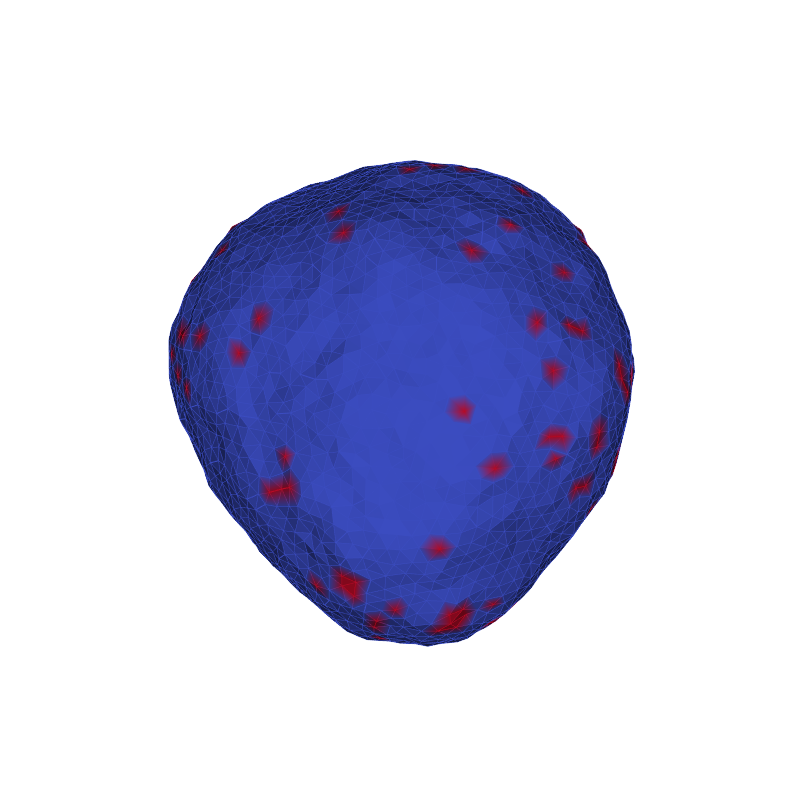}};
	\node[inner sep=0pt] at (axis cs:0.06971538215542053,1.1111111111111112) {\includegraphics[width=0.16\textwidth]{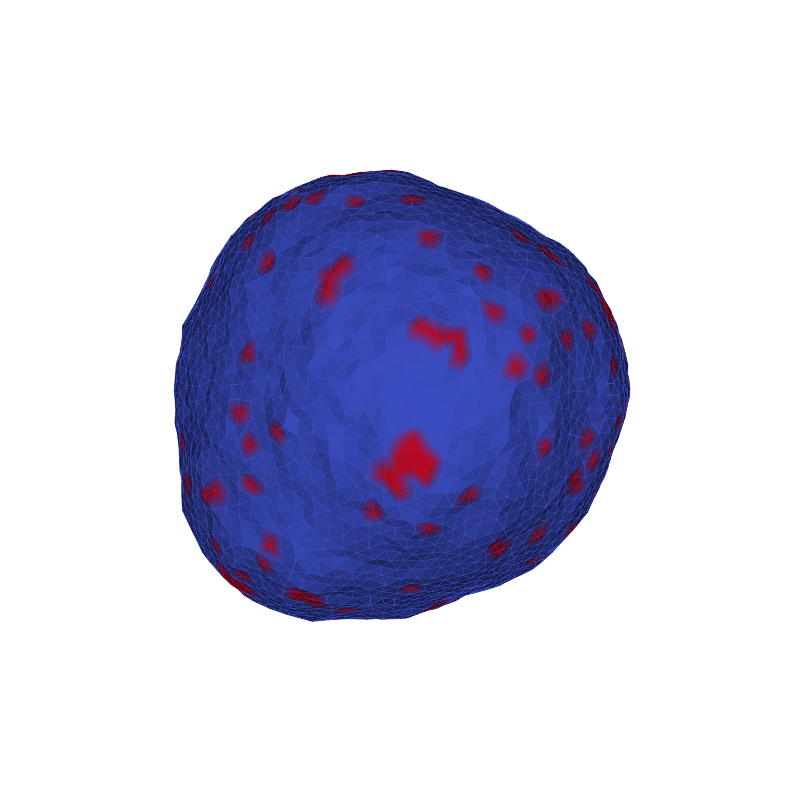}};
	\node[inner sep=0pt] at (axis cs:0.08986248800767509,1.1111111111111112) {\includegraphics[width=0.16\textwidth]{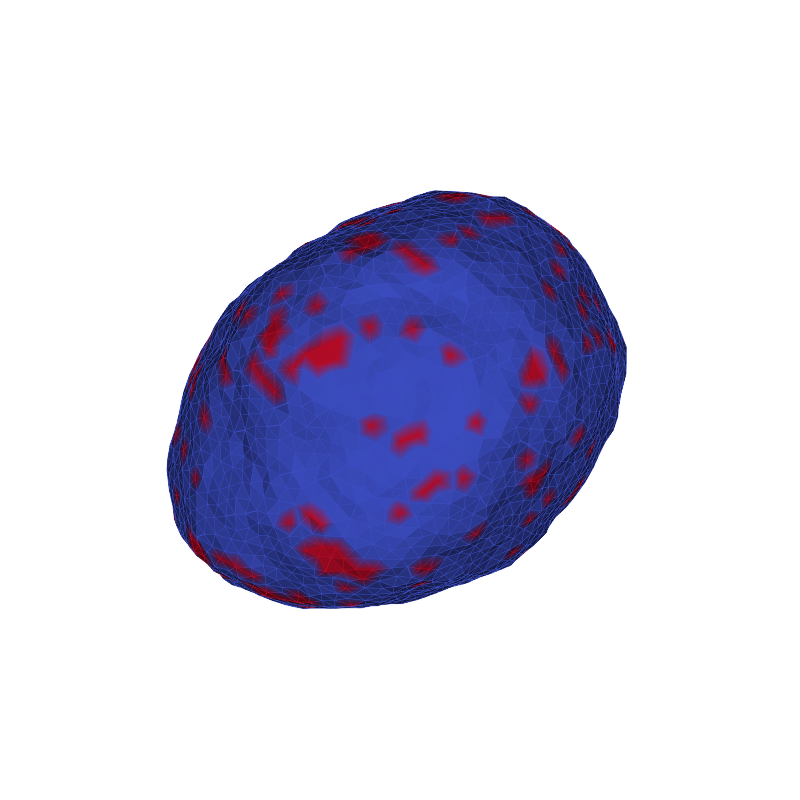}};
	\node[inner sep=0pt] at (axis cs:0.10968979852894148,1.1111111111111112) {\includegraphics[width=0.16\textwidth]{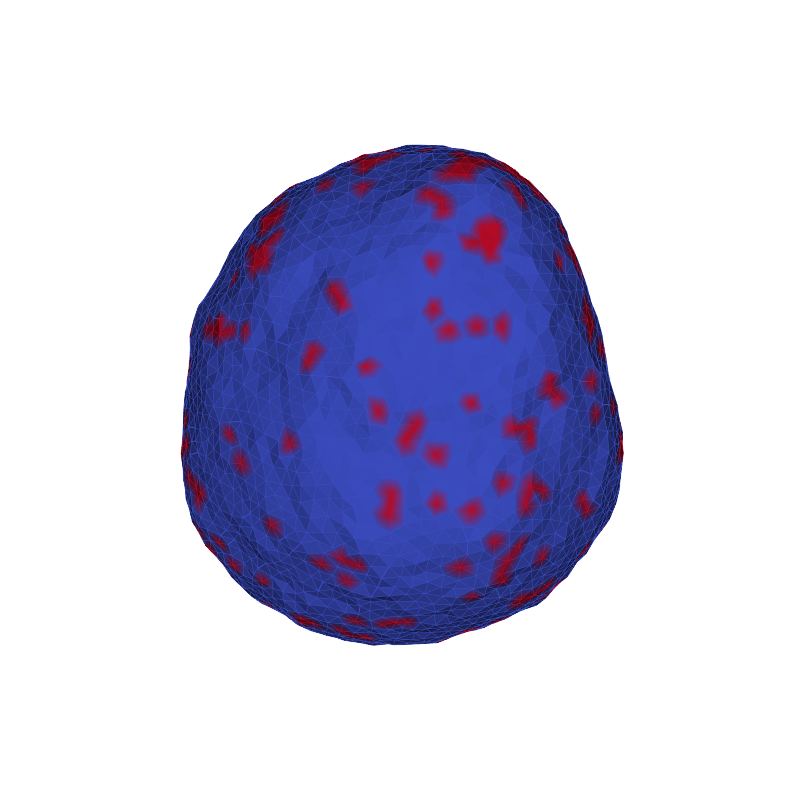}};
	\node[inner sep=0pt] at (axis cs:0.12983690438119602,1.1111111111111112) {\includegraphics[width=0.16\textwidth]{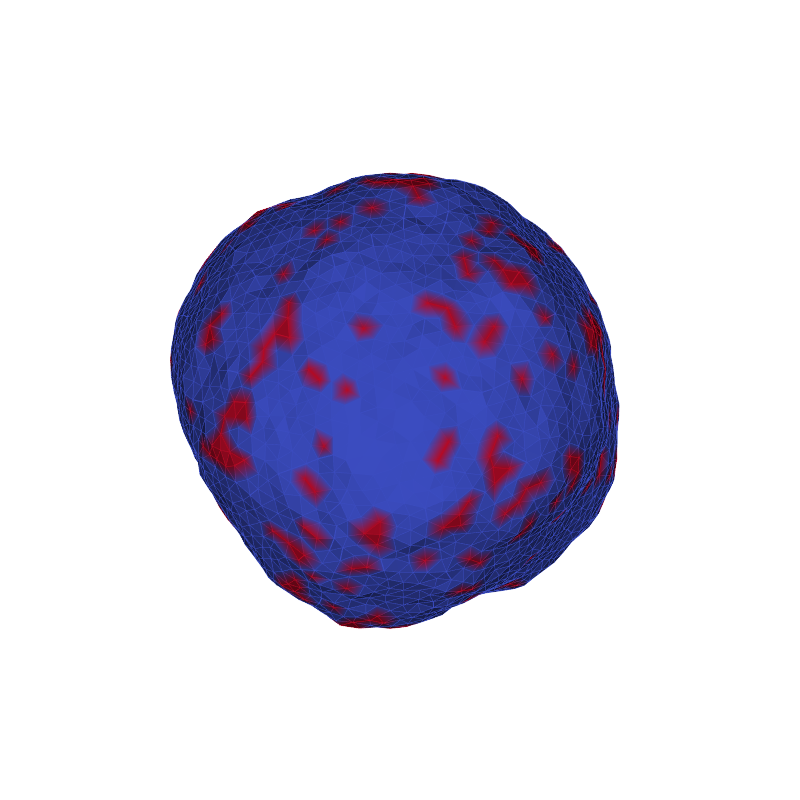}};
	\node[inner sep=0pt] at (axis cs:0.1499840102334506,1.1111111111111112) {\includegraphics[width=0.16\textwidth]{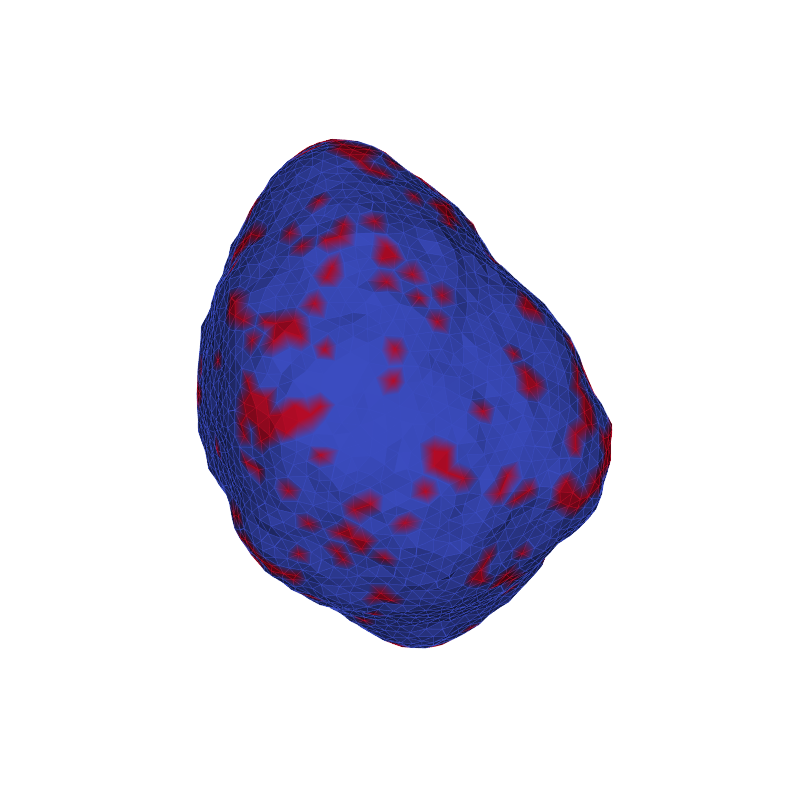}};
	\node[inner sep=0pt] at (axis cs:0.16981132075471697,1.1111111111111112) {\includegraphics[width=0.16\textwidth]{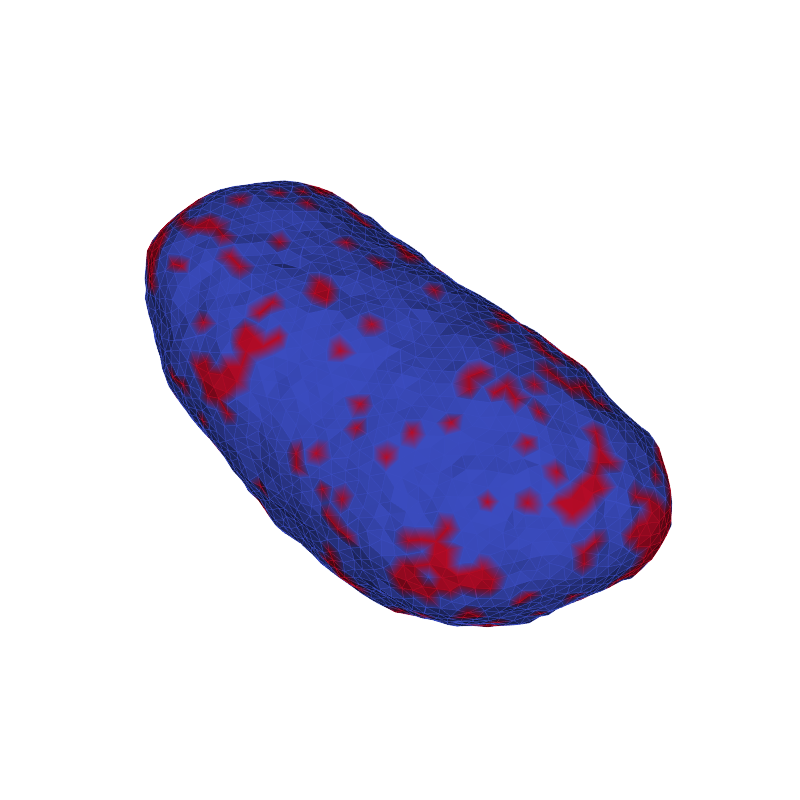}};
	\node[inner sep=0pt] at (axis cs:0.04988807163415414,1.3333333333333333) {\includegraphics[width=0.16\textwidth]{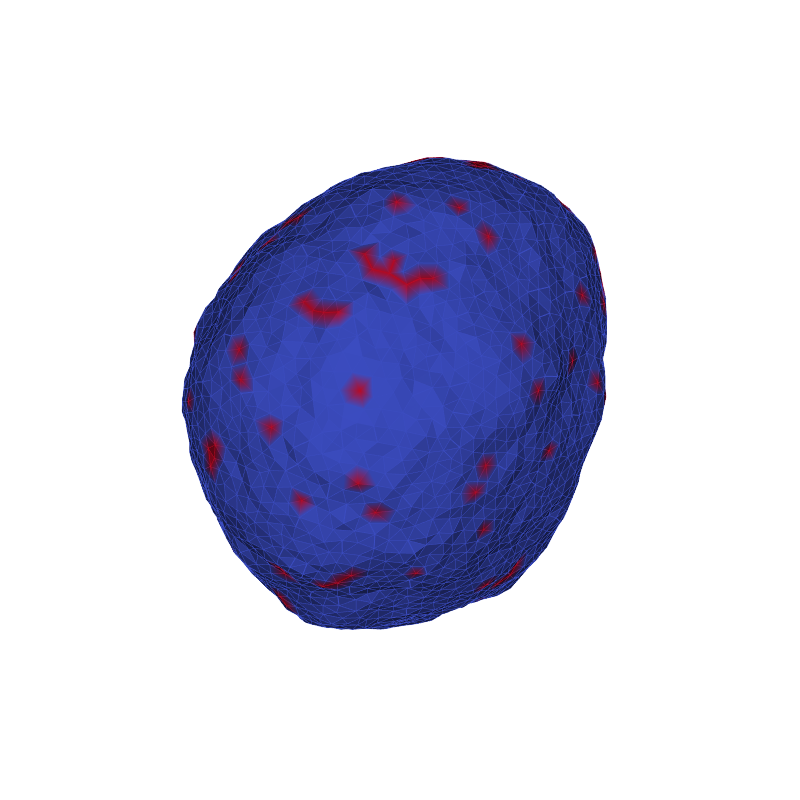}};
	\node[inner sep=0pt] at (axis cs:0.06971538215542053,1.3333333333333333) {\includegraphics[width=0.16\textwidth]{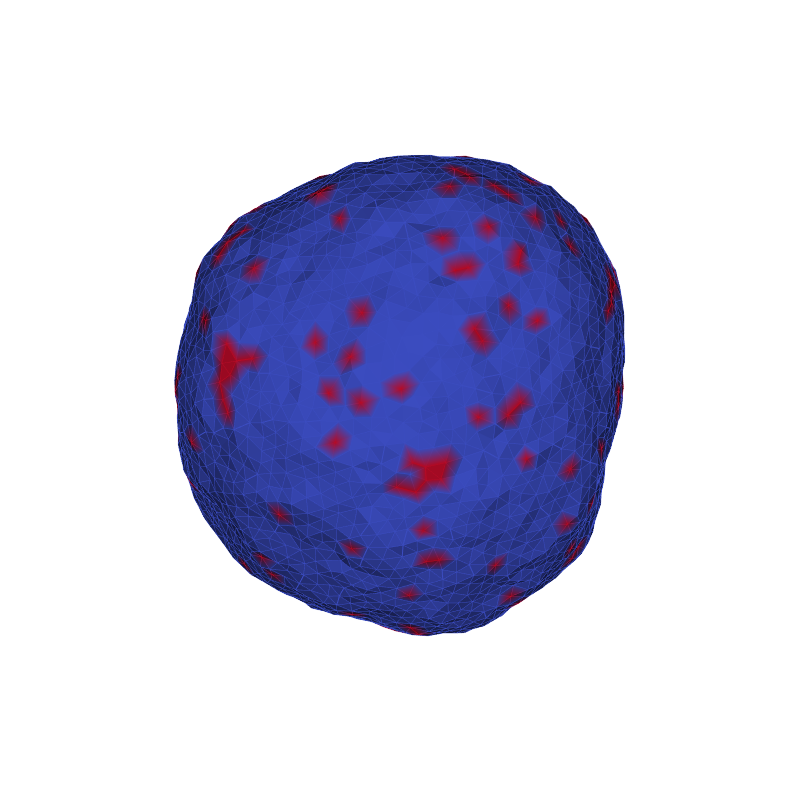}};
	\node[inner sep=0pt] at (axis cs:0.08986248800767509,1.3333333333333333) {\includegraphics[width=0.16\textwidth]{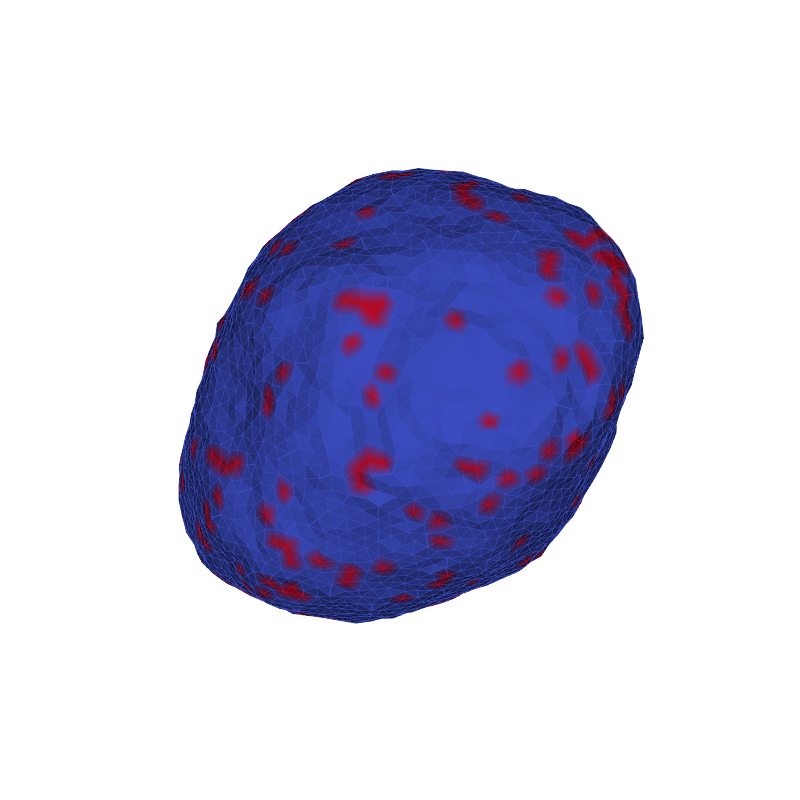}};
	\node[inner sep=0pt] at (axis cs:0.10968979852894148,1.3333333333333333) {\includegraphics[width=0.16\textwidth]{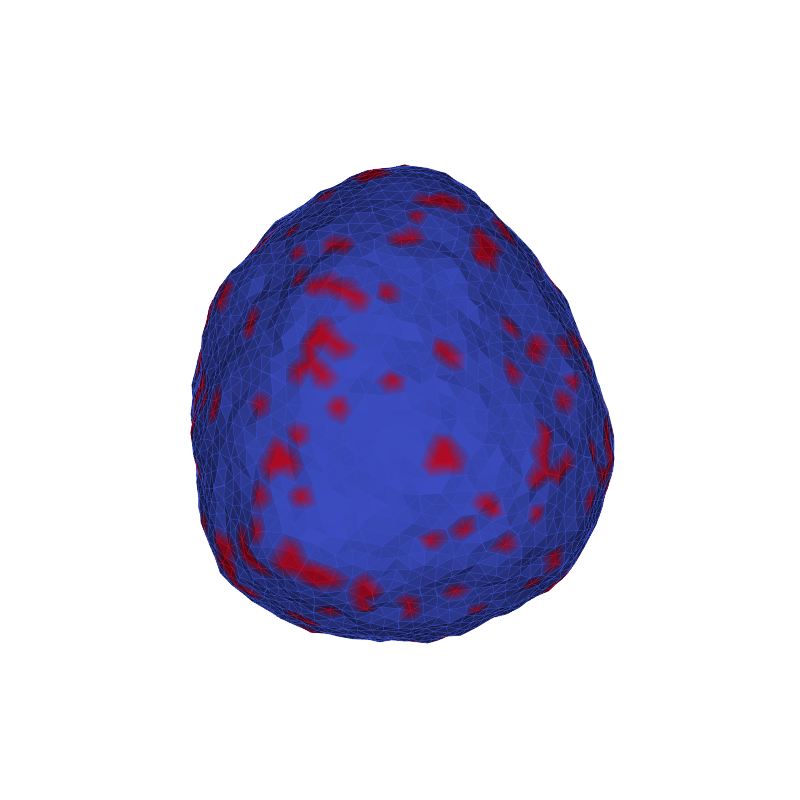}};
	\node[inner sep=0pt] at (axis cs:0.12983690438119602,1.3333333333333333) {\includegraphics[width=0.16\textwidth]{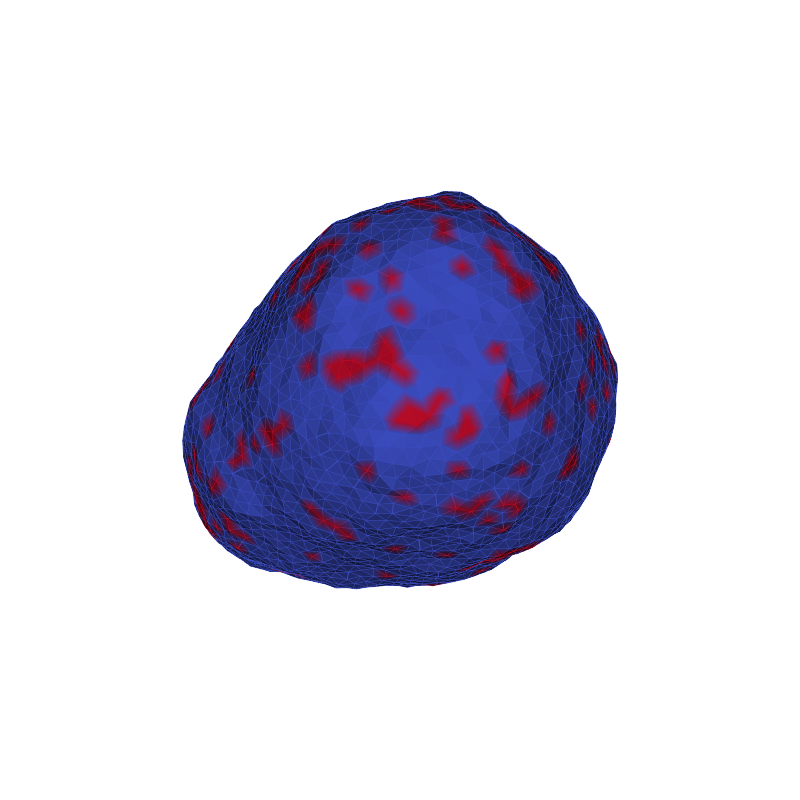}};
	\node[inner sep=0pt] at (axis cs:0.1499840102334506,1.3333333333333333) {\includegraphics[width=0.16\textwidth]{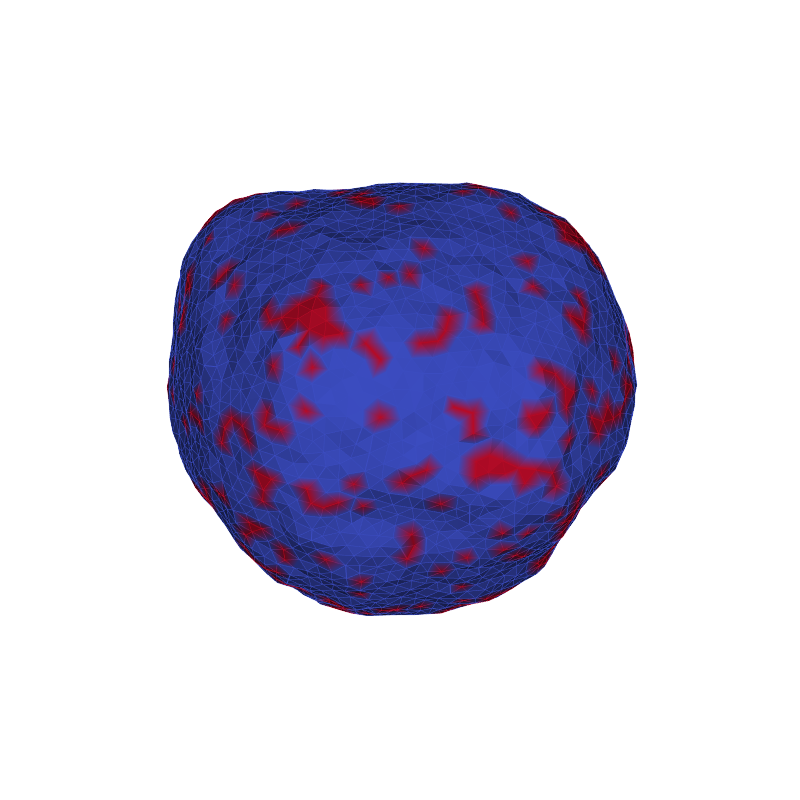}};
	\node[inner sep=0pt] at (axis cs:0.16981132075471697,1.3333333333333333) {\includegraphics[width=0.16\textwidth]{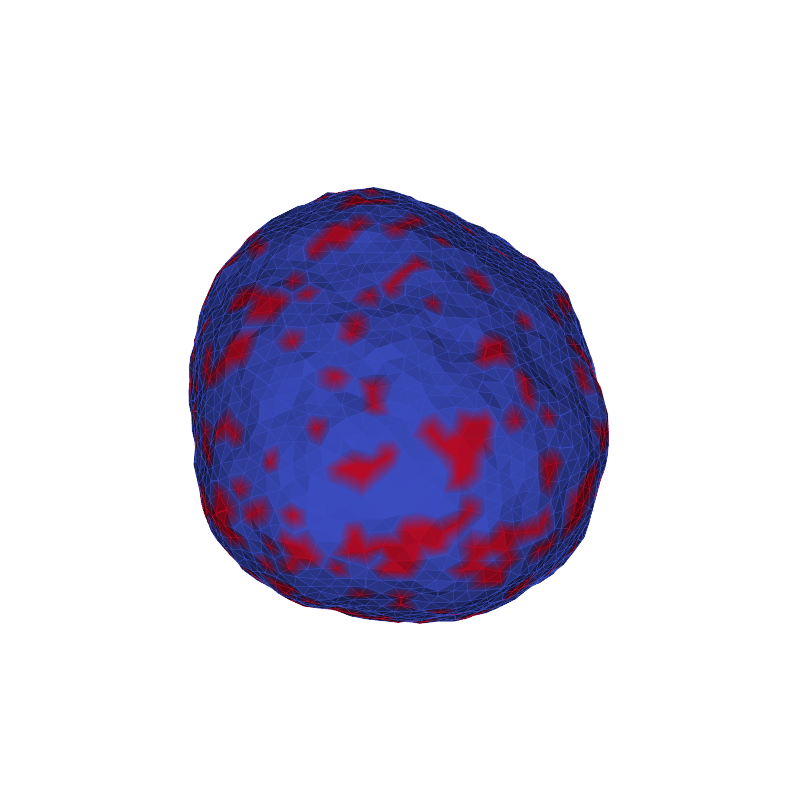}};
	\node[inner sep=0pt] at (axis cs:0.04988807163415414,1.5384615384615385) {\includegraphics[width=0.16\textwidth]{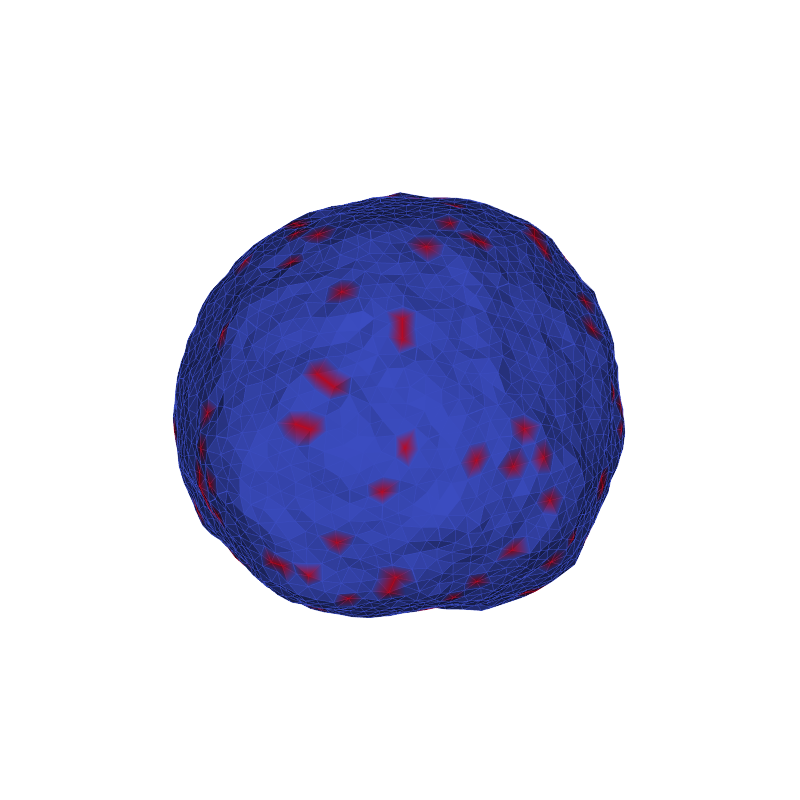}};
	\node[inner sep=0pt] at (axis cs:0.06971538215542053,1.5384615384615385) {\includegraphics[width=0.16\textwidth]{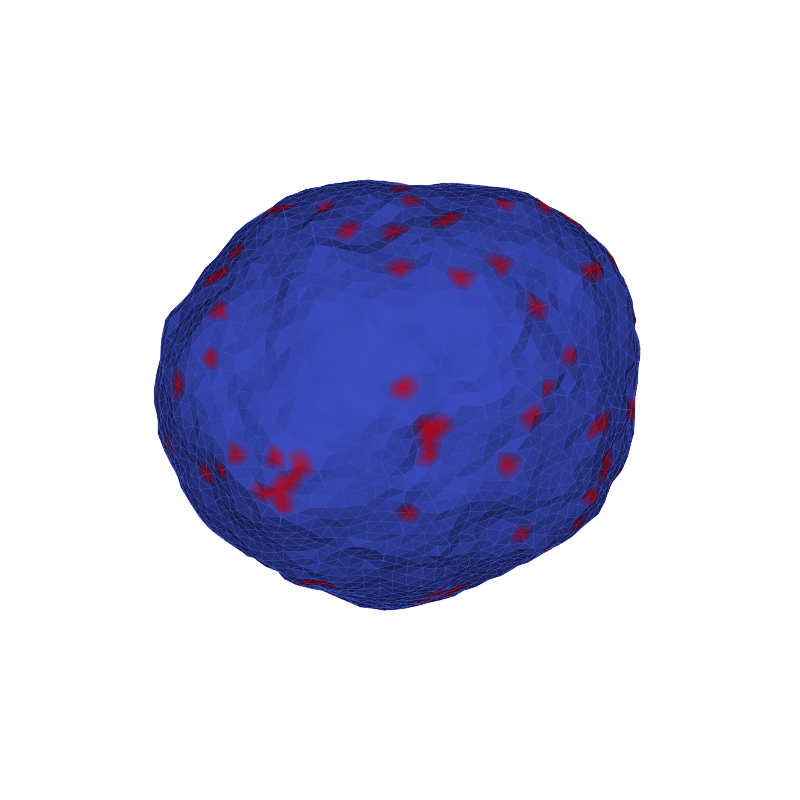}};
	\node[inner sep=0pt] at (axis cs:0.08986248800767509,1.5384615384615385) {\includegraphics[width=0.16\textwidth]{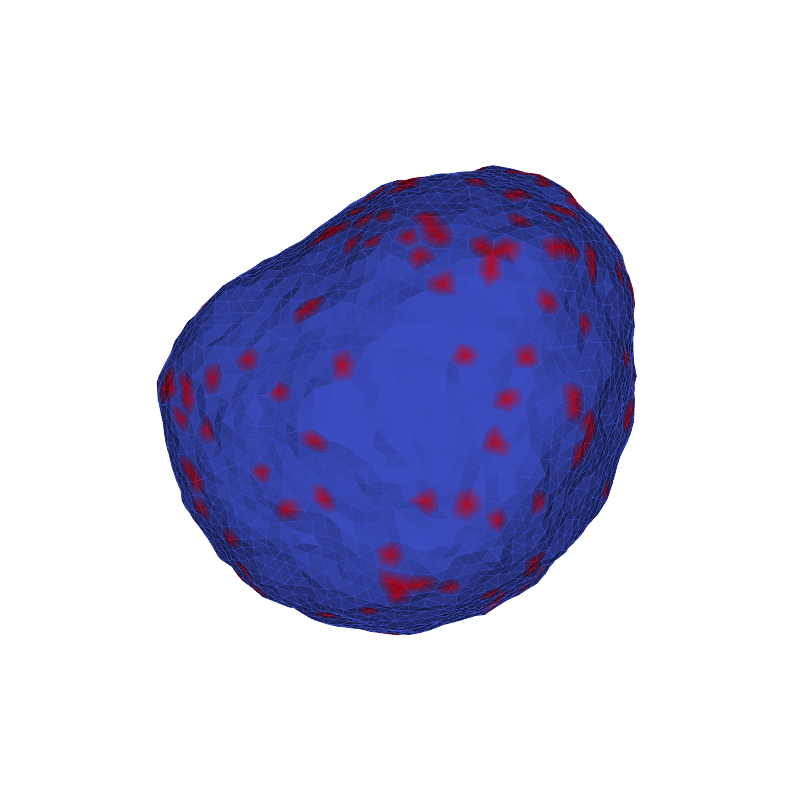}};
	\node[inner sep=0pt] at (axis cs:0.10968979852894148,1.5384615384615385) {\includegraphics[width=0.16\textwidth]{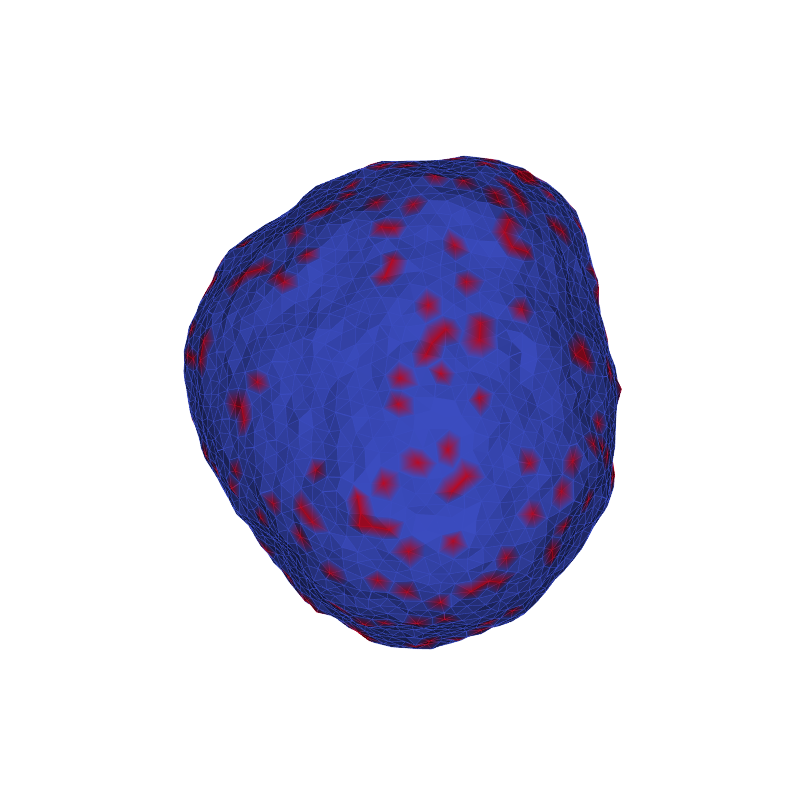}};
	\node[inner sep=0pt] at (axis cs:0.12983690438119602,1.5384615384615385) {\includegraphics[width=0.16\textwidth]{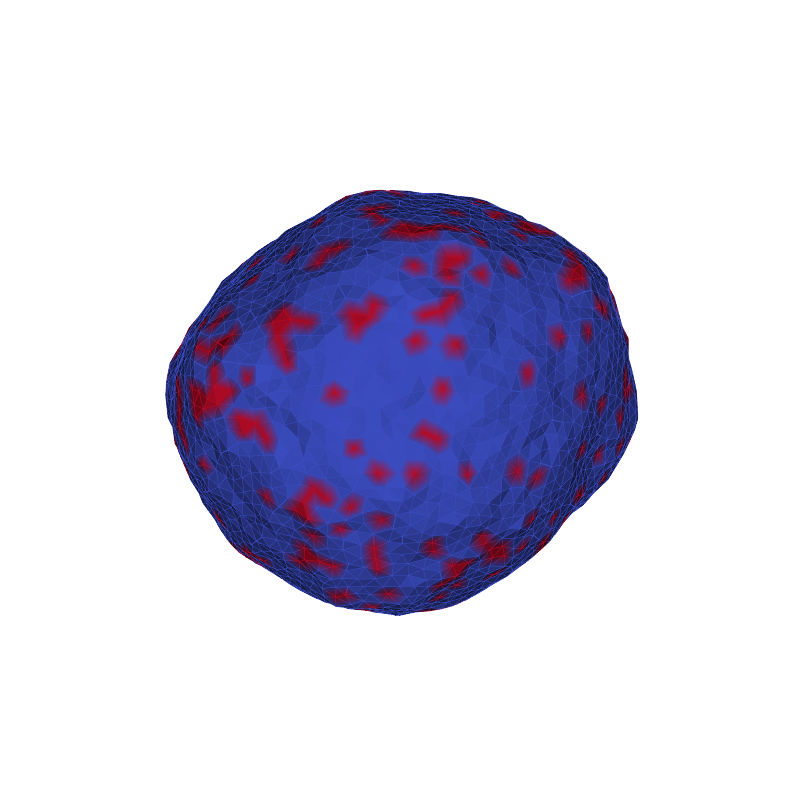}};
	\node[inner sep=0pt] at (axis cs:0.1499840102334506,1.5384615384615385) {\includegraphics[width=0.16\textwidth]{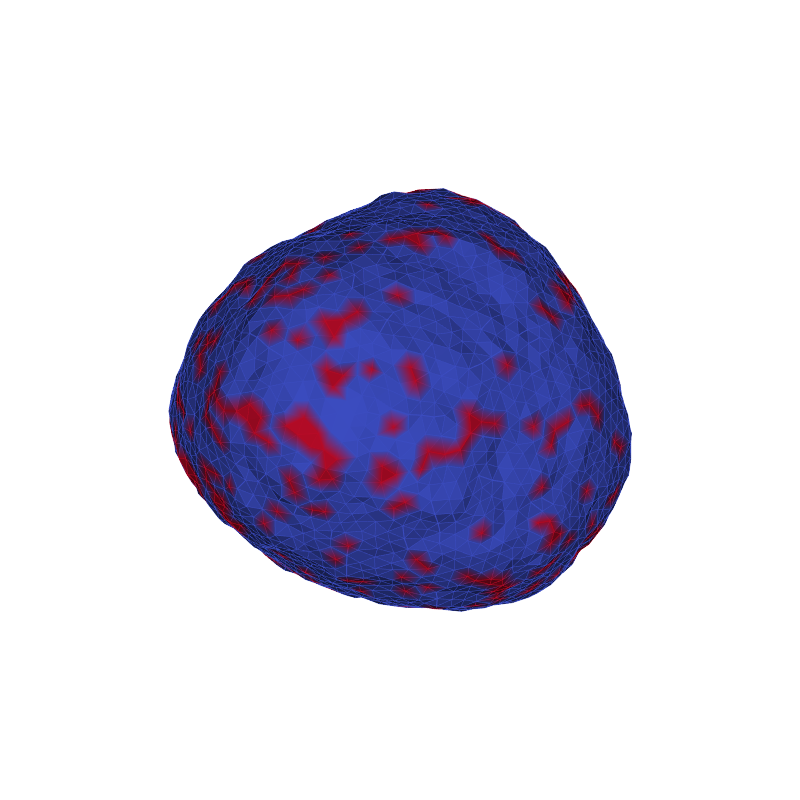}};
	\node[inner sep=0pt] at (axis cs:0.16981132075471697,1.5384615384615385) {\includegraphics[width=0.16\textwidth]{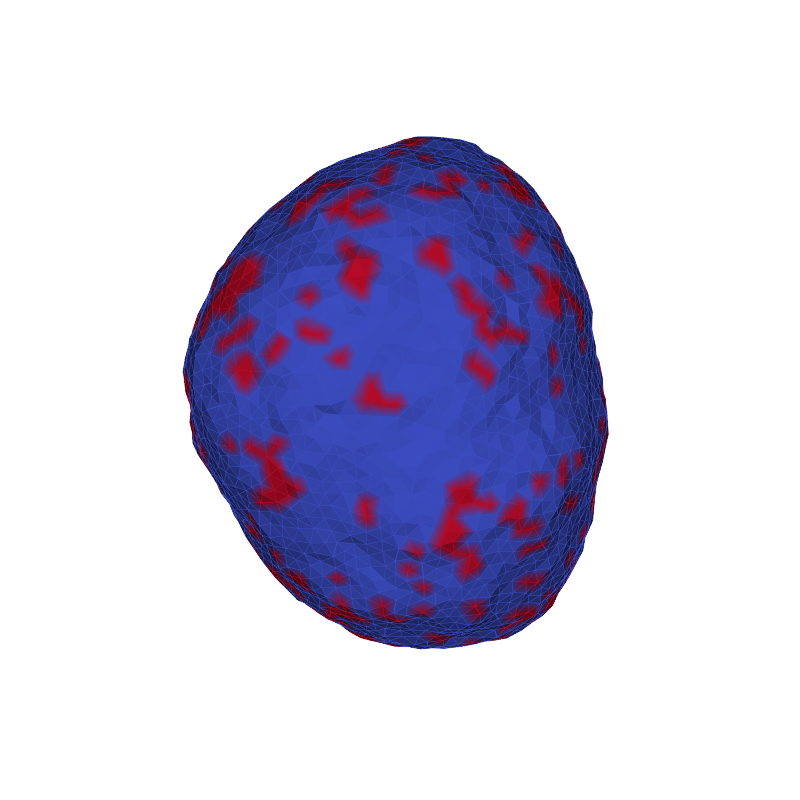}};

	\addplot[domain=0.045:0.14]{12*x*(1-x)};
	
	\addplot [mark=*,mark options={solid, scale=1.0}, style={dashed}, color=green!80
] table[y error minus index=2, y error plus index=3] {./figs/data/hydra_demixing_f1.dat};

	\addplot [mark=*,mark options={solid, scale=1.0}, style={dashed}, color=red!80
] table[y error minus index=2, y error plus index=3] {./figs/data/hydra_phase_transition_f1_error_bars.dat};

\end{axis}
\end{tikzpicture}

%% file: figs/asph_c0.tex
\begin{tikzpicture}[scale=0.5]
\begin{axis}[
    font=\huge,
    title={\bf (c)},
    xlabel={$T/T_0$},
    ylabel={$Asph$},
    legend entries={$\rho=11\%$},
]
\input ./figs/asph_c0_addplots.tex
	
\draw [dashed,thick,red] (axis cs:1.02,-0.05) -- (axis cs:1.02,1.05);

\node[anchor=center] (3zdesne) at (axis cs:1.35,0.4) {\includegraphics[width=0.8\columnwidth]{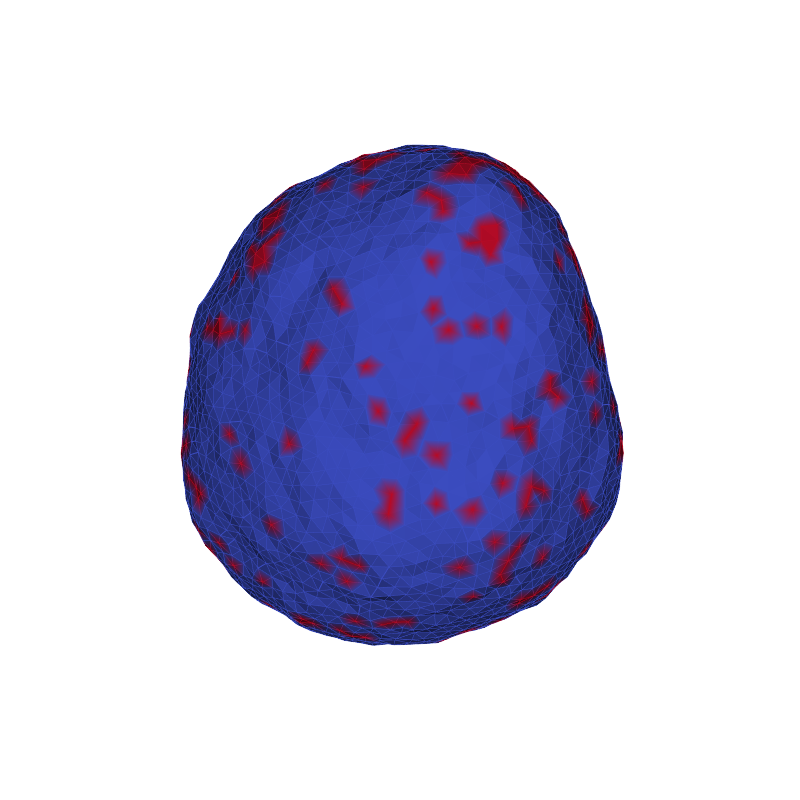}}; 
\draw[->] ([xshift=-2.6em,yshift=-2.8em]3zdesne.center) -- (axis cs:1.15,0.08);

\node[anchor=center] (4zdesne) at (axis cs:0.85,0.1) {\includegraphics[width=0.8\columnwidth]{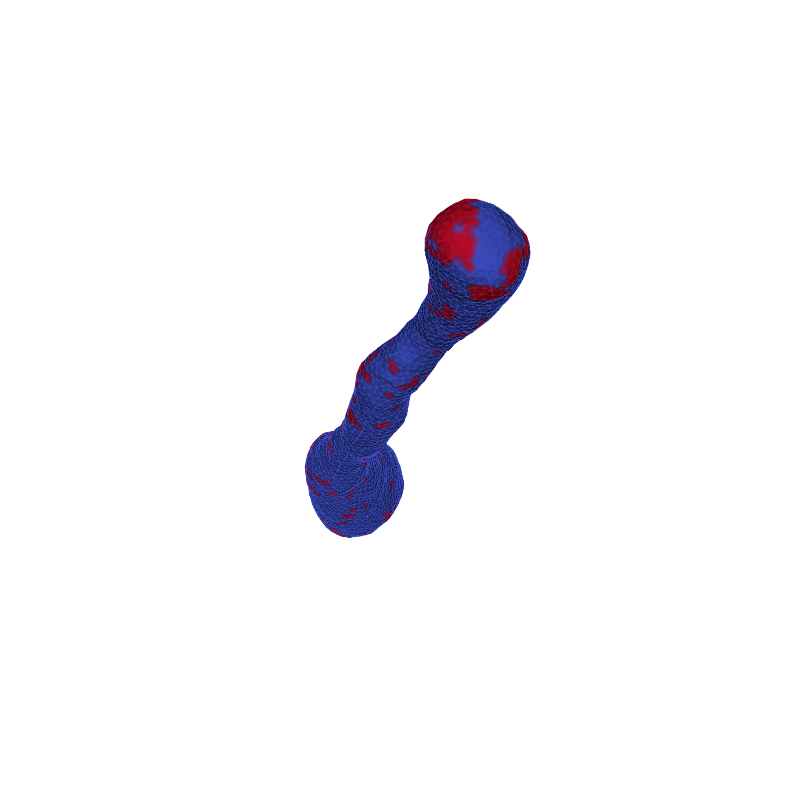}}; 
\draw[->] ([xshift=0.8em,yshift=3.0em]4zdesne.center) -- (axis cs:0.9,0.45);

\node[anchor=center] (5zdesne) at (axis cs:0.6,0.7) {\includegraphics[width=0.8\columnwidth]{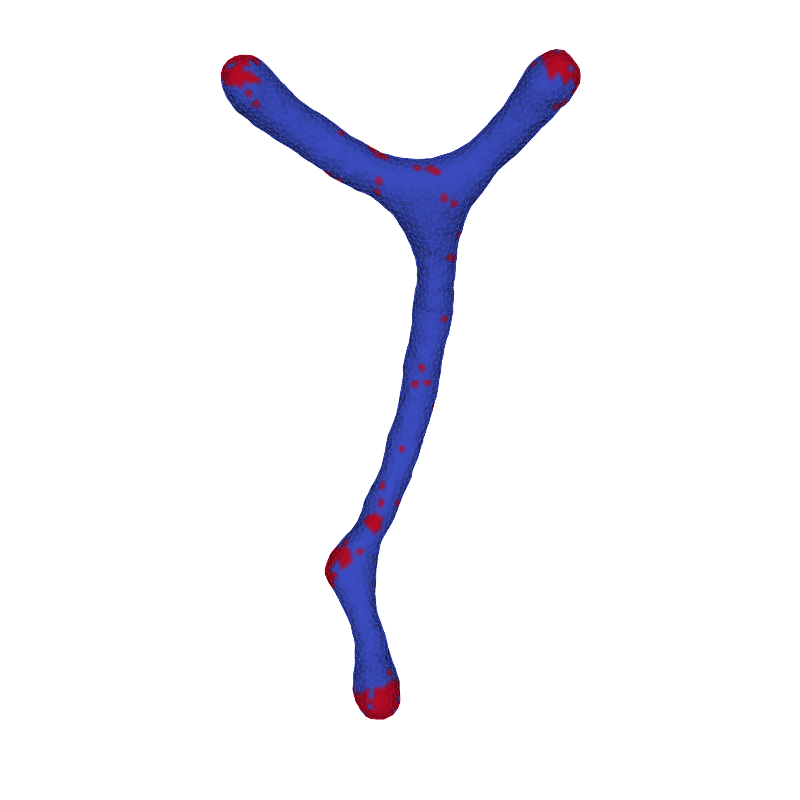}}; 
\draw[->] ([xshift=0.5em,yshift=-1.8em]5zdesne.center) -- (axis cs:0.67,0.37);

\end{axis}
\end{tikzpicture}


%% file: figs/pancaketohydra.tex
\begin{tikzpicture}

\definecolor{color0}{rgb}{0.12156862745098,0.466666666666667,0.705882352941177}
\definecolor{color3}{rgb}{0.83921568627451,0.152941176470588,0.156862745098039}
\definecolor{color2}{rgb}{0.172549019607843,0.627450980392157,0.172549019607843}
\definecolor{color1}{rgb}{1,0.498039215686275,0.0549019607843137}

\begin{axis}[
	width=\columnwidth,
	height=0.3\textheight,
	ylabel near ticks,
	title={{\small \bf (d)}},
	xmin=-0.05,
	xmax=1.05,
	ymin=-14.3088351791494,
	ymax=300,
xtick={-0.2,0,0.2,0.4,0.6,0.8,1,1.2},
xticklabels={,0.0,0.2,0.4,0.6,0.8,1.0,},
tick align=outside,
tick pos=left,
x grid style={lightgray!92.02614379084967!black},
y grid style={lightgray!92.02614379084967!black},
xlabel={$c_0\times l_\mr{min}$},
ylabel={$\left<\bar{N}_{\mathrm vc}\right>$},
]
\addlegendimage{no markers, color0}
\addlegendimage{no markers, color1}
\addlegendimage{no markers, color2}
\addlegendimage{no markers, color3}
\path [draw=color0, semithick] (axis cs:0,0)
--(axis cs:0,0);

\path [draw=color0, semithick] (axis cs:0.111111111111111,1.41627080859486)
--(axis cs:0.111111111111111,1.51327047522439);

\path [draw=color0, semithick] (axis cs:0.222222222222222,1.44991246711183)
--(axis cs:0.222222222222222,1.55587005642738);

\path [draw=color0, semithick] (axis cs:0.333333333333333,1.46430473357079)
--(axis cs:0.333333333333333,1.57876564125319);

\path [draw=color0, semithick] (axis cs:0.444444444444444,1.45683860963317)
--(axis cs:0.444444444444444,1.55845717740272);

\path [draw=color0, semithick] (axis cs:0.555555555555556,1.4660813430849)
--(axis cs:0.555555555555556,1.57997783615679);

\path [draw=color0, semithick] (axis cs:0.666666666666667,1.43019273078011)
--(axis cs:0.666666666666667,1.52964182540602);

\path [draw=color0, semithick] (axis cs:0.777777777777778,1.41441101273927)
--(axis cs:0.777777777777778,1.50547604173725);

\path [draw=color0, semithick] (axis cs:0.888888888888889,1.38711355952408)
--(axis cs:0.888888888888889,1.4771138636);

\path [draw=color0, semithick] (axis cs:1,1.37436128130337)
--(axis cs:1,1.46565820792223);











\path [draw=color2, semithick] (axis cs:0,0)
--(axis cs:0,0);

\path [draw=color2, semithick] (axis cs:0.111111111111111,7.03211640935452)
--(axis cs:0.111111111111111,8.80103468232213);

\path [draw=color2, semithick] (axis cs:0.222222222222222,6.00479634470419)
--(axis cs:0.222222222222222,6.97964109200243);

\path [draw=color2, semithick] (axis cs:0.333333333333333,6.62222649019813)
--(axis cs:0.333333333333333,8.60434727633812);

\path [draw=color2, semithick] (axis cs:0.444444444444444,6.88736112470845)
--(axis cs:0.444444444444444,8.6477968970504);

\path [draw=color2, semithick] (axis cs:0.555555555555556,7.60341290950434)
--(axis cs:0.555555555555556,10.2236329591144);

\path [draw=color2, semithick] (axis cs:0.666666666666667,28.1854741497254)
--(axis cs:0.666666666666667,56.7398462088376);

\path [draw=color2, semithick] (axis cs:0.777777777777778,17.8102893106268)
--(axis cs:0.777777777777778,37.4355251124671);

\path [draw=color2, semithick] (axis cs:0.888888888888889,14.2908455291156)
--(axis cs:0.888888888888889,27.0750511870695);

\path [draw=color2, semithick] (axis cs:1,12.3896155571421)
--(axis cs:1,18.7972677672071);

\path [draw=color3, semithick] (axis cs:0,0)
--(axis cs:0,0);

\path [draw=color3, semithick] (axis cs:0.111111111111111,23.17690781117)
--(axis cs:0.111111111111111,29.5846176470319);

\path [draw=color3, semithick] (axis cs:0.222222222222222,24.0720354771306)
--(axis cs:0.222222222222222,31.6980810786992);

\path [draw=color3, semithick] (axis cs:0.333333333333333,17.8287329374244)
--(axis cs:0.333333333333333,24.3127299719142);

\path [draw=color3, semithick] (axis cs:0.444444444444444,17.1732387921032)
--(axis cs:0.444444444444444,22.6922488039904);

\path [draw=color3, semithick] (axis cs:0.555555555555556,14.154495158655)
--(axis cs:0.555555555555556,19.7909335167072);

\path [draw=color3, semithick] (axis cs:0.666666666666667,109.645296417011)
--(axis cs:0.666666666666667,286.176703582989);

\path [draw=color3, semithick] (axis cs:0.777777777777778,98.1725437871247)
--(axis cs:0.777777777777778,266.379289546209);

\path [draw=color3, semithick] (axis cs:0.888888888888889,92.7699744680654)
--(axis cs:0.888888888888889,279.213525531935);

\path [draw=color3, semithick] (axis cs:1,78.6479227465289)
--(axis cs:1,152.133910586804);

\addplot [semithick, color0, mark=*, mark size=2, mark options={solid}, forget plot]
table {%
0 0
0.111111111111111 1.46477064190963
0.222222222222222 1.50289126176961
0.333333333333333 1.52153518741199
0.444444444444444 1.50764789351794
0.555555555555556 1.52302958962084
0.666666666666667 1.47991727809307
0.777777777777778 1.45994352723826
0.888888888888889 1.43211371156204
1 1.4200097446128
};
\addplot [semithick, color2, mark=*, mark size=2, mark options={solid}, forget plot]
table {%
0 0
0.111111111111111 7.91657554583832
0.222222222222222 6.49221871835331
0.333333333333333 7.61328688326812
0.444444444444444 7.76757901087943
0.555555555555556 8.91352293430938
0.666666666666667 42.4626601792815
0.777777777777778 27.622907211547
0.888888888888889 20.6829483580925
1 15.5934416621746
};
\addplot [semithick, color3, mark=*, mark size=2, mark options={solid}, forget plot]
table {%
0 0
0.111111111111111 26.380762729101
0.222222222222222 27.8850582779149
0.333333333333333 21.0707314546693
0.444444444444444 19.9327437980468
0.555555555555556 16.9727143376811
0.666666666666667 197.911
0.777777777777778 182.275916666667
0.888888888888889 185.99175
1 115.390916666667
};




\node at (axis cs:0.5,1)[
  scale=0.6,
  anchor=base,
  text=black,
  rotate=0.0
]{ };
\node at (axis cs:0,1)[
  scale=0.6,
  anchor=base west,
  text=black,
  rotate=0.0
]{ };
\node at (axis cs:1,1)[
  scale=0.6,
  anchor=base east,
  text=black,
  rotate=0.0
]{ };

\node[anchor=center] (c01) at (axis cs:0.07,80) {\includegraphics[width=0.30\columnwidth]{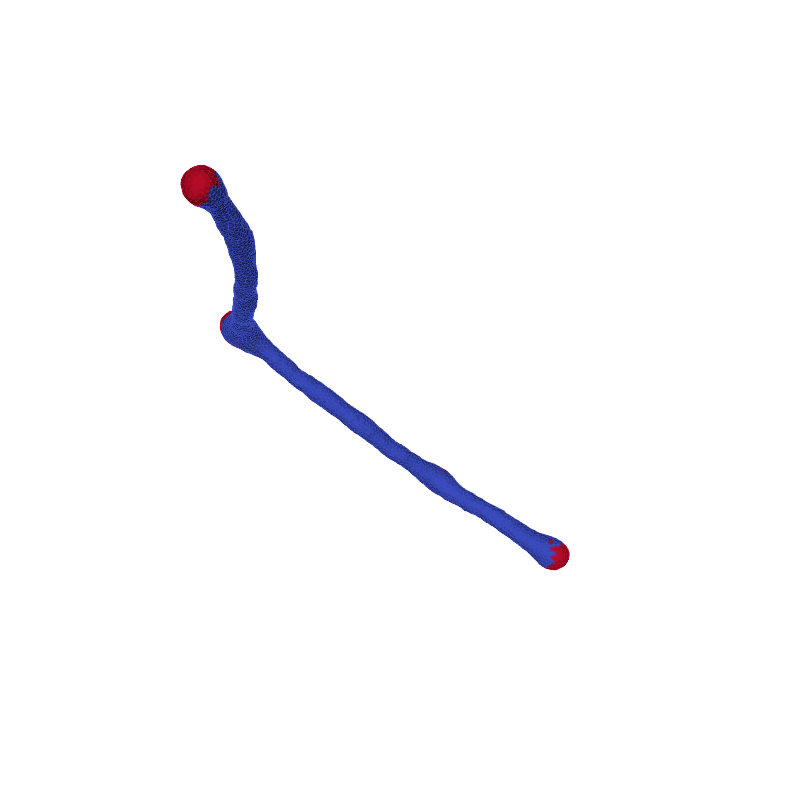}};
\draw[draw=color3,->] ([yshift=-0.5em,xshift=-0.6em]c01.center) -- (axis cs:0.005,9);

\node[anchor=center] (c01) at (axis cs:0.15,250) {\includegraphics[width=0.30\columnwidth]{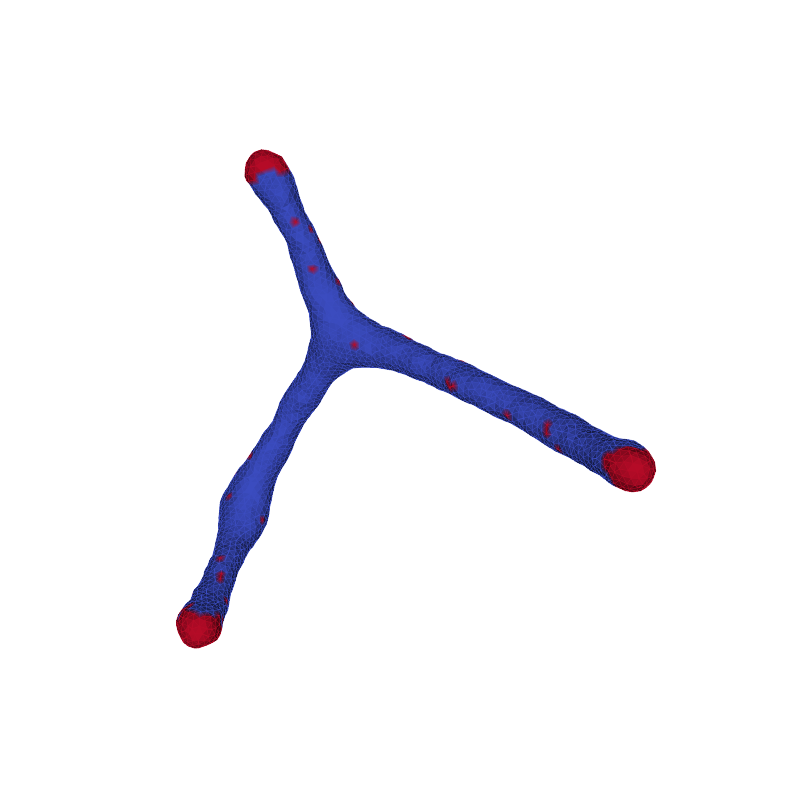}};
\draw[draw=color2,->] ([yshift=-0.5em,xshift=-0.6em]c01.center) -- (axis cs:0.1111,15);

\node[anchor=center] (c01) at (axis cs:0.222222,120) {\includegraphics[width=0.30\columnwidth]{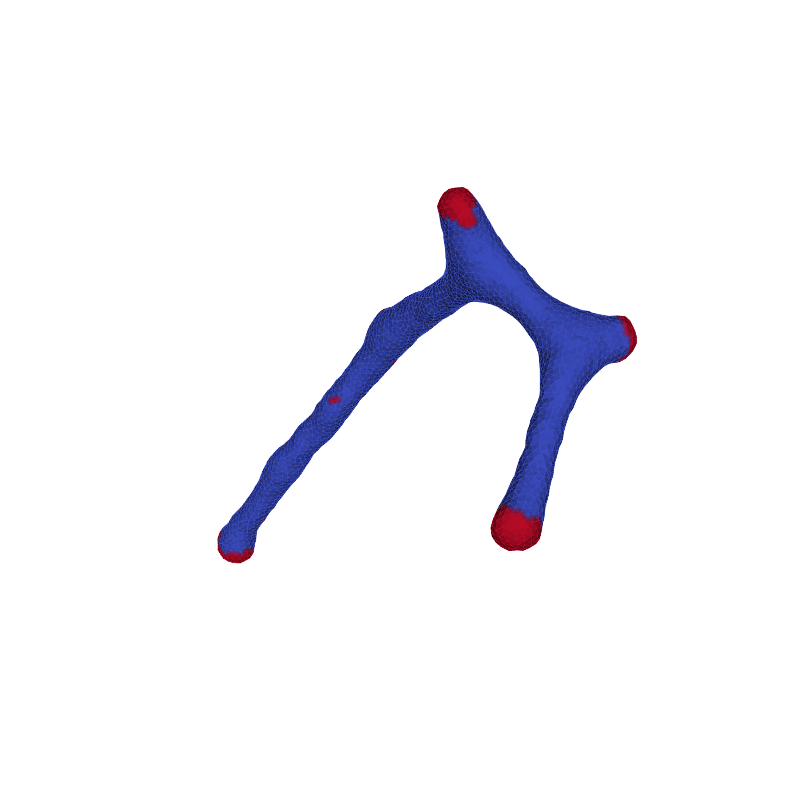}};
\draw[draw=color3,->] ([yshift=-0.5em,xshift=-0.0em]c01.center) -- (axis cs:0.222222,40);


\node[anchor=center] (c01) at (axis cs:0.45,85) {\includegraphics[width=0.30\columnwidth]{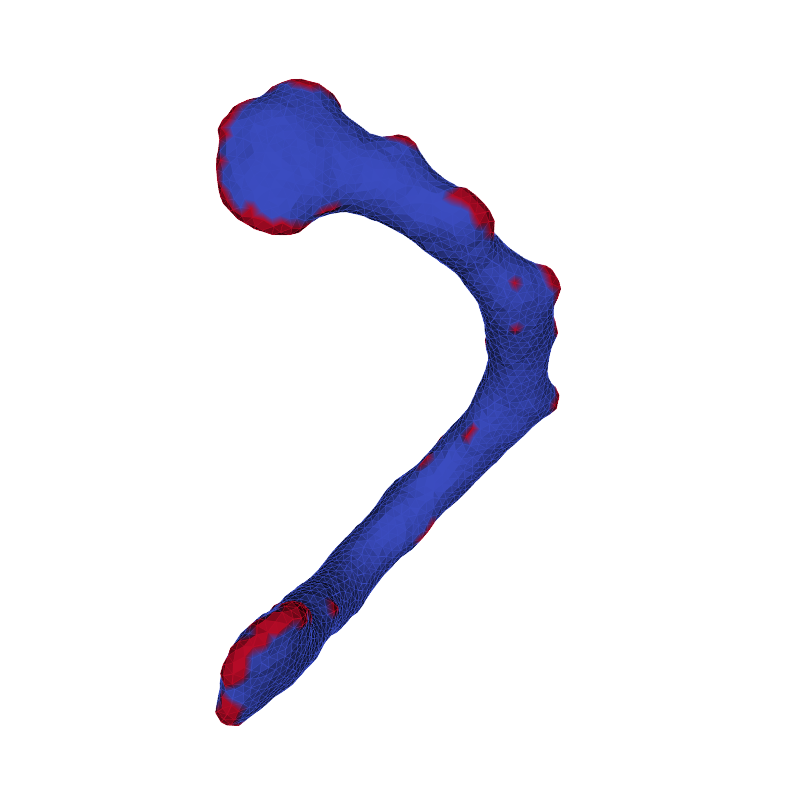}};
\draw[draw=color3,->] ([yshift=-1.5em,xshift=0.6em]c01.center) -- (axis cs:0.54,30);

\node[anchor=center] (c01) at (axis cs:0.48,220) {\includegraphics[width=0.30\columnwidth]{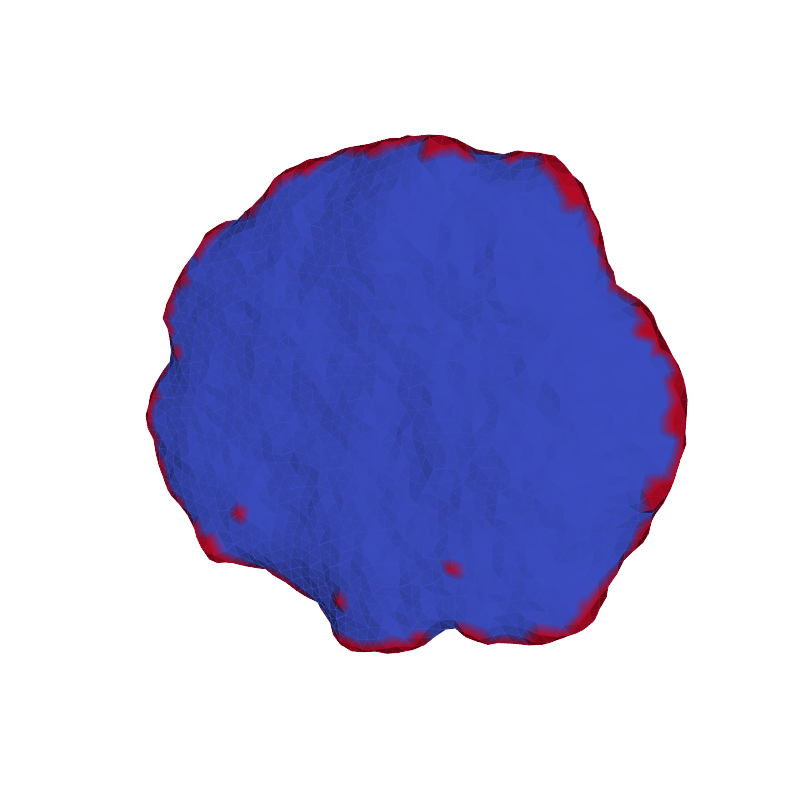}};
\draw[draw=color2,->] ([yshift=-2.3em,xshift=0.5em]c01.center) -- (axis cs:0.65,50);

\node[anchor=center] (c01) at (axis cs:0.9,80) {\includegraphics[width=0.30\columnwidth]{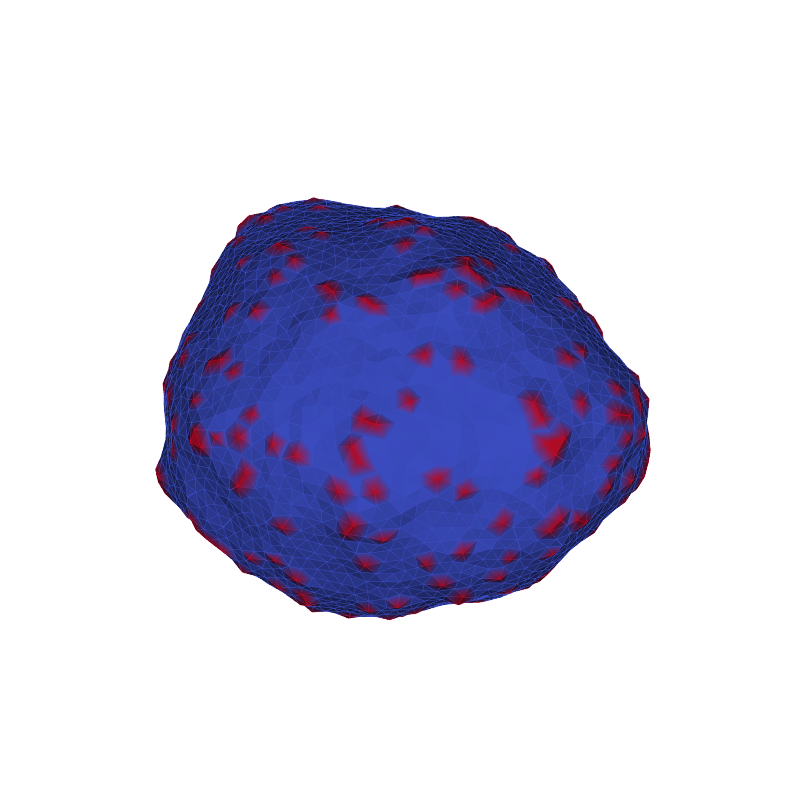}};
\draw[draw=color0,->] ([yshift=-2.0em,xshift=-2.0em]c01.center) -- (axis cs:0.68,10);

\node[anchor=center] (c01) at (axis cs:0.98,232) {\includegraphics[width=0.30\columnwidth]{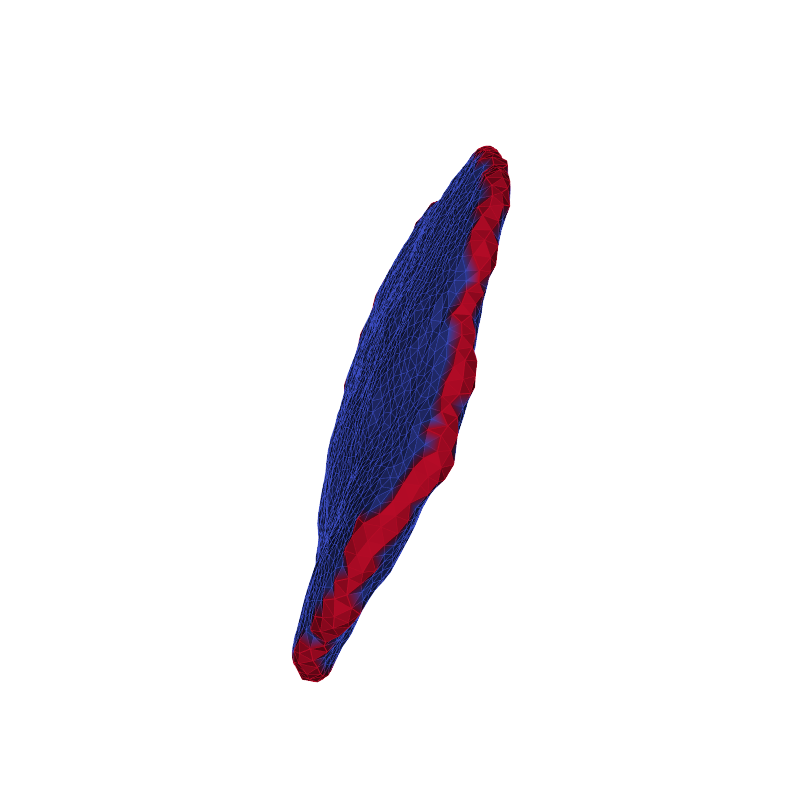}};
\draw[draw=color3,->] ([yshift=-1.0em,xshift=-1.5em]c01.center) -- (axis cs:0.69,200);

\end{axis}

\end{tikzpicture}

%% file: figs/hydra_radius_dependence.tex
\begin{tikzpicture}
\begin{axis}[
	xlabel={$\kappa/F \times l_{min}$},
	scaled x ticks=manual:{}{\pgfmathparse{(#1)/1000}},
	tick label style={/pgf/number format/fixed},
	ylabel={$R_c$},
	grid=none,
	tick label style={/pgf/number format/fixed},
]
	\addplot[style={solid}, color=black!80,domain=2:80]{(1.5783*2*x)^(1/3)};

	\addplot [mark=*,mark options={solid, scale=0.5}, style={dashed}, color=red!80, error bars/.cd, y dir=both, y explicit, error bar style={solid}] table[y error minus index=2, y error plus index=3] {./figs/data/hydra_radius_vs_kapadivf.dat};

\end{axis}
\end{tikzpicture}

%% file: figs/eamcsvstdivtc_w_comparison_F1only.tex
\begin{tikzpicture}
\begin{axis}[
    xlabel={$T/T^\mr{(c)}$},
    ylabel={$\left<\bar{N}_{\mr vc}\right>$},
    legend entries={$w=1.0~kT_0$,$w=1.5~kT_0$},
    legend pos=north east,
     xmax=1,
]
\input ./figs/eamcsvstdivtc_w_comparison_f1_addplots.tex
\coordinate (insetPosition) at (rel axis cs:0.364,0.25);
\end{axis}





\end{tikzpicture}

%% file: figs/hysteresis.tex
\begin{tikzpicture}

\definecolor{color0}{rgb}{0.12156862745098,0.466666666666667,0.705882352941177}
\definecolor{color1}{rgb}{1,0.498039215686275,0.0549019607843137}

\begin{axis}[
xmin=0.6, xmax=1.01875,
ymin=-0.0, ymax=46.1727948805248,
xtick={0.6,0.65,0.7,0.75,0.8,0.85,0.9,0.95,1,1.05},
xticklabels={0.6,,0.7,,0.8,,0.9,,1.0,},
tick align=outside,
tick pos=left,
xlabel={$T/T_0$},
ylabel={$\left<\bar{N}_{\mathrm vc}\right>$},
x grid style={lightgray!92.02614379084967!black},
y grid style={lightgray!92.02614379084967!black},
legend cell align={left},
legend style={draw=white!80.0!black},
legend entries={{Decreasing $T$},{Increasing $T$}},
]
\addlegendimage{no markers, color0}
\addlegendimage{no markers, color1}
\path [draw=color0, semithick] (axis cs:0.625,22.9026973991419)
--(axis cs:0.625,41.1661594172444);

\path [draw=color0, semithick] (axis cs:0.652173913043478,23.7262163366798)
--(axis cs:0.652173913043478,44.0894216868034);

\path [draw=color0, semithick] (axis cs:0.681818181818182,14.2538935555627)
--(axis cs:0.681818181818182,24.3626743655686);

\path [draw=color0, semithick] (axis cs:0.714285714285714,11.4822340735021)
--(axis cs:0.714285714285714,20.6456045472538);

\path [draw=color0, semithick] (axis cs:0.75,7.77953546803685)
--(axis cs:0.75,12.5991436863409);

\path [draw=color0, semithick] (axis cs:0.789473684210526,4.39400355744229)
--(axis cs:0.789473684210526,5.90439493598144);

\path [draw=color0, semithick] (axis cs:0.833333333333333,3.02492974079973)
--(axis cs:0.833333333333333,3.66621556710617);

\path [draw=color0, semithick] (axis cs:0.882352941176471,2.80762542012871)
--(axis cs:0.882352941176471,3.21627652696107);

\path [draw=color0, semithick] (axis cs:0.9375,2.66408239785828)
--(axis cs:0.9375,3.06637294366482);

\path [draw=color0, semithick] (axis cs:1,2.42195781237483)
--(axis cs:1,2.79257220750951);

\path [draw=color1, semithick] (axis cs:0.625,21.802194594375)
--(axis cs:0.625,32.6271346347035);

\path [draw=color1, semithick] (axis cs:0.652173913043478,18.2172736560384)
--(axis cs:0.652173913043478,34.013373259619);

\path [draw=color1, semithick] (axis cs:0.681818181818182,15.5016583957821)
--(axis cs:0.681818181818182,25.3856635337873);

\path [draw=color1, semithick] (axis cs:0.714285714285714,13.1145775552433)
--(axis cs:0.714285714285714,22.2687045254184);

\path [draw=color1, semithick] (axis cs:0.75,9.29657693401385)
--(axis cs:0.75,14.6396856626815);

\path [draw=color1, semithick] (axis cs:0.789473684210526,3.72178414230876)
--(axis cs:0.789473684210526,4.56067249006965);

\path [draw=color1, semithick] (axis cs:0.833333333333333,3.15666559608011)
--(axis cs:0.833333333333333,3.81021246567882);

\path [draw=color1, semithick] (axis cs:0.882352941176471,2.79558626572555)
--(axis cs:0.882352941176471,3.25700251304639);

\path [draw=color1, semithick] (axis cs:0.9375,2.66602140161476)
--(axis cs:0.9375,3.04510754730469);

\path [draw=color1, semithick] (axis cs:1,2.46797015857616)
--(axis cs:1,2.79986212518071);

\addplot [semithick, color0, mark=*, mark size=2, mark options={solid}, forget plot]
table {%
0.625 32.0344284081931
0.652173913043478 33.9078190117416
0.681818181818182 19.3082839605656
0.714285714285714 16.0639193103779
0.75 10.1893395771889
0.789473684210526 5.14919924671187
0.833333333333333 3.34557265395295
0.882352941176471 3.01195097354489
0.9375 2.86522767076155
1 2.60726500994217
};
\addplot [semithick, color1, mark=*, mark size=2, mark options={solid}, forget plot]
table {%
0.625 27.2146646145392
0.652173913043478 26.1153234578287
0.681818181818182 20.4436609647847
0.714285714285714 17.6916410403309
0.75 11.9681312983477
0.789473684210526 4.1412283161892
0.833333333333333 3.48343903087946
0.882352941176471 3.02629438938597
0.9375 2.85556447445972
1 2.63391614187844
};




\node at (axis cs:0.5,1)[
  scale=0.6,
  anchor=base,
  text=black,
  rotate=0.0
]{ };
\node at (axis cs:0,1)[
  scale=0.6,
  anchor=base west,
  text=black,
  rotate=0.0
]{ };
\node at (axis cs:1,1)[
  scale=0.6,
  anchor=base east,
  text=black,
  rotate=0.0
]{ };
\end{axis}

\end{tikzpicture}

%% file: figs/eamcsvsrhoSI.tex
\begin{tikzpicture}

\begin{axis}[
	scaled x ticks=manual:{}{\pgfmathparse{(#1)*100}},
	xlabel={$\rho[\%]$},
	ylabel={$\left<\bar{N}_{\mr vc}\right>$},
	xticklabel style={/pgf/number format/.cd,fixed,precision=5}]
	\input ./figs/eamcsvsnc_f1_addplots_mix.tex
\end{axis}

\end{tikzpicture}

%% file: figs/pancaketranstempvsn.tex
\begin{tikzpicture}
\begin{axis}[
	xlabel={$N\times 1000$},
	scaled x ticks=manual:{}{\pgfmathparse{(#1)/1000}},
	tick label style={/pgf/number format/fixed},
	ylabel={$T^\mr{(c)}/T_0$},
	grid=none,
	tick label style={/pgf/number format/fixed},
]
	\addplot[style={solid}, color=black!80,domain=800:3200]{12*0.095*(1-0.095)*(1+sqrt((1-1/(0.095*0.3544*sqrt(x)))/12))};
	\addplot [mark=*,mark options={solid, scale=0.5}, style={dashed}, color=red!80, error bars/.cd, y dir=both, y explicit, error bar style={solid}] table[y error minus index=2, y error plus index=3] {./figs/data/pancake_phase_transition_vs_N.dat};

	\addplot [mark=*,mark options={solid, scale=0.5}, style={dashed}, color=green!80, error bars/.cd, y dir=both, y explicit, error bar style={solid}] table[y error minus index=2, y error plus index=3] {./figs/data/buddingtransitionvsn.dat};

\node[above] at (axis cs:2100,1.5) {Mixed};
\node[above] at (axis cs:2100,0.9) {Budded};
\node[above] at (axis cs:2100,0.60) {Pancake};

\end{axis}
\end{tikzpicture}

%% file: snapspressure001_fig.tex
\begin{tikzpicture}
\begin{axis}[
	width=0.5\textwidth,
	xlabel={$\rho[\%]$},
	ylabel near ticks,
	ylabel=$T/T_0$,
	ymin=0.5,
	ymax=1.1,
	xmin=0.025,xmax=0.17,
	scaled x ticks=manual:{}{\pgfmathparse{(#1)*100}},
	tick label style={/pgf/number format/fixed}]

	\node[inner sep=0pt] at (axis cs:0.04988807163415414,0.6060606060606061) {\includegraphics[width=0.15\textwidth]{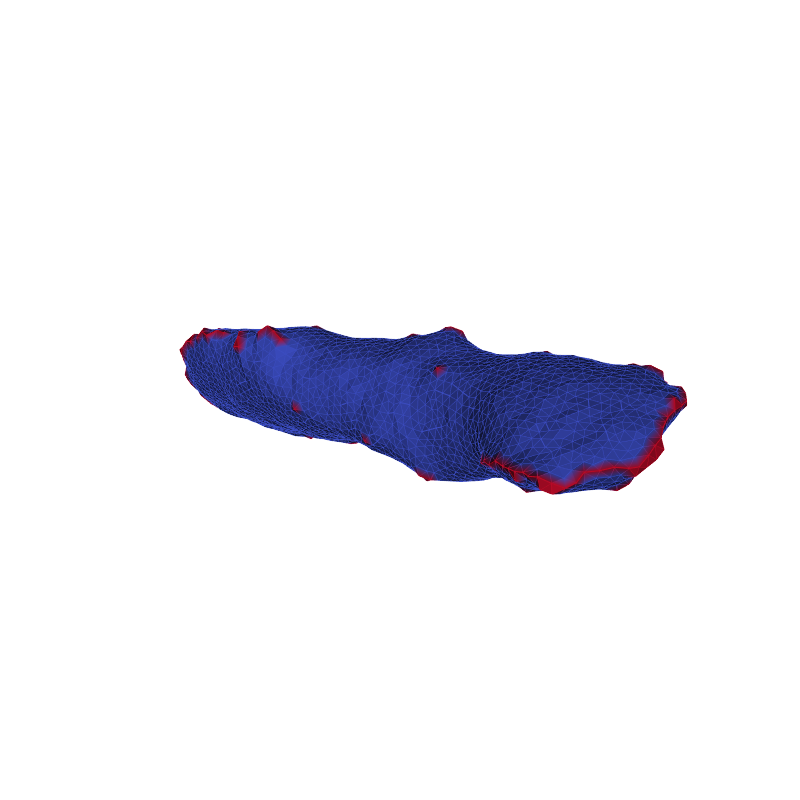}};
	\node[inner sep=0pt] at (axis cs:0.09977614326830828,0.6060606060606061) {\includegraphics[width=0.15\textwidth]{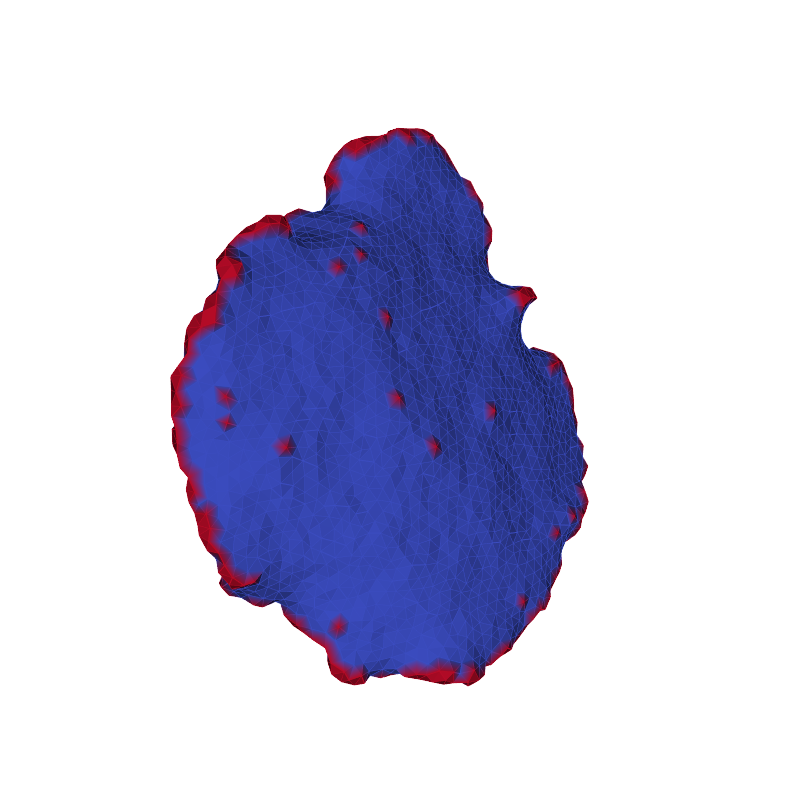}};
	\node[inner sep=0pt] at (axis cs:0.1499840102334506,0.6060606060606061) {\includegraphics[width=0.15\textwidth]{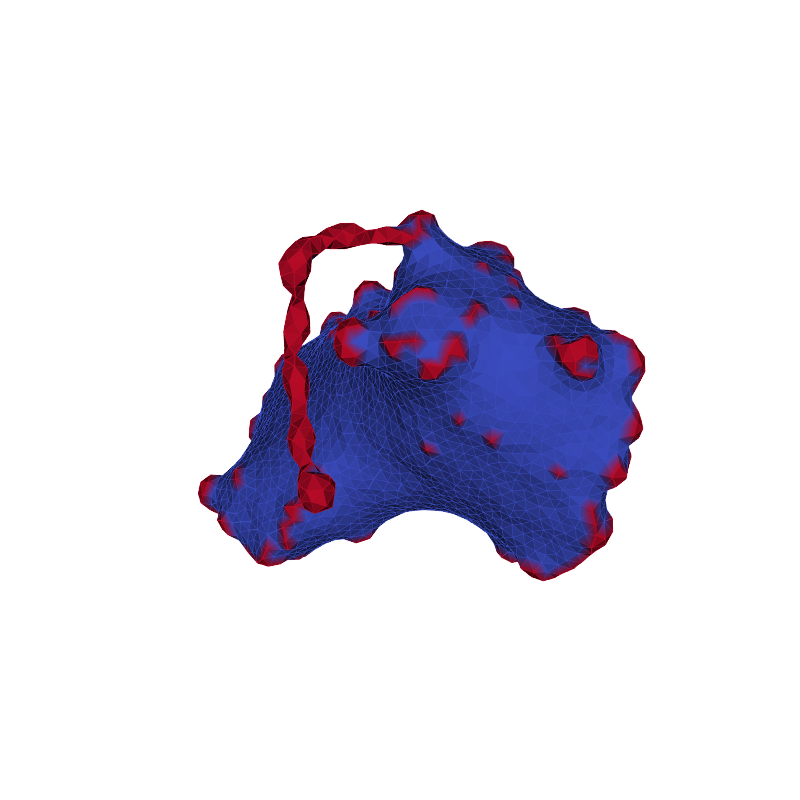}};
	\node[inner sep=0pt] at (axis cs:0.04988807163415414,0.8) {\includegraphics[width=0.15\textwidth]{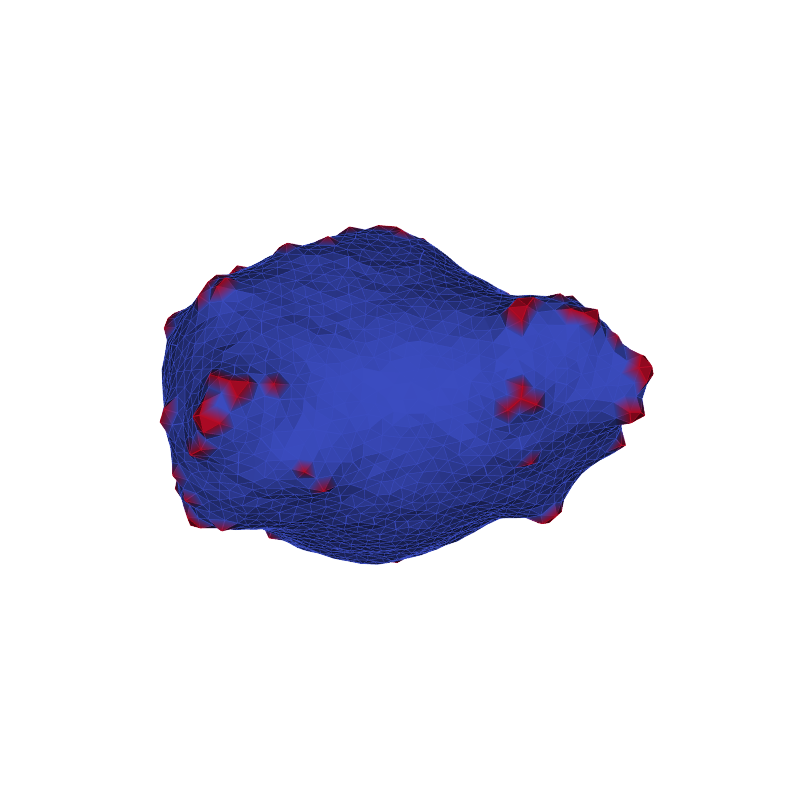}};
	\node[inner sep=0pt] at (axis cs:0.09977614326830828,0.8) {\includegraphics[width=0.15\textwidth]{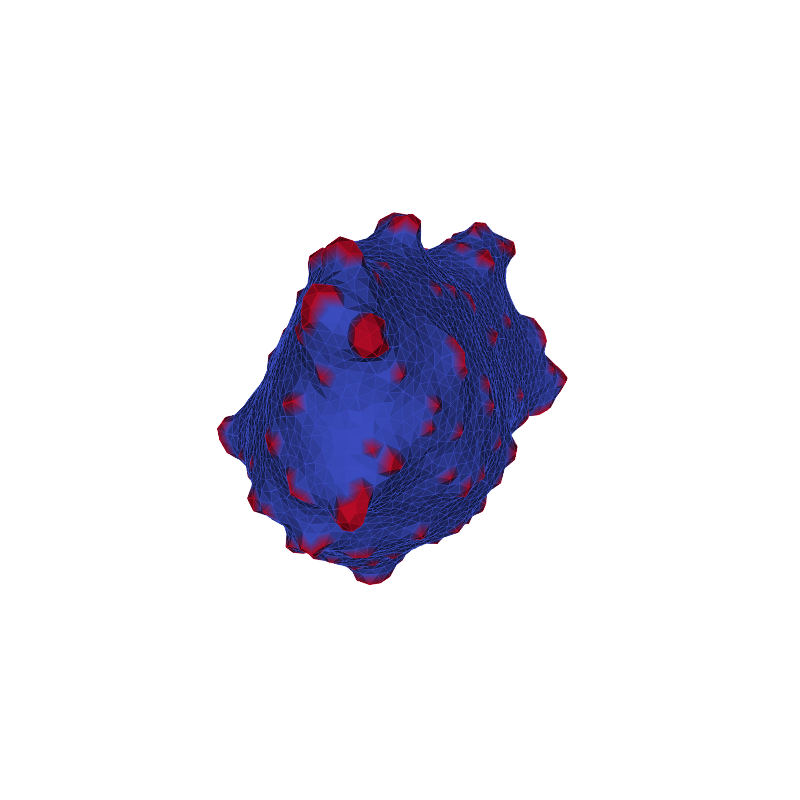}};
	\node[inner sep=0pt] at (axis cs:0.1499840102334506,0.8) {\includegraphics[width=0.15\textwidth]{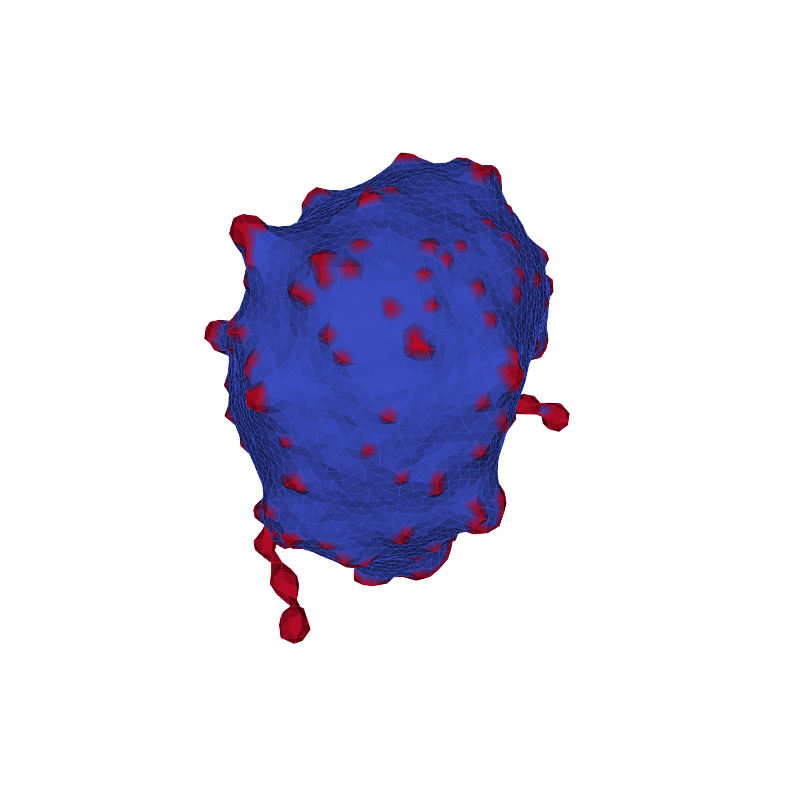}};
	\node[inner sep=0pt] at (axis cs:0.04988807163415414,1.0) {\includegraphics[width=0.15\textwidth]{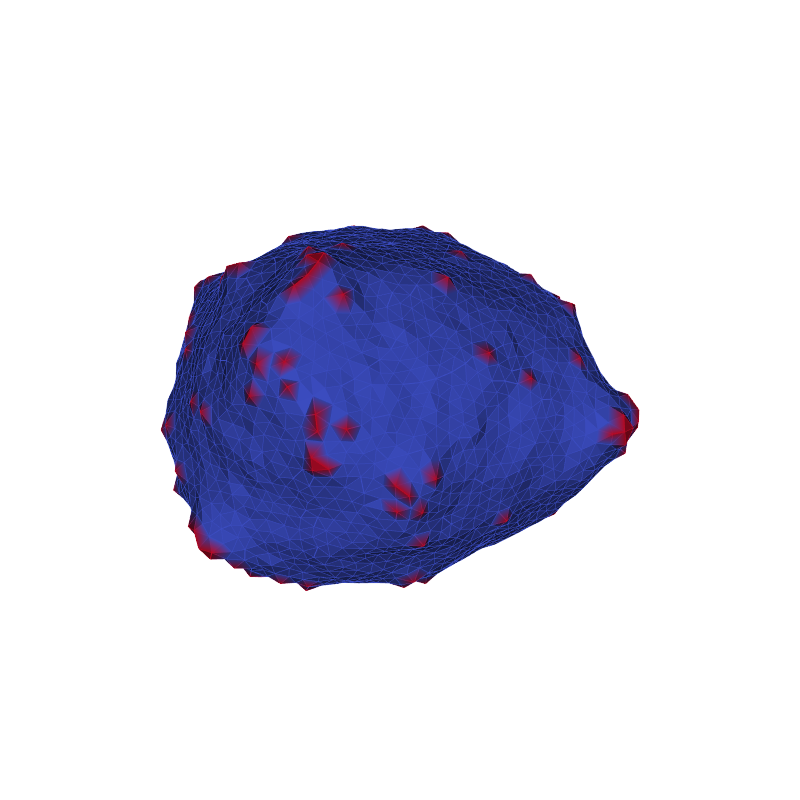}};
	\node[inner sep=0pt] at (axis cs:0.09977614326830828,1.0) {\includegraphics[width=0.15\textwidth]{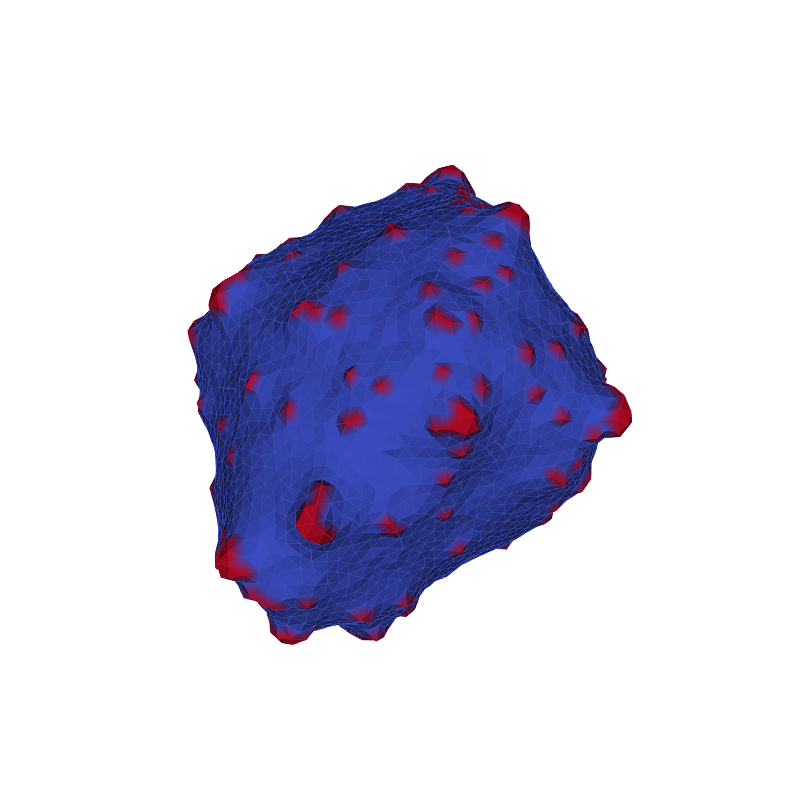}};
	\node[inner sep=0pt] at (axis cs:0.1499840102334506,1.0) {\includegraphics[width=0.15\textwidth,angle=40]{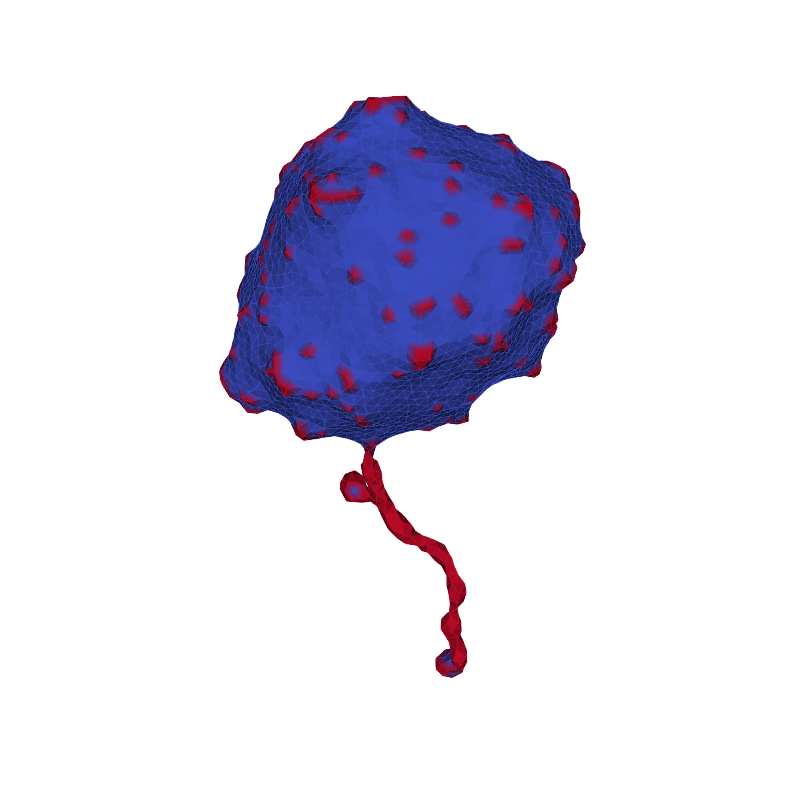}};


\end{axis}
\end{tikzpicture}

%% file: snapspressure_fig.tex
\begin{tikzpicture}
\begin{axis}[
	width=0.5\textwidth,
	xlabel={$\rho[\%]$},
	ylabel near ticks,
	ylabel=$T/T_0$,
	ymin=0.5,
	ymax=1.1,
	xmin=0.025,xmax=0.17,
	scaled x ticks=manual:{}{\pgfmathparse{(#1)*100}},
	tick label style={/pgf/number format/fixed}]

	\node[inner sep=0pt] at (axis cs:0.04988807163415414,0.6060606060606061) {\includegraphics[width=0.15\textwidth]{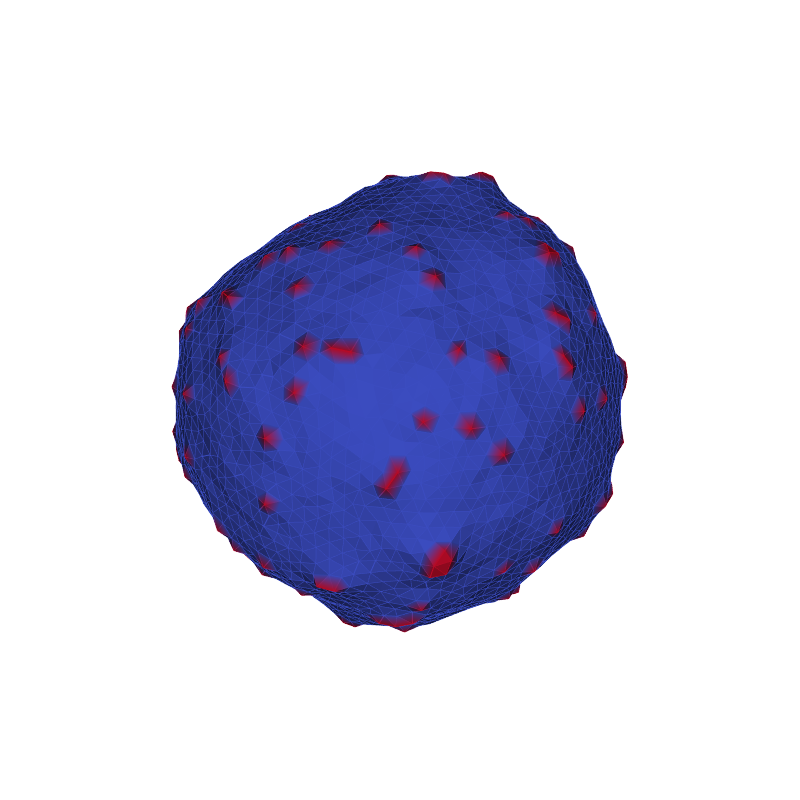}};
	\node[inner sep=0pt] at (axis cs:0.09977614326830828,0.6060606060606061) {\includegraphics[width=0.15\textwidth]{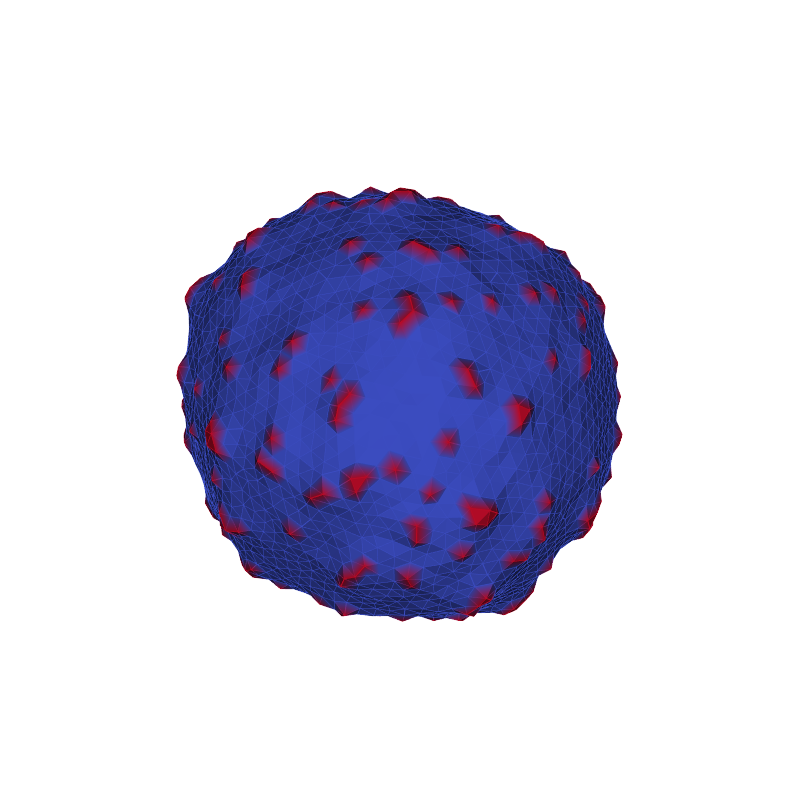}};
	\node[inner sep=0pt] at (axis cs:0.1499840102334506,0.6060606060606061) {\includegraphics[width=0.15\textwidth]{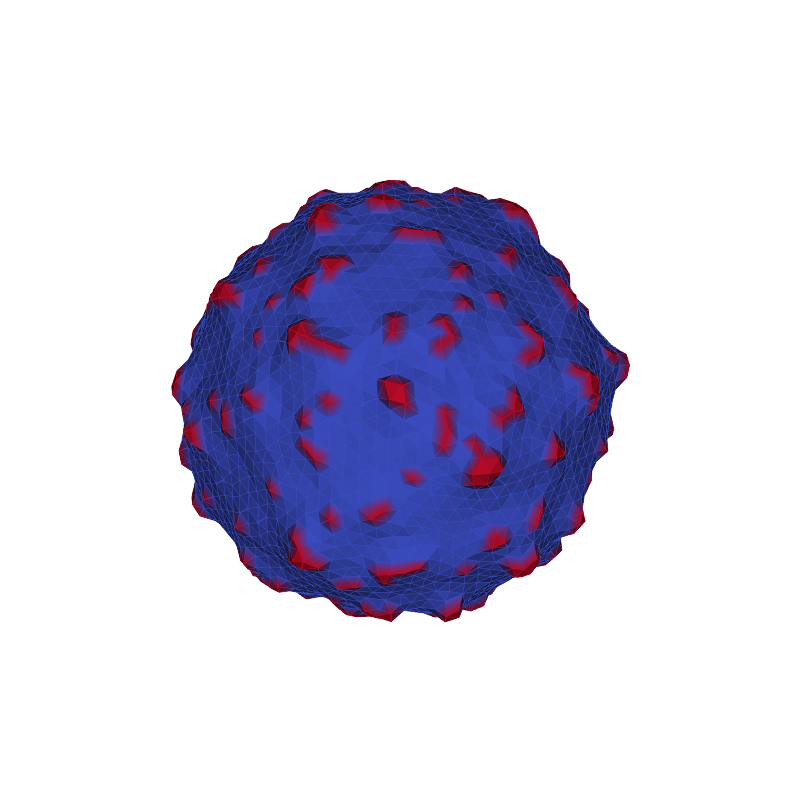}};
	\node[inner sep=0pt] at (axis cs:0.04988807163415414,0.8) {\includegraphics[width=0.15\textwidth]{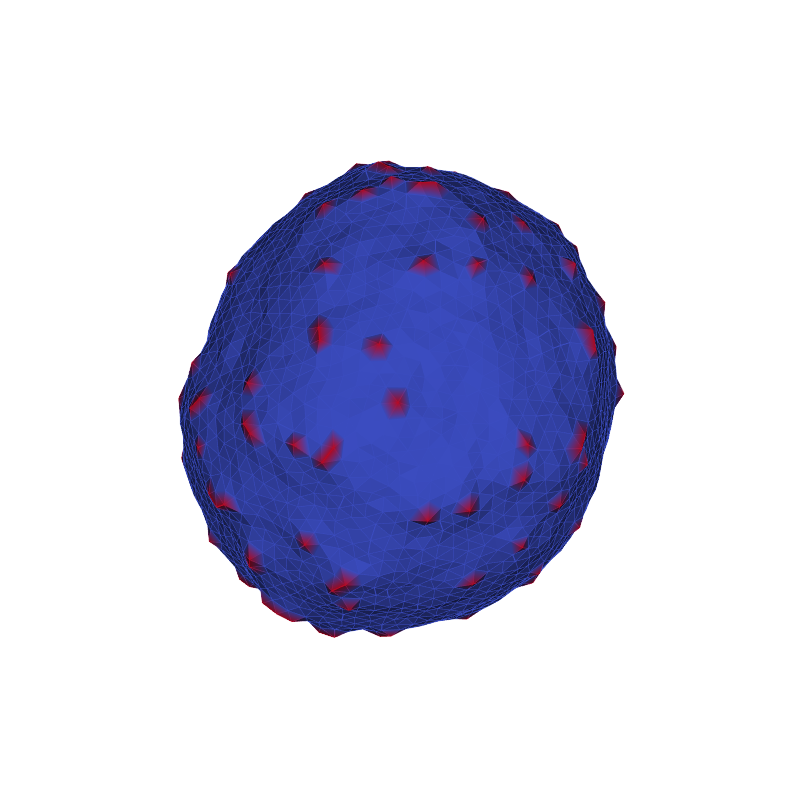}};
	\node[inner sep=0pt] at (axis cs:0.09977614326830828,0.8) {\includegraphics[width=0.15\textwidth]{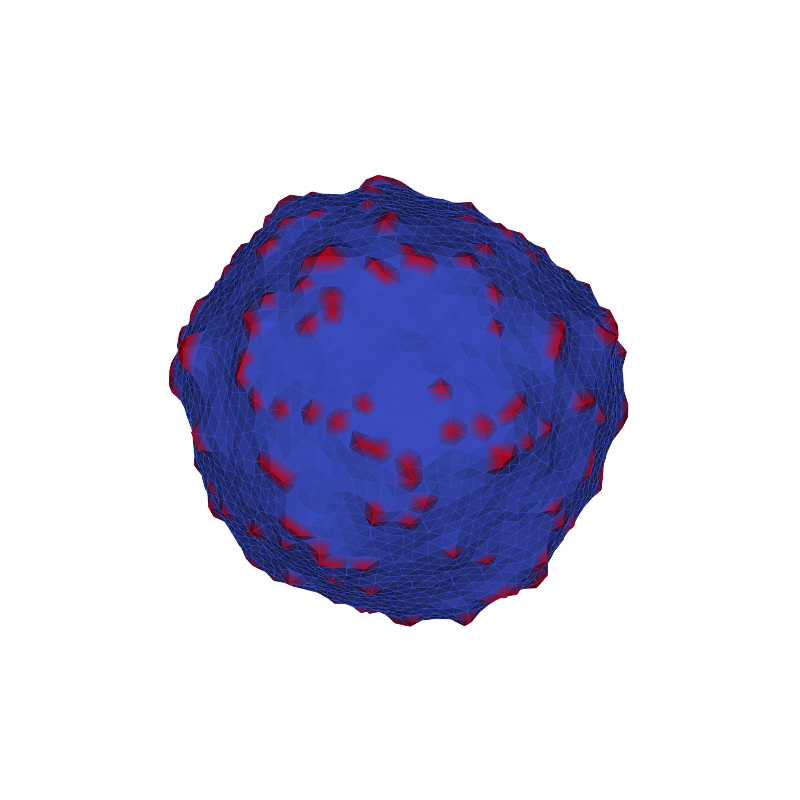}};
	\node[inner sep=0pt] at (axis cs:0.1499840102334506,0.8) {\includegraphics[width=0.15\textwidth]{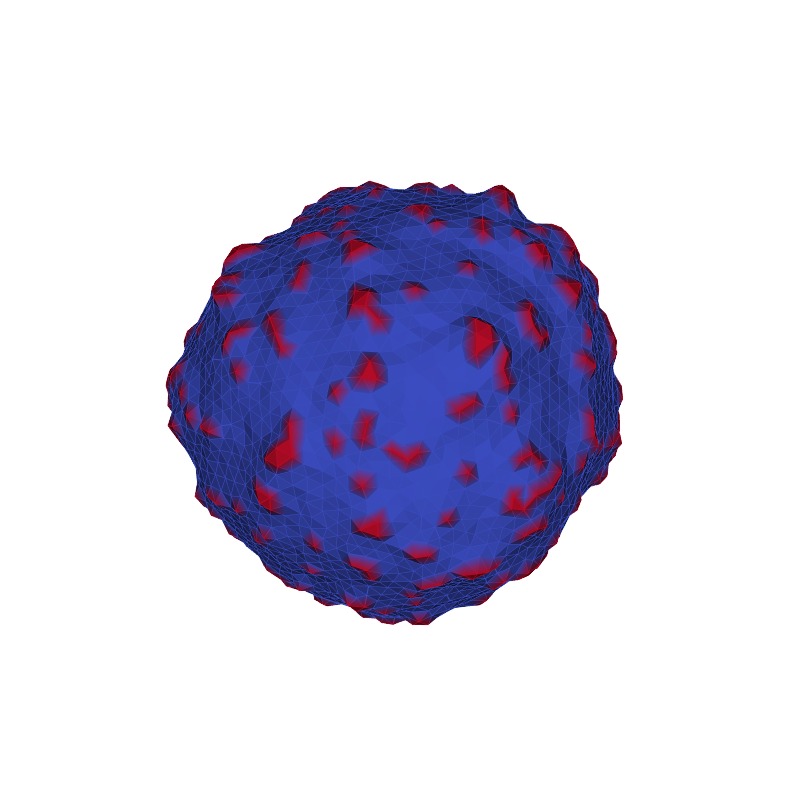}};
	\node[inner sep=0pt] at (axis cs:0.04988807163415414,1.0) {\includegraphics[width=0.15\textwidth]{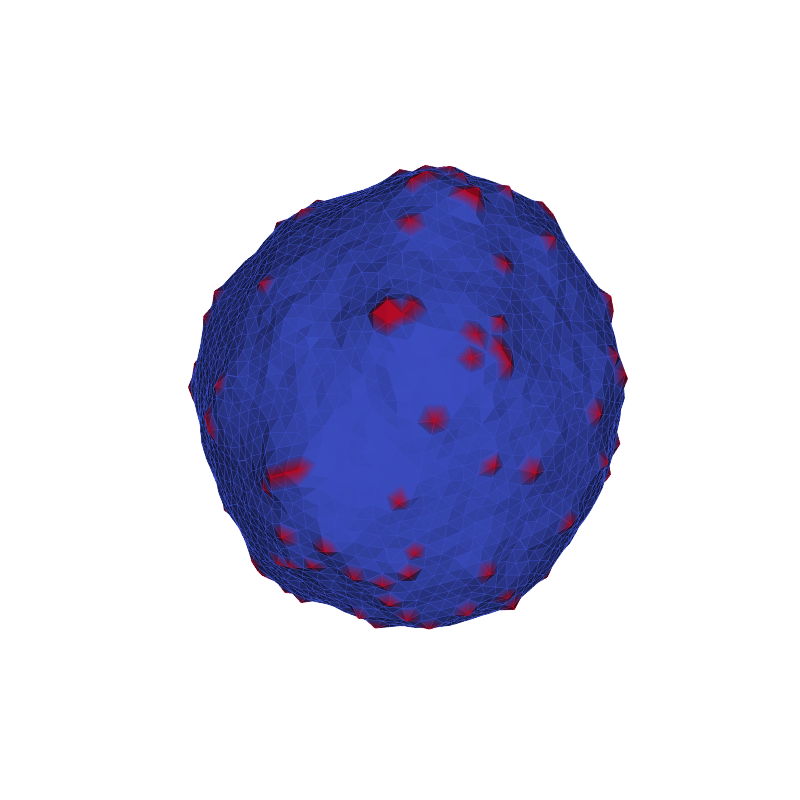}};
	\node[inner sep=0pt] at (axis cs:0.09977614326830828,1.0) {\includegraphics[width=0.15\textwidth]{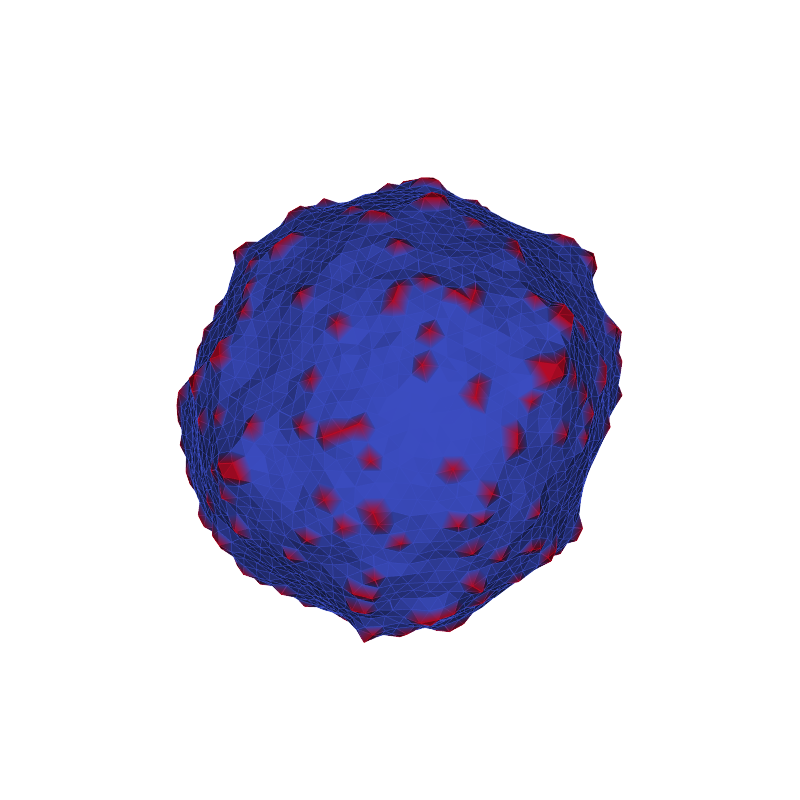}};
	\node[inner sep=0pt] at (axis cs:0.1499840102334506,1.0) {\includegraphics[width=0.15\textwidth]{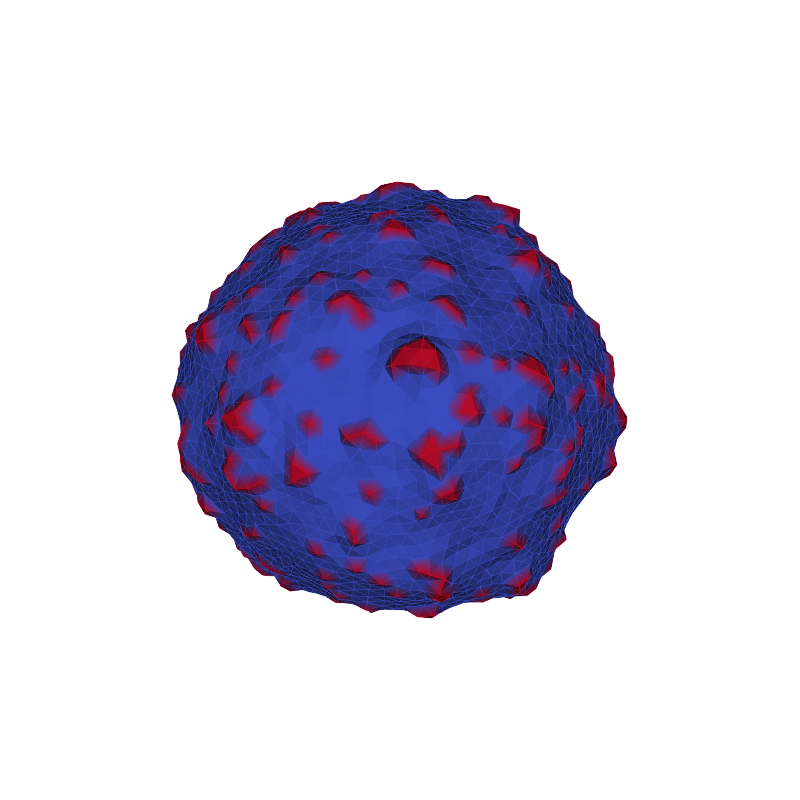}};


\end{axis}
\end{tikzpicture}